\documentclass{gNST2e}

\usepackage{amsthm, amssymb, amsmath, graphicx, subfigure, psfrag, natbib}
\RequirePackage[colorlinks, citecolor=blue, urlcolor=blue, breaklinks=true]{hyperref}
\usepackage{epsfig}
\usepackage{epstopdf}
\usepackage{url}
\usepackage[usenames, dvipsnames]{color}

\theoremstyle{plain}
\newtheorem{theorem}{Theorem}[section]

\theoremstyle{remark}

\theoremstyle{definition}
\newtheorem{definition}{Definition}
\def\T{{ \mathrm{\scriptscriptstyle T} }}

\newcommand{\btheta} {\mbox{\boldmath $\theta$}}
\newcommand{\bbeta} {\mbox{\boldmath $\beta$}}

\newcommand{\bSigma} {\mbox{\boldmath $\Sigma$}}

\newcommand{\bc} {\mbox{\boldmath $c$}}
\newcommand{\be} {\mbox{\boldmath $e$}}

\newcommand{\bY} {\mathbf {Y}}

\newcommand{\bG} {\mathbf {G}}
\newcommand{\bI} {\mathbf {I}}
\newcommand{\bK} {\mathbf {K}}

\newcommand{\bS} {\mathbf {S}}
\newcommand{\bT} {\mathbf {T}}
\newcommand{\bW} {\mathbf {W}}

\newcommand{\diag}{\textrm{diag}}

\newcommand{\mathC}{\mathbb{C}}
\newcommand{\mathR}{\mathbb{R}}

\def \calD {\mathcal{D}}
\def \calF {\mathcal{F}}

\newcommand{\ISE}{{\rm ISE}}

\newcommand{\bz}{\boldsymbol{z}}

\begin{document}

\title{An alternative local polynomial estimator for the error-in-variables problem}

\author{
\name{Xianzheng Huang\textsuperscript{a}$^{\ast}$\thanks{$^\ast$Corresponding author. Email: huang@stat.sc.edu}
and Haiming Zhou\textsuperscript{b}}
\affil{\textsuperscript{a}Department of Statistics, University of South Carolina, Columbia, South Carolina, U.S.A.;
\textsuperscript{b}Division of Statistics, Northern Illinois University, DeKalb, Illinois, U.S.A.}
\received{v4.0 released June 2015}
}

\maketitle 

\begin{abstract}
We consider the problem of estimating a regression function when a covariate is measured with error. Using the local polynomial estimator of  \citet{Delaigle09} as a benchmark, we propose an alternative way of solving the problem without transforming the kernel function. The asymptotic properties of the alternative estimator are rigorously studied. A detailed implementing algorithm and a computationally efficient bandwidth selection procedure are also provided. The proposed estimator is compared with the existing local polynomial estimator via extensive simulations and an application to the motorcycle crash data. The results show that the new estimator can be less biased than the existing estimator and is numerically more stable. 
\end{abstract}


\begin{classcode}
62G05; 62G08; 62G20
\end{classcode}

\section{Introduction}
\label{s:intro}
The error-in-covariates problem has received great attention among researchers who study nonparametric inference for regression functions over the past two decades. \citet{Schennach04a, Schennach04b} proposed an estimator of the regression function when the error-prone covariate is measured twice. Her estimator does not require a known measurement error distribution. \citet{Zwanzig07} proposed a local least square estimator of the regression function, assuming a uniformly distributed error-prone covariate with normal measurement error. Many more existing methods are developed under the assumption of a known measurement error distribution and an unknown true covariate distribution. Among these works, many follow the theme of deconvolution kernel pioneered in the density estimation problem in the presence of measurement error \citep{Carroll88, Stefanski90}. In particular, starting from the well-known Nadaraya-Watson kernel estimator developed for error-free case \citep{Nadaraya, Watson}, \citet{Fan93} formulated the local constant estimator of a regression function using the deconvolution kernel technique. Generalization of this estimator to local polynomial estimators of higher orders was achieved by \citet{Delaigle09} via introducing a complex transform of the kernel function. This transform is the key step that allows for the extension from the zero-order to a higher-order local polynomial estimator in error-in-variables problems. 

In this study, we propose a new estimator motivated by an identity that relates the Fourier transform of the functions to be estimated to the Fourier transform of the counterpart naive functions. Here, a naive estimate refers to an estimate that results from replacing the unobserved true covariate one would use in the absence of measurement error with the error-contaminated observed covariate. This identity and the new estimator are presented in Section 2, following a brief review of the estimator in \citet{Delaigle09}, which we refer to as the DFC estimator henceforth. Sections~\ref{s:bias}, \ref{s:variance}, and \ref{s:normality} are devoted to studying the asymptotic distribution of the new estimator. The finite sample performance of our estimator is demonstrated in comparison with the DFC estimator in Section~\ref{s:simulation}. We summarize our contribution and findings, discuss some practical issues in Section~\ref{s:discussion}. All appendices referenced in this article are provided in the Supplementary Materials. 

\section{Existing and proposed estimators}
\label{s:methods}
Denote by $\{(Y_j, W_j), \, j=1, \ldots, n\}$ a random sample of size $n$ from a regression model with additive measurement error in the covariate specified as follow,  
\begin{equation}
E(Y_j|X_j) = m(X_j), \,\, W_j  =  X_j+U_j, \label{eq:regmodel}
\end{equation}
where $X_j$ is the unobserved true covariate following a distribution with probability density function (pdf) $f_{\hbox {\tiny $X$}}(x)$, $U_j$ is the measurement error, assumed to be independent of $(X_j, Y_j)$ and follow a known distribution with pdf $f_{\hbox {\tiny $U$}}(u)$, $W_j$ is the error-contaminated observed covariate following a distribution with pdf $f_{\hbox {\tiny $W$}}(w)$, for $j=1, \ldots, n$. The problem of interest in this study is to estimate the regression function, $m(x)$, based on the observed data. The index $j$ is often suppressed in the sequel when a generic observation or random variable is referenced. 

\subsection{The DFC estimator} \label{s:theirs}
In the absence of measurement error, the well-known local polynomial estimator of order $p$ for $m(x)$ is given by \citep[][Chapter 3]{Fanbook}
\begin{equation}
\hat m(x)=\be_1^\T \bS_n^{-1}\bT_n, 
\label{eq:mxnome}
\end{equation}
where $\be_1$ is a $(p+1)\times 1$ vector with 1 in the first entry and 0 in the remaining $p$ entries,  
\begin{equation}
\bS_n=
\begin{bmatrix}
S_{n,0}(x) & \ldots & S_{n,p}(x) \\
\vdots     & \ddots & \vdots \\
S_{n,p}(x) & \ldots & S_{n, 2p}(x) 
\end{bmatrix}, \nonumber 
\end{equation}
and $\bT_n=(T_{n,0}(x), \ldots, T_{n,p}(x))^\T$, in which
\begin{equation}
\left\{
\begin{array}{l}
S_{n,\ell}(x) = n^{-1}\displaystyle{\sum_{j=1}^n\left(\frac{X_j-x}{h} \right)^\ell K_h(X_j-x)},  \textrm{ for $\ell=0, 1, \ldots, 2p$},  \\
T_{n,\ell}(x) = n^{-1}\displaystyle{\sum_{j=1}^nY_j\left(\frac{X_j-x}{h} \right)^\ell K_h(X_j-x)}, \textrm{ for $\ell=0, 1, \ldots, p$},
\end{array}
\right. \label{eq:SnTn}
\end{equation}
and $K_h(x)=h^{-1}K(x/h)$ with $K(\cdot)$ being a symmetric kernel function and $h$ being the bandwidth. 

In the presence of measurement error, one could replace $X_j$ with $W_j$ for $j=1, \ldots, n$ in the above local polynomial estimator, yielding a naive estimator of $m(x)$, denoted by $\hat m^*(x)$. Clearly, $\hat m^*(x)$ is merely a sensible estimator of the naive regression function $m^*(x)=E(Y|W=x)$. Following the rationale behind the corrected score method \citep[][Section 7.4]{LASbook}, \citet{Delaigle09} sought some function, denoted by $L_{\ell}(\cdot)$, that satisfies 
\begin{equation}
E\left\{(W_j-x)^\ell L_{\ell,h}(W_j-x)|X_j\right\}=(X_j-x)^\ell K_{h}(X_j-x), \textrm{ for $\ell=0, 1, \ldots, 2p$,}
\label{eq:corrected}
\end{equation}
where $L_{\ell, h}(x)=h^{-1}L_{\ell}(x/h)$. The authors derived such function via solving the Fourier transform version of (\ref{eq:corrected}), and showed that $L_{\ell}(x)=x^{-\ell}K_{\hbox {\tiny $U$}, \ell}(x)$, where 
\begin{equation}
K_{\hbox {\tiny $U$}, \ell}(x)=i^{-\ell}\frac{1}{2\pi}\int e^{-itx} \frac{\phi_{\hbox {\tiny $K$}}^{(\ell)}(t)}{\phi_{\hbox {\tiny $U$}}(-t/h)} dt,  \textrm{ for $\ell=0, 1, \ldots, 2p$}, \label{eq:KUl}
\end{equation}
in which $i=\sqrt{-1}$, $\phi^{(\ell)}_{\hbox {\tiny $K$}}(t)$ is the $\ell$-th derivative of $\phi_{\hbox {\tiny $K$}}(t)=\int e^{itx}K(x)dx$, and $\phi_{\hbox {\tiny $U$}}(x)$ is the characteristic function of $U$. Throughout this article, $\phi_{g}$ denotes the Fourier transform (characteristic function) of $g$ if $g$ is a function (random variable). All integrals in this article integrate over either the entire real line or a subset of it that guarantees the existence of relevant integrals, and we will make remarks on such subset whenever it is needed for clarity. The DFC estimator is given by $\hat m_{\hbox {\tiny DFC}}(x)=\be_1^\T \hat\bS_n^{-1}\hat\bT_n$, where $\hat \bS_n$ and $\hat\bT_n$ are similarly defined as $\bS_n$ and $\bT_n$ in (\ref{eq:mxnome}) but with the elements in the matrices given by 
\begin{equation}
\left\{
\begin{array}{l}
\hat S_{n,\ell}(x) = n^{-1}\displaystyle{\sum_{j=1}^n\left(\frac{W_j-x}{h} \right)^\ell L_{\ell,h}(W_j-x)}, \textrm{ for $\ell=0, 1, \ldots, 2p$,}\\
\hat T_{n,\ell}(x) = n^{-1}\displaystyle{\sum_{j=1}^nY_j\left(\frac{W_j-x}{h} \right)^\ell L_{\ell, h}(W_j-x)}, \textrm{ for $\ell=0, 1, \ldots, p$}.
\end{array}
\right. \nonumber 
\end{equation}

The transform of $K$ defined in (\ref{eq:KUl}) is a natural extension of the transform used in the deconvolution density estimator \citep{Stefanski90} and the local constant estimator \citep{Fan93} of $m(x)$ under the setting of (\ref{eq:regmodel}). In particular, the estimator in \citet{Fan93} is a special case of the DFC estimator with $p=0$. 

\subsection{The proposed estimator} \label{s:ours}
Deviating from the theme of deconvolution kernel and its extension in (\ref{eq:KUl}), we propose a new estimator that more directly exploits the naive inference as a whole. This direct use of the naive inference is motivated by the following result proved in \citet{Delaigle14}, $m^*(w)f_{\hbox {\tiny $W$}}(w)=(m f_{\hbox {\tiny $X$}})*f_{\hbox {\tiny $U$}}(w)$, where $(m f_{\hbox {\tiny $X$}})*f_{\hbox {\tiny $U$}}(w)$ is the convolution given by $\int m(x) f_{\hbox {\tiny $X$}}(x)f_{\hbox {\tiny $U$}}(w-x)dx$. Applying Fourier transform on both sides of this identity, one has 
\begin{equation}
\phi_{m^* f_{\hbox {\tiny $W$}}}(t)=\phi_{m f_{\hbox {\tiny $X$}}}(t) \phi_{\hbox {\tiny $U$}}(t),
\label{eq:transf}
\end{equation}
where $\phi_{m^* f_{\hbox {\tiny $W$}}}(t)$ is the Fourier transform of $m^*(w)f_{\hbox {\tiny $W$}}(w)$ and 
$\phi_{m f_{\hbox {\tiny $X$}}}(t)$ is the Fourier transform of $m(x)f_{\hbox {\tiny $X$}}(x)$.
Immediately following (\ref{eq:transf}), by the Fourier inversion theorem, one has 
$m(x)f_{\hbox {\tiny $X$}}(x)=(2\pi)^{-1}\int e^{-itx}\phi_{m^*f_{\hbox {\tiny $W$}}}(t)/\phi_{\hbox {\tiny $U$}}(t)dt$. This motivates our local polynomial estimator of order $p$ for $m(x)$ given by, assuming the relevant Fourier transforms well defined,
\begin{equation}
\hat m_{\hbox {\tiny HZ}}(x)= \left\{\hat f_{\hbox {\tiny $X$}}(x)\right\}^{-1}\frac{1}{2\pi}\int e^{-itx}\frac{\phi_{\hat m^* \hat f_{\hbox {\tiny $W$}}}(t)}{\phi_{\hbox {\tiny $U$}}(t)}dt,
\label{eq:mz}
\end{equation}
where $\hat f_{\hbox {\tiny $X$}}(x)$ is the deconvolution kernel density estimator of $f_{\hbox {\tiny $X$}}(x)$ in \citet{Stefanski90}, and $\phi_{\hat m^* \hat f_{\hbox {\tiny $W$}}}(t)$ is the Fourier transform of $\hat m^*(w) \hat f_{\hbox {\tiny $W$}}(w)$, in which $\hat m^*(w)$ is the $p$-th order local polynomial estimator of $m^*(w)$, and $\hat f_{\hbox {\tiny $W$}}(w)$ is the regular kernel density estimator of $f_{\hbox {\tiny $W$}}(w)$ \citep[][Section 2.7.1]{Fanbook}, i.e., the naive estimator of $f_{\hbox {\tiny $X$}}(\cdot)$. Note that, although we consider a scalar covariate for notational simplicity in this article, the estimators on the right-hand side of (\ref{eq:mz}) have their multivariate counterparts to account for multivariate covariates. Hence, with multivariate (inverse) Fourier transform used in (\ref{eq:mz}), the proposed estimator becomes applicable to regression models with multiple covariates. Moreover, if some of these covariates are measured without error, one may reflect this in $\phi_{\hbox {\tiny $U$}}(t)$ by viewing that the elements in the multivariate $U$ corresponding to the error-free covariates follow a degenerate distribution with all probability mass on zero. 

By its appearance, the new estimator in (\ref{eq:mz}) results from applying an integral transform similar to that in (\ref{eq:KUl}) on the naive product $\hat m^*(\cdot) \hat f_{\hbox {\tiny $W$}}(\cdot)$ rather than on $K$. It can be shown (via straightforward algebra omitted here) that, when $p=0$, this new estimator is the same as the DFC estimator, both reducing to the local constant estimator in \citet{Fan93}. Other than this special case, $\hat m_{\hbox {\tiny HZ}}(x)$ differs from $\hat m_{\hbox {\tiny DFC}}(x)$ in general. 

\subsection{Preamble for asymptotic analyses}
\label{s:prep}
The majority of the theoretical development presented in \citet{Delaigle09} revolves around properties of the transformed kernel, $K_{\hbox {\tiny $U$},\ell}(x)$, which is not surprising as $K_{\hbox {\tiny $U$},\ell}(x)$ is everywhere in the building blocks of their estimator. Because of the close tie between our proposed estimator and the naive estimators, much of our theoretical development builds upon well established results for kernel-based estimators in the absence of measurement error. This can be better appreciated by interchanging the order of the two integrals in (\ref{eq:mz}), assuming that $\phi_{\hat m^* \hat f_{\hbox {\tiny $W$}}}(t)$ is compactly supported on $I_t$ (to allow the interchange), 
$\hat m_{\hbox {\tiny HZ}}(x)\hat f_{\hbox {\tiny $X$}}(x)  = \int \hat m^*(w) \hat f_{\hbox {\tiny $W$}}(w) (2\pi)^{-1} \int_{I_t} e^{-it(x-w)}/\phi_{\hbox {\tiny $U$}}(t)\, dt dw$. This identity can be re-expressed more succinctly as
\begin{equation}
\mathcal{B}(x) = \int \mathcal{A}(w) D(x-w)\, dw=(\mathcal{A}*D)(x), \label{eq:convo} 
\end{equation}
where $\mathcal{A}(w)=\hat m^*(w) \hat f_{\hbox {\tiny $W$}}(w)$, $\mathcal{B}(x)=\hat m_{\hbox {\tiny HZ}}(x)\hat f_{\hbox {\tiny $X$}}(x)$, and $D(s)=(2\pi)^{-1}\int_{I_t} e^{-its}/\phi_{\hbox {\tiny $U$}}(t)dt$. Note that $\mathcal{A}(w)$ is a random process depending on the native estimators $\hat m^*(w)$ and $\hat f_{\hbox {\tiny $W$}}(w)$, and $\mathcal{B}(x)$ results from convoluting $\mathcal{A}(w)$ and the non-random function $D(s)$. A natural question is, given the asymptotic properties of $\mathcal{A}(w)$, what can be deduced from the convolution of $\mathcal{A}$ and $D$. More specifically, we are interested to know how the moments of $\mathcal{A}$ compare with those of $\mathcal{B}$, and whether a  Gaussian process on $\mathcal{A}(w)$ implies another Gaussian process on $\mathcal{B}(x)$. These questions about random process convolution are of mathematical interest in their own rights besides being the key to understanding $\hat m_{\hbox {\tiny HZ}}(x)$. 

Here we provide two definitions of smoothness of a distribution \citep{Fan91a, Fan91b, Fan91c} and two sets of conditions to be referenced later. 
\begin{definition}
\label{df:ordsm}
The distribution of $U$ is ordinary smooth of order $b$ if
$$\lim_{t\to +\infty} t^b \phi_{\hbox {\tiny $U$}}(t)=c \textrm{ and } \lim_{t\to +\infty} t^{b+1} \phi'_{\hbox {\tiny $U$}}(t)=-cb$$ 
for some positive constants $b$ and $c$. 
\end{definition}
\begin{definition}
\label{df:supsm}
The distribution of $U$ is super smooth of order $b$ if
$$d_0|t|^{b_0}\exp(-|t|^b/d_2)\le |\phi_{\hbox {\tiny $U$}}(t)| \le d_1|t|^{b_1}\exp(-|t|^b/d_2)  \textrm { as $|t|\to \infty$}$$ 
for some positive constants $d_0$, $d_1$, $d_2$, $b$, $b_0$ and $b_1$.
\end{definition}
\begin{description}
\item[Condition O:] 
For $\ell =0, \ldots, 2p+1$, $\|\phi^{(\ell)}_{\hbox {\tiny $K$}}(t)\|_{\infty}<\infty$ and $\int (|t|^b+|t|^{b-1})|\phi^{(\ell)}_{\hbox {\tiny $K$}}(t)|dt<\infty$. For $0\le \ell_1, \ell_2 \le 2p$, $\int |t|^{2b} |\phi^{(\ell_1)}_{\hbox {\tiny $K$}}(t)||\phi^{(\ell_2)}_{\hbox {\tiny $K$}}(t)|  dt < \infty$. And, $\|\phi'_{\hbox {\tiny $U$}}(t)\|_{\infty}<\infty$. 
\item[Condition S:] 
For $\ell=0, \ldots, 2p$, $\|\phi^{(\ell)}_{\hbox {\tiny $K$}}(t)\|_\infty<\infty$, and $\phi_{\hbox {\tiny $K$}}(t)$ is supported on $[-1, 1]$. 
\end{description}

In addition, we assume $f_{\hbox {\tiny $X$}}(x)>0$ and $\phi_{\hbox {\tiny $U$}}(t)$ is an even function that never vanishes. We reach the convolution form in (\ref{eq:convo}) under the assumption that $\phi_{\hat m^* \hat f_{\hbox {\tiny $W$}}}(t)$ is compactly supported on $I_t$, where $I_t$ is a region that guarantees $D(s)$ well defined. This assumption can be easily satisfied by choosing a kernel of which the Fourier transform has a finite support. Even without this assumption the asymptotic properties presented in the following three sections still hold, although some of the proof need to be revised to use the estimator of its original form in (\ref{eq:mz}). While acknowledging the overlap between the regularity conditions needed in our asymptotic analyses and those required for the DFC estimator, we also assume existence of the Fourier transform of $m^*(\cdot)f_{\hbox {\tiny $W$}}(\cdot)$ and that of $m(\cdot)f_{\hbox {\tiny $X$}}(\cdot)$ in (\ref{eq:transf}). We next dissect the asymptotic bias, variance and normality of $\hat m_{\hbox {\tiny HZ}}(x)$.

\section{Asymptotic bias}
\label{s:bias}
We provide the derivations of the asymptotic bias of $\hat m_{\hbox {\tiny HZ}}(x)$ for $p\ge 0$ in Appendix A. To better apprehend the distinction between our bias results and those of $\hat m_{\hbox {\tiny DFC}}(x)$, we present a brief derivation of the bias when $p=1$ in this section. 

\subsection{Dominating bias when $p=1$}
\label{s:bias1}
Define $\mu_\ell=\int u^{\ell}K(u)\,du$, for $\ell=0, 1, \ldots, 2p$. Let $A(w)=m^*(w)f_{\hbox {\tiny $W$}}(w)$ and $B(x)=m(x)f_{\hbox {\tiny $X$}}(x)$ be the non-random counterparts of $\mathcal{A}(w)$ and $\mathcal{B}(x)$ in (\ref{eq:convo}), respectively. Then, like (\ref{eq:convo}), we have $B(x)=(A*D)(x)$. 

By Theorem 2.1 in \citet{Stefanski90}, the deconvolution density estimator $\hat f_{\hbox {\tiny $X$}}(x)$ is a consistent estimator of $f_{\hbox {\tiny $X$}}(x)$. Noting that $ \hat f_{\hbox {\tiny $X$}}(x) /f_{\hbox {\tiny $X$}}(x)$ converges to one in probability, we derive the dominating bias via elaborating $E[\{\hat m_{\hbox {\tiny HZ}}(x)-m(x)\} \hat f_{\hbox {\tiny $X$}}(x) /f_{\hbox {\tiny $X$}}(x)|\mathbb{W}]$, which is equal to 
\begin{equation}
\left\{f_{\hbox {\tiny $X$}}(x)\right\}^{-1} \left[E\left\{\mathcal{B}(x)|\mathbb{W}\right\}-m(x)\hat f_{\hbox {\tiny $X$}}(x)\right], \label{eq:expdiff}
\end{equation}
where $\mathbb{W}=(W_1, \ldots, W_n)$,  and 
\begin{equation}
\hat f_{\hbox {\tiny $X$}}(x) = f_{\hbox {\tiny $X$}} (x)+ \mu_2h^2 f^{(2)}_{\hbox {\tiny $X$}} (x)/2+o_{\hbox {\tiny $P$}}(h^2). \label{eq:expfxhat}
\end{equation}
To derive $E\{\mathcal{B}(x)|\mathbb{W}\}$ in (\ref{eq:expdiff}), we invoke the following two results for kernel-based estimators in the absence of measurement error \citep[][Chapter 3]{Fanbook}, 
\begin{eqnarray}
E\left\{\hat m^*(w)|\mathbb{W}\right\} & = & m^*(w)+\mu_2m^{*(2)}(w)h^2/2+o_{\hbox {\tiny $P$}}(h^2), \nonumber \\
\hat f_{\hbox {\tiny $W$}}(w) &= & f_{\hbox {\tiny $W$}}(w)+ \mu_2f^{(2)}_{\hbox {\tiny $W$}}(w)h^2/2+o_{\hbox {\tiny $P$}}(h^2).\nonumber 
\end{eqnarray}
Following these results, one can show that 
\begin{equation}
E\left\{\mathcal{A}(w)|\mathbb{W}\right\}= A(w)+\mu_2 M(w)h^2/2+o_{\hbox {\tiny $P$}}(h^2), 
\label{eq:inside}
\end{equation}
where $M(w)=m^*(w)f^{(2)}_{\hbox {\tiny $W$}}(w)+m^{*(2)}(w)f_{\hbox {\tiny $W$}}(w)$. Then, assuming interchangeability of expectation and integration, (\ref{eq:convo}) and (\ref{eq:inside}) imply
\begin{equation}
E\{\mathcal{B}(x)|\mathbb{W}\}  = \left\{ E(\mathcal{A}|\mathbb{W})*D\right\}(x)   =  B(x)+\mu_2 h^2 (M*D)(x)/2+o_{\hbox {\tiny $P$}}(h^2). \label{eq:expBx}
\end{equation}
Finally, by (\ref{eq:expfxhat}) and (\ref{eq:expBx}), (\ref{eq:expdiff}) reduces to 
\begin{equation}
\frac{\mu_2 h^2}{2 f_{\hbox {\tiny $X$}}(x)}\left\{(M*D)(x)-m(x) f^{(2)}_{\hbox {\tiny $X$}}(x)\right\}+o_{\hbox {\tiny $P$}}(h^2), \label{eq:dombias} 
\end{equation}
which reveals the dominating bias of $\hat m_{\hbox {\tiny HZ}}(x)$ of order $h^2$.  

Different from \citet{Delaigle09}, we directly use the existing results associated with estimators in the absence of measurement error for deriving the asymptotic bias. 

\subsection{Comparison with the bias of the DFC estimator}
\label{s:toy}
By Theorem 3.2 in \citet{Delaigle09}, the dominating bias of $\hat m_{\hbox {\tiny DFC}}(x)$ is the same as that of $\hat m(x)$, which is 
$\mu_2 h^2 m^{(2)}(x)/2$ when $p=1$. To make the comparison of dominating bias more tractable, we consider regression functions in the form of a polynomial of order $r$, $m(x)=\sum_{k=0}^r \beta_k x^k$. Furthermore, we set $X\sim N(0, 1)$ and $U\sim N(0, \sigma^2_u)$, resulting in a reliability ratio \citep[][Section 3.2.1]{LASbook} of  $\lambda=1/(1+\sigma_u^2)$. 

Under this setting, the dominating bias in (\ref{eq:dombias}) can be derived explicitly. Instead of directly comparing the dominating bias associated with the two estimators, we focus on studying the number of $x$'s at which each dominating bias is zero. Note that $m^{(2)}(x)$ is a polynomial of order $r-2$ provided that $r\ge 2$, and thus the dominating bias of $\hat m_{\hbox {\tiny DFC}}(x)$ is zero at no more than $r-2$ $x$'s. In contrast, we show in Appendix A that the dominating bias in (\ref{eq:dombias}) reduces to a polynomial of order $r$, suggesting that the dominating bias of $\hat m_{\hbox {\tiny HZ}}(x)$ can be zero at $r$ $x$'s. Suppose that the bias of each estimator is continuous in $x$, which is a realistic assumption in many applications. Then having two more roots to the equation, $\textrm{dominating bias}=0$, for $\hat m_{\hbox {\tiny HZ}}(x)$ indicates that the proposed estimator can have two more regions in the support of $m(x)$ within which $\hat m_{\hbox {\tiny HZ}}(x)$ is less biased than $\hat m_{\hbox {\tiny DFC}}(x)$, where each region is a neighborhood of some root. For example, when $r=2$, clearly the dominating bias of $\hat m_{\hbox {\tiny DFC}}(x)$ can never be zero. It is shown in Appendix A that, the dominating bias of $\hat m_{\hbox {\tiny HZ}}(x)$ is zero at the roots of the equation $2(\lambda-1)\beta_2x^2+(\lambda-1)\beta_1x+(2\lambda^2-2\lambda+1)\beta_2=0$. With $\lambda\in (0, 1)$, one can easily show that this quadratic equation has two roots. 

\section{Asymptotic variance} \label{s:variance}
Because 
\begin{equation}
\textrm{Var}\{\hat m_{\hbox {\tiny HZ}}(x)|\mathbb{W}\}=\textrm{Var}\left\{\mathcal{B}(x) |\mathbb{W}\right\}f^{-2}_{\hbox {\tiny $X$}}(x)\left\{1+o_{\hbox {\tiny $P$}}(1) \right\}, \label{eq:reallyend}
\end{equation}
we focus on deriving $\textrm{Var}\{\mathcal{B}(x)|\mathbb{W}\}$ in order to study the asymptotic variance of $\hat m_{\hbox {\tiny HZ}}(x)$. Detailed derivations are provided in Appendix B, which consists of five steps. In what follows, we provide a sketch of the derivations, where we highlight the connection between our results and the counterpart results in the absence of measurement error, and how our derivations differ from and relate to those in \citet{Delaigle09}.   

\subsection{Derivations of $\textrm{Var}\{\mathcal{B}(x)|\mathbb{W}\}$}
\label{s:varBx}
First, we deduce from (\ref{eq:convo}) that $\textrm{Var}\{\mathcal{B}(x)|\mathbb{W}\}$ can be formulated as an iterative convolution of the covariance of $\mathcal{A}(w)$ as follows, 
\begin{equation}
\textrm{Var}\{\mathcal{B}(x)|\mathbb{W}\} =  \int D(x-w_1) \int D(x-w_2) \textrm{Cov}\left\{\mathcal{A}(w_1), \mathcal{A}(w_2) |\mathbb{W}\right\} dw_2 dw_1. \label{eq:varB}
\end{equation}
Since $\hat f_{\hbox {\tiny $W$}}(w) /f_{\hbox {\tiny $W$}}(w)$ converges to 1 in probability under regularity conditions, 
\begin{equation}
\textrm{Cov}\left\{\mathcal{A}(w_1), \, \mathcal{A}(w_2)|\mathbb{W} \right\}= \textrm{Cov}\{\hat m^*(w_1), \,  \hat m^*(w_2)|\mathbb{W}\}f_{\hbox {\tiny $W$}}(w_1)f_{\hbox {\tiny $W$}}(w_2)\{1+o_{\hbox {\tiny $P$}}(1)\}.
\label{eq:covAA}
\end{equation}

Second, we view $\hat m^*(w)$ as a weighted least squares estimator \citep[][page 58]{Fanbook}, and show that 
\begin{equation}
\textrm{Cov}\{\hat m^*(w_1), \, \hat m^*(w_2)|\mathbb{W}\} = \be_1^\T (\bG_1^\T \bW_1 \bG_1)^{-1} (\bG_1^\T \bSigma_{12} \bG_2)(\bG_2^\T \bW_2 \bG_2)^{-1} \be_1, \label{eq:covmm}
\end{equation}
where $\bSigma_{12} = \diag \{K_h(W_1-w_1) K_h(W_1-w_2)\nu^2(W_1), \ldots, K_h(W_n-w_1) K_h(W_n-w_2)\nu^2(W_n)\}$, $\nu^2(w)=\textrm{Var}(Y|W=w)$, and, for $k=1, 2$, $\bW_k = \diag\{ K_h(W_1-w_k), \ldots, K_h(W_n-w_k)\}$, 
\begin{equation}
\bG_k = 
\begin{bmatrix}
1 & (W_1-w_k) & \ldots & (W_1-w_k)^p \cr 
\vdots & \vdots & \ddots & \vdots \cr
1 & (W_n-w_k) & \ldots & (W_n-w_k)^p
\end{bmatrix}. \nonumber \\
\end{equation}
Then we approximate the random quantities on the right hand side of (\ref{eq:covmm}) to establish that
\begin{eqnarray}
& & \textrm{Cov}\{\hat m^*(w_1), \, \hat m^*(w_2) |\mathbb{W}\} \nonumber \\
& = &\frac{\nu^2\left\{(w_1+w_2)/2\right\}f_{\hbox {\tiny $W$}}\left\{(w_1+w_2)/2\right\}}{nh f_{\hbox {\tiny $W$}}(w_1)f_{\hbox {\tiny $W$}}(w_2)}\be_1^\T \bS^{-1}\bS^*_{\hbox {\tiny $W$},h}\bS^{-1}\be_1\left\{1+o_{\hbox {\tiny $P$}}\left(\frac{1}{nh}\right)\right\}, \nonumber \\
\label{eq:covmm2}
\end{eqnarray}
where $\bS=(\mu_{\ell_1+\ell_2})_{0\le \ell_1, \ell_2\le p}$ and $\bS^*_{\hbox {\tiny $W$},h}=(\xi_{\ell_1, \ell_2}((w_1-w_2)/2,h))_{0\le \ell_1, \ell_2\le p}$, in which, for $\ell_1, \ell_2=0, 1, \ldots, p$, 
\begin{equation}
\xi_{\ell_1, \ell_2}(w, h)=\int (u-w/h)^{\ell_1} (u+w/h)^{\ell_2} K(u-w/h) K(u+w/h)du. \label{eq:tauh}
\end{equation}
The result in (\ref{eq:covmm2}) is a counterpart result of $\textrm{Var}\{\hat m(x)|\mathbb{X}\}$, where $\mathbb{X}=(X_1, \ldots, X_n)$ \citep[][equation (3.7)]{Fanbook}. 

Third, substituting (\ref{eq:covmm2}) in (\ref{eq:covAA}) gives 
\begin{equation}
\textrm{Cov}\left\{\mathcal{A}(w_1), \, \mathcal{A}(w_2) |\mathbb{W}\right\}=\frac{\gamma\left\{(w_1+w_2)/2\right\}}{nh}\be_1^\T \bS^{-1}\bS^*_{\hbox {\tiny $W$},h}\bS^{-1}\be_1\left\{1+o_{\hbox {\tiny $P$}}\left(\frac{1}{nh}\right)\right\}, 
\label{eq:covAA2}
\end{equation}
where $\gamma(w)=\nu^2(w)f_{\hbox {\tiny $W$}}(w)$. And plugging (\ref{eq:covAA2}) in (\ref{eq:varB}) yields 
\begin{eqnarray}
\textrm{Var}\left\{\mathcal{B}(x) |\mathbb{W}\right\} 
& = & \int D(x-w_1) \int D(x-w_2) \times \nonumber \\
& & \left[\frac{\gamma\left\{(w_1+w_2)/2\right\}}{nh}\be_1^\T \bS^{-1}\bS^*_{\hbox {\tiny $W$},h}\bS^{-1}\be_1\left\{1+o_{\hbox {\tiny $P$}}\left(\frac{1}{nh}\right)\right\}\right] dw_2 dw_1. \nonumber \\
\label{eq:varB2}
\end{eqnarray}
Note that, among the matrices in (\ref{eq:varB2}), only $\bS^*_{\hbox {\tiny $W$},h}$ depends on $w_1$ and $w_2$, of which the entries are $\xi_{\ell_1, \ell_2}(w,h)$ in (\ref{eq:tauh}). 

The fourth step is to derive
\begin{equation}
\int D(x-w_1) \int D(x-w_2) \gamma\left(\frac{w_1+w_2}{2}\right) \xi_{\ell_1, \ell_2}\left(\frac{w_1-w_2}{2},h\right) \, dw_2 dw_1,
\label{eq:varB2en}
\end{equation}
which is equal to 
\begin{equation}
\{\gamma(x)+O(h)\} \int K_{\hbox {\tiny $U$}, \ell_1}(v) K_{\hbox {\tiny $U$}, \ell_2}(v) \, dv. \label{eq:KK}
\end{equation}
Define $\kappa_{\ell_1, \ell_2}(h)= \int K_{\hbox {\tiny $U$}, \ell_1}(v) K_{\hbox {\tiny $U$}, \ell_2}(v)\, dv$ to highlight the dependence of this integral on $h$ (since $K_{\hbox {\tiny $U$}, \ell}(v)$ depends on $h$ according to (\ref{eq:KUl})), and define matrix $\bK(h)=(\kappa_{\ell_1, \ell_2}(h))_{0\le \ell_1, \ell_2\le p}$. To this end, we can conclude that, by (\ref{eq:varB2}) and (\ref{eq:KK}), 
\begin{equation}
\textrm{Var}\left\{\mathcal{B}(x) |\mathbb{W}\right\} = \frac{\gamma(x)}{nh}\be_1^\T \bS^{-1} \bK(h) \bS^{-1}\be_1 \left\{1+o_{\hbox {\tiny $P$}}\left(\frac{1}{nh}\right)\right\}. 
\label{eq:varB3}
\end{equation}
This is where the path of our derivations meets that of \citet{Delaigle09}, as now we need to incorporate the properties of $\kappa_{\ell_1, \ell_2}(h)$ as $n\to \infty$ (and thus $h\to 0$), for an ordinary smooth $U$ and for a super smooth $U$, respectively. These properties are thoroughly studied in \citet{Delaigle09} and summarized in their Lemmas B.4, B.6, B.9, which are restated in Appendix B for completeness. Equipped with these lemmas, we are ready to move on to the fifth step of the derivations. 

By Lemma B.4, for an ordinary smooth $U$, under Condition O, $\kappa_{\ell_1, \ell_2}(h)=h^{-2b}\eta_{\ell_1, \ell_2}+o\left(h^{-2b}\right)$ as $n\to \infty$, where 
\begin{equation}
\eta_{\ell_1, \ell_2}=i^{-\ell_1-\ell_2}(-1)^{-\ell_2}c^{-2}(2\pi)^{-1}\int |t|^{2b} \phi^{(\ell_1)}_{\hbox {\tiny $K$}}(t)\phi^{(\ell_2)}_{\hbox {\tiny $K$}}(t)\, dt, \nonumber 
\end{equation}
in which $b$ and $c$ are constants in Definition~\ref{df:ordsm}. Define $\bS^*=(\eta_{\ell_1, \ell_2})_{0\le \ell_1, \ell_2\le p}$, then  $\bK(h)=h^{-2b}\bS^*+o\left(h^{-2b}\right)$, and thus (\ref{eq:varB3}) implies (\ref{eq:ourthmo}) in Theorem~\ref{thm:var} below. For a super smooth $U$, by Lemma B.9, under Condition S, $|\kappa_{\ell_1, \ell_2}(h)|\le Ch^{2b_2}\exp(2h^{-b}/d_2)$, where $b_3=b_0I(b_0<0.5)$, $b_0$, $b$ and $d_2$ are constants in Definition~\ref{df:supsm}, and $C$ is some generic non-negative finite constant appearing in Lemma B.8 in \citet{Delaigle09}. This leads to (\ref{eq:ourthms}) in Theorem~\ref{thm:var} below, which serves as a recap of our findings in this subsection. 
\begin{theorem}
\label{thm:var}
When $U$ is ordinary smooth of order $b$, under Condition O, if $nh^{2b+1}\to \infty$, then 
\begin{equation}
\textrm{Var}\left\{\hat m_{\hbox {\tiny HZ}}(x) |\mathbb{W}\right\}=\be_1^\T \bS^{-1} \bS^* \bS^{-1}\be_1 \frac{\gamma(x)}{f^2_{\hbox {\tiny $X$}}(x)nh^{2b+1}}+o_{\hbox {\tiny $P$}}\left(\frac{1}{nh^{2b+1}} \right).
\label{eq:ourthmo}
\end{equation}
When $U$ is super smooth of order $b$, under Condition S, if $n\exp(2h^b/d_2)h^{1-2b_3}\to \infty$, then $\textrm{Var}\left\{\hat m_{\hbox {\tiny HZ}}(x) |\mathbb{W}\right\}$ is bounded from above by 
\begin{equation}
\be_1^\T \bS^{-1}\bS^{-1}\be_1\frac{C\gamma(x)h^{2b_3-1}}{f^2_{\hbox {\tiny $X$}}(x)n\exp(2h^b/d_2)}+o_{\hbox {\tiny $P$}}\left\{\frac{h^{2b_3-1}}{n\exp(2h^b/d_2)} \right\}.
\label{eq:ourthms}
\end{equation}
\end{theorem} 

\subsection{Comparison with the variance of the DFC estimator}
\label{s:comparevar}
By Theorem 3.1 in \citet{Delaigle09}, when the distribution of $U$ is ordinary smooth, under Condition O,  if $nh^{2b+1}\to \infty$, then 
\begin{equation}
\textrm{Var}\left\{\hat m_{\hbox {\tiny DFC}}(x) \right\}=\be_1^\T \bS^{-1} \bS^* \bS^{-1} \be_1 \frac{(\tau^2f_{\hbox {\tiny $X$}})*f_{\hbox {\tiny $U$}}(x)}{f^2_{\hbox {\tiny $X$}}(x)n h^{2b+1}}+o\left(\frac{1}{n h^{2b+1}}  \right), 
\label{eq:theirthmo}
\end{equation}
where $\tau^2(x)=\textrm{Var}(Y|X=x)$. Note that the asymptotic variance results in Theorem~\ref{thm:var}, as well as the asymptotic bias results in Section~\ref{s:bias}, are conditional on $\mathbb{W}$ whereas (\ref{eq:theirthmo}) is an unconditional variance. The conditional arguments in our moment analysis originate from the direct use of asymptotic moments of the local polynomial estimator of a regression function in the absence of measurement error, which are conditional moments given $\mathbb{X}$ \citep{Ruppert94}. As pointed out in \citet[][Remark 1, page 1351]{Ruppert94}, because the dominating terms in these conditional moments are free of $\mathbb{W}$, they still have the interpretation of unconditional dominating moments. Once this is clear, one can see that the difference between the dominating variance in (\ref{eq:theirthmo}) and that in (\ref{eq:ourthmo}) lies in the distinction between $(\tau^2f_{\hbox {\tiny $X$}})*f_{\hbox {\tiny $U$}}(x)$ and $\gamma(x)$. It is shown in Appendix B that $\gamma(x) = (\tau^2f_{\hbox {\tiny $X$}})*f_{\hbox {\tiny $U$}}(x)+f_{\hbox {\tiny $W$}}(x) \textrm{Var}\{m(X)|W=x\}\ge (\tau^2f_{\hbox {\tiny $X$}})*f_{\hbox {\tiny $U$}}(x)$. Hence, for an ordinary smooth $U$, the dominating variance of $\hat m_{\hbox {\tiny HZ}}(x)$ is greater than or equal to that of $\hat m_{\hbox {\tiny DFC}}(x)$.  In Section~\ref{s:simulation}, we will see how this large sample comparison takes effect in the comparison of finite sample variances associated with the two estimators. 

\section{Asymptotic normality} \label{s:normality}
Under the conditions stated in Theorem~\ref{thm:var}, we show the asymptotic normality of $\hat m_{\hbox {\tiny HZ}}(x)$ in Appendix C. The logic behind the proof is similar to that in \citet{Delaigle09}. More specifically, we first approximate $\mathcal{B}(x)-B(x)$ via an average, $n^{-1}\sum_{j=1}^n \tilde U_{n, j}(x)$, where $\{\tilde U_{n, j}(x)\}_{j=1}^n$ is a set of independent and identically distributed (i.i.d.) random variables at each fixed $x$. Then we show that, for some positive constant $\eta$, 
\begin{equation}
\lim_{n\to \infty} \frac{E|\tilde U_{n,1}|^{2+\eta}}{n^{\eta/2}\{E(\tilde U_{n,1}^2)\}^{(2+\eta)/2}}=0,\nonumber
\end{equation}
which is a sufficient condition for 
$$\frac{\sum_{j=1}^n \tilde U_{n,j} -n E(\tilde U_{n,j})}{\sqrt{n\textrm{Var}(\tilde U_{n,j})}}\stackrel{\mathcal{L}}{\to} N(0, 1).$$ 
This in turn leads to the asymptotic normality of $\mathcal{B}(x)-B(x)$, and further suggests the asymptotic normality of $\hat m_{\hbox {\tiny HZ}}(x)$. 

To this end, we have answered the questions raised in Section~\ref{s:prep} regarding the properties of a random process $\mathcal{B}(x)$ resulting from the convolution of another random process $\mathcal{A}(w)$ and the non-random function $D(s)$. We now see that the first two moments of $\mathcal{B}(x)$ are closely related to the the first two moments of $\mathcal{A}(w)$ via similar convolutions. Also, if $\mathcal{A}(w)$ is asymptotically Gaussian, then under mild regularity conditions, $\mathcal{B}(x)$ is also asymptotically Gaussian, and many of these conditions can be satisfied by choosing an appropriate kernel function in $\mathcal{A}(w)$. 

\section{Implementation and finite sample performance}
\label{s:simulation}
After a thorough investigation of asymptotic properties of the proposed estimator, we are now in the position to look into its finite sample performance. By the construction of $\hat m_{\hbox {\tiny HZ}}(x)$, we need to evaluate continuous Fourier transforms (CFT) and inverse CFTs. In this section we first describe the algorithm for these evaluations, then discuss bandwidth selection. Finally, we present experiments to compare our estimator with the DFC estimator under four settings where we simulate data from the true models with our design of $m(x)$, and under another setting where error-prone data are simulated from a motorcycle-crash data set with the underlying $m(x)$ unknown. 

\subsection{Numerical evaluations}
\label{s:fft}
For an integrable function that maps the real line onto the complex space, $f:\mathR\rightarrow\mathC$, define the CFT of $f$ as
\begin{equation}
\calF[f](t) = \int_{-\infty}^\infty f(s) e^{-its} ds, \quad \forall t\in\mathR.
\label{eq:cft}
\end{equation} 
In our study, we first approximate the CFT via a discrete Fourier transform (DFT), then we use the fast Fourier transform algorithm \citep[FFT,][]{Bailey94} to evaluate the corresponding DFT. For a sequence of $G$ complex values $\bz=\{z_0, \ldots, z_{G-1}\}$, the DFT is defined as 
$D_k[\bz] = \sum_{g=0}^{G-1} z_g e^{-i2\pi kg/G}$, for $k=0, \ldots, G-1$, which can be easily evaluated using FFT in standard statistical software. The approximation of CFT using DFT is sketched next. 

To prepare for the approximation, one first specifies a sequence of input values and then specifies a sequence of output values accordingly. More specifically, let $\{s_g=(g-G/2)\alpha_1, \, g=0, 1, \ldots, G-1\}$ be the input values for the CFT, where $G/2$ is an even integer, $\alpha_1=a/G$ is the increment, and $a$ is chosen such that (\ref{eq:cft}) can be well approximated by $\int_{-a/2}^{a/2} f(s) e^{-its} ds$. With the input values specified, the corresponding output values are $\{t_k = (k-G/2)\alpha_2, \, k=0, 1, \ldots, G-1\}$, where $\alpha_2 = 2\pi/(G\alpha_1)$. With the input and output values ready, we approximate the CFT as follows, for $k=0, 1, \ldots, G-1$, 
\begin{eqnarray*}
\calF[f](t_k) &\approx & \int_{-a/2}^{a/2} f(s) e^{-it_k s} ds \\
& \approx & \sum_{g=0}^{G-1} f(s_g) e^{-it_k s_g} \alpha_1 \\
& = & \alpha_1 \sum_{g=0}^{G-1} f(s_g) e^{-i (k-G/2)\alpha_2 (g-G/2)\alpha_1} \\
& = & \alpha_1 e^{i(k-G/2)\pi} \sum_{g=0}^{G-1} f(s_g)e^{i\pi g} e^{-i2\pi k g/G} \\
& = & \alpha_1(-1)^k D_k\left[ \left\{(-1)^g f(s_g) \right\} \right].
\end{eqnarray*}

This approximation converges to the truth very rapidly provided that the Fourier coefficients of $f$ rapidly decrease \citep{Davis.Rabinowitz1984}. The values of $\alpha_1$ and $\alpha_2$ determine the resolution of the input and output results, respectively. Comparable resolutions in $s$ and $t$ are typically desired, which can be achieved by setting $\alpha_1=\alpha_2=\sqrt{2\pi/G}$. A larger $G$ tends to yield a more accurate approximation of the CFT. \cite{Bailey94} computed the CFT of the standard normal density function using $G=2^{16}$ and achieved the root-mean-squared error of order $10^{-16}$. In the simulations presented in this article, we set $G=2^{16}$, resulting in $\alpha_1=\alpha_2\approx 0.01$ and $a\approx641.7$. In additional simulation studies where we used a larger $G$, we found the results  essentially unchanged. This algorithm can be similarly applied to approximate the inverse CFT.

\subsection{Bandwidth selection}
\label{s:band}
It has been well acknowledged that the choice of bandwidth is crucial in kernel-based nonparametric estimation. In our study, we adopt the method of cross-validation (CV) in conjunction with simulation extrapolation \citep[SIMEX,][Chapter 5]{LASbook} as proposed by \citet{Delaigle08}. To implement this method, one first randomly divides the observed data, $\{(Y_j, W_j)\}_{j=1}^n$, into $\delta$ subsamples of (nearly) equal size. Denote by $\calD_{k}$ the $k$th subsample, and $I_k$ the set of subject indices corresponding to the observations in $\calD_k$, for $k=1,\ldots, \delta$. Then one carries out two rounds of $\delta$-fold cross validation using further contaminated data. In the first round, one generates further contaminated data according to $W_{b,j}^*=W_j+U_{b,j}^*$, for $b=1, \ldots, B$ and $j=1, \ldots, n$, where $\{U_{b,j}^*, b=1, \ldots, B\}_{j=1}^n$ are i.i.d. according to $f_{\hbox {\tiny $U$}}(u)$. Viewing $\mathbb{W}$ as the ``unobserved true" covariate values, and $m^*(x)=E(Y|W=x)$ as the target regression function to be estimated using the ``observed" data, $\{(Y_j, W^*_{b,j})\}_{j=1}^n$, for $b=1, \ldots, B$, one may use the proposed method to estimate $m^*(x)$. Denote this estimator by $\hat{m}_{\hbox {\tiny HZ}}^*(x)$. Now one carries out the $\delta$-fold cross validation to choose a bandwidth for estimating $m^*(x)$ that minimizes 
\begin{equation*}
\textrm{CV}_1(h) = \frac{1}{nB}\sum_{b=1}^B\sum_{k=1}^\delta \sum_{j \in I_k} \left\{Y_{j}-\hat{m}_{\hbox {\tiny HZ},b}^{*(-k)}(W_j)\right\}^2 w(W_{i,j}),
\end{equation*}
where $\hat{m}_{\hbox {\tiny HZ},b}^{*(-k)}(x)$ is the estimate $\hat{m}_{\hbox {\tiny HZ}}^*(x)$ computed using the further contaminated data excluding $\calD_k$, for $k=1, \ldots, \delta$, and $w(\cdot)$ is a suitable weight function. Define $\hat h_1=\textrm{argmin}_{h>0}\textrm{CV}_1(h)$. In the second round of $\delta$-fold cross validation, another set of further contaminated data is produced according to $W_{b,j}^{**}=W_{b,j}^*+U_{b,j}^{**}$, where $\{U_{b,j}^{**}, b=1, \ldots, B\}_{j=1}^n$ are i.i.d. according to $f_{\hbox {\tiny $U$}}(u)$, for $b=1, \ldots, B$ and $j=1, \ldots, n$, also independent of $\{U_{b,j}^*, b=1, \ldots, B\}_{j=1}^n$. Similar to the first round, one views $\mathbb{W}^*=\{W_{b,j}^*, b=1, \ldots, B\}_{j=1}^n$ as the ``unobserved true" covariate values, and considers estimating another target regression function $m^{**}(x)=E(Y|W^*=x)$ using the proposed method based on the ``observed" data $\{(Y_j, W^{**}_{b,j})\}_{j=1}^n$, for $b=1, \ldots, B$. Denote this estimator by $\hat {m}_{\hbox {\tiny HZ}}^{**}(x)$. To select a bandwidth for estimating $m^{**}(x)$, one minimizes the following criterion with respect to $h$, 
\begin{equation*}
\textrm{CV}_2(h) = \frac{1}{nB}\sum_{b=1}^B\sum_{k=1}^{\delta}\sum_{j\in I_k} \left\{Y_j-\hat{m}_{\hbox {\tiny HZ},b}^{**(-k)}(W_{b,j}^{*})\right\}^2 w(W_{b,j}^{*}),
\end{equation*}
where $\hat{m}_{\hbox {\tiny HZ},b}^{**(-k)}(x)$ is the estimate $\hat {m}_{\hbox {\tiny HZ}}^{**}(x)$ computed using the data $\{(Y_j, W^{**}_{b,j})\}_{j=1}^n$ excluding $\calD_k$, for $k=1, \ldots, \delta$. Define $\hat{h}_2={\rm argmin}_{h>0}~\textrm{CV}_2(h)$. Finally, one sets $\hat{h}=\hat{h}_1^2/\hat{h}_2$ as the bandwidth used in $\hat m_{\hbox {\tiny HZ}}(x)$ for estimating $m(x)$ based on the original observed data $\{(Y_j, W_j)\}_{j=1}^n$. 

This bandwidth selection procedure can be computationally cumbersome because, first, in search of $\hat h_1$ and $\hat h_2$, one needs to evaluate $\textrm{CV}_1(h)$ and $\textrm{CV}_2(h)$ on a fine grid of candidate bandwidths; second, as recommended in most SIMEX applications, one needs a $B$ not too small in order to control the Monte Carlo variability when generating further contaminated data. To reduce the computational burden, we propose a procedure to refine the search region of $h$. Take the first round of cross validation described above as an example. Recall that, during this round, $\mathbb{W}$ is viewed as the unobserved true covariate values whereas $\mathbb{W}^*$ is the error-contaminated version of the true covariate values. To narrow down the search region of $h$ when minimizing $\textrm{CV}_1(h)$, we first find an initial bandwidth, $\tilde h_1$. In particular, we obtain $\tilde h_1$ by minimizing the following approximated mean integrated squared error (MISE) for the deconvolution kernel density estimator of $f_{\hbox {\tiny $W$}}(w)$ using $\mathbb{W}^*$ \citep{Stefanski90}, 
\begin{equation}
\textrm{MISE}(h) = \frac{1}{2\pi nh}\int \frac{|\phi_{\hbox {\tiny $K$}}(t)|^2}{|\phi_{\hbox {\tiny $U$}}(t/h)|^2}dt + \frac{h^4}{4}\int \left\{f''_{\hbox {\tiny $W$}}(w)\right\}^2 dw \int x^2 K(x)dx,
\label{eq:mise}
\end{equation}
where $\int \{f''_{\hbox {\tiny $W$}}(w)\}^2 dw$ can be easily estimated using $\mathbb{W}$. After $\tilde h_1$ is found, we search for $\hat h_1$ across $L$ grid points within $[0.2\tilde{h}_1, \, 2\tilde{h}_1]$. This strategy is motivated by the theoretical finding that the deconvolution kernel regression estimators have the same optimal rates as the deconvolution kernel density estimators. In our extensive trial-and-error simulation experiments under the model settings described in Section~\ref{s:4cases}, we considered a wider search region that encompasses $[0.2\tilde{h}_1, \, 2\tilde{h}_1]$, and we observed all selected $h$ indeed fell in the above refined search region. Similarly, in the second round of cross validation where we search for $\hat h_2$ across $L$ grid points within $[0.2\tilde{h}_2, \, 2\tilde{h}_2]$, where $\tilde{h}_2$ is chosen by minimizing (\ref{eq:mise}), but, different from the first round, now $\int \{f''_{\hbox {\tiny $W$}}(w)\}^2 dw$ there is replaced by $\int \{f''_{\hbox {\tiny $W^*$}}(w)\}^2 dw$, which can be easily estimated using $\mathbb{W}^*$. 

One may legitimately question our choice of the multiplicative factors, 0.2 and 2, in the recommended refined search region of $h$. For a given application, the safe and conservative way to choose $h$ usually involves some trial-and-error. If the optimal $h$ found within this refined region is too close to one of the boundaries, one may consider pushing that end of the region out slightly and adjusting the search region accordingly. Using the refined search region of $h$ at each round of cross validation, we also observe in simulations that one can even use a much smaller $B$ without noticeably compromising the quality of $\hat m_{\hbox {\tiny HZ}}(x)$. This refined bandwidth selection procedure and the algorithm for approximating CFT and inverse CFT described in Section~\ref{s:fft} are implemented in an {\tt R} package called {\tt lpme} created and maintained by the second author, which provides both $\hat m_{\hbox {\tiny HZ}}(x)$ and $\hat m_{\hbox {\tiny DFC}}(x)$.

\subsection{Simulation study}
\label{s:4cases}
In the simulation experiments, we compare realizations of $\hat m_{\hbox {\tiny HZ}}(x)$ and $\hat m_{\hbox {\tiny DFC}}(x)$ (with $p=1$) obtained under the following four model configurations:
\begin{enumerate}
	\item[(C1)] $[Y|X=x] \sim N(m(x), 0.2^2)$, where $m(x)=2x\exp(-10x^4/81)$, $X=0.8X_1 + 0.2X_2$, $X_1\sim f_{\hbox {\tiny $X_1$}}(x)=0.1875x^2I_{[-2,2]}(x)$, $X_2\sim \textrm{uniform}(-1,1)$, and $U\sim{\rm Laplace}(0,\sigma_u/\sqrt{2})$. 
	\item[(C2)] $[Y|X=x] \sim N(m(x), 0.5^2)$, where $m(x)=(x+x^2)/4$, $X\sim N(0,1)$, and $U\sim N(0, \sigma_u^2)$.
	\item[(C3)] $[Y|X=x]\sim N(m(x), 0.2^2)$, where $m(x)=x^6/30-5x^4/6+9x^2/2 +x$, $X\sim\mathrm{uniform}(-2,2)$, and $U\sim \mathrm{Laplace}(0, \sigma_u/\sqrt{2})$.
	\item[(C4)] $[Y|X=x]\sim N(m(x), 0.2^2)$, where $m(x)=\cos(x^2)+\sin(x)$, $X\sim\mathrm{uniform}(-2,2)$, and $U\sim \mathrm{Laplace}(0, 	\sigma_u/\sqrt{2})$.
 \end{enumerate}
Among these configurations, (C1) is considered in \citet{Delaigle09}; (C2) creates a scenario where the dominating bias of $\hat m_{\hbox {\tiny DFC}}(x)$ never vanishes since $m(x)$ is a second-order polynomial; (C3), with $m(x)$ being a higher order polynomial, results in zero dominating bias for $\hat m_{\hbox {\tiny DFC}}(x)$ within the support of $X$ at $\pm 1$; and (C4) has $m(x)$ out of the polynomial family yet it can be expanded as a polynomial of infinite order. Besides the model configuration, we also vary the reliability ratio $\lambda=\textrm{Var}(X)/\{\textrm{Var}(X)+\sigma_u^2\}$ from 0.7 to 0.95 at increments of 0.05 when generating $\mathbb{W}$. Under (C2), although the measurement errors are simulated from a normal distribution, we computed the estimates of $m(x)$ assuming a normal $U$ first, and then we repeated the estimation assuming a Laplace $U$. This exercise allows us to observe the effects of a misspecified distribution for $U$ on the estimates. Under each simulation setting, $500$ Monte Carol (MC) replicates of sample size $n=500$ are generated from the true model of $(Y,W)$. For both estimation methods, we used the kernel of which the Fourier transform is given by $\phi_{\hbox {\tiny $K$}}(t) = (1-t^2)^8 I_{[-1,1]}(t)$. 

Denote by $\hat m_{[\cdot]}(x)$ one of the two estimates under comparison generically. For the majority of the simulation experiments, in order to  mitigate the confounding effect of a data-driven bandwidth selection method on the quality of $\hat m_{[\cdot]}(x)$, we computed $\hat m_{[\cdot]}(x)$ using the theoretical optimal bandwidth obtained via minimizing an approximate of the integrated squared error (ISE), $\ISE=\int_{x_{\hbox {\tiny L}}}^{x_{\hbox {\tiny U}}}\{\hat{m}_{[\cdot]}(x)-m(x)\}^2 dx$, where $[x_{\hbox {\tiny L}}, \, x_{\hbox {\tiny U}}]$ is the interval of the true covariate value of interest. This approximated $\ISE$ is given by $\sum_{k=0}^{\mathcal{M}}\{\hat{m}_{[\cdot]}(x_k)-m(x_k)\}^2\Delta$, where $\Delta$ is the partition resolution, $\mathcal{M}$ is the largest integer no greater than $(x_{\hbox {\tiny U}}-x_{\hbox {\tiny L}})/\Delta$, and $x_k=x_{\hbox {\tiny L}}+k\Delta$, for $k=0,\ldots, \mathcal{M}$. For a small portion of the presented simulation experiments, we used the CV-SIMEX bandwidth selection strategy described in Section~\ref{s:band} to select a bandwidth for each of the two estimators. Note that, when choosing a bandwidth for $\hat m_{\hbox {\tiny DFC}}(x)$, one should change $\hat m^*_{\hbox {\tiny HZ}}(x)$ and $\hat m^{**}_{\hbox {\tiny HZ}}(x)$ in Section~\ref{s:band} to the counterpart estimates $\hat m^*_{\hbox {\tiny DFC}}(x)$ and $\hat m^{**}_{\hbox {\tiny DFC}}(x)$, respectively. 

\begin{figure}
	\centering
	\setlength{\linewidth}{4.5cm}
	\subfigure[]{ \includegraphics[width=\linewidth]{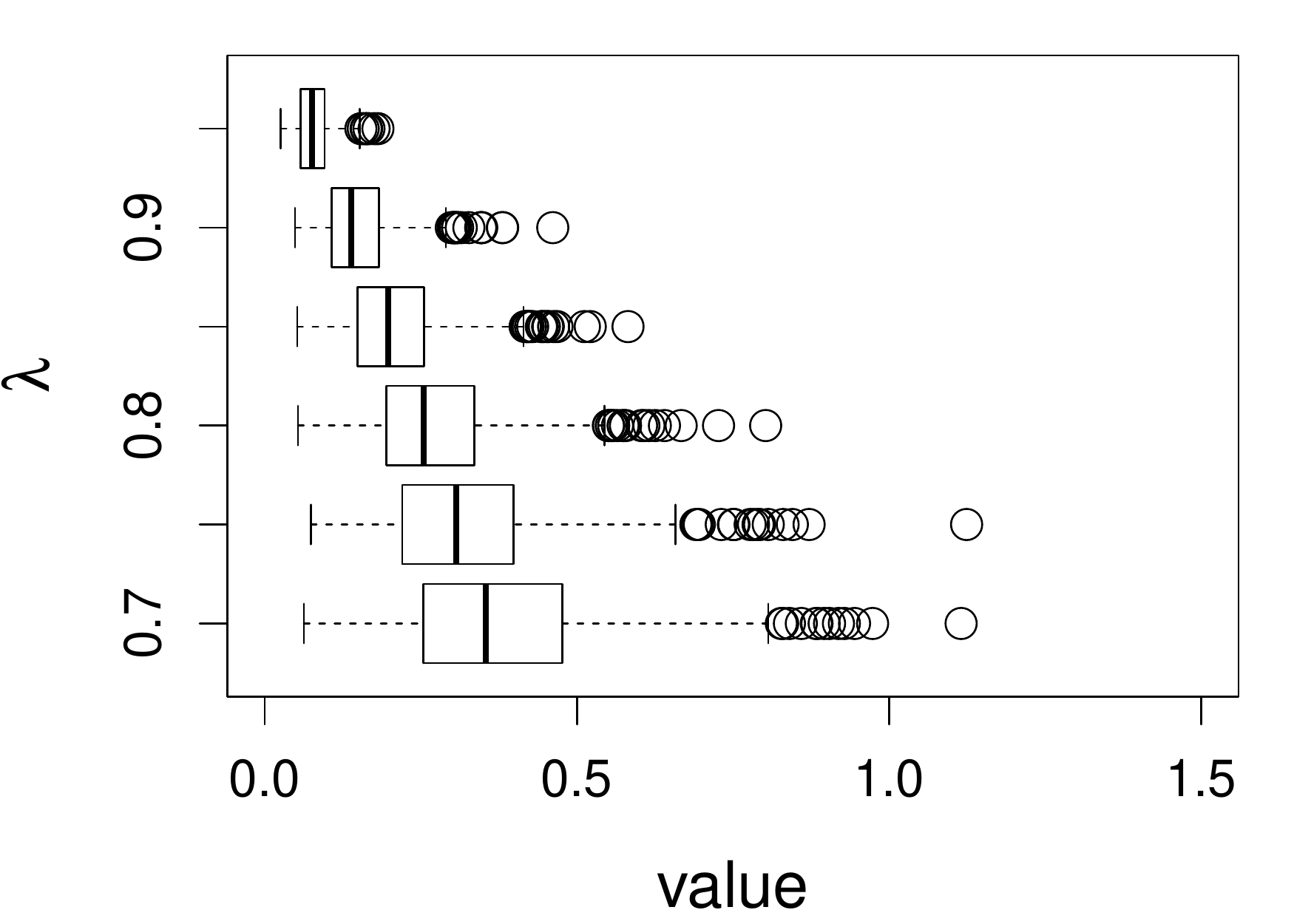} }
	\subfigure[]{ \includegraphics[width=\linewidth]{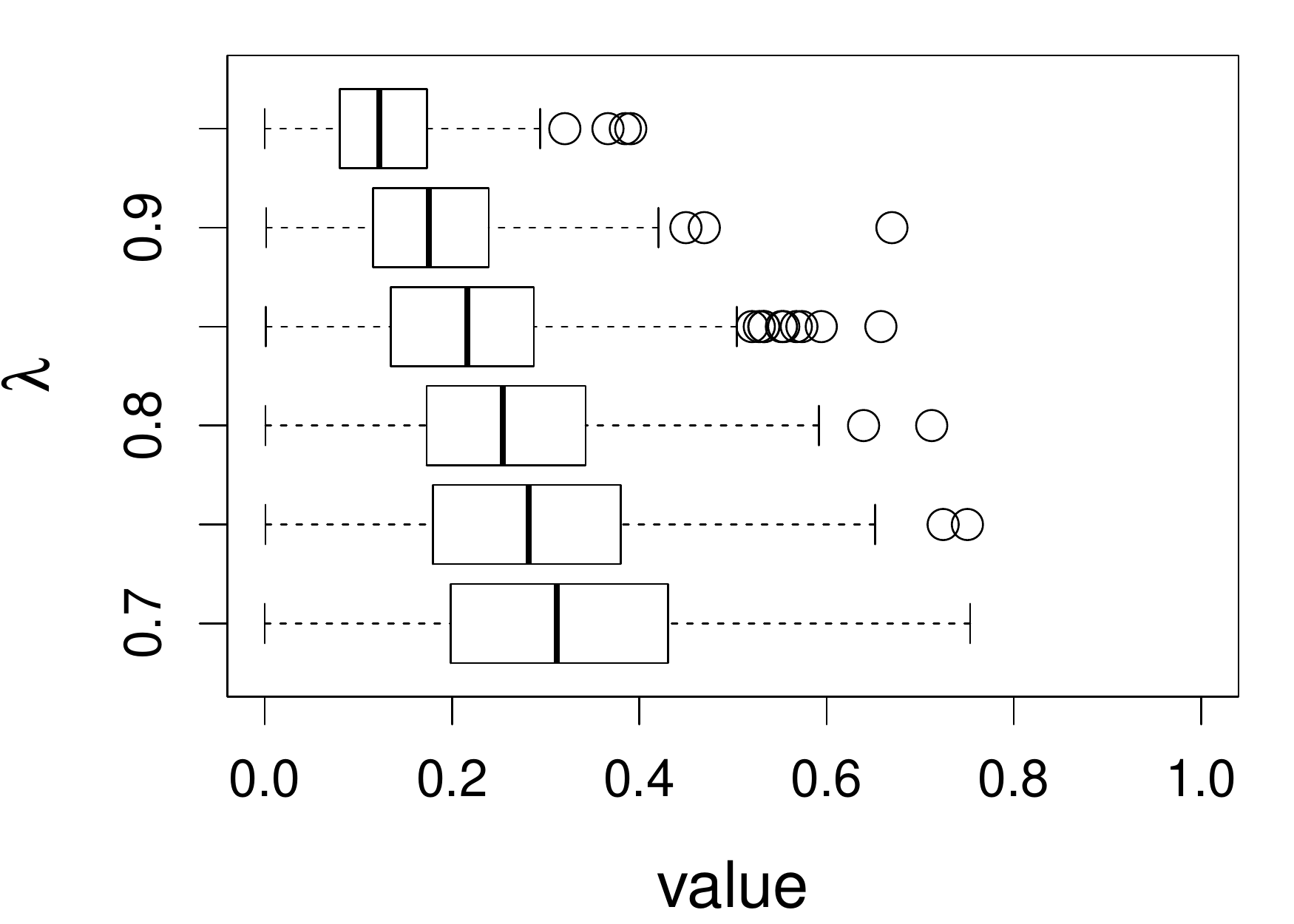} }
	\subfigure[]{ \includegraphics[width=\linewidth]{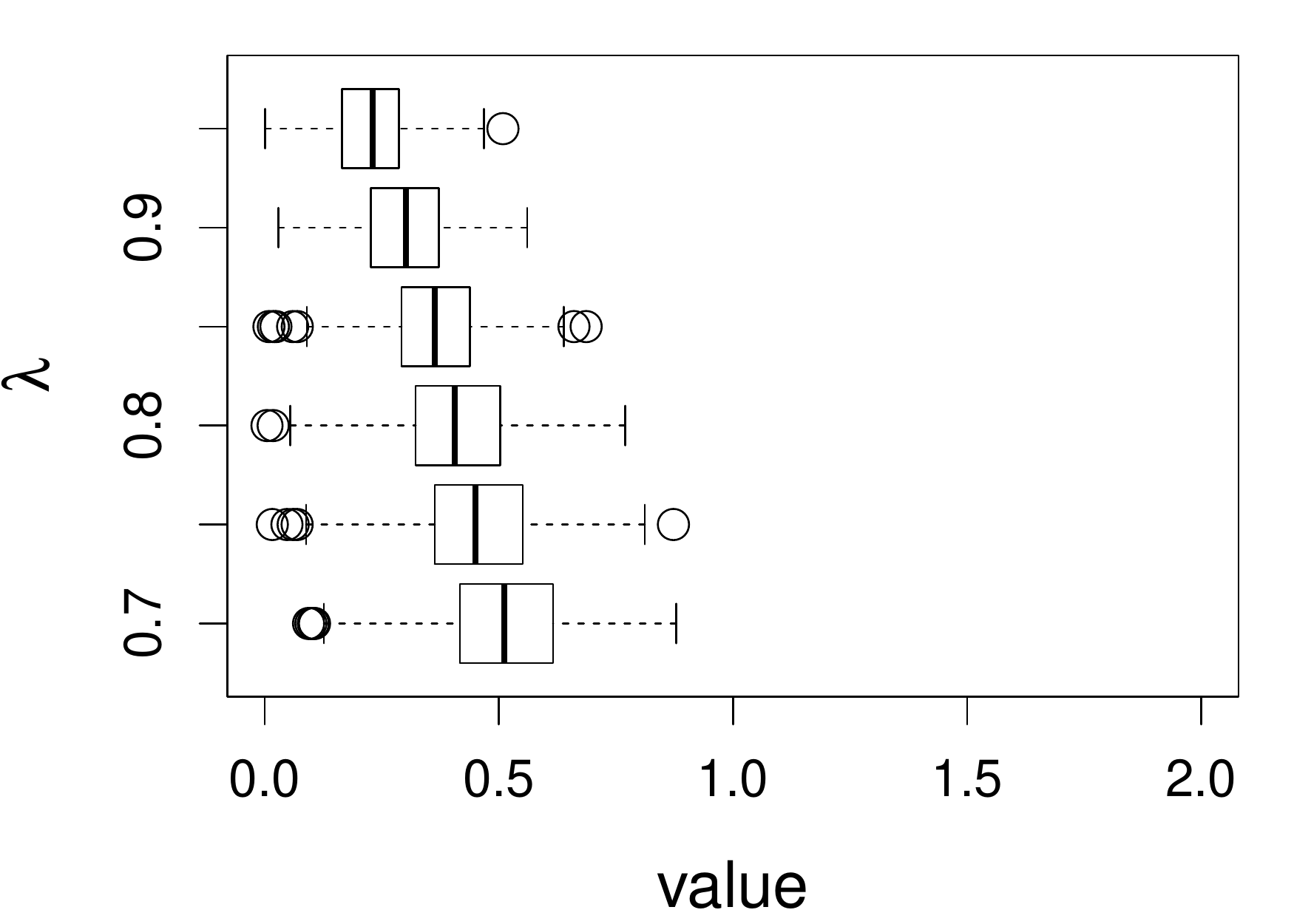} }\\
	\subfigure[]{ \includegraphics[width=\linewidth]{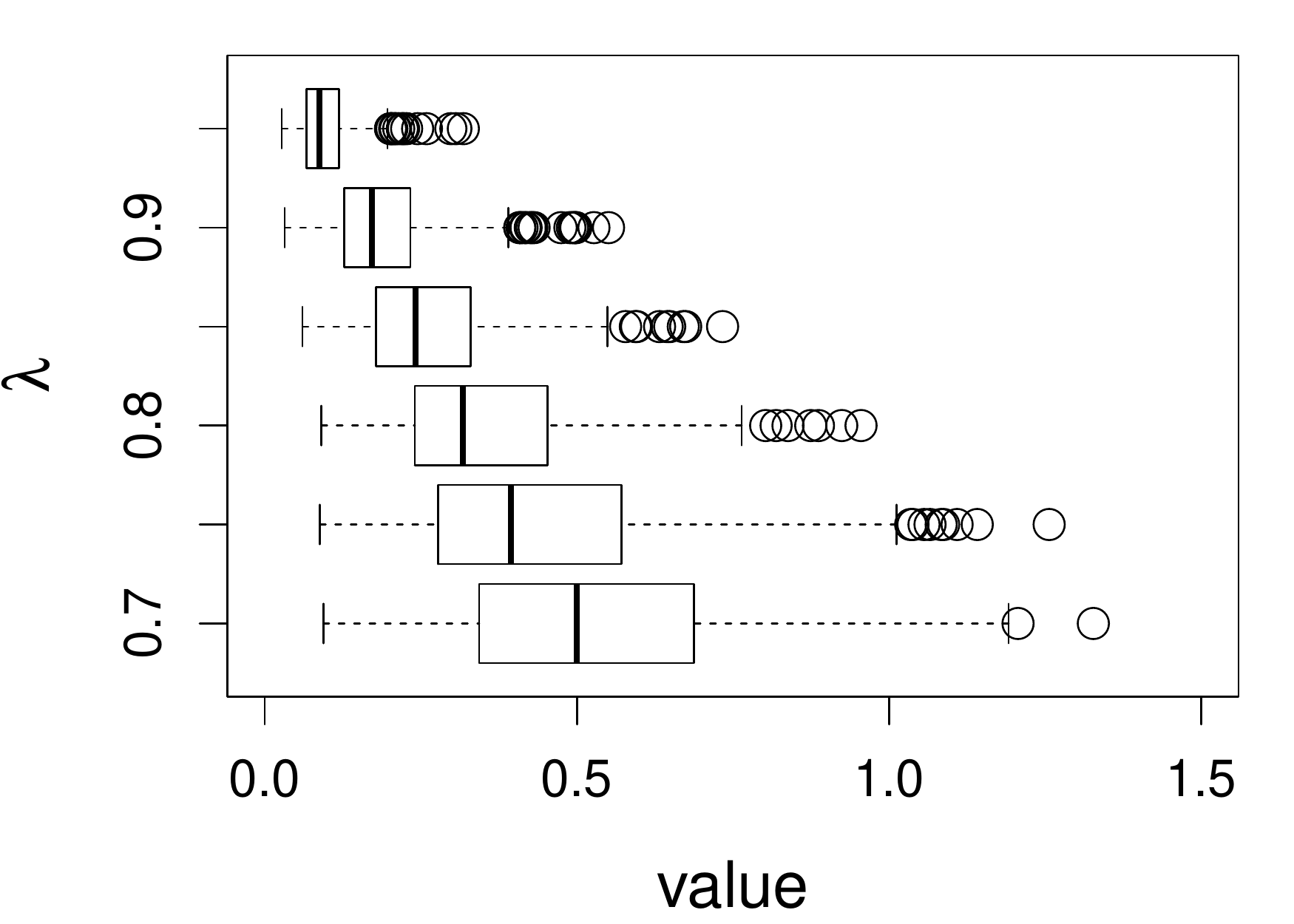} }
	\subfigure[]{ \includegraphics[width=\linewidth]{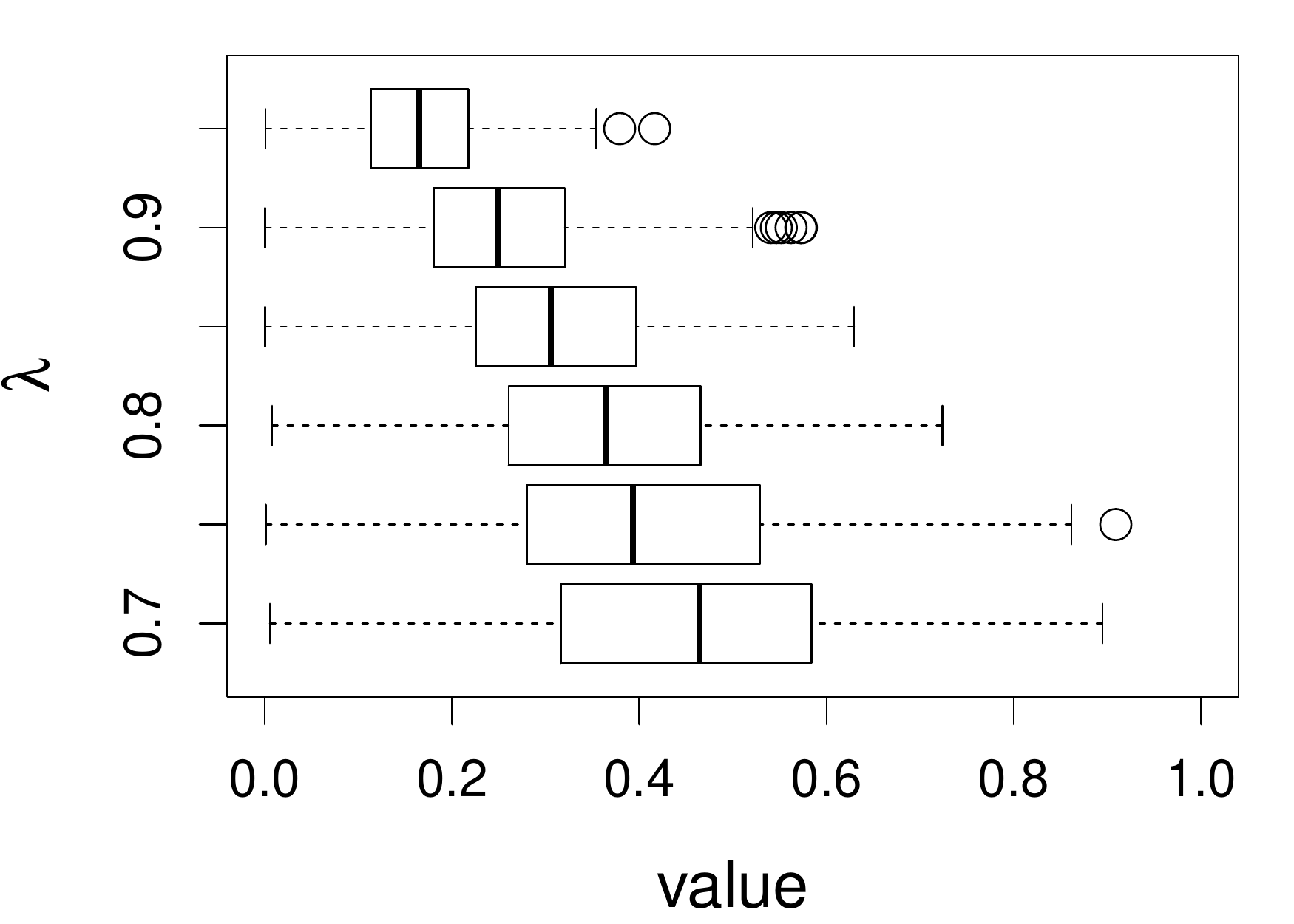} }
	\subfigure[]{ \includegraphics[width=\linewidth]{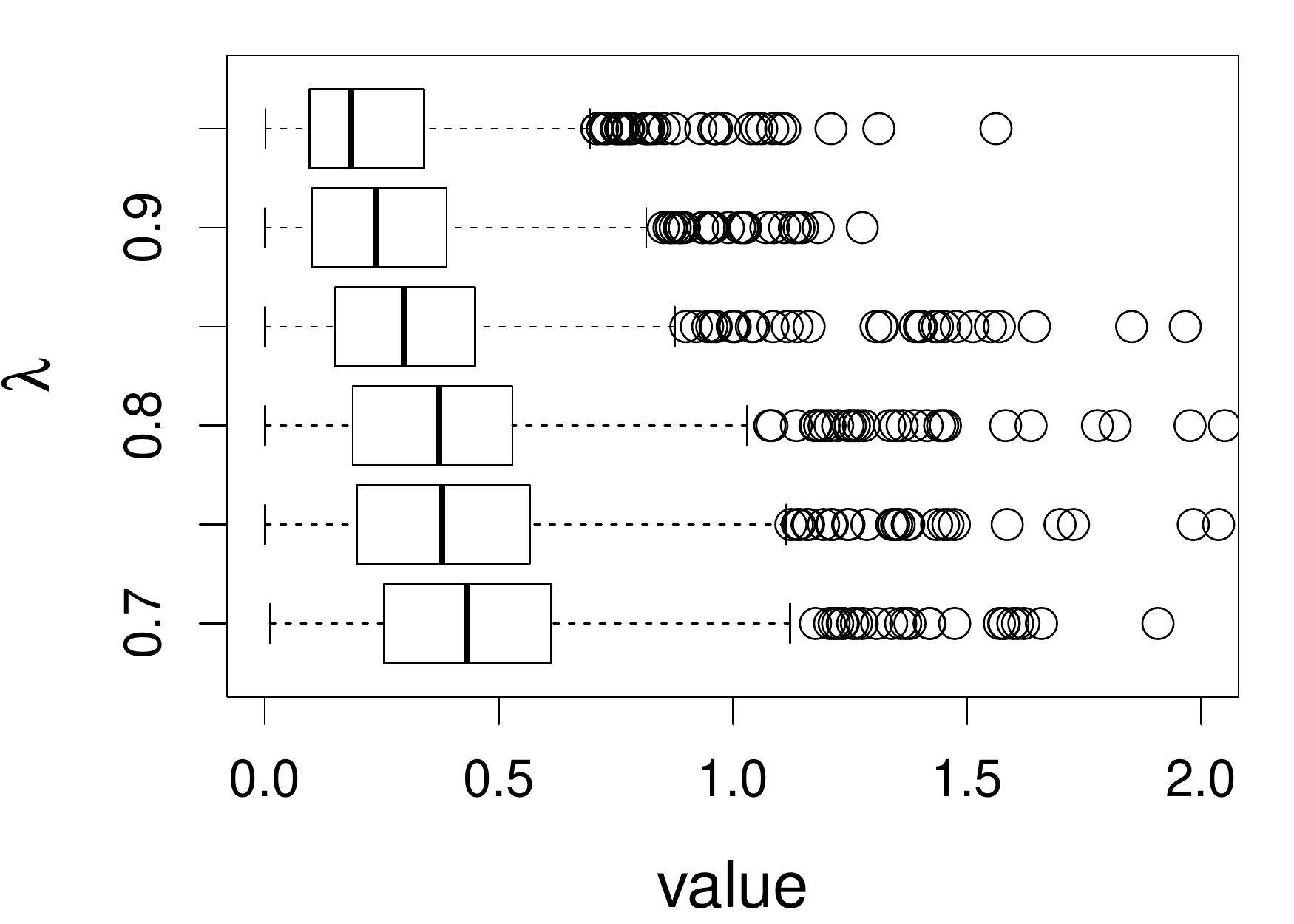} }\\
	\subfigure[]{ \includegraphics[width=\linewidth]{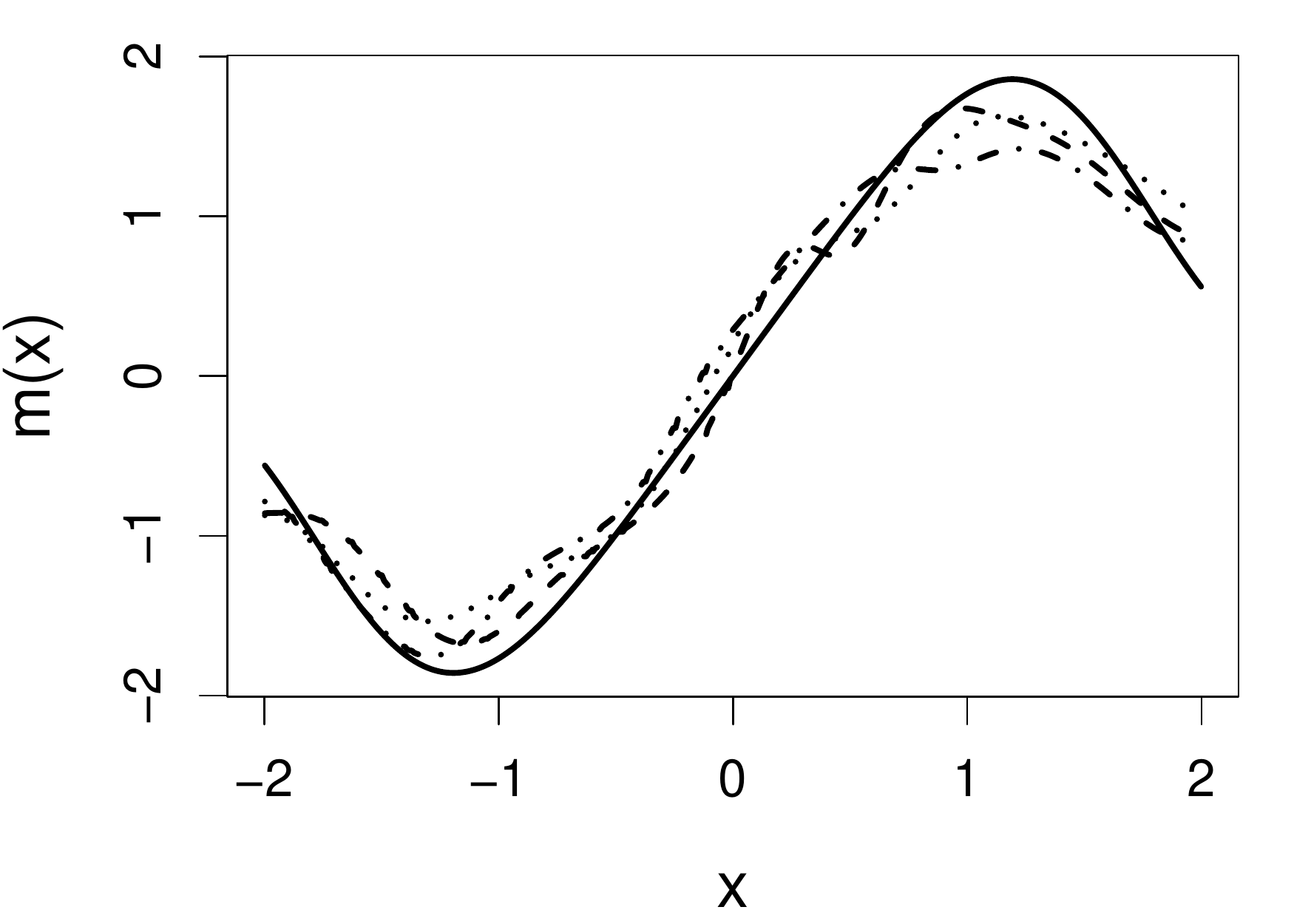} }
	\subfigure[]{ \includegraphics[width=\linewidth]{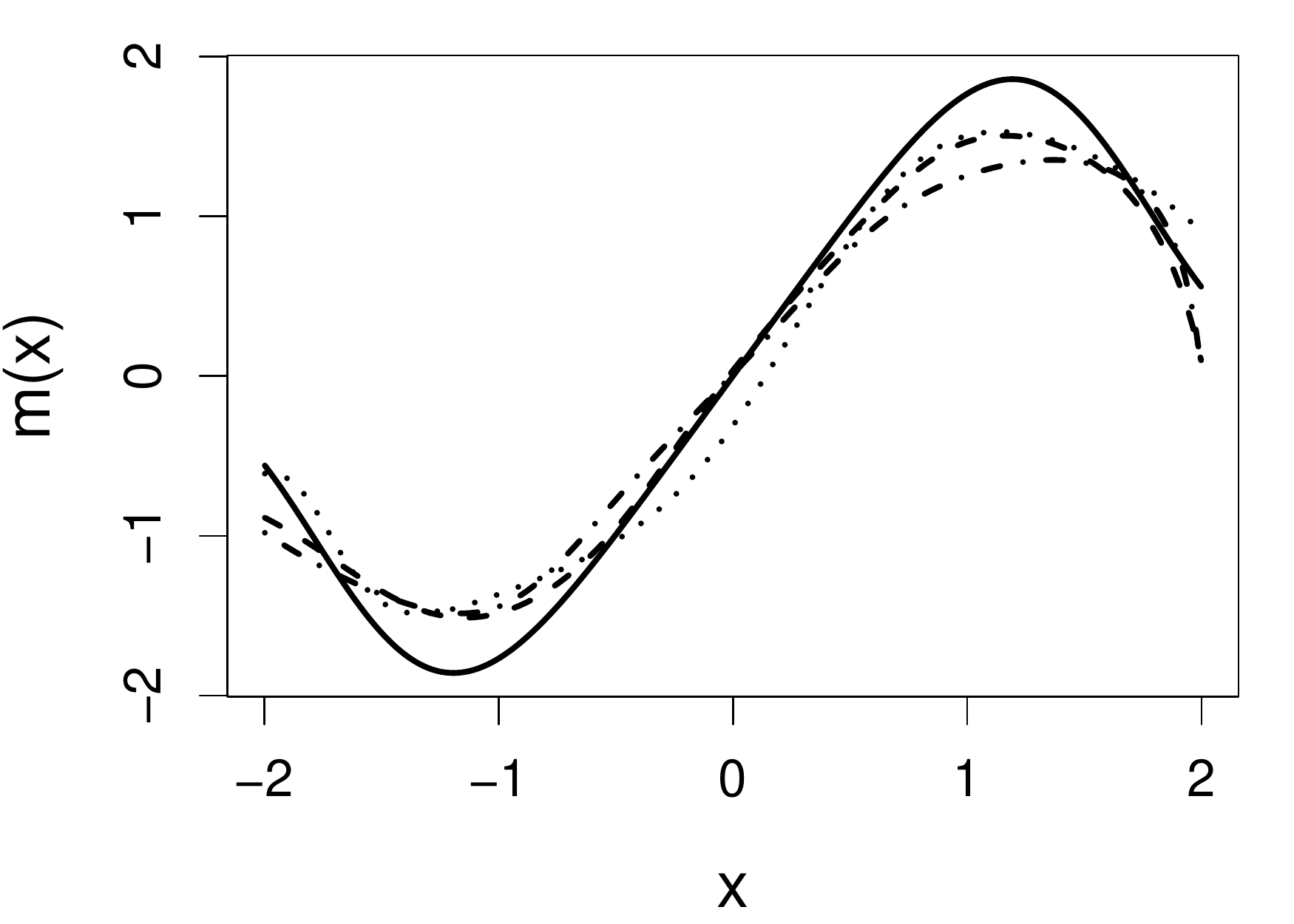} }
	\subfigure[]{ \includegraphics[width=\linewidth]{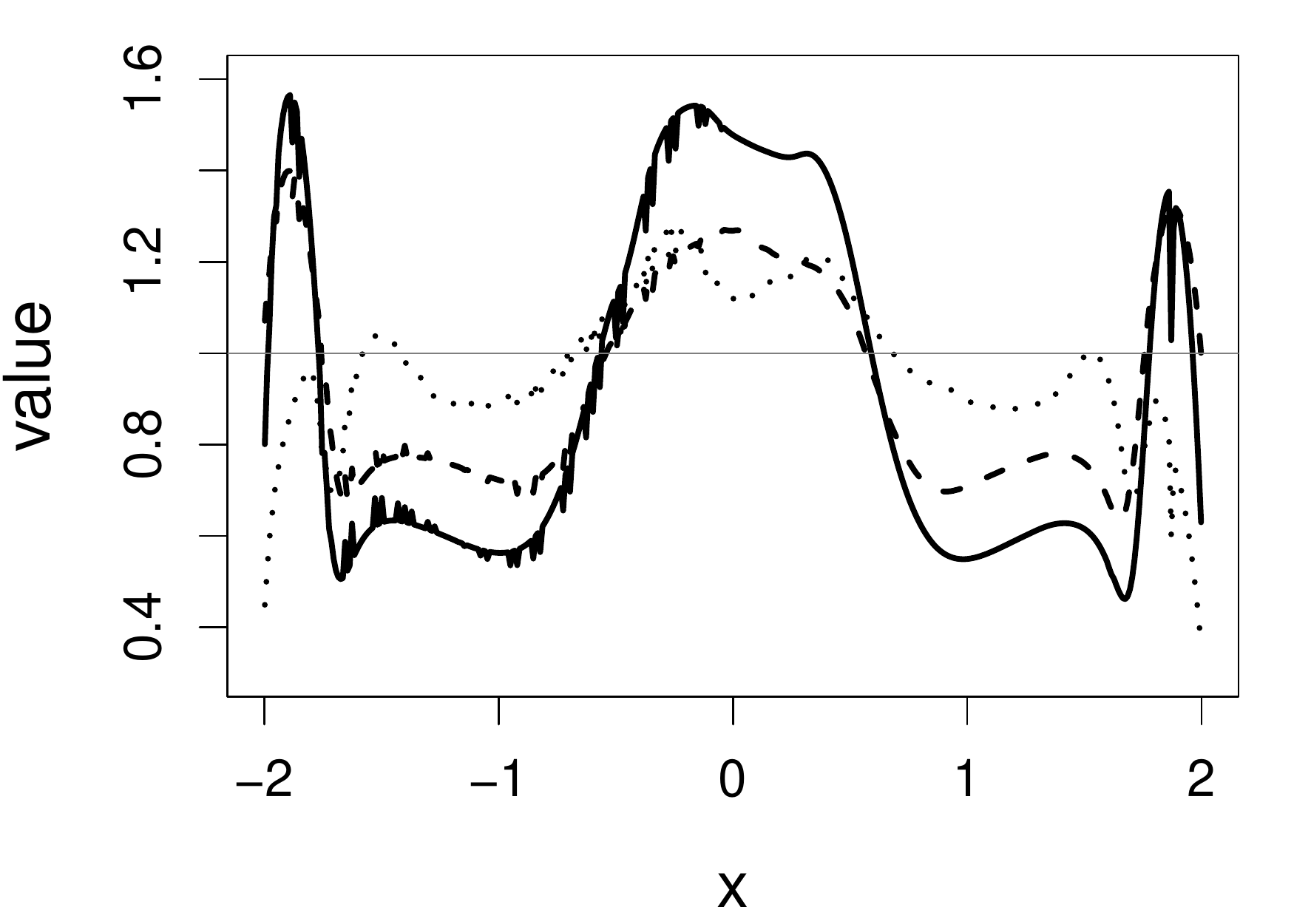} }\\
	\caption{Simulation results under (C1) using the theoretical optimal $h$. Panels (a) \& (d): boxplots of ISEs versus $\lambda$ for $\hat m_{\hbox {\tiny HZ}}(x)$ and $\hat m_{\hbox {\tiny DFC}}(x)$, respectively. Panels (b) \& (e): boxplots of PAE(1) versus $\lambda$ for $\hat m_{\hbox {\tiny HZ}}(1)$ and $\hat m_{\hbox {\tiny DFC}}(1)$, respectively. Panels (c) \& (f): boxplots of PAE(2) versus $\lambda$ for $\hat m_{\hbox {\tiny HZ}}(2)$ and $\hat m_{\hbox {\tiny DFC}}(2)$, respectively. Panels (g) \& (h): quantile curves when $\lambda=0.85$ for $\hat m_{\hbox {\tiny HZ}}(x)$ and $\hat m_{\hbox {\tiny DFC}}(x)$, respectively, based on ISEs (dashed lines for the first quartile, dotted lines for the second quartile, and dot-dashed lines for the third quartile, solid lines for the truth). Panel (i): PmAER (dashed line), PsdAER (dotted line), and PMSER (solid line) versus $x$ when $\lambda=0.85$; the horizontal reference line highlights the value 1.} 
	\label{Sim1LapLap500:box}
\end{figure}

\begin{figure}
	\centering
	\setlength{\linewidth}{4.5cm}
	\subfigure[]{ \includegraphics[width=\linewidth]{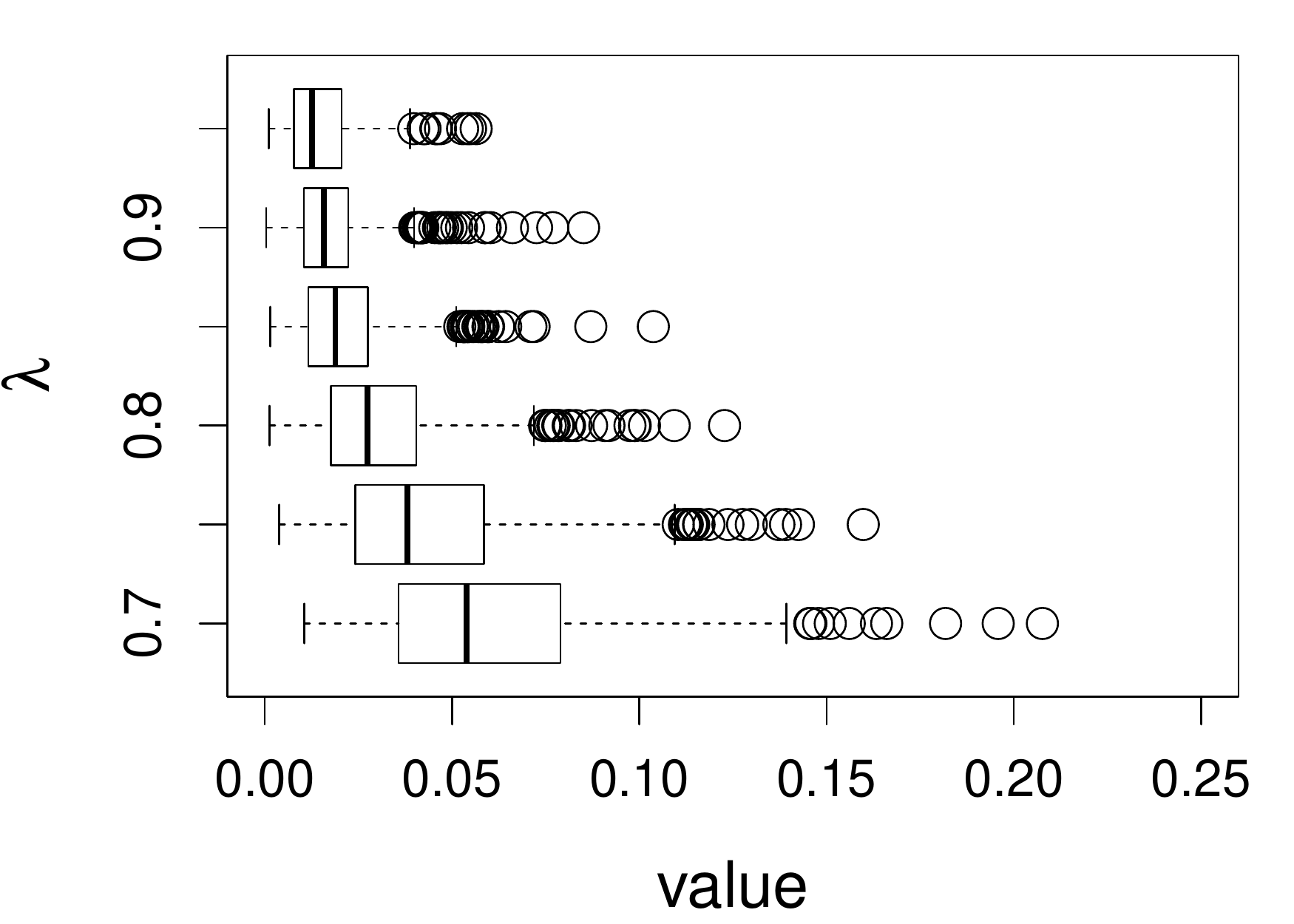} }
	\subfigure[]{ \includegraphics[width=\linewidth]{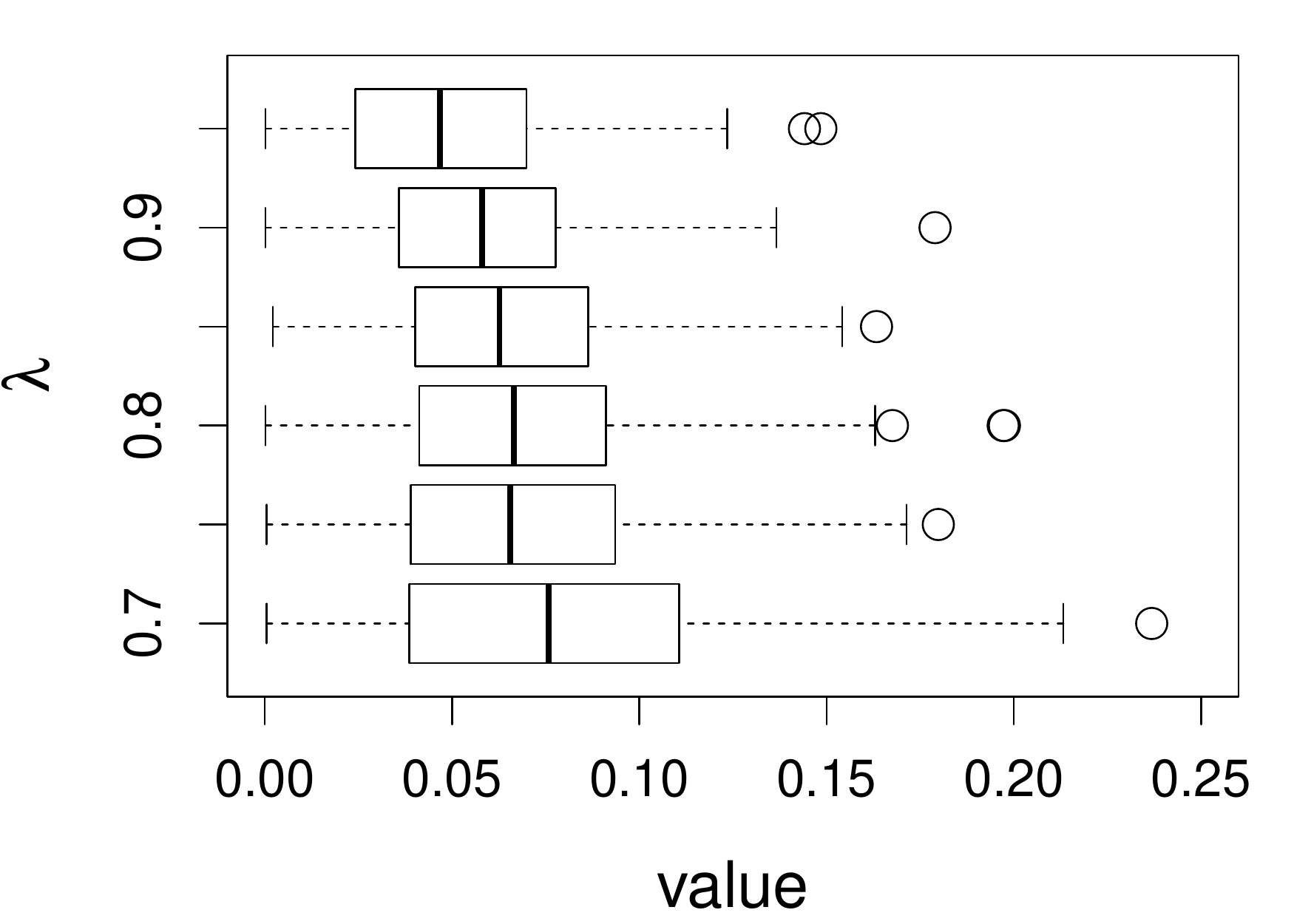} }
	\subfigure[]{ \includegraphics[width=\linewidth]{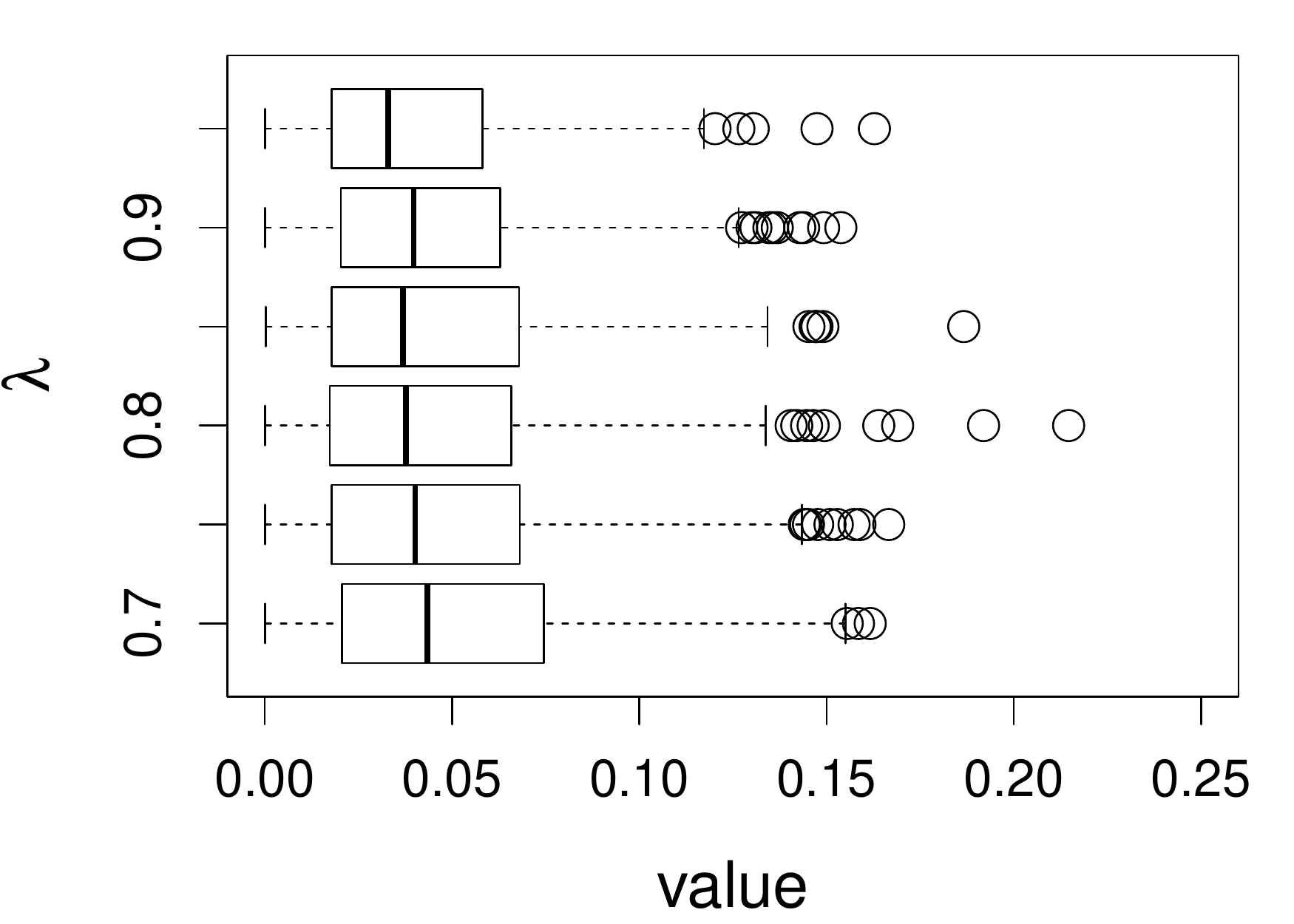} }\\
	\subfigure[]{ \includegraphics[width=\linewidth]{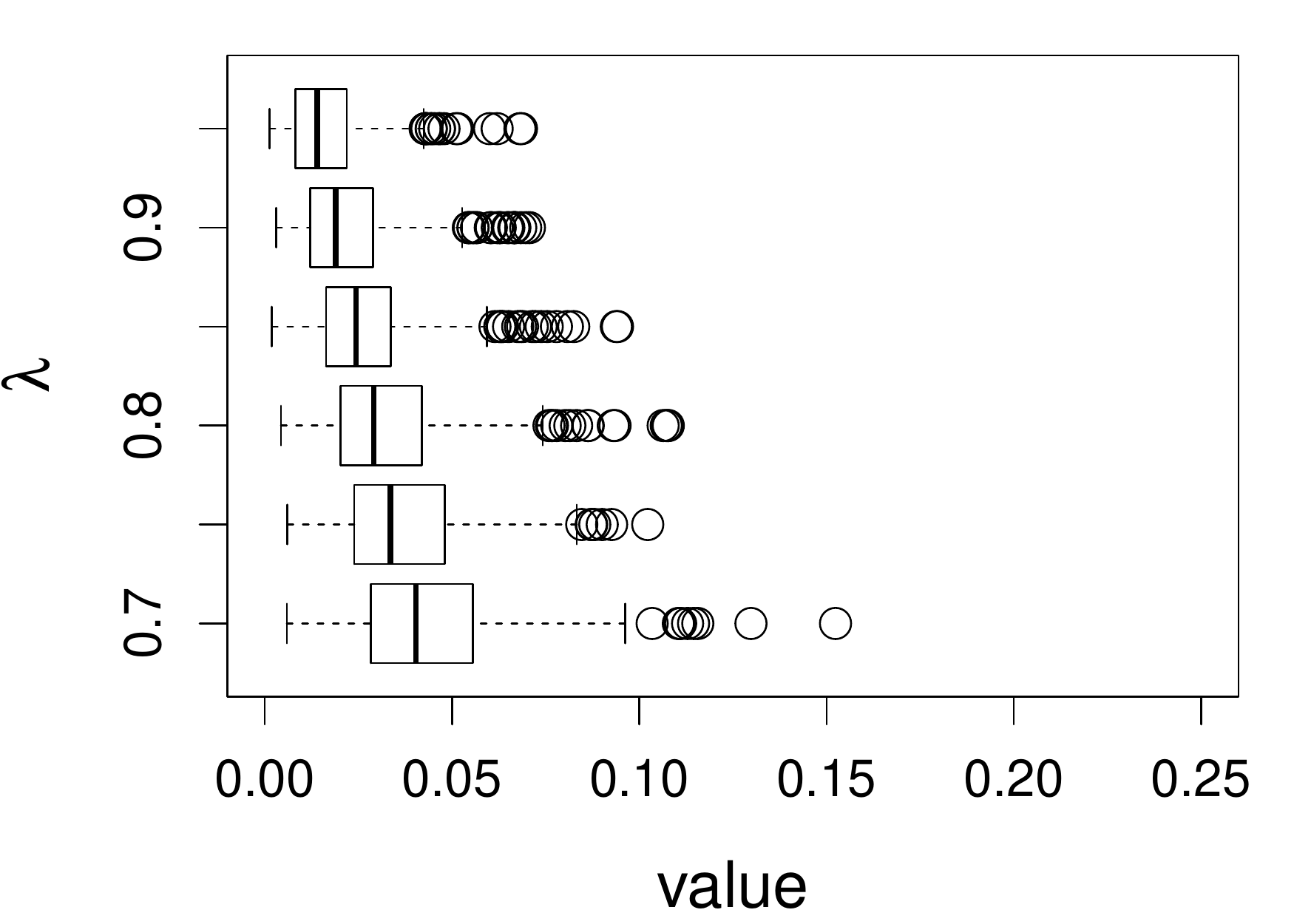} }
	\subfigure[]{ \includegraphics[width=\linewidth]{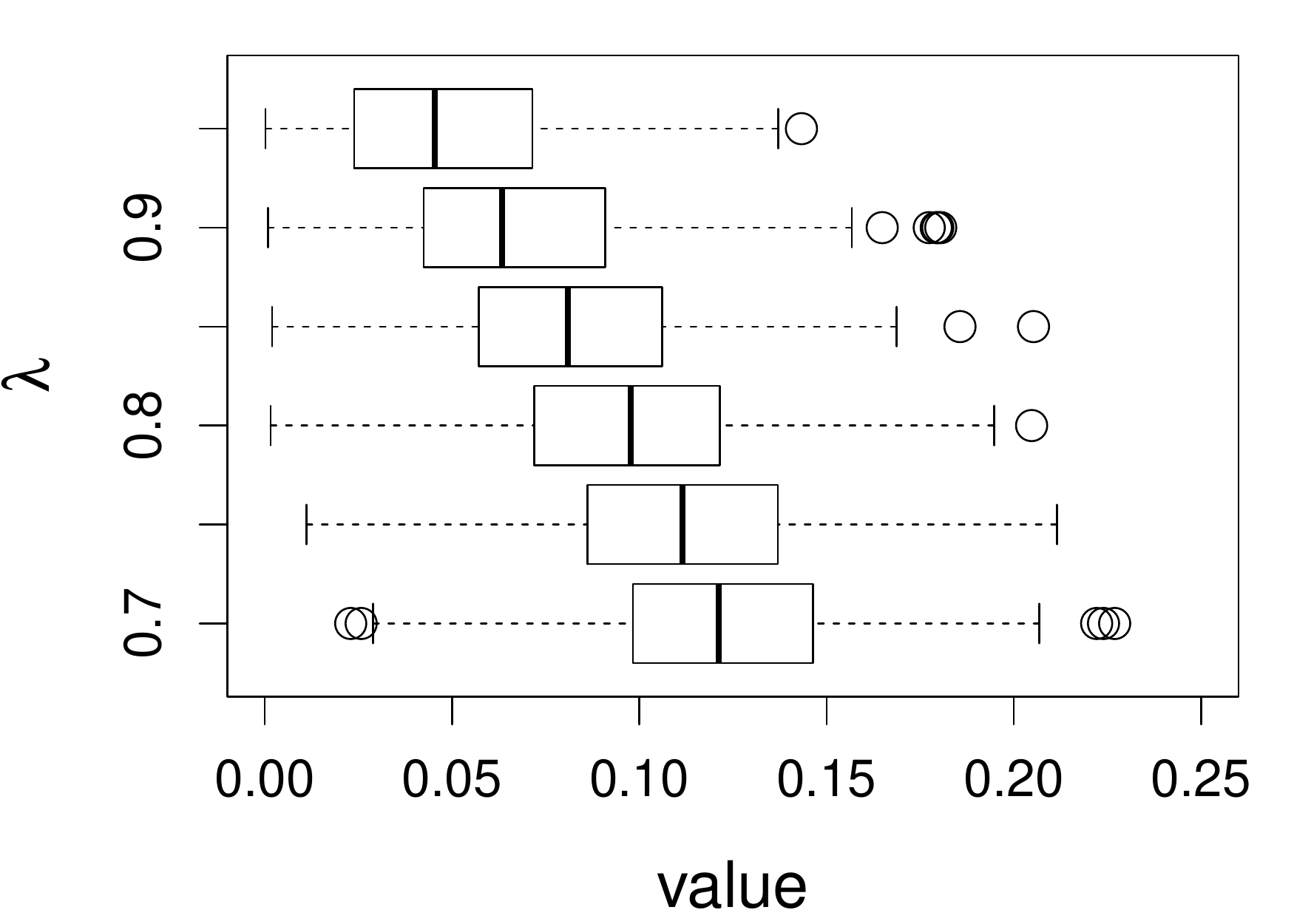} }
	\subfigure[]{ \includegraphics[width=\linewidth]{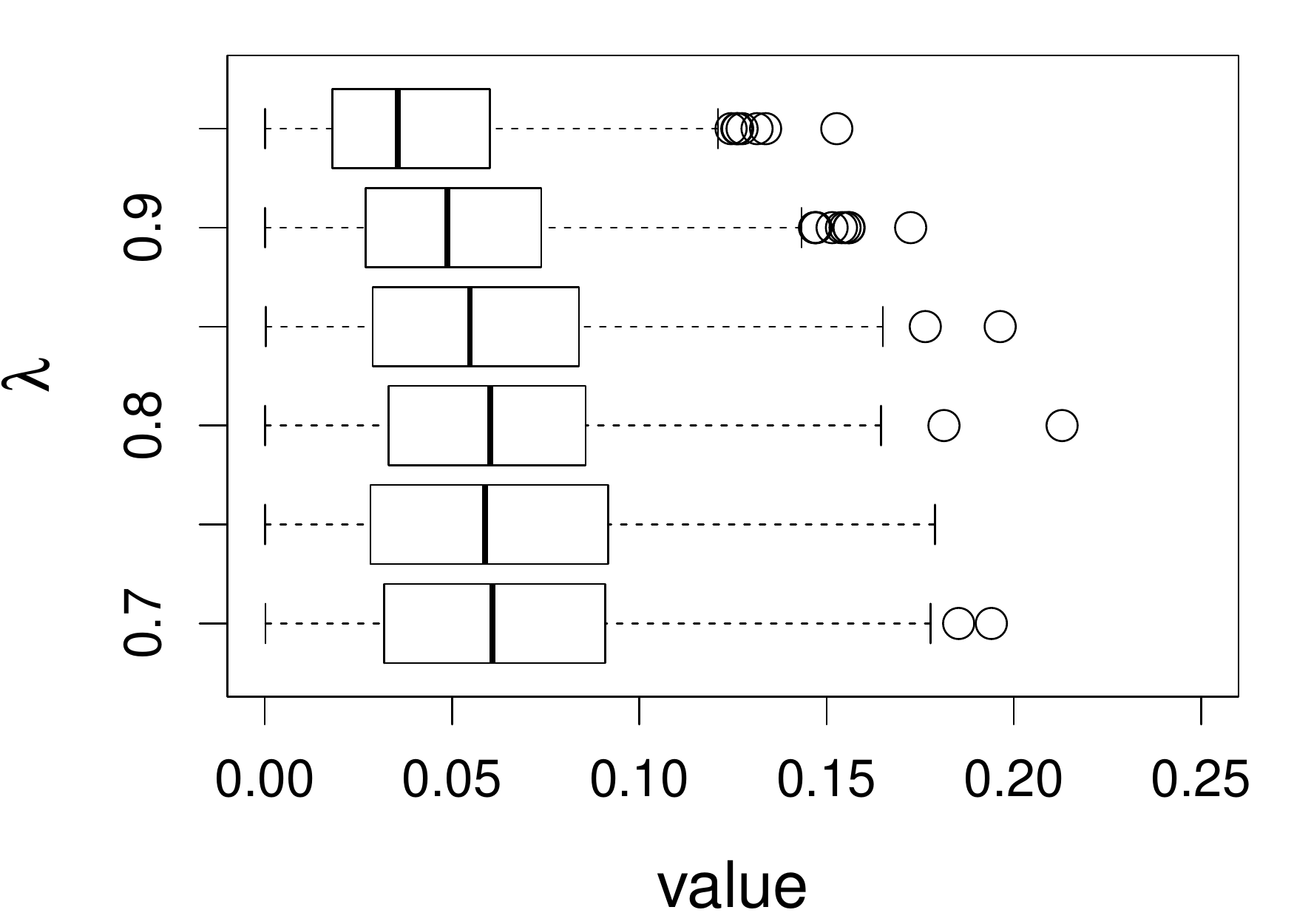} }\\
	\subfigure[]{ \includegraphics[width=\linewidth]{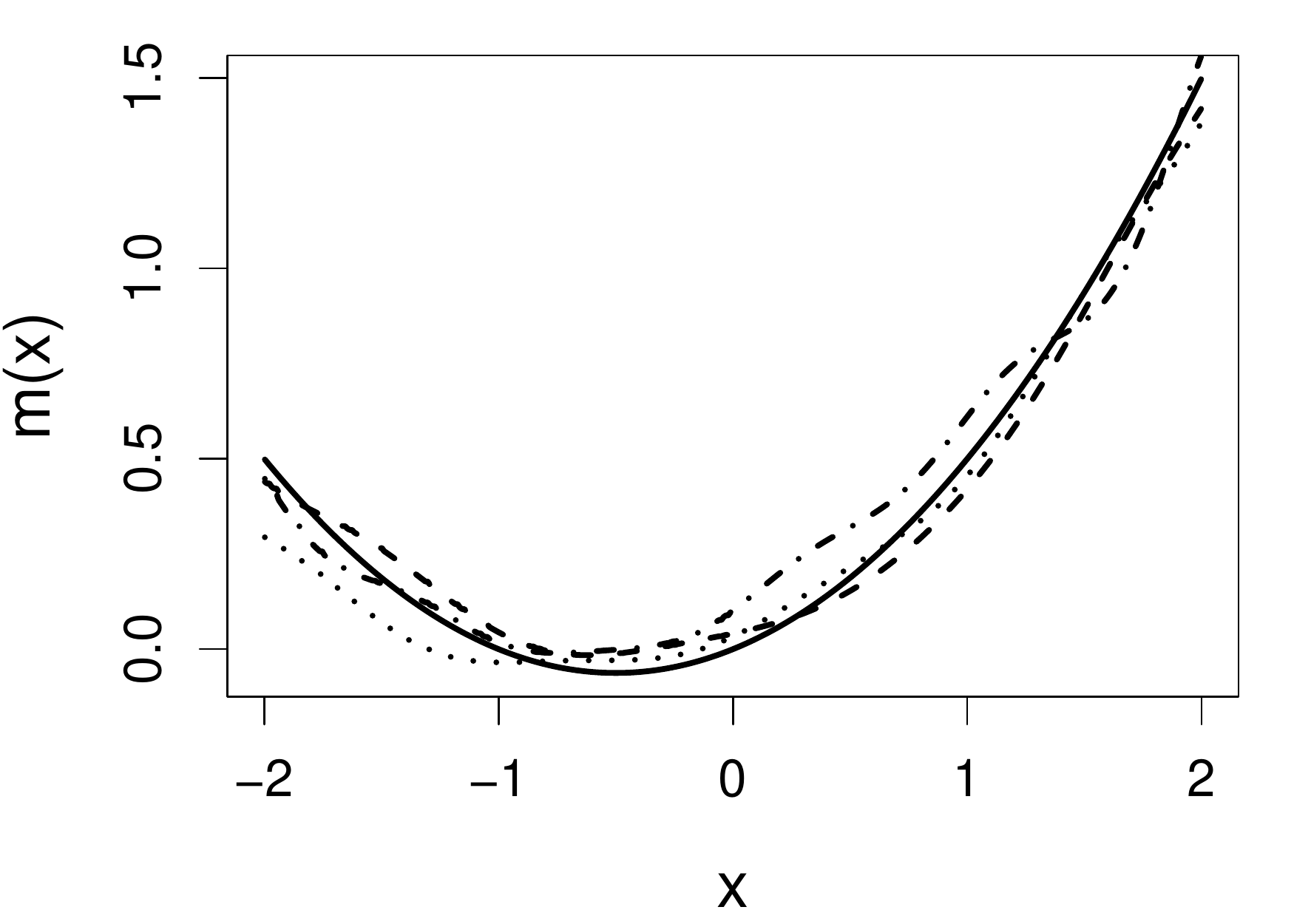} }
	\subfigure[]{ \includegraphics[width=\linewidth]{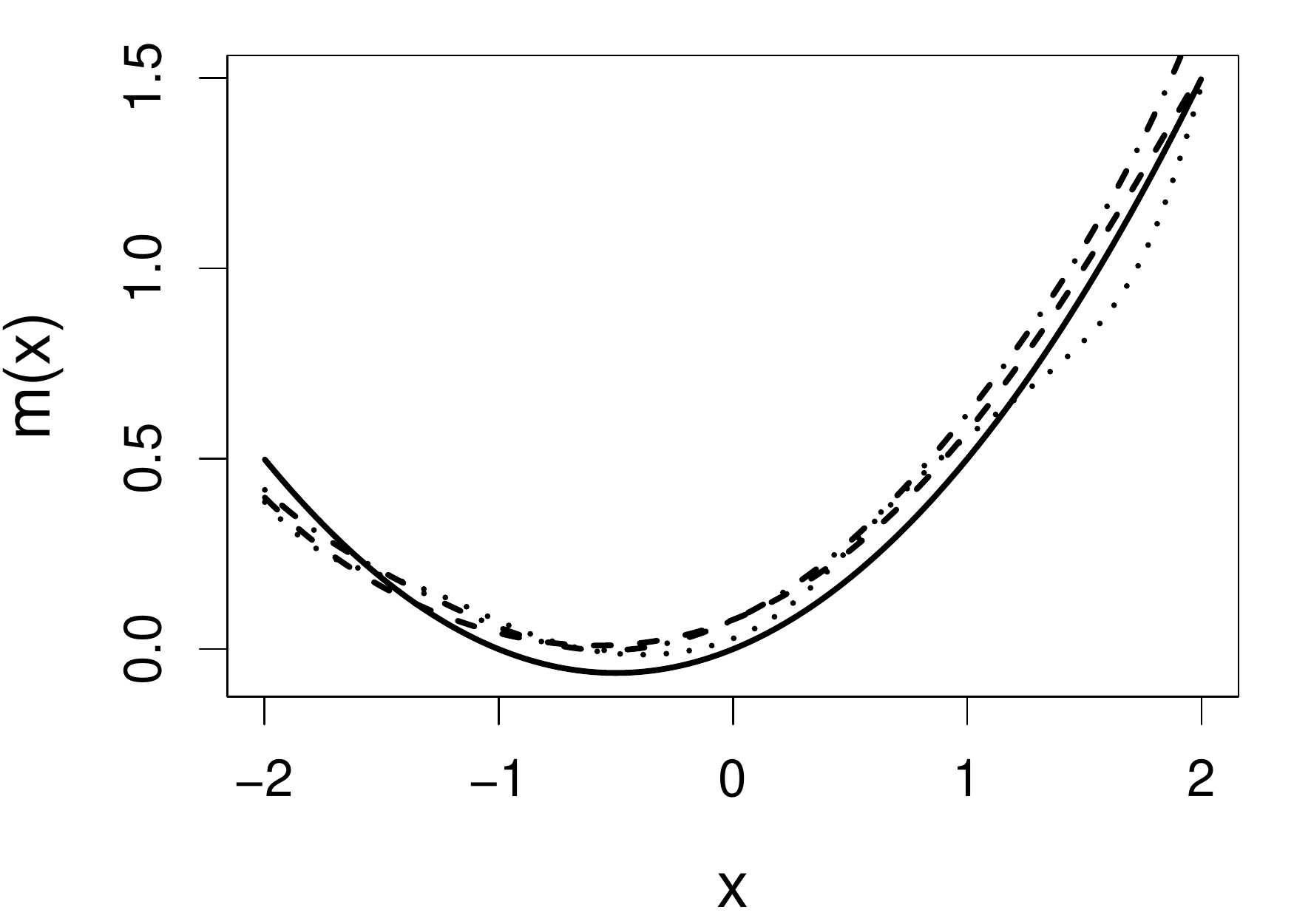} }
	\subfigure[]{ \includegraphics[width=\linewidth]{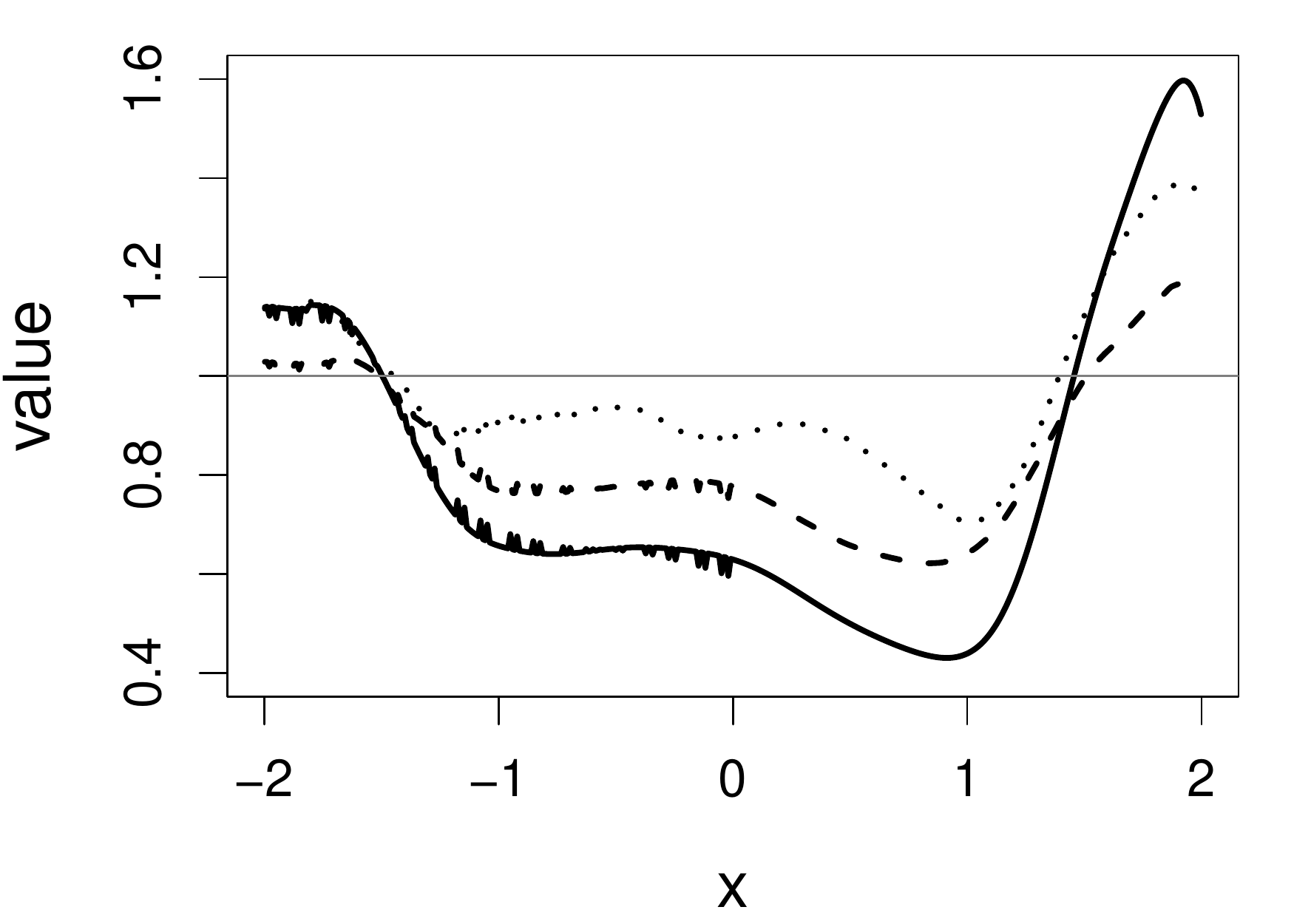} }\\
	\caption{Simulation results under (C2) using the theoretical optimal $h$. Panels (a) \& (d): boxplots of ISEs versus $\lambda$ for $\hat m_{\hbox {\tiny HZ}}(x)$ and $\hat m_{\hbox {\tiny DFC}}(x)$, respectively. Panels (b) \& (e): boxplots of PAE(0) versus $\lambda$ for $\hat m_{\hbox {\tiny HZ}}(0)$ and $\hat m_{\hbox {\tiny DFC}}(0)$, respectively. Panels (c) \& (f): boxplots of PAE($-1$) versus $\lambda$ for $\hat m_{\hbox {\tiny HZ}}(-1)$ and $\hat m_{\hbox {\tiny DFC}}(-1)$, respectively. Panels (g) \& (h): quantile curves when $\lambda=0.85$ for $\hat m_{\hbox {\tiny HZ}}(x)$ and $\hat m_{\hbox {\tiny DFC}}(x)$, respectively, based on ISEs (dashed lines for the first quartile, dotted lines for the second quartile, and dot-dashed lines for the third quartile, solid lines for the truth). Panel (i): PmAER (dashed line), PsdAER (dotted line), and PMSER (solid line) versus $x$ when $\lambda=0.85$; the horizontal reference line highlights the value 1.}
	\label{Sim2GauGau500:box}
\end{figure}

\begin{figure}
	\centering
	\setlength{\linewidth}{4.5cm}
	\subfigure[]{ \includegraphics[width=\linewidth]{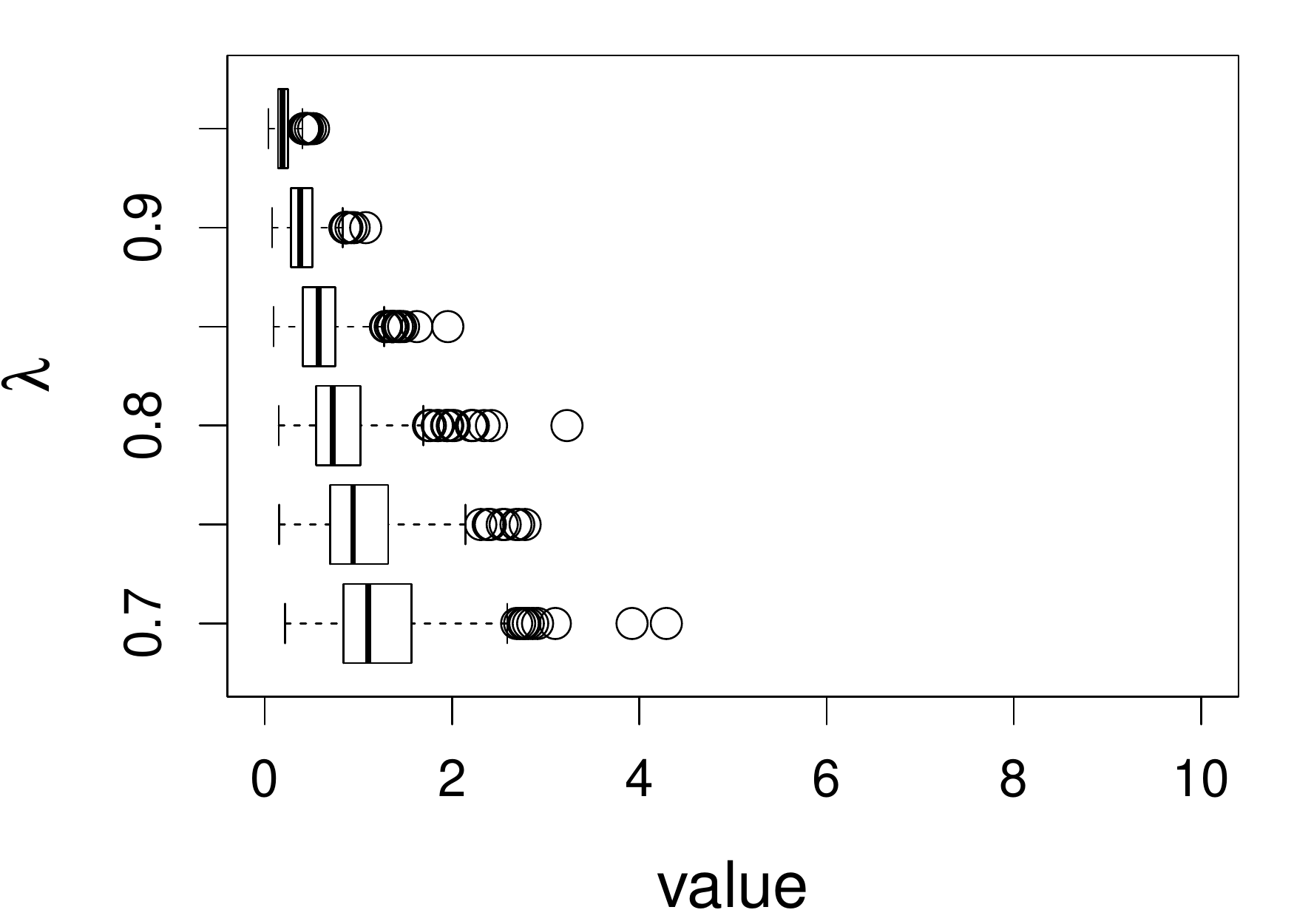} }
	\subfigure[]{ \includegraphics[width=\linewidth]{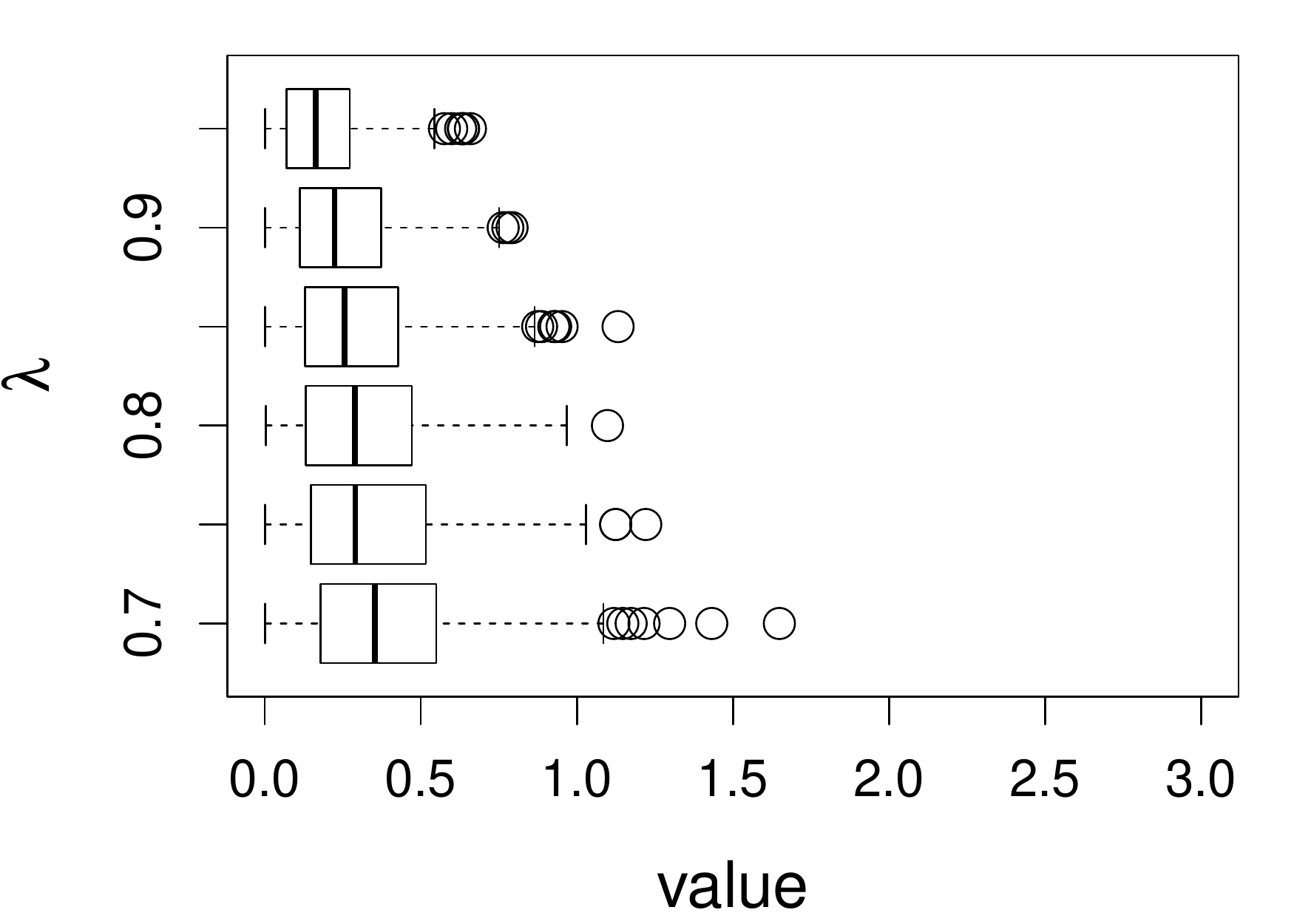} }
	\subfigure[]{ \includegraphics[width=\linewidth]{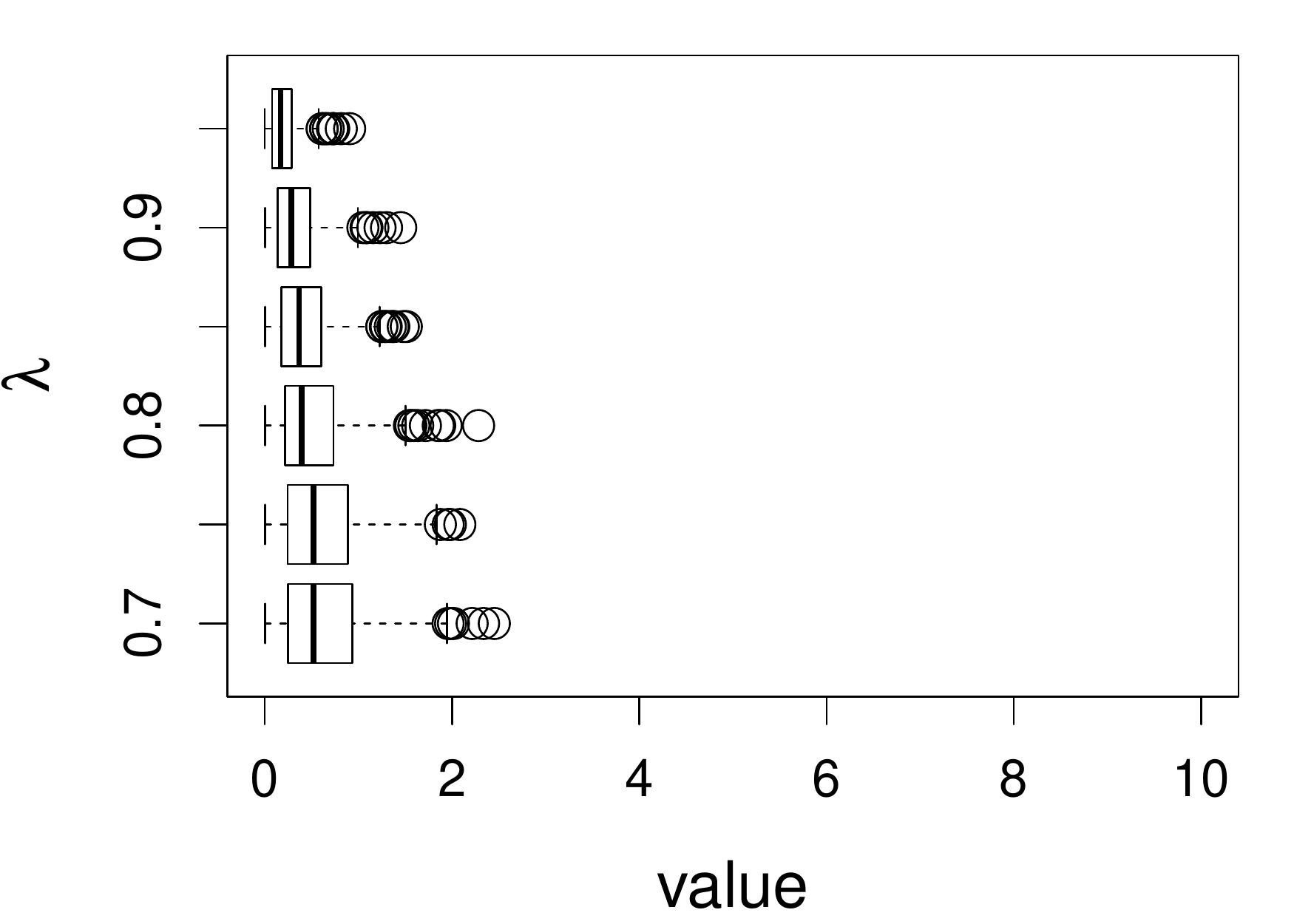} }\\
	\subfigure[]{ \includegraphics[width=\linewidth]{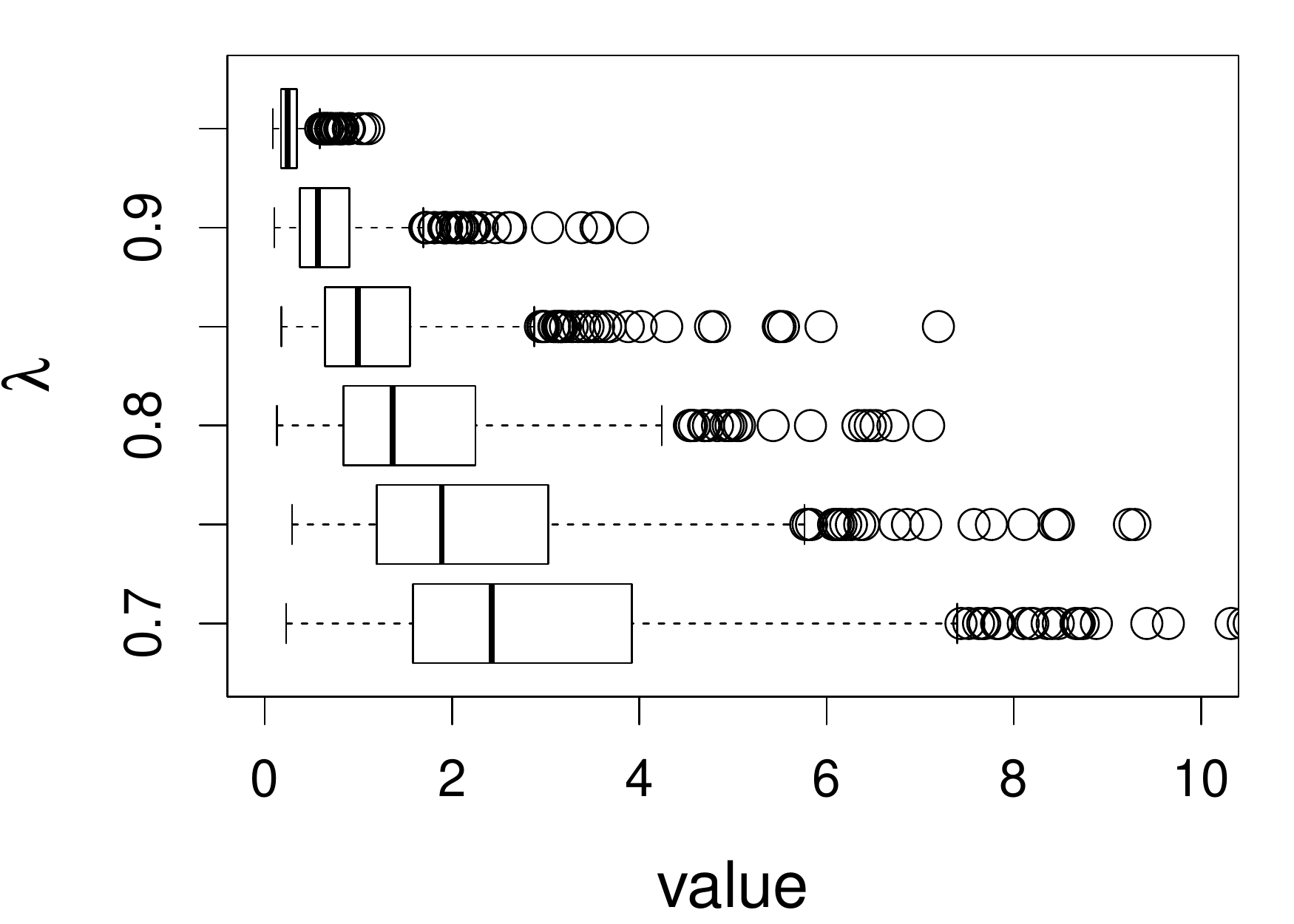} }
	\subfigure[]{ \includegraphics[width=\linewidth]{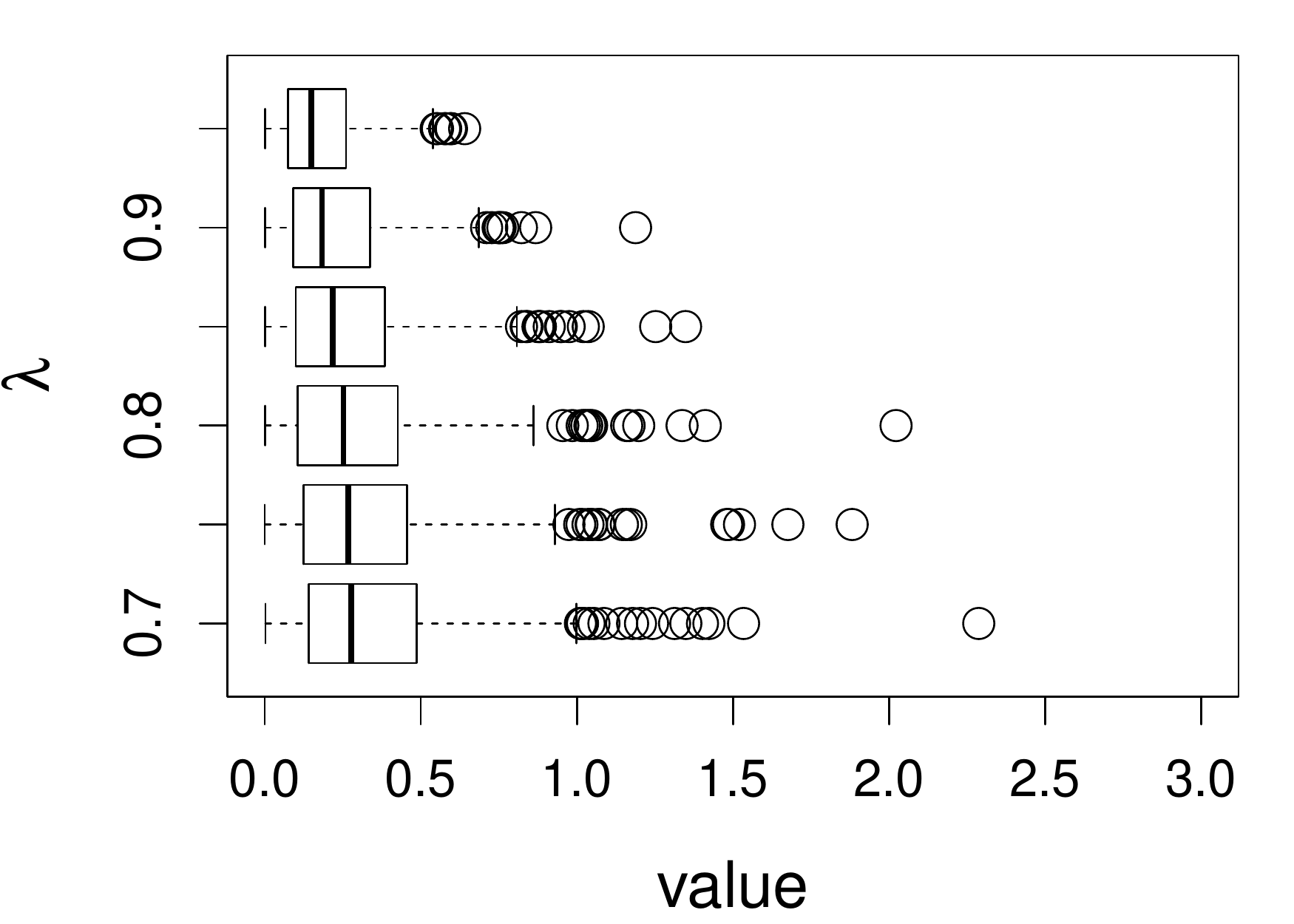} }
	\subfigure[]{ \includegraphics[width=\linewidth]{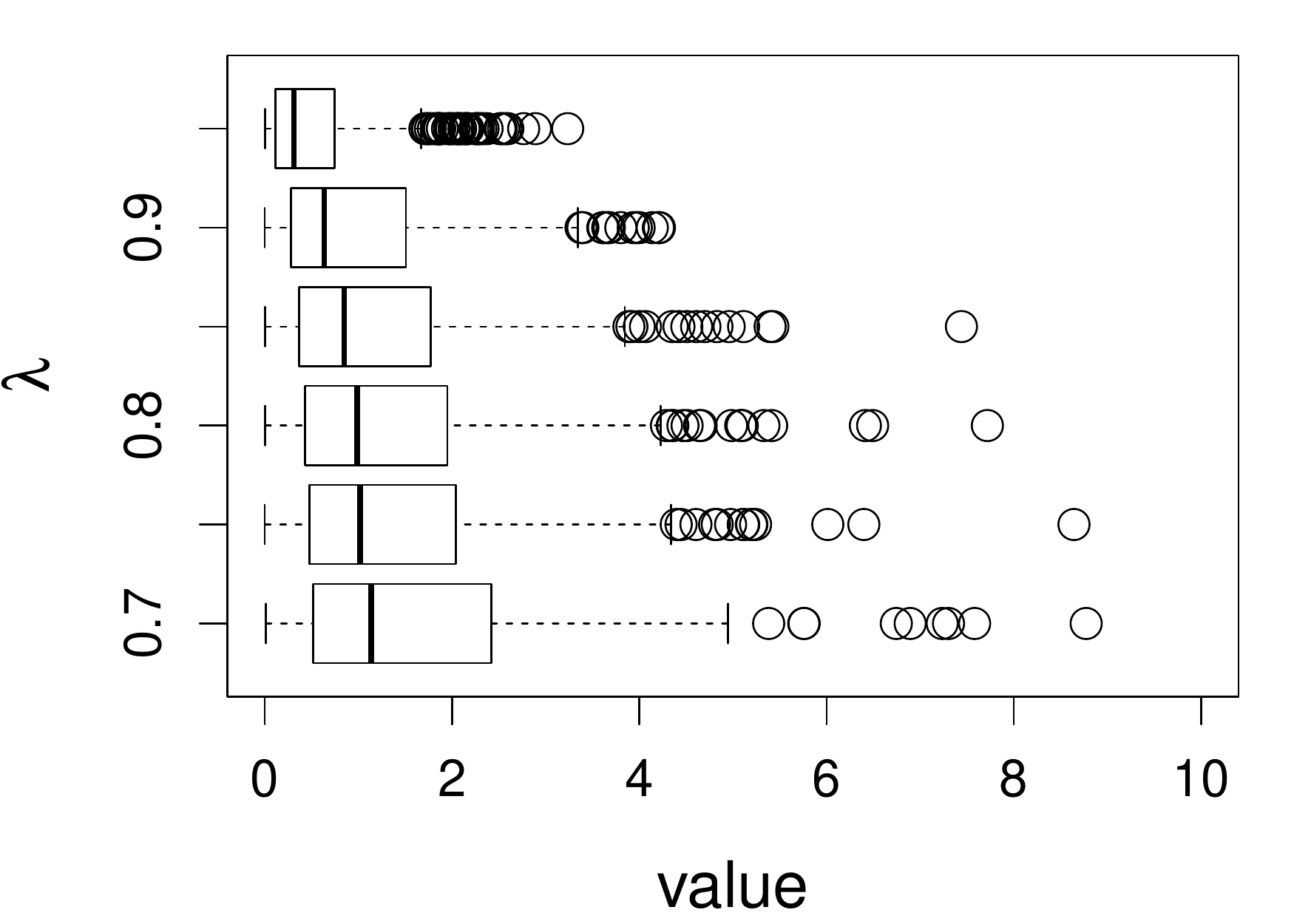} }\\
	\subfigure[]{ \includegraphics[width=\linewidth]{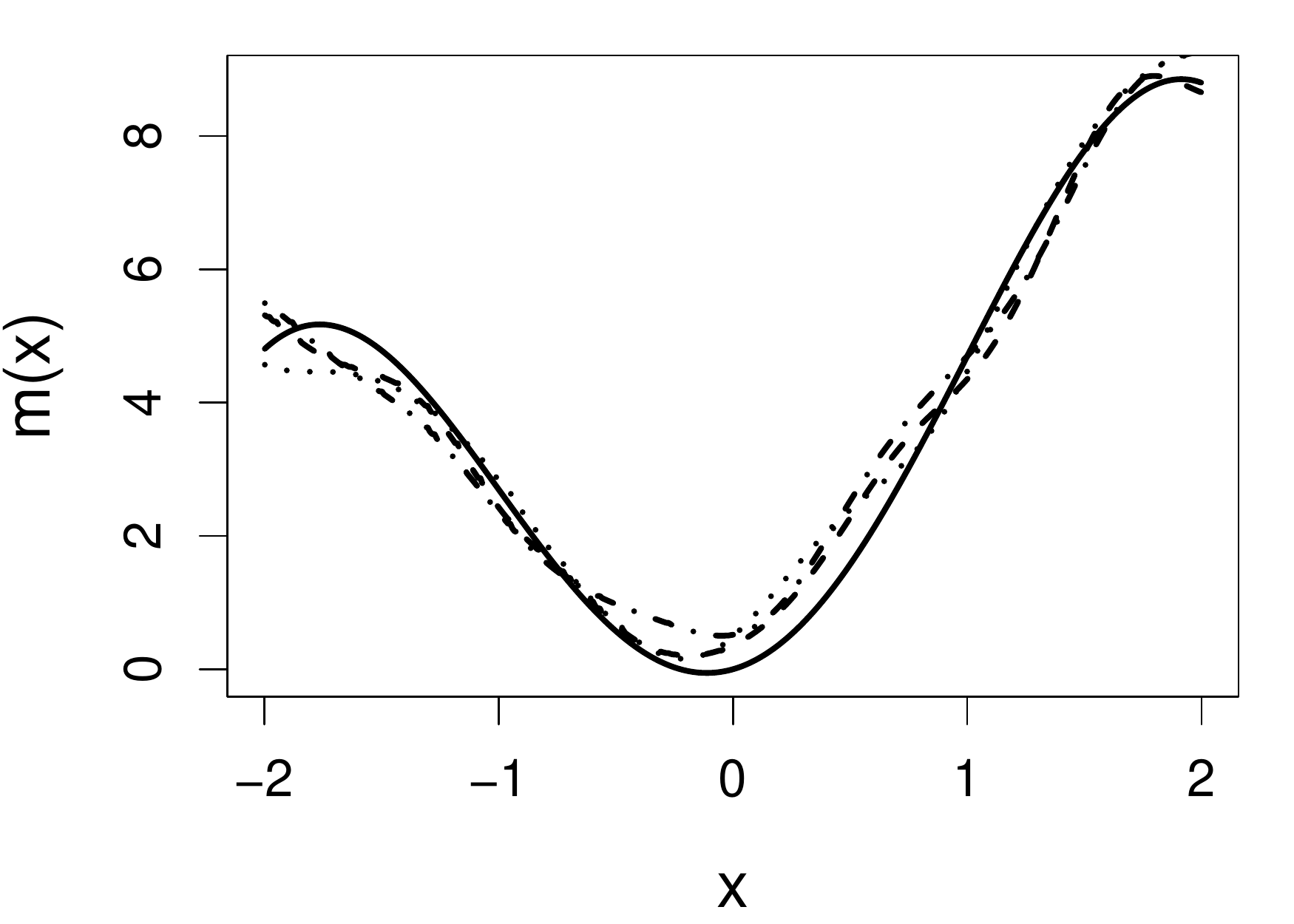} }
	\subfigure[]{ \includegraphics[width=\linewidth]{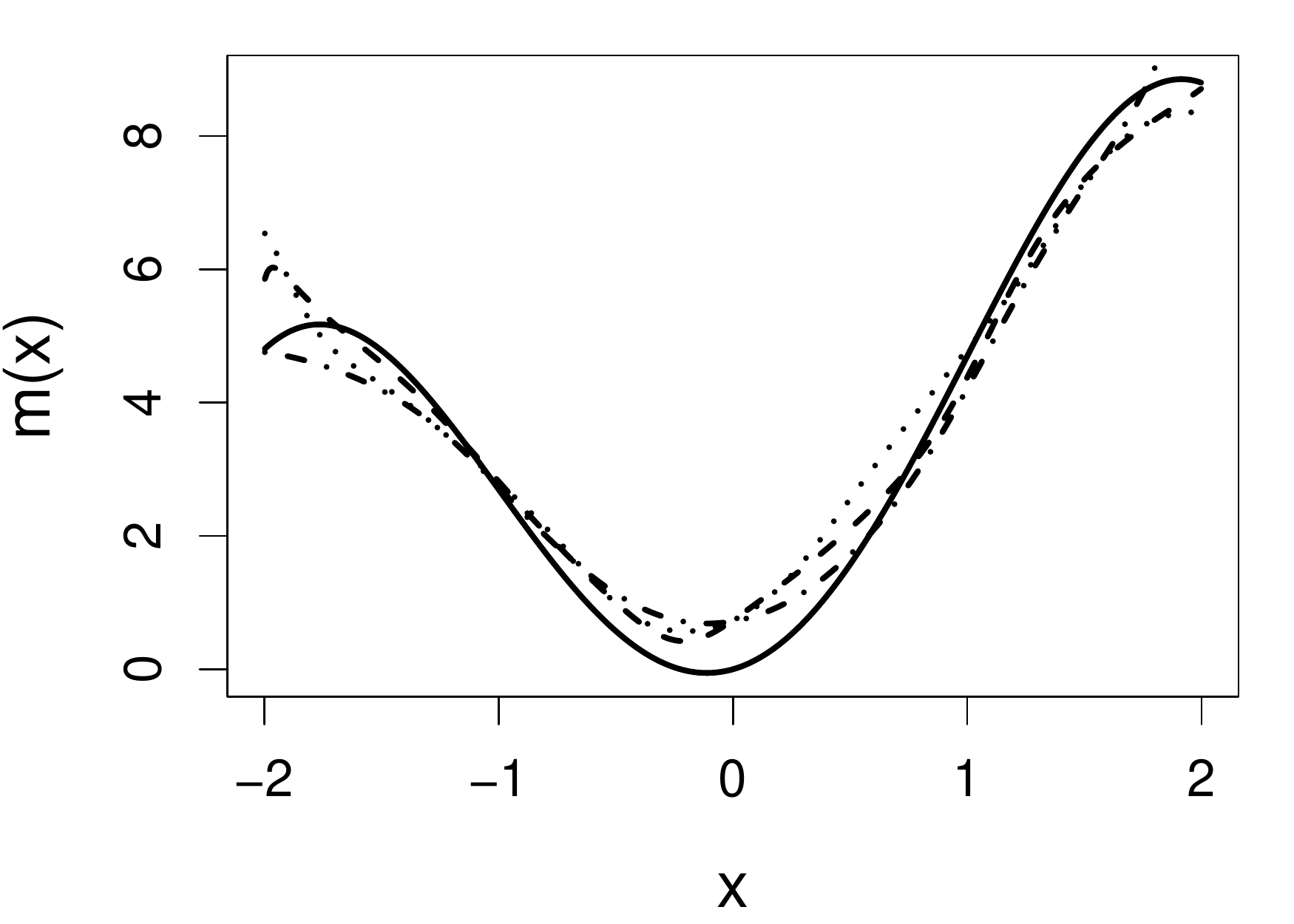} }
	\subfigure[]{ \includegraphics[width=\linewidth]{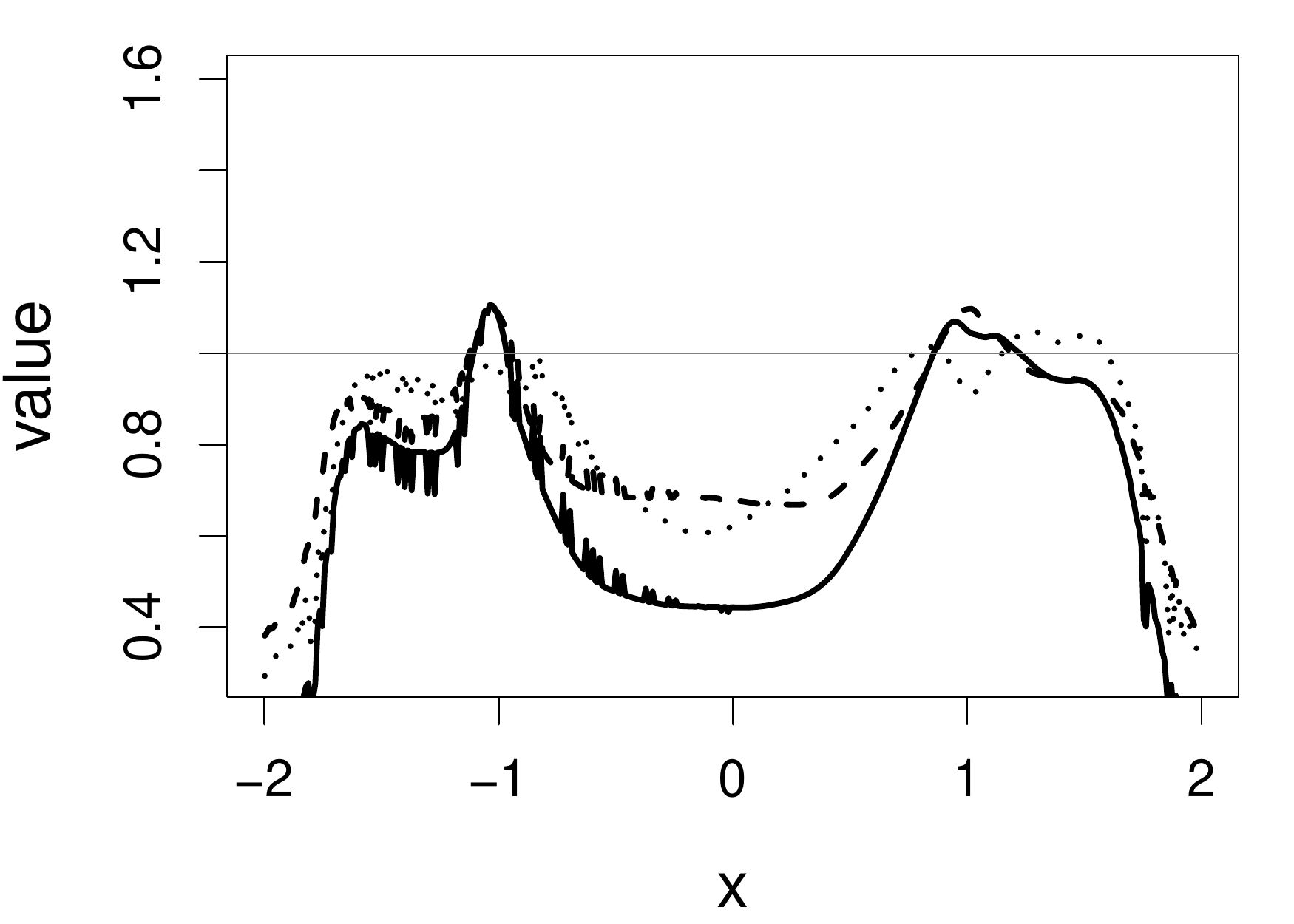} }\\
	\caption{Simulation results under (C3) using the theoretical optimal $h$. Panels (a) \& (d): boxplots of ISEs versus $\lambda$ for $\hat{m}_{\hbox {\tiny HZ}}(x)$ and $\hat{m}_{\hbox {\tiny DFC}}(x)$, respectively. Panels (b) \& (e): boxplots of PAE(1) versus $\lambda$ for  $\hat{m}_{\hbox {\tiny HZ}}(1)$ and $\hat{m}_{\hbox {\tiny DFC}}(1)$, respectively. Panels (c) \& (f): boxplots of PAE(2) versus $\lambda$ for  $\hat{m}_{\hbox {\tiny HZ}}(2)$ and $\hat{m}_{\hbox {\tiny DFC}}(2)$, respectively. Panels (g) \& (h): quantile curves when $\lambda=0.8$ for $\hat{m}_{\hbox {\tiny HZ}}(x)$ and $\hat{m}_{\hbox {\tiny DFC}}(x)$, respectively, based on ISEs (dashed lines for the first quartile, dotted lines for the second quartile, dot-dashed lines for the third quartile, and solid lines for the truth). Panel (i): PmAER (dashed line), PsdAER (dotted line), and PMSER (solid line) versus $x$ when $\lambda=0.8$; the horizontal reference line highlights the value 1.}
	\label{Sim4LapLap:box}
\end{figure}

\begin{figure}
	\centering
	\setlength{\linewidth}{4.5cm}
	\subfigure[]{ \includegraphics[width=\linewidth]{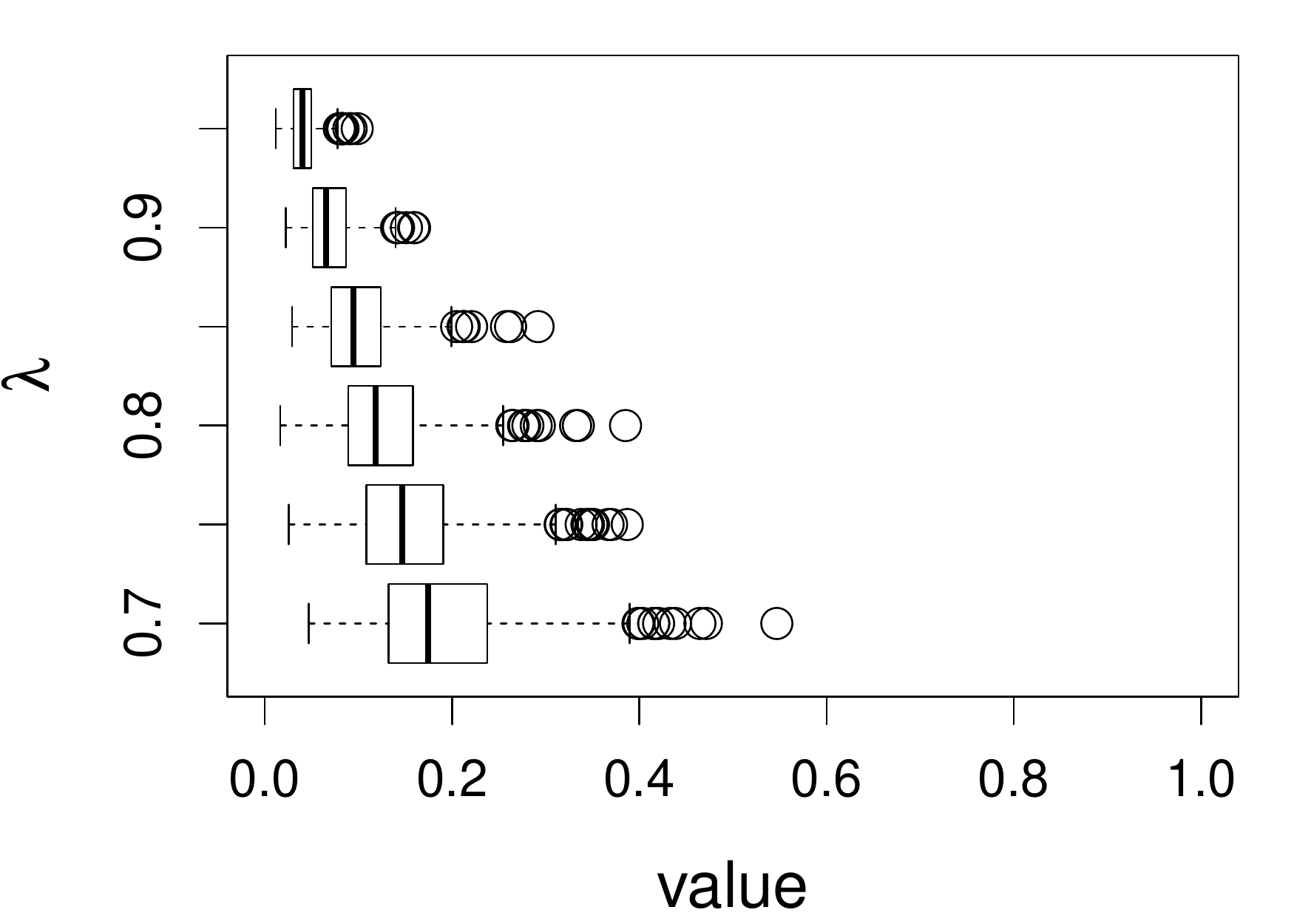} }
	\subfigure[]{ \includegraphics[width=\linewidth]{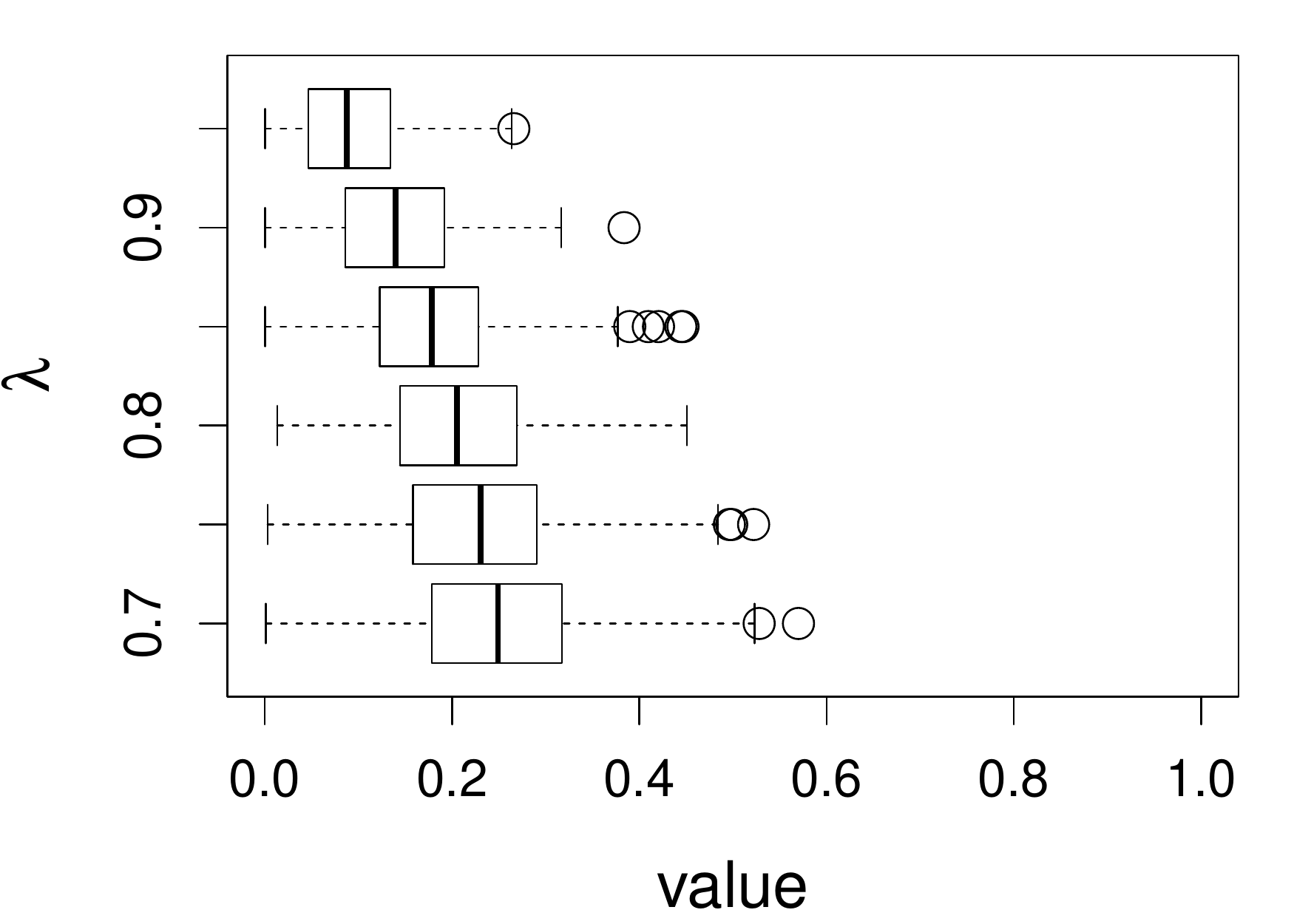} }
	\subfigure[]{ \includegraphics[width=\linewidth]{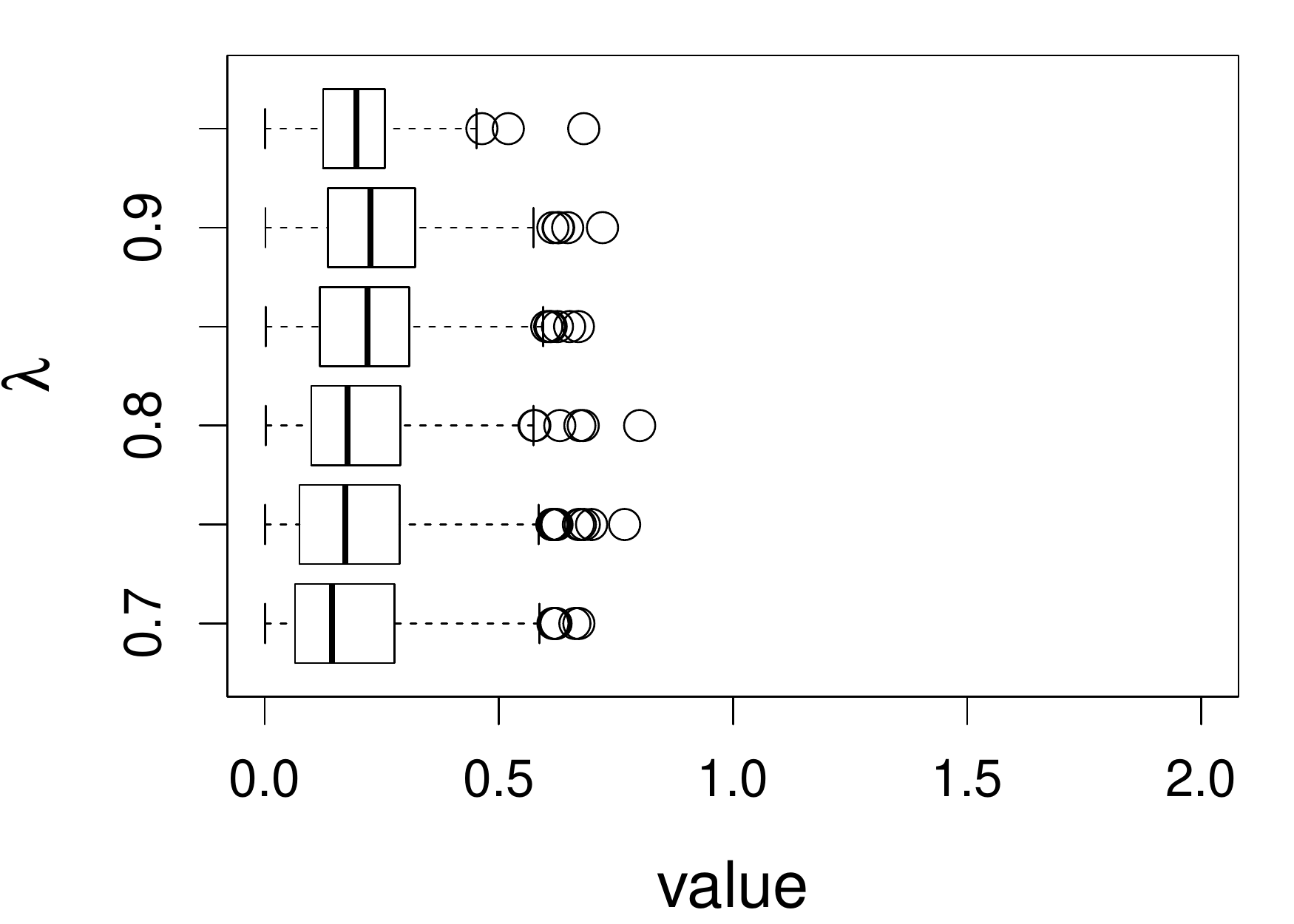} }\\
	\subfigure[]{ \includegraphics[width=\linewidth]{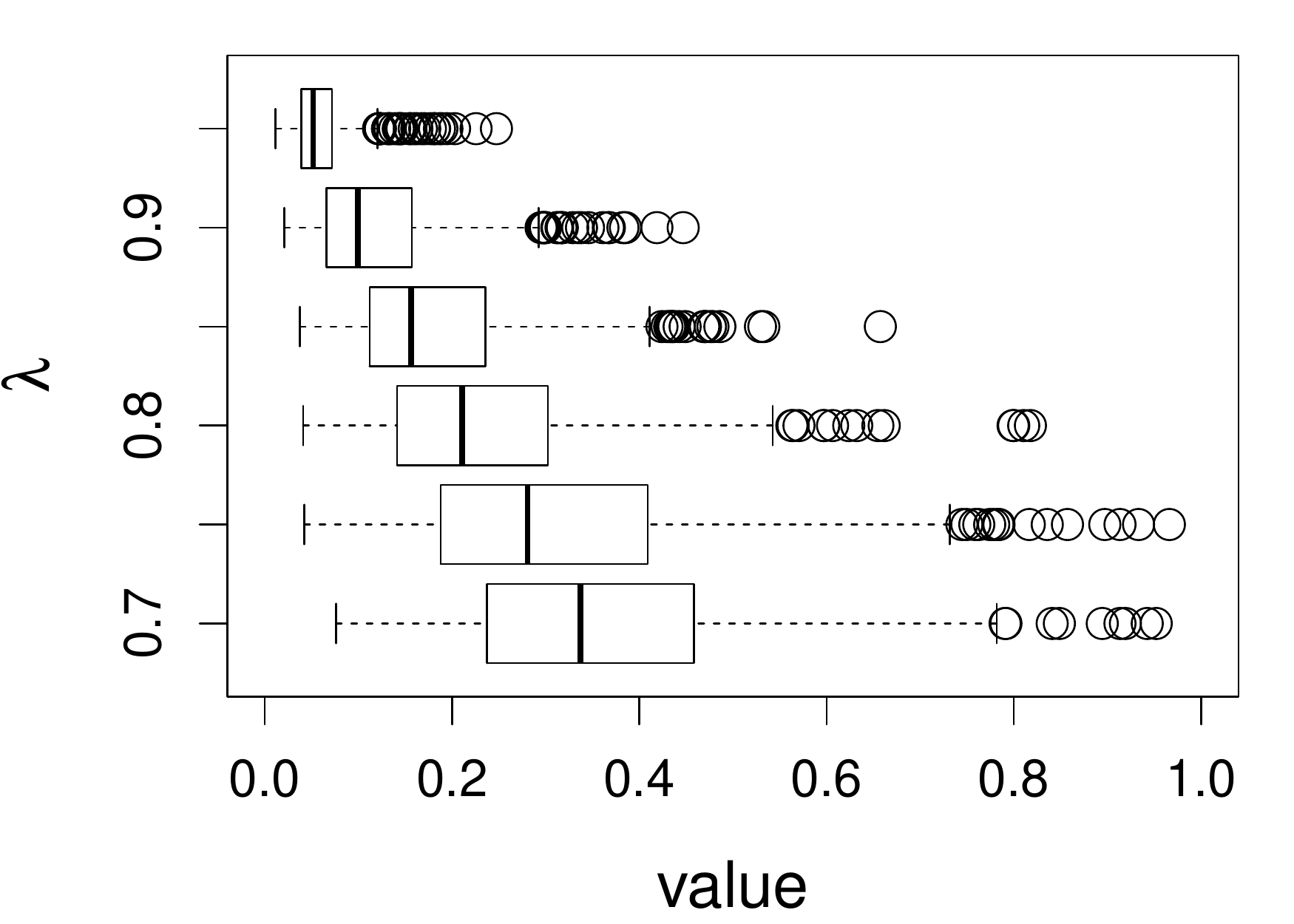} }
	\subfigure[]{ \includegraphics[width=\linewidth]{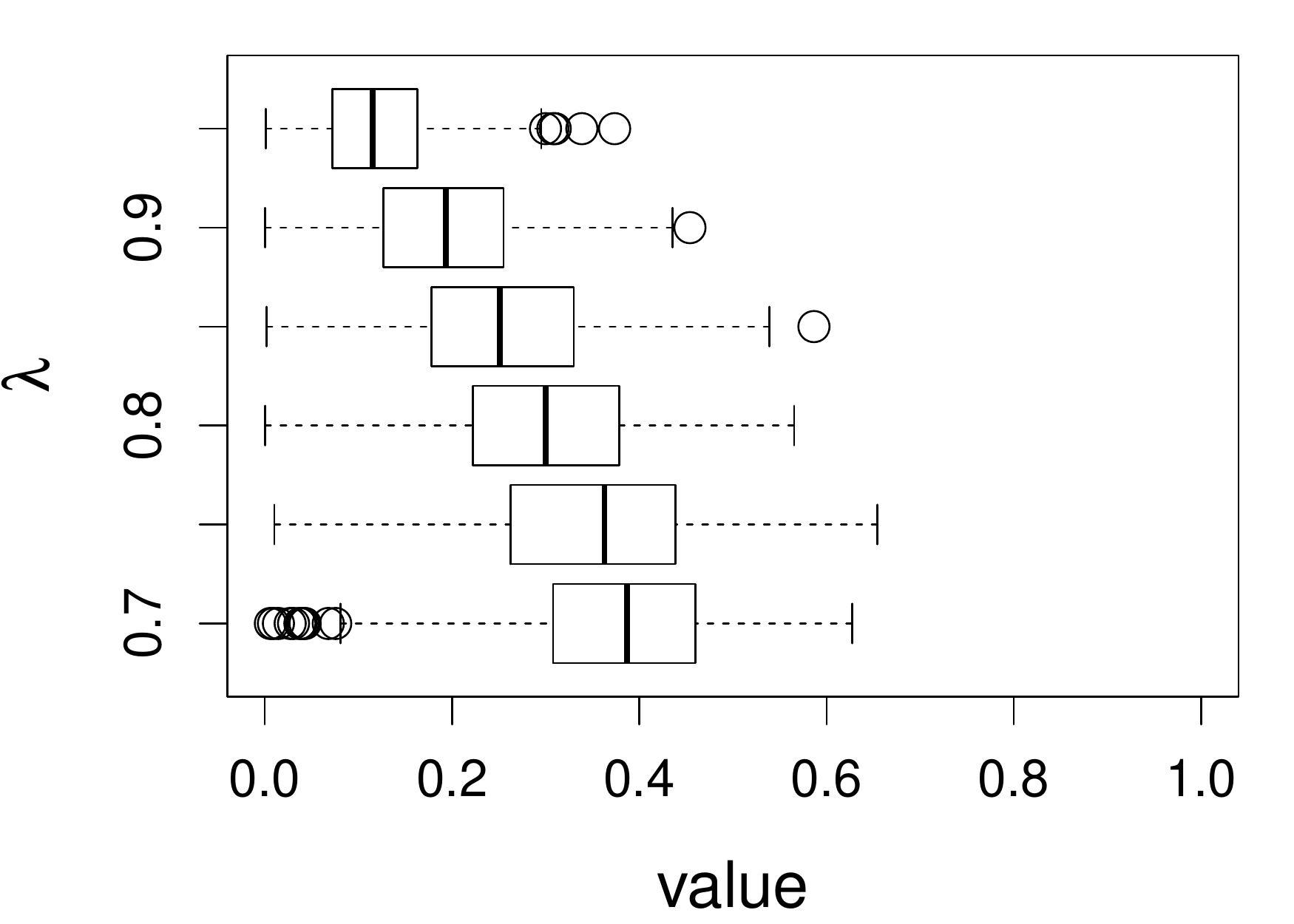} }
	\subfigure[]{ \includegraphics[width=\linewidth]{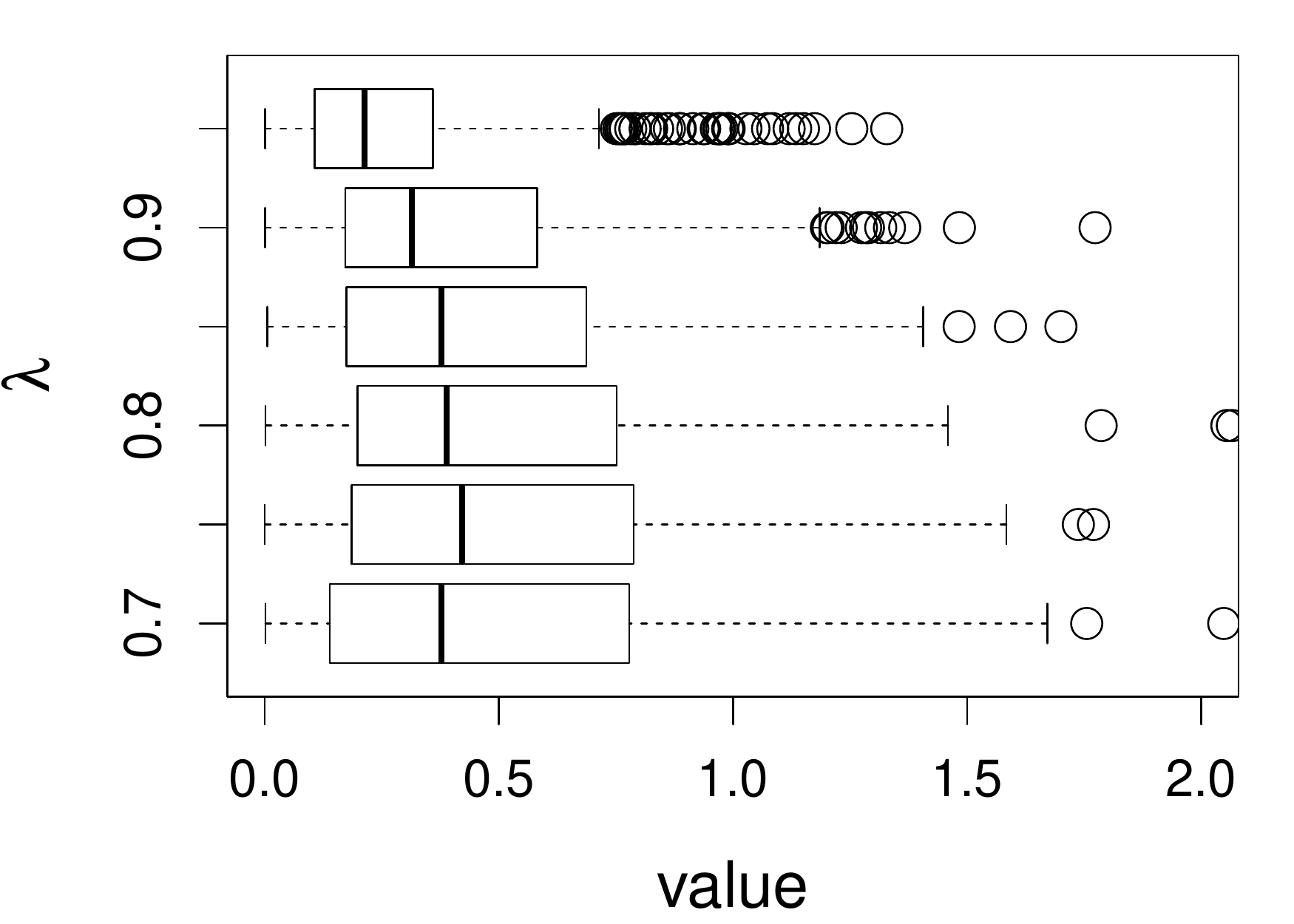} }\\
	\subfigure[]{ \includegraphics[width=\linewidth]{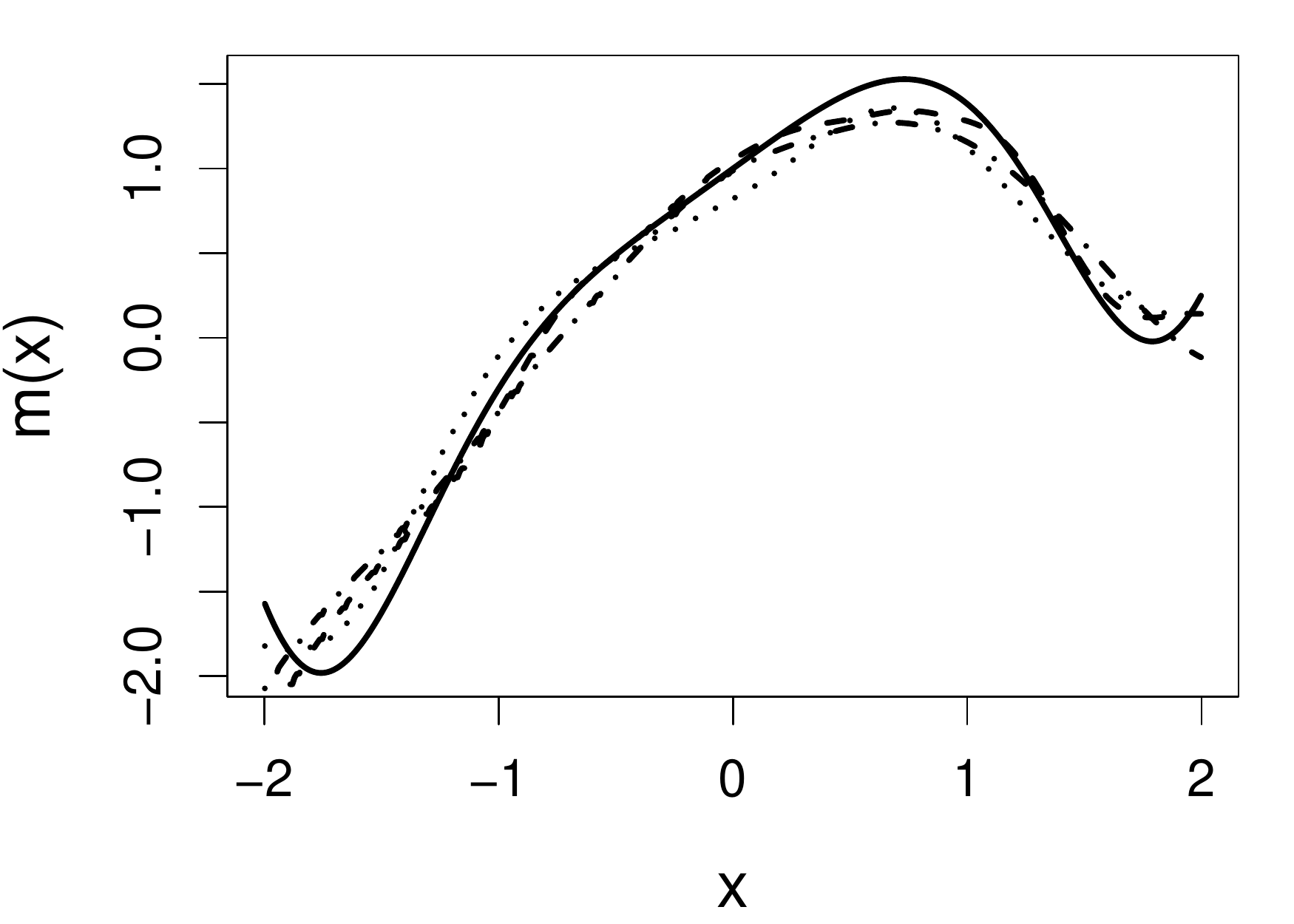} }
	\subfigure[]{ \includegraphics[width=\linewidth]{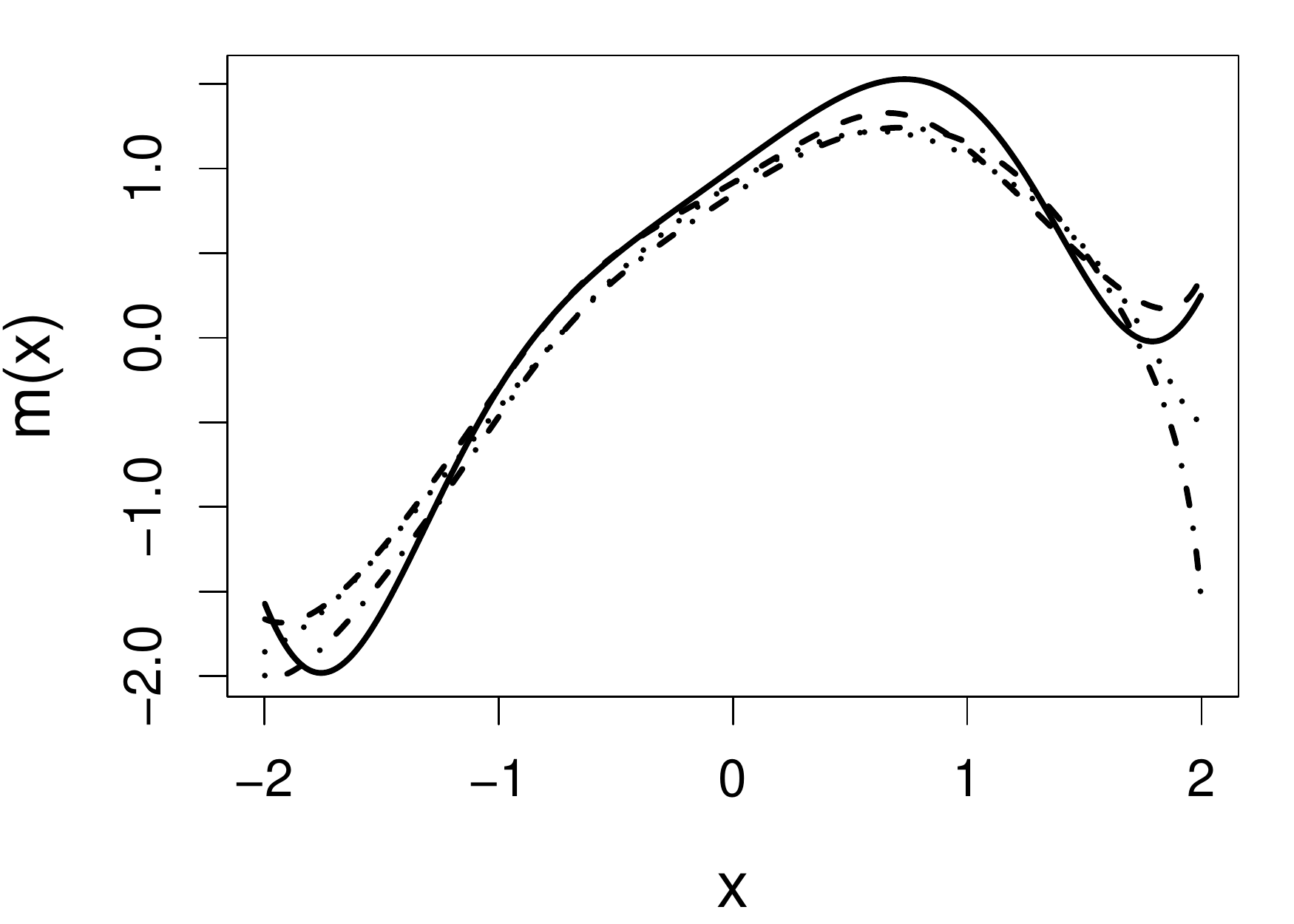} }
	\subfigure[]{ \includegraphics[width=\linewidth]{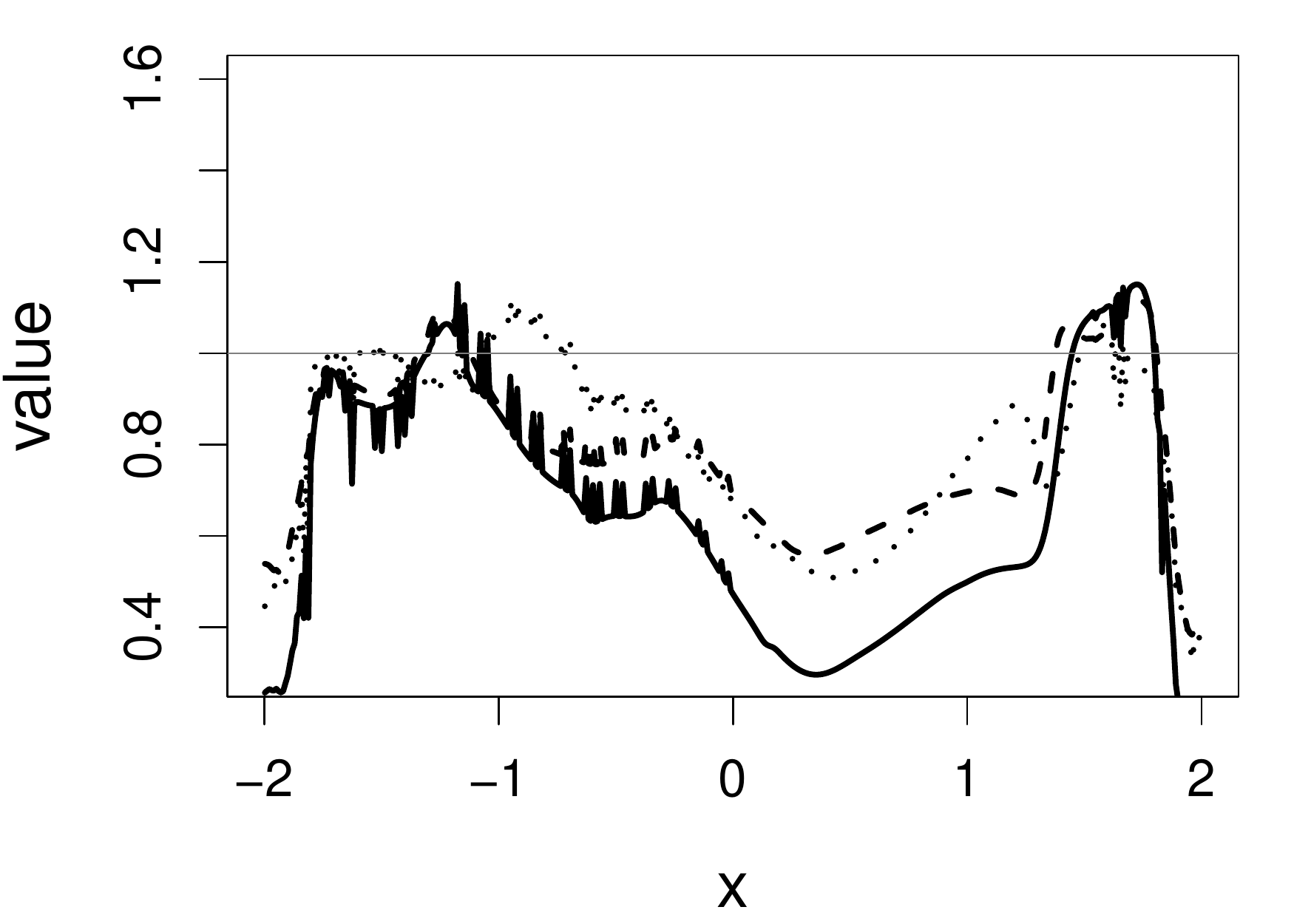} }\\
	\caption{Simulation results under (C4) using the theoretical optimal $h$. Panels (a) \& (d): boxplots of ISEs versus $\lambda$ for $\hat{m}_{\hbox {\tiny HZ}}(x)$ and $\hat{m}_{\hbox {\tiny DFC}}(x)$, respectively. Panels (b) \& (e): boxplots of PAE(1) versus $\lambda$ for  $\hat{m}_{\hbox {\tiny HZ}}(1)$ and $\hat{m}_{\hbox  {\tiny DFC}}(1)$, respectively. Panels (c) \& (f): boxplots of PAE(2) versus $\lambda$ for  $\hat{m}_{\hbox {\tiny HZ}}(2)$ and $\hat{m}_{\hbox {\tiny DFC}}(2)$, respectively. Panels (g) \& (h): quantile curves when $\lambda=0.8$ for $\hat{m}_{\hbox {\tiny HZ}}(x)$ and $\hat{m}_{\hbox {\tiny DFC}}(x)$, respectively, based on ISEs (dashed lines for the first quartile, dotted lines for the second quartile, dot-dashed lines for the third quartile, and solid lines for the truth). Panel (i): PmAER(dashed line), PsdAER (dotted line), and PMSER (solid line) versus $x$ when $\lambda=0.8$; the horizontal reference line highlights the value 1.}
	\label{Sim3LapLap:box}
\end{figure}

\begin{figure}
	\centering
	\setlength{\linewidth}{4.5cm}
	\subfigure[]{ \includegraphics[width=\linewidth]{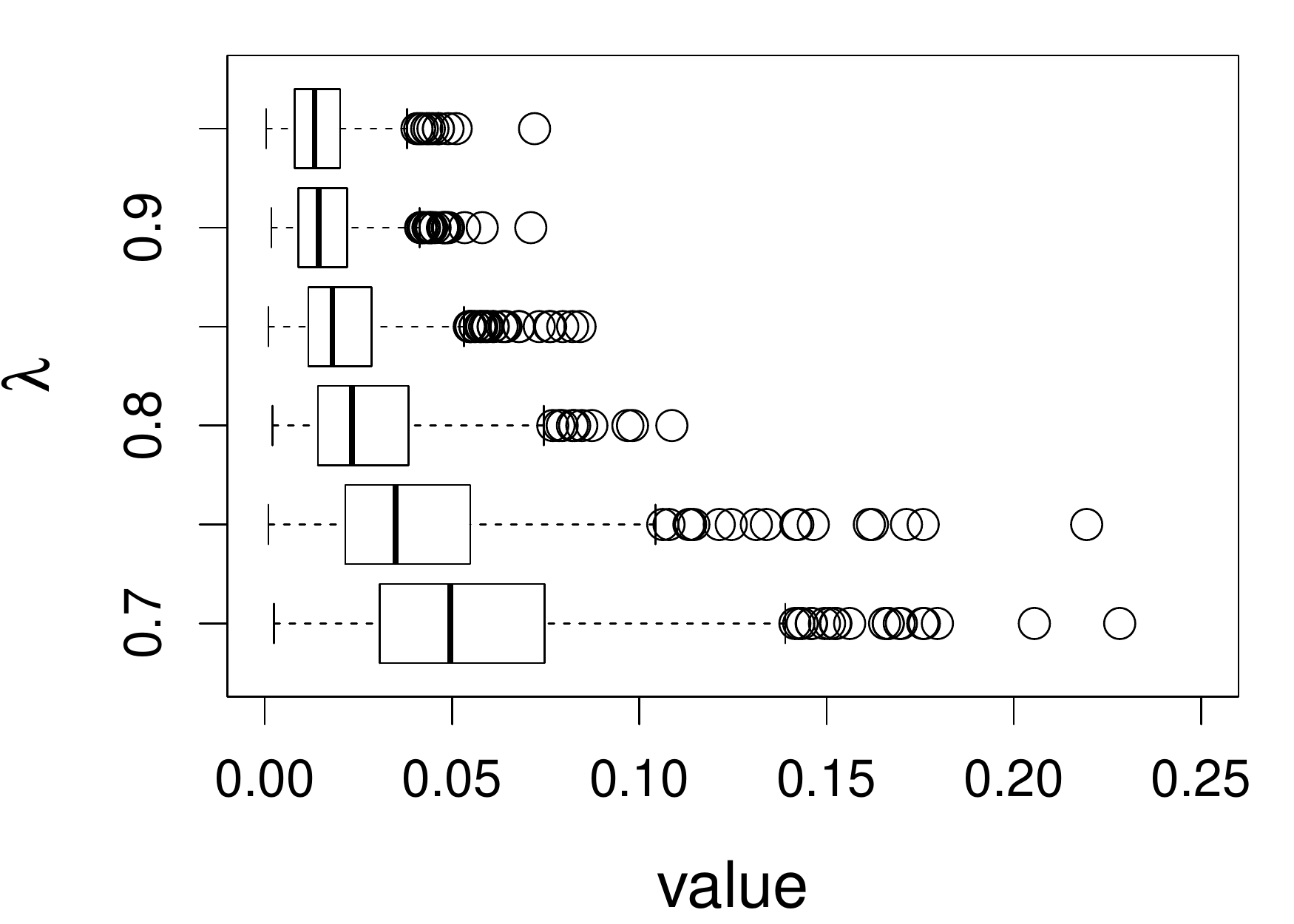} }
	\subfigure[]{ \includegraphics[width=\linewidth]{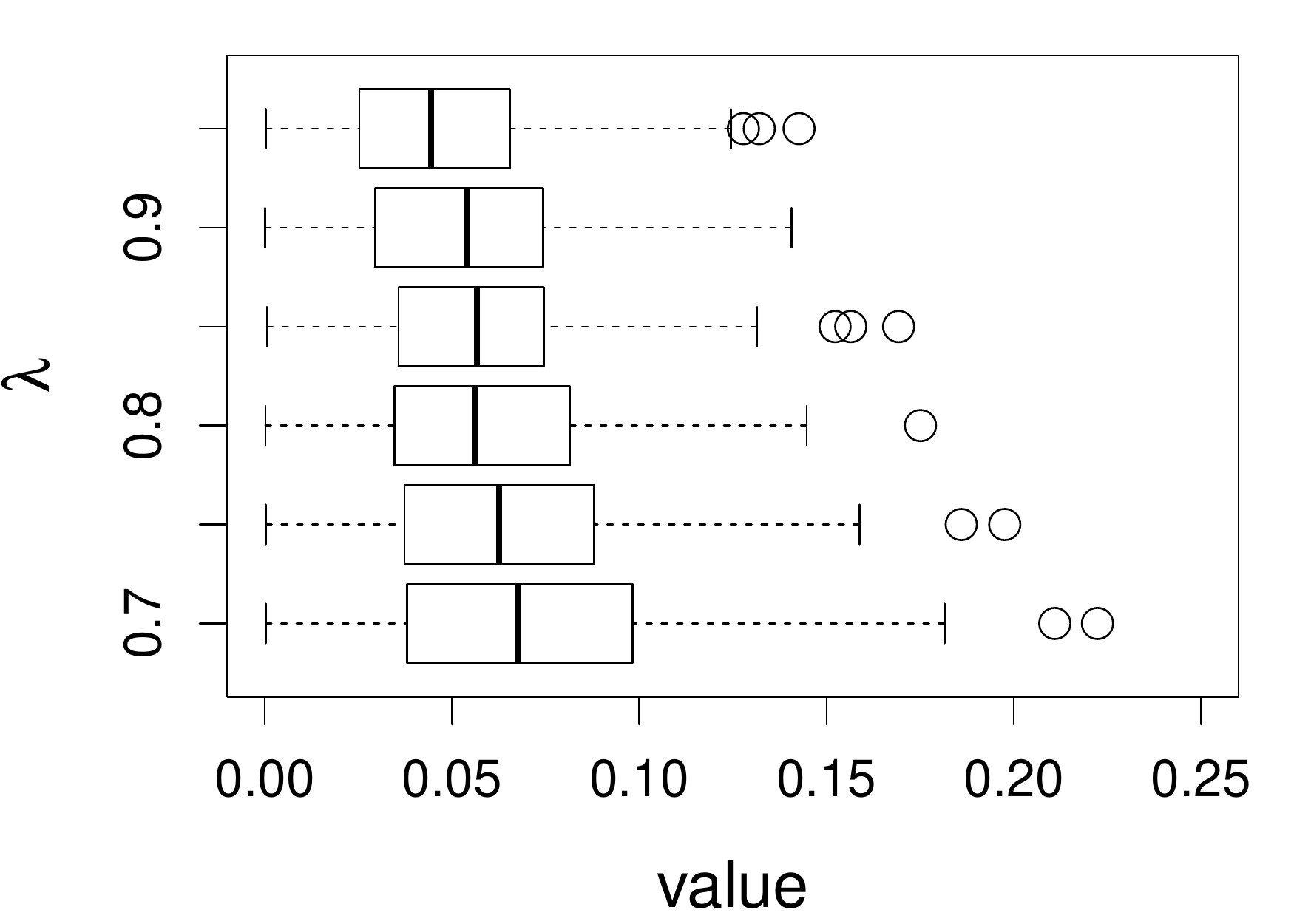} }
	\subfigure[]{ \includegraphics[width=\linewidth]{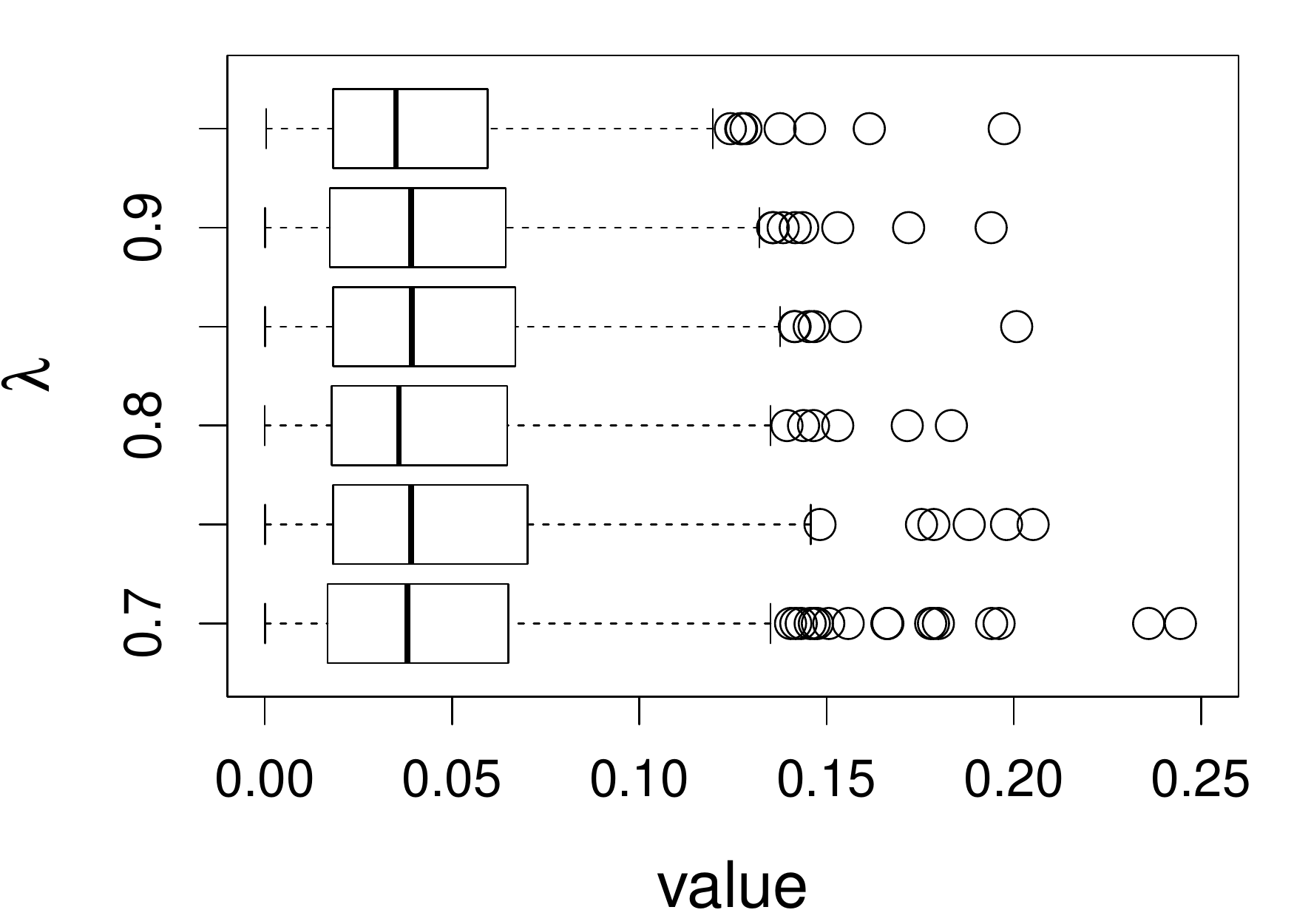} }\\
	\subfigure[]{ \includegraphics[width=\linewidth]{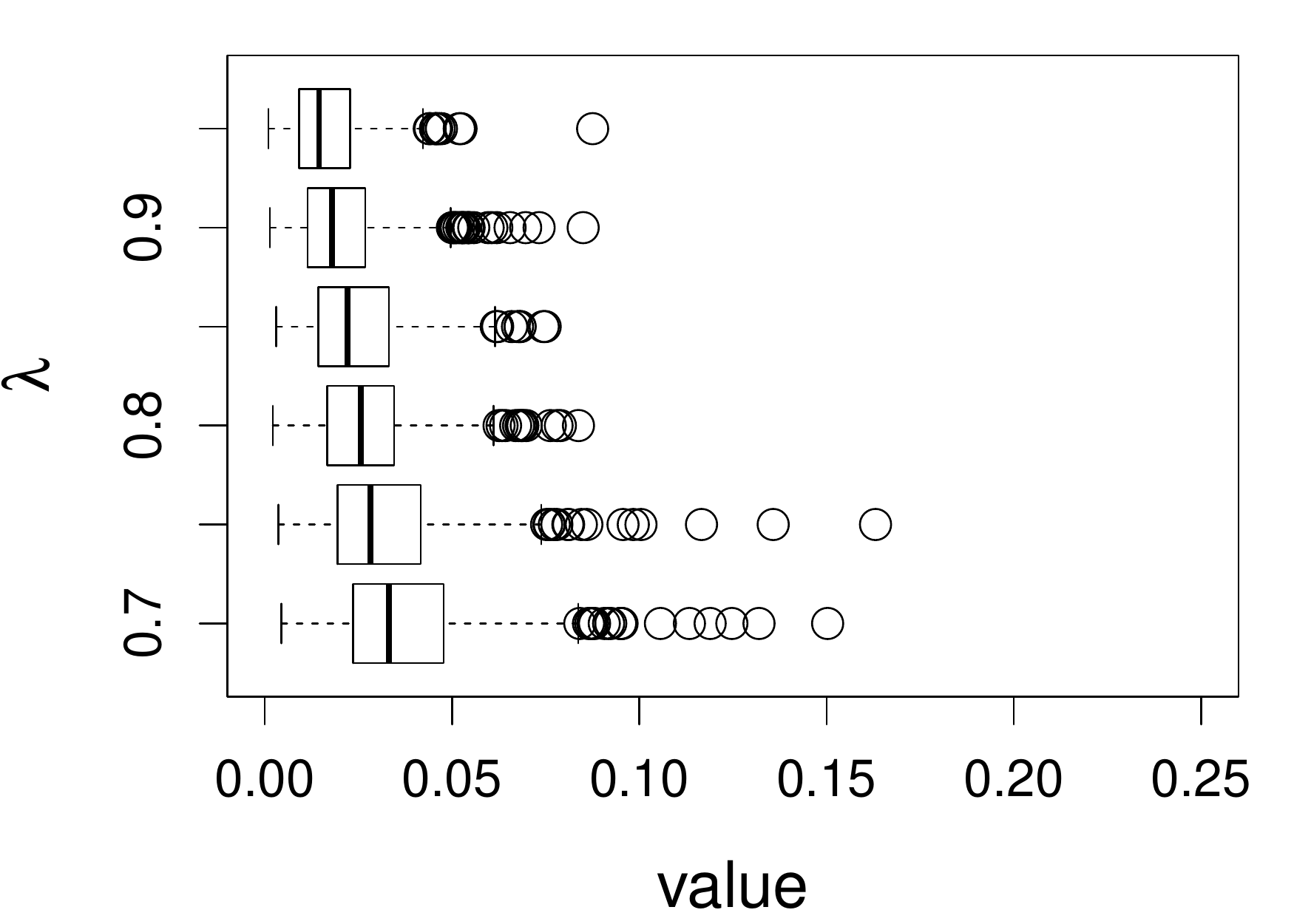} }
	\subfigure[]{ \includegraphics[width=\linewidth]{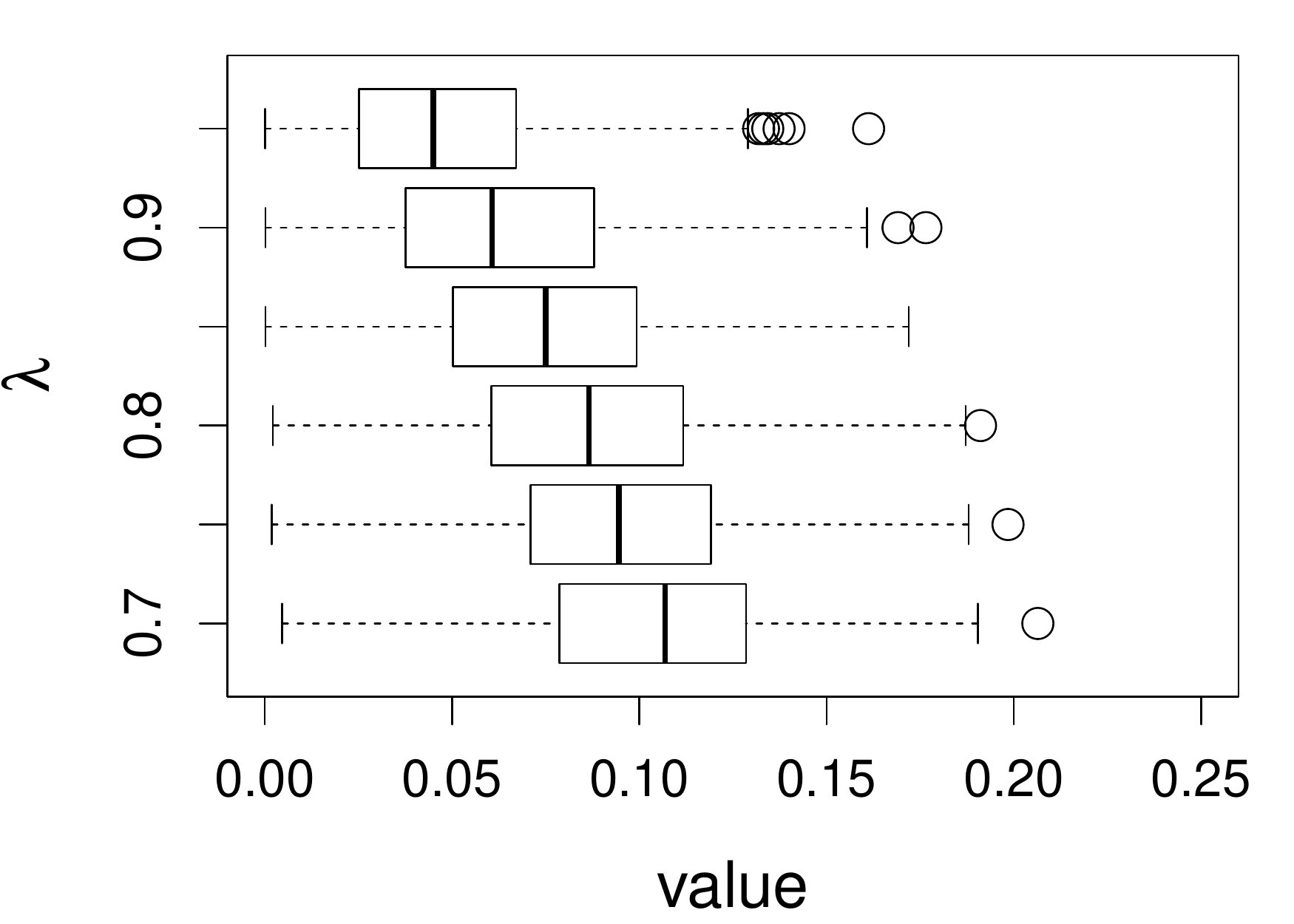} }
	\subfigure[]{ \includegraphics[width=\linewidth]{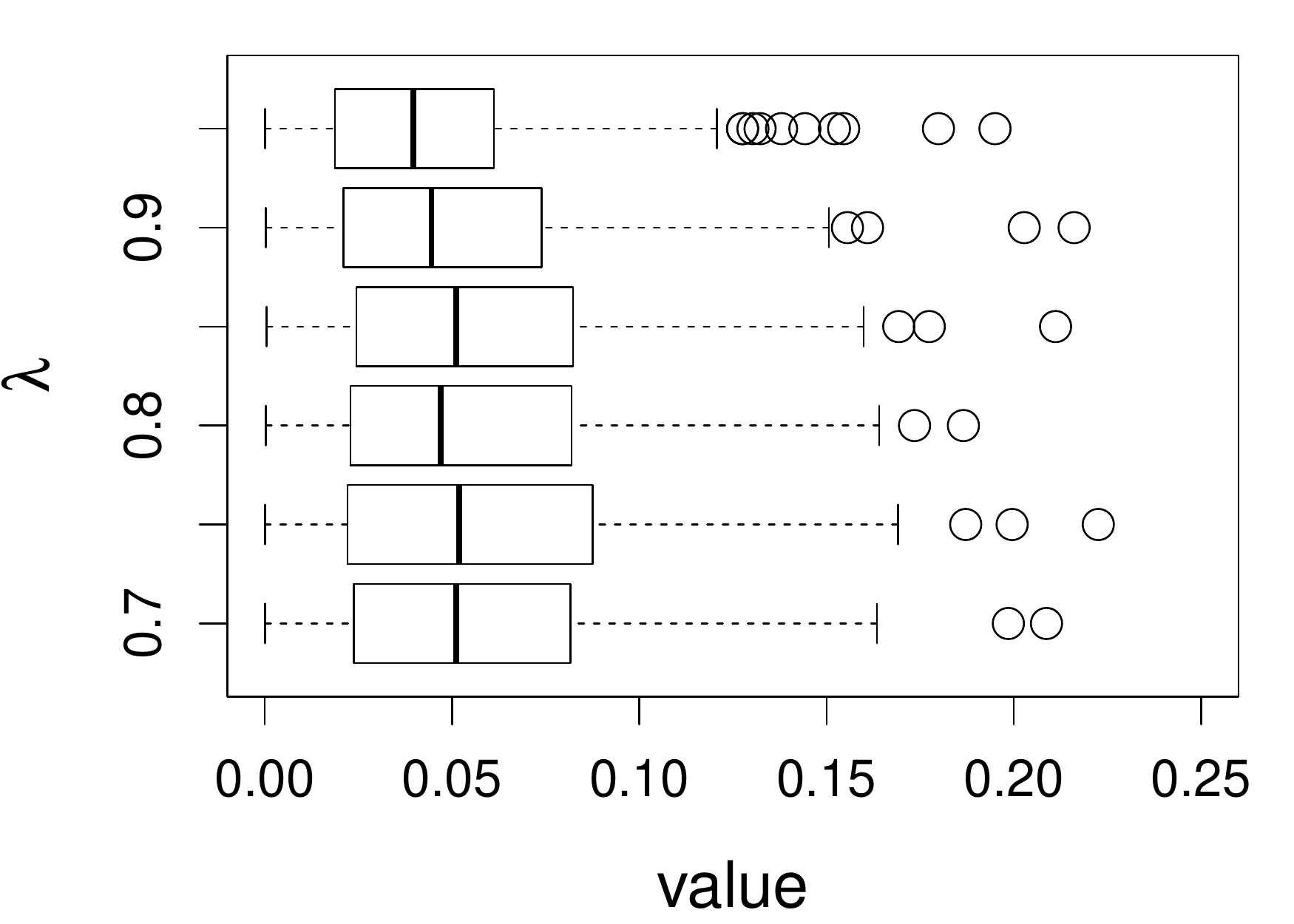} }\\
	\subfigure[]{ \includegraphics[width=\linewidth]{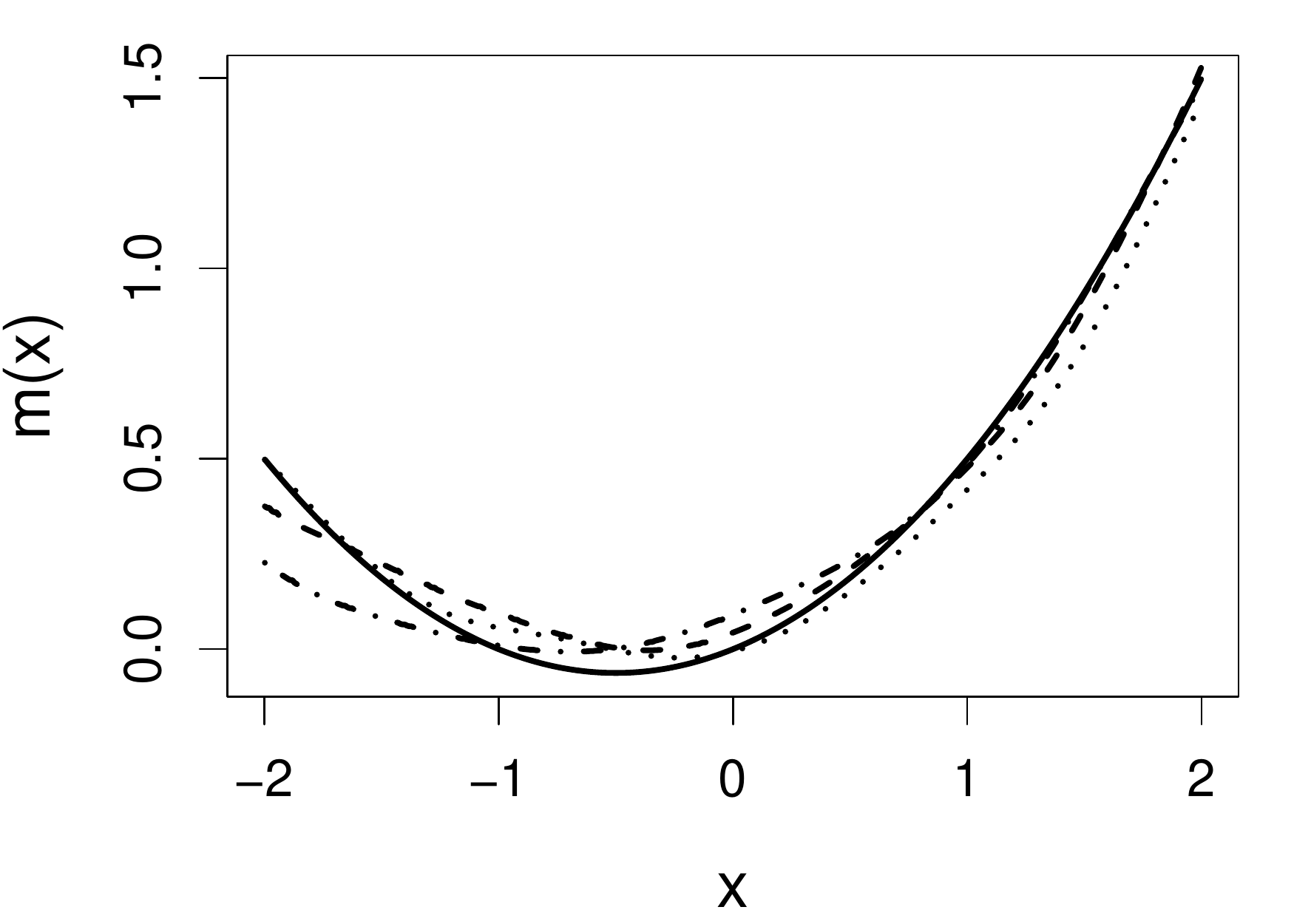} }
	\subfigure[]{ \includegraphics[width=\linewidth]{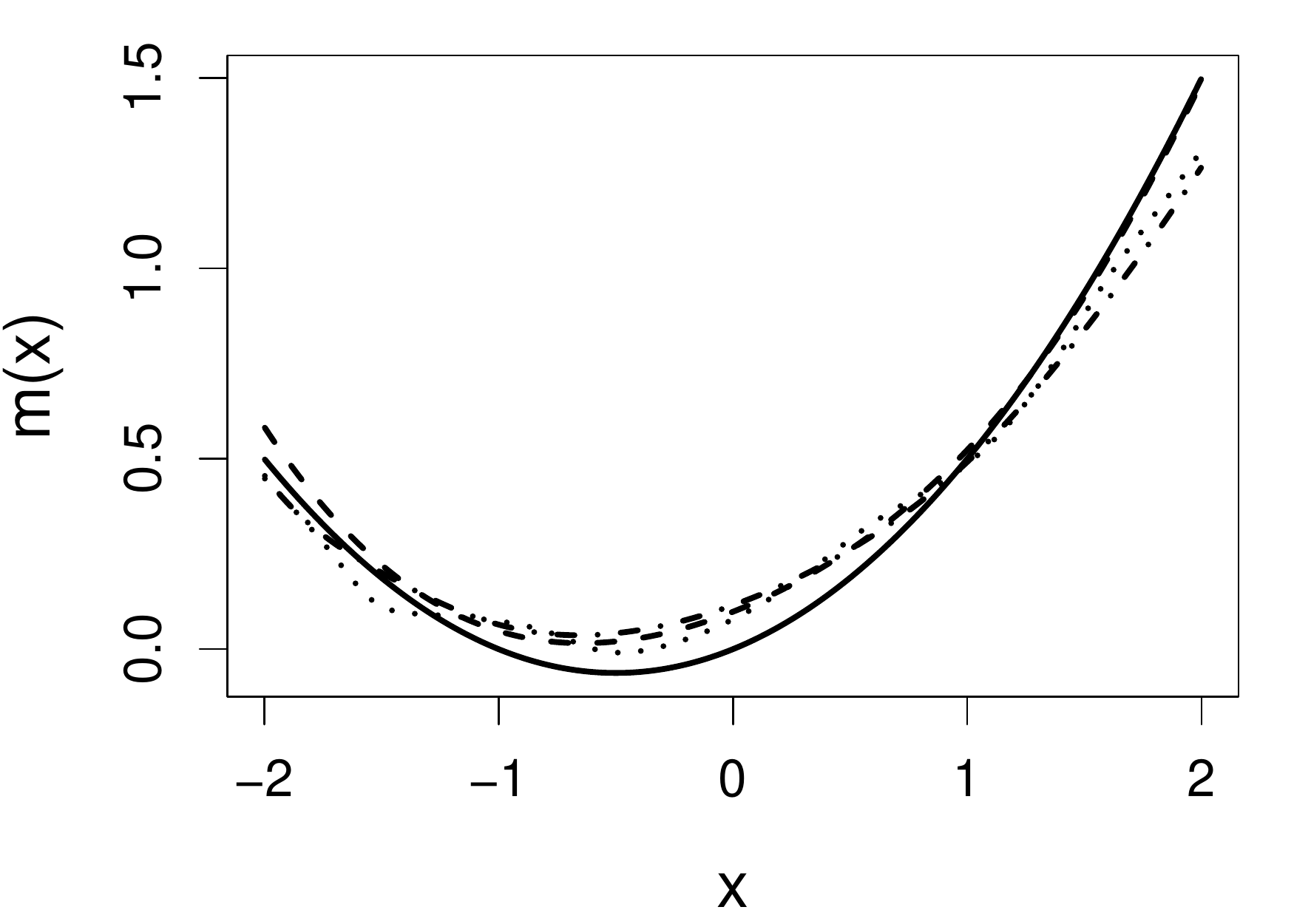} }
	\subfigure[]{ \includegraphics[width=\linewidth]{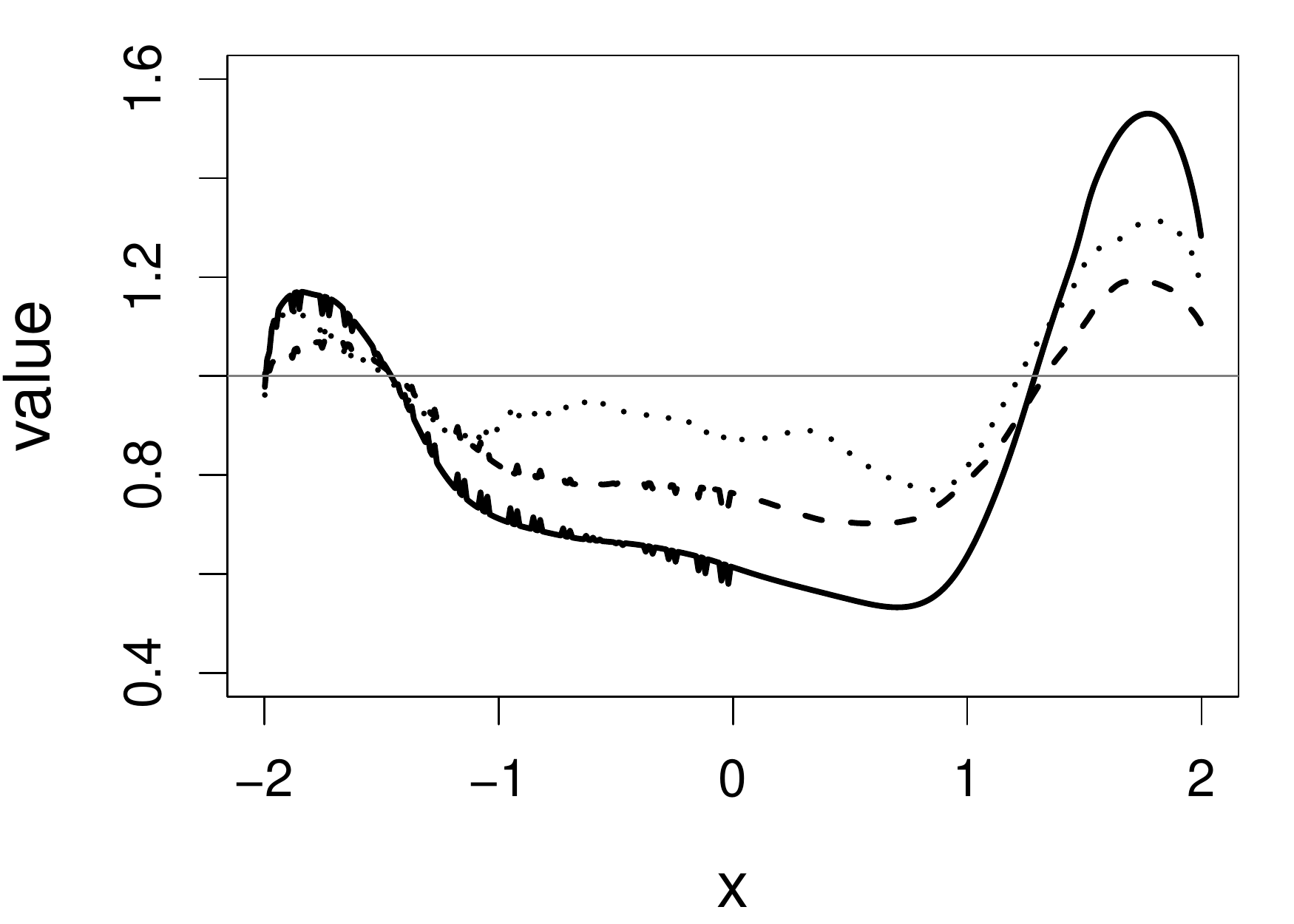} }\\
	\caption{Simulation results under (C2) using the theoretical optimal $h$, with $U$-distribution misspecified as Laplace. Panels (a) \& (d): boxplots of ISEs versus $\lambda$ for $\hat m_{\hbox {\tiny HZ}}(x)$ and $\hat m_{\hbox {\tiny DFC}}(x)$, respectively. Panels (b) \& (e): boxplots of PAE(0) versus $\lambda$ for $\hat m_{\hbox {\tiny HZ}}(0)$ and $\hat m_{\hbox {\tiny DFC}}(0)$, respectively. Panels (c) \& (f): boxplots of PAE($-1$) versus $\lambda$ for $\hat m_{\hbox {\tiny HZ}}(-1)$ and $\hat m_{\hbox {\tiny DFC}}(-1)$, respectively. Panels (g) \& (h): quantile curves when $\lambda=0.85$ for $\hat m_{\hbox {\tiny HZ}}(x)$ and $\hat m_{\hbox {\tiny DFC}}(x)$, respectively, based on ISEs (dashed lines for the first quartile, dotted lines for the second quartile, and dot-dashed lines for the third quartile, solid lines for the truth). Panel (i): PmAER (dashed line), PsdAER (dotted line), and PMSER (solid line) versus $x$ when $\lambda=0.85$; the horizontal reference line highlights the value 1.}
	\label{Sim2GauLap500:box}
\end{figure}

\begin{figure}
	\centering
	\setlength{\linewidth}{4.5cm}
	\subfigure[]{ \includegraphics[width=\linewidth]{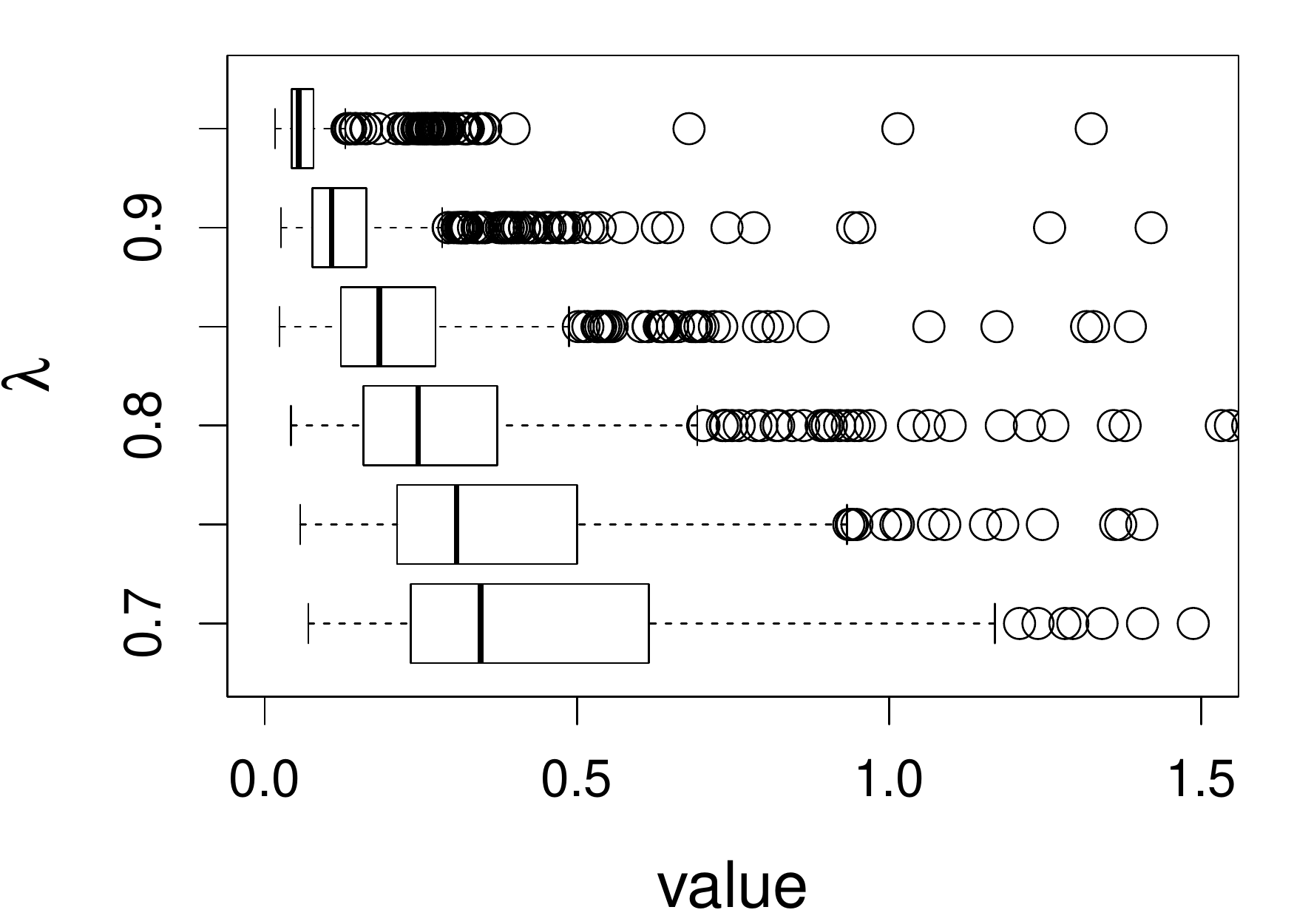} }
	\subfigure[]{ \includegraphics[width=\linewidth]{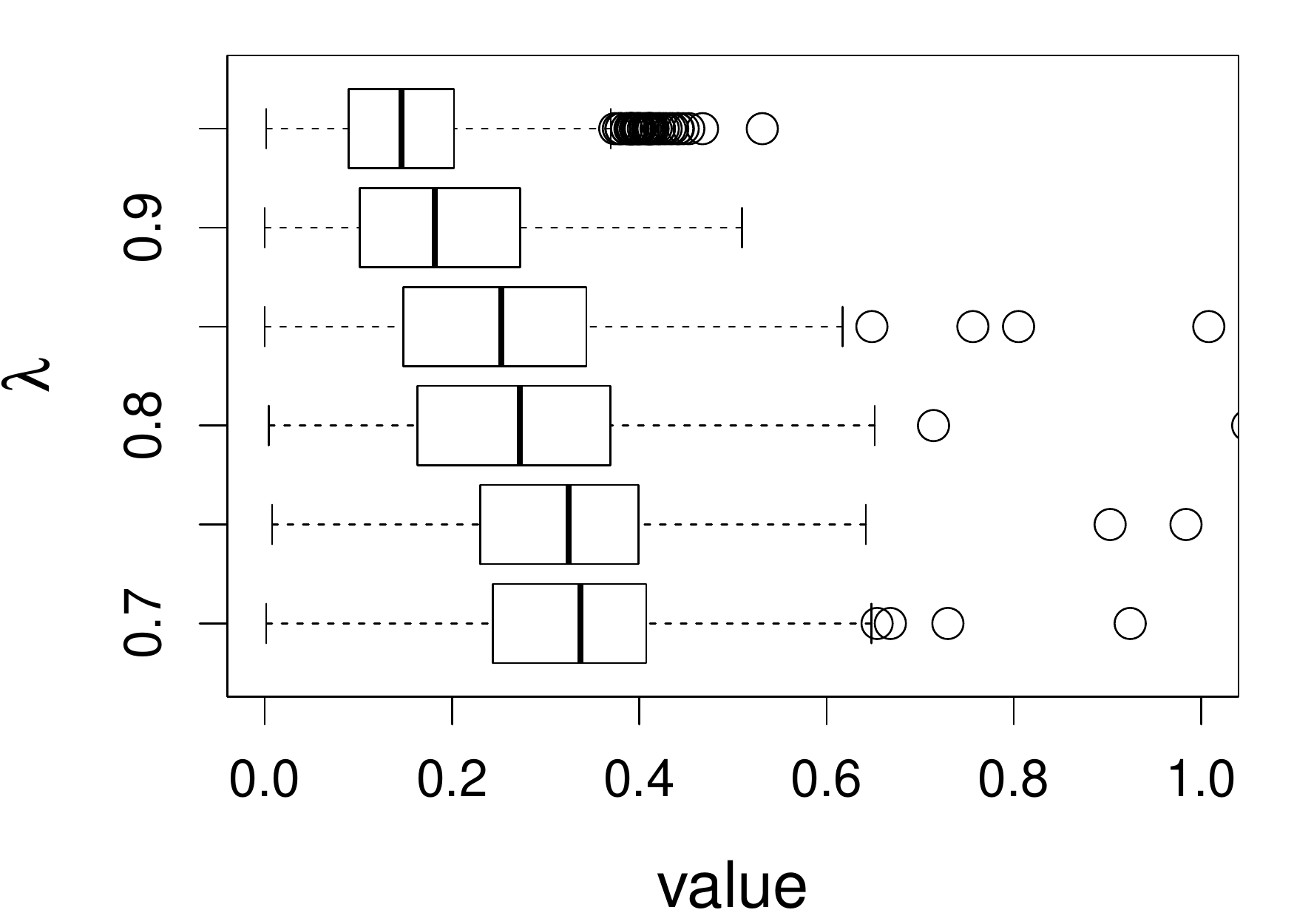} }
	\subfigure[]{ \includegraphics[width=\linewidth]{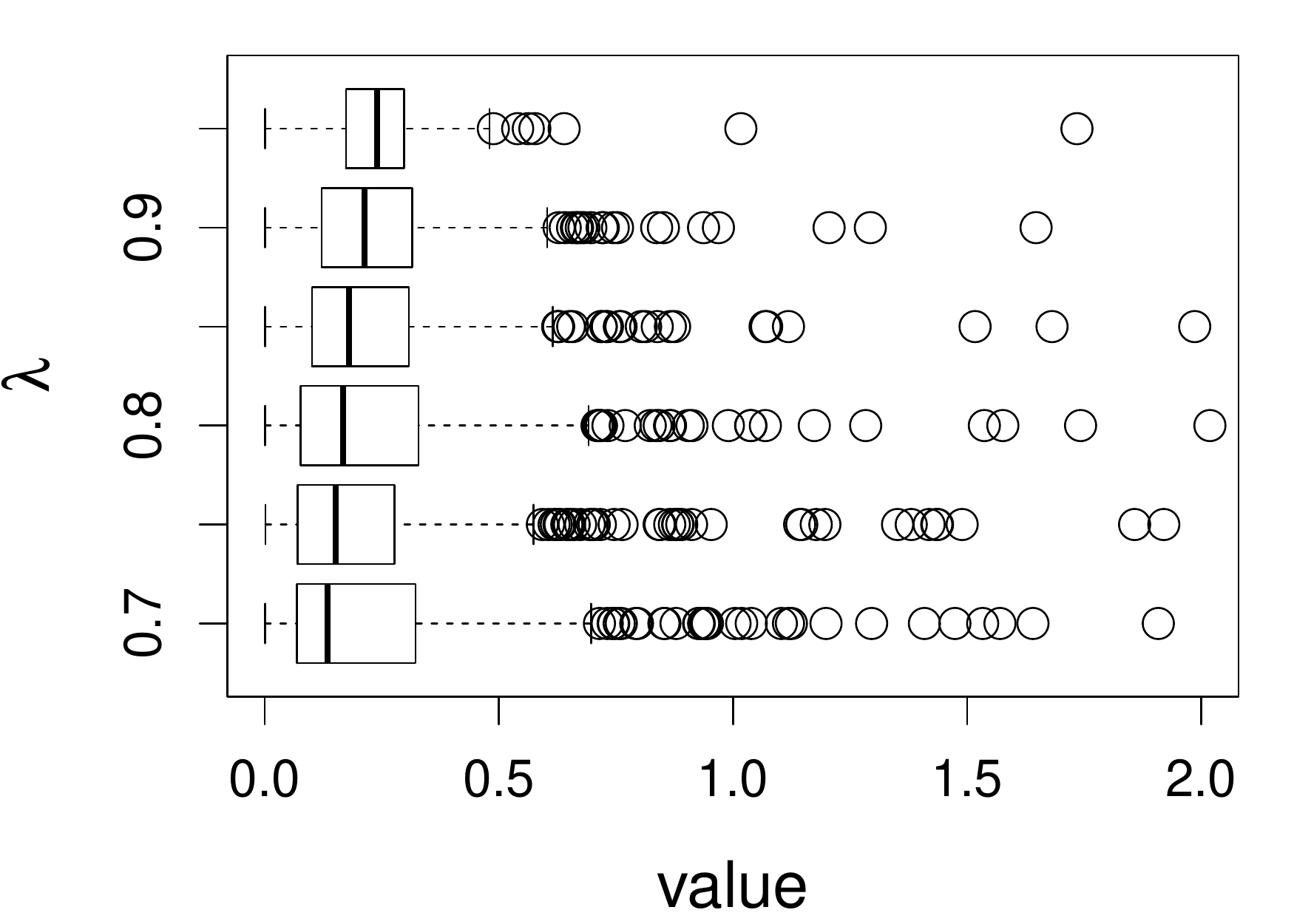} }\\
	\subfigure[]{ \includegraphics[width=\linewidth]{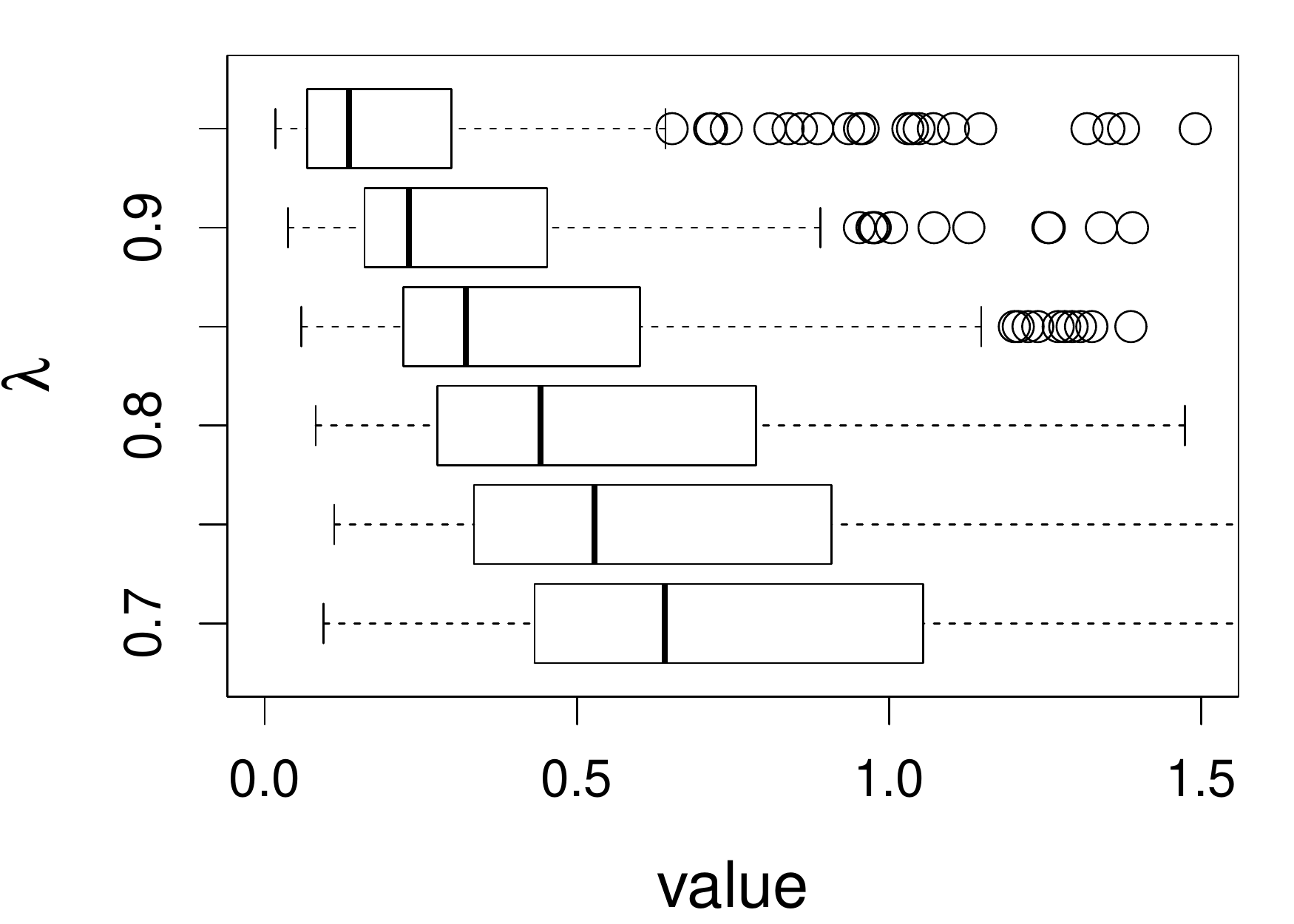} }
	\subfigure[]{ \includegraphics[width=\linewidth]{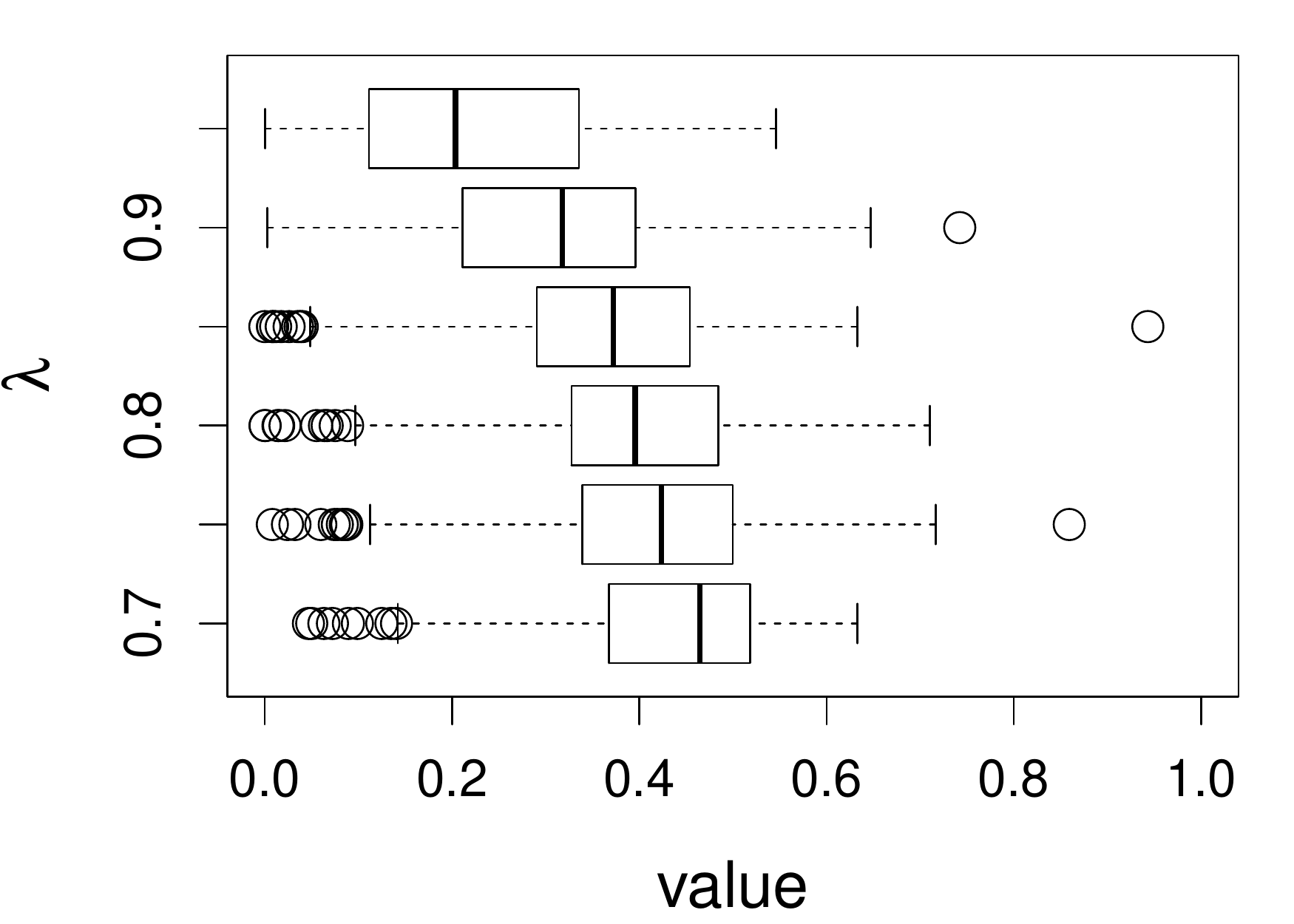} }
	\subfigure[]{ \includegraphics[width=\linewidth]{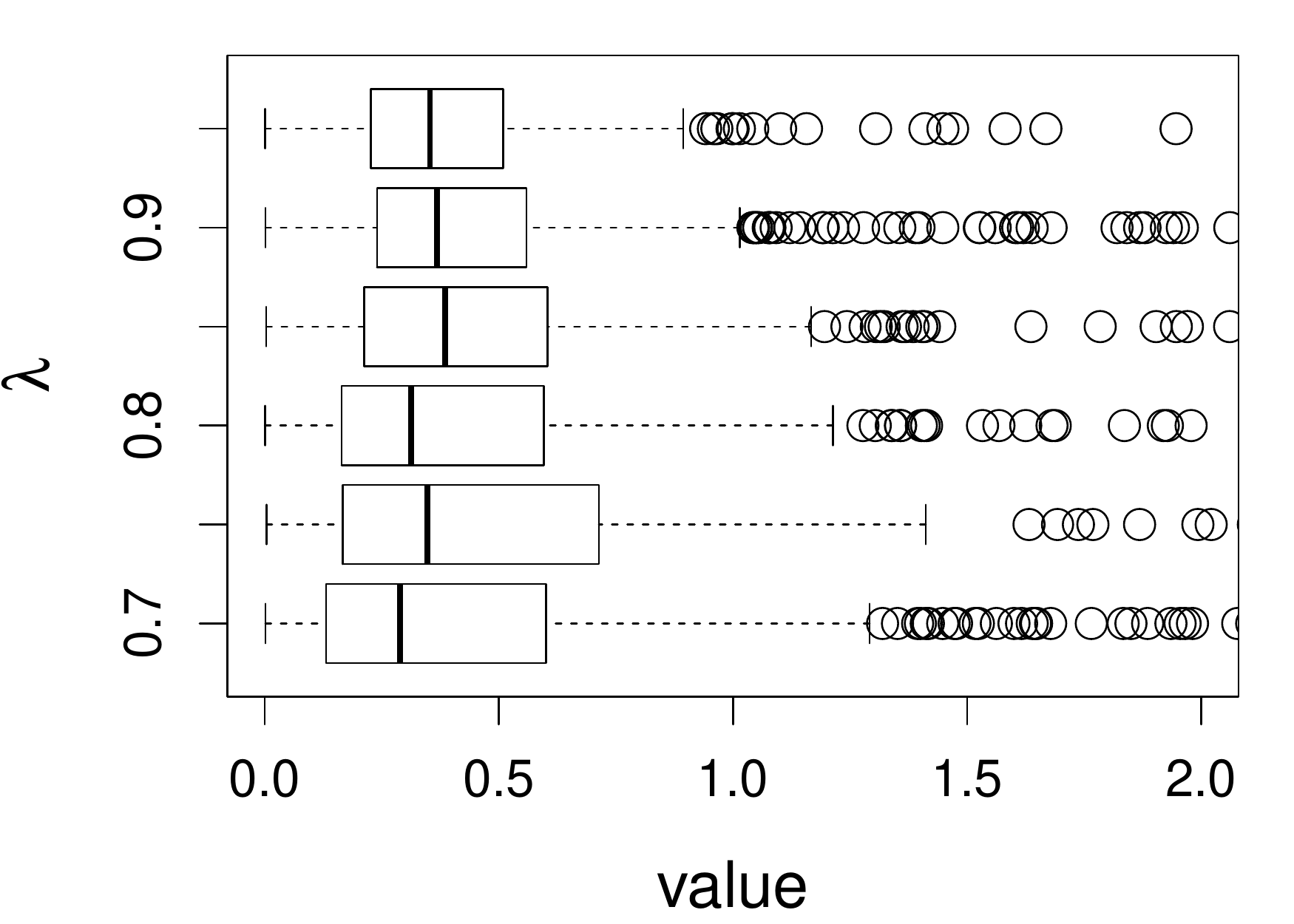} }\\
	\subfigure[]{ \includegraphics[width=\linewidth]{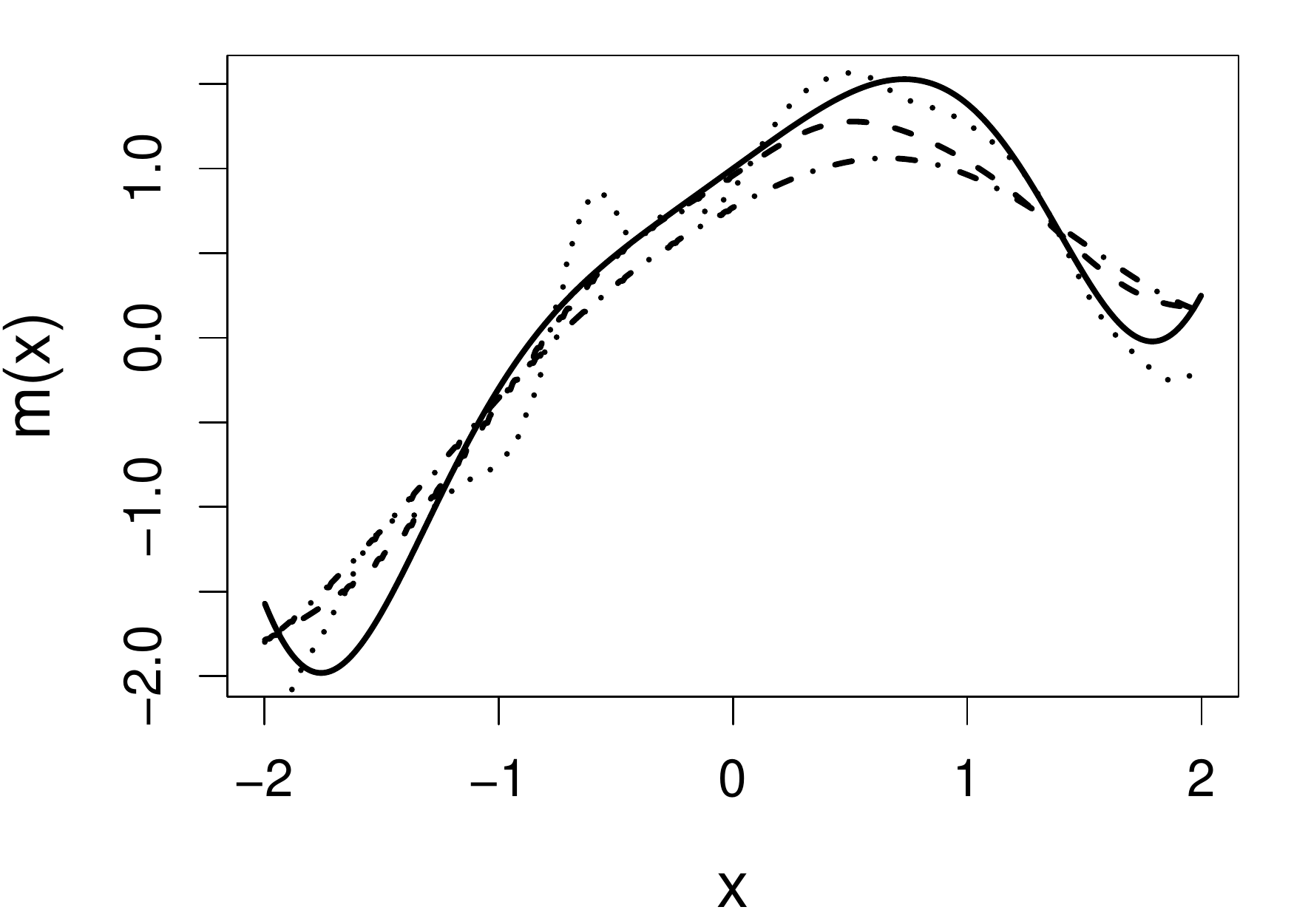} }
	\subfigure[]{ \includegraphics[width=\linewidth]{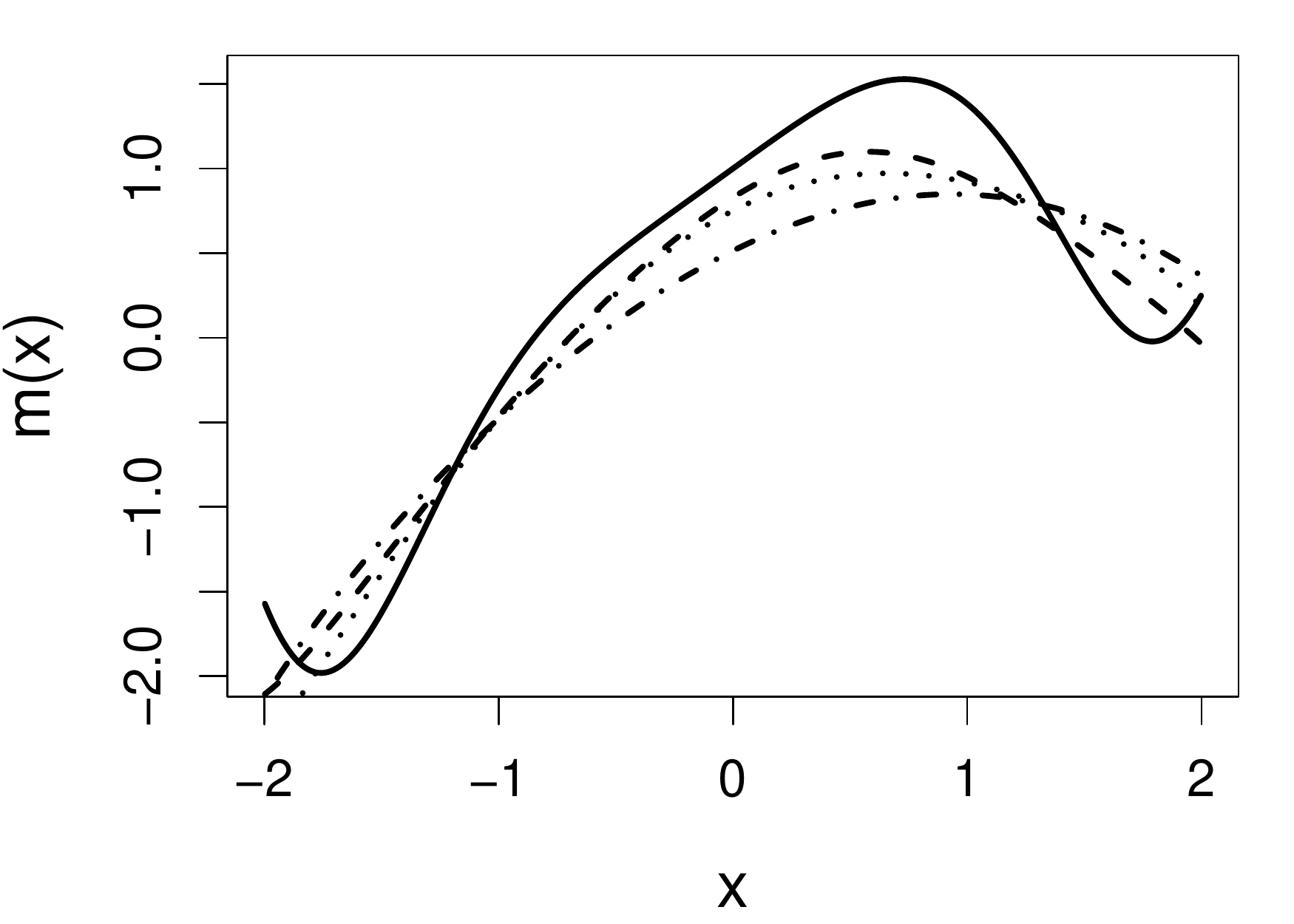} }
	\subfigure[]{ \includegraphics[width=\linewidth]{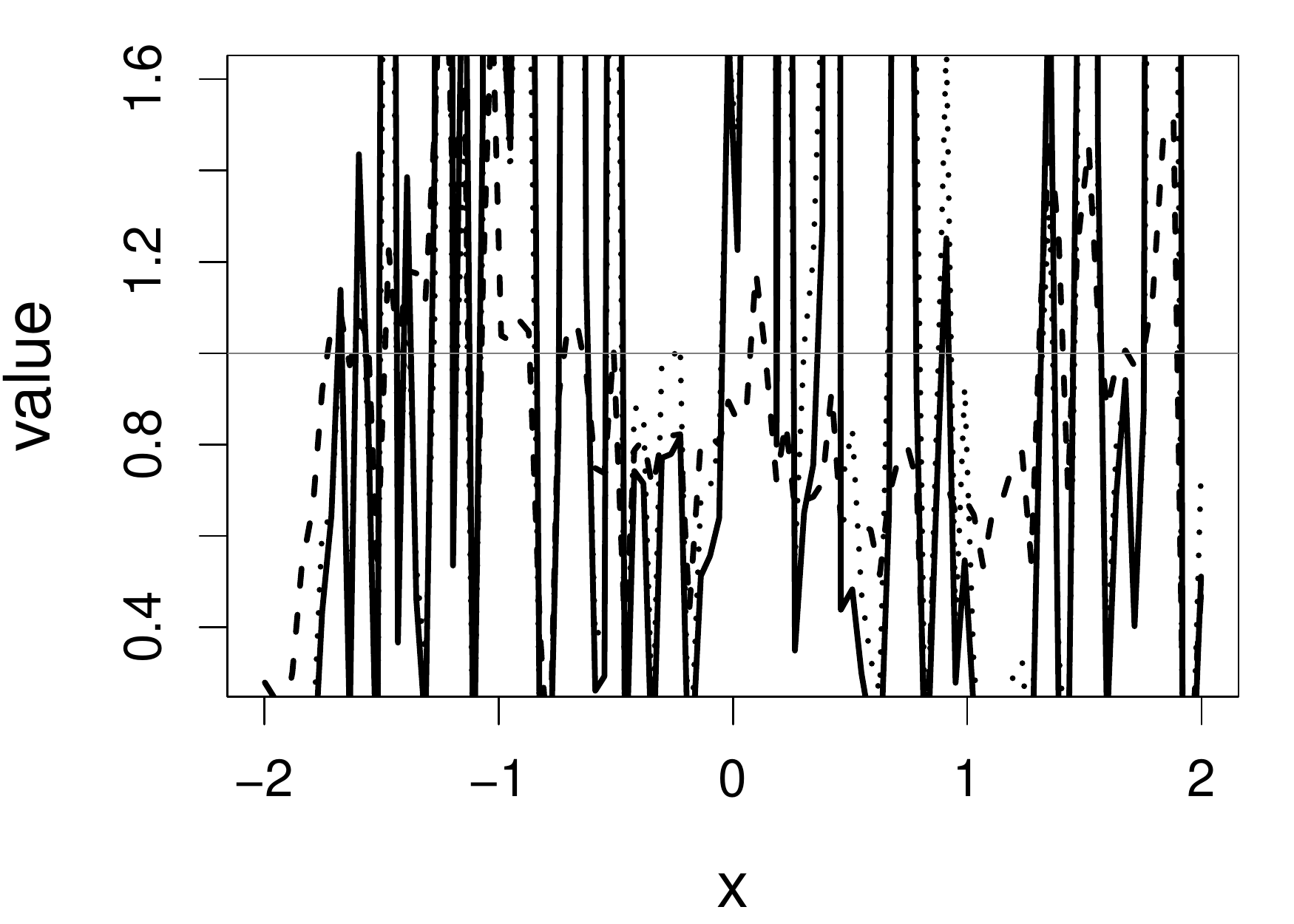} }\\
	\caption{Simulation results under (C4) using CV-SIMEX bandwidth selection. Panels (a) \& (d): boxplots of ISEs versus $\lambda$ for $\hat m_{\hbox {\tiny HZ}}(x)$ and $\hat m_{\hbox {\tiny DFC}}(x)$, respectively. Panels (b) \& (e): boxplots of PAE(1) versus $\lambda$ for $\hat m_{\hbox {\tiny HZ}}(1)$ and $\hat m_{\hbox {\tiny DFC}}(1)$, respectively. Panels (c) \& (f): boxplots of PAE(2) versus $\lambda$ for $\hat m_{\hbox {\tiny HZ}}(2)$ and $\hat m_{\hbox {\tiny DFC}}(2)$, respectively. Panels (g) \& (h): quantile curves when $\lambda=0.8$ for $\hat m_{\hbox {\tiny HZ}}(x)$ and $\hat m_{\hbox {\tiny DFC}}(x)$, respectively, based on ISEs (dashed lines for the first quartile, dotted lines for the second quartile, and dot-dashed lines for the third quartile, solid lines for the truth). Panel (i): PmAER (dashed line), PsdAER (dotted line), and PMSER (solid line) versus $x$ when $\lambda=0.8$; the horizontal reference line highlights the value 1.} 
	\label{Sim3LapLapS500:box}
\end{figure}

We compare the performance of $\hat m_{\hbox {\tiny HZ}}(x)$ and $\hat m_{\hbox {\tiny DFC}}(x)$ with regard to the quality of the entire regression curve estimation over $[x_{\hbox {\tiny L}}, \, x_{\hbox {\tiny U}}]$, as well as the quality of the estimation of $m(x)$ at individual $x$'s. The quantity used to monitor the overall regression curve estimation is the approximated ISE. The quantities used to assess the quality of $\hat m_{[\cdot]}(x)$ at a particular point $x=x_0$ are based on the pointwise absolute error (PAE), $\textrm{PAE}(x_0)=|\hat m_{[\cdot]}(x_0)-m(x_0)|$. Specifically, we compute the following three summary statistics: first, the pointwise mean absolute error ratio (PmAER) defined by 
$${\rm PmAER}(x_k) = \frac{\textrm{MC average of }\left|\hat{m}_{\hbox {\tiny HZ}}(x_k)-m(x_k)\right|} { \text{MC average of  }\left|\hat{m}_{\hbox {\tiny DFC}}(x_k)-m(x_k)\right|};$$
second, the pointwise standard deviation of absolute error ratio (PsdAER) defined by 
$${\rm PsdAER}(x_k) = \frac{\text{MC standard deviation of }\left|\hat{m}_{\hbox {\tiny HZ}}(x_k)-m(x_k)\right|} { \text{MC standard deviation of }\left|\hat{m}_{\hbox {\tiny DFC}}(x_k)-m(x_k)\right|};$$
and third, the pointwise mean squared error ratio (PMSER) defined by 
$${\rm PMSER}(x_k) = \frac{\textrm{MC average of }\left|\hat{m}_{\hbox {\tiny HZ}}(x_k)-m(x_k)\right|^2} { \text{MC average of  }\left|\hat{m}_{\hbox {\tiny DFC}}(x_k)-m(x_k)\right|^2}.$$ These quantities are presented in Figures~\ref{Sim1LapLap500:box}--\ref{Sim3LapLap:box} for (C1)--(C4), respectively. Figure~\ref{Sim2GauLap500:box} shows the counterpart results of Figure~\ref{Sim2GauGau500:box} under (C2) when it is (incorrectly) assumed that $U$ follows a Laplace distribution. These five figures depict results obtained when the theoretical optimal $h$ is used. Lastly, Figure~\ref{Sim3LapLapS500:box} is the counterpart of Figure~\ref{Sim3LapLap:box} under (C4) with $h$ chosen by the CV-SIMEX bandwidth selection procedure with $B=10$ and $L=10$. Very similar performance of the two estimates is observed when larger values of $B$ or $L$ are used in this round of experiment.  

When the theoretical optimal bandwidth is used, as in Figures~\ref{Sim1LapLap500:box}--\ref{Sim2GauLap500:box}, $\hat m_{\hbox {\tiny HZ}}(x)$ outperforms $\hat m_{\hbox {\tiny DFC}}(x)$ over the majority region of each considered range of $x$ in regard to both accuracy and precision. Even though it is shown in Section~\ref{s:comparevar} that the dominating variance of $\hat m_{\hbox {\tiny HZ}}(x)$ is higher than that of $\hat m_{\hbox {\tiny DFC}}(x)$ when the distribution of $U$ is ordinary smooth (e.g., a Laplace distribution), this large sample trend does not take effect for the majority region of $x$ in these finite sample experiments. The regions where $\hat m_{\hbox {\tiny DFC}}(x)$ performs better than $\hat m_{\hbox {\tiny HZ}}(x)$ in regard to bias, variance, and MSE are usually neighborhoods of the inflection points of $m(x)$. For instance, under (C3) (see panel (i) in Figure~\ref{Sim4LapLap:box}), $\hat m_{\hbox {\tiny DFC}}(x)$ is less biased than $\hat m_{\hbox {\tiny HZ}}(x)$ at the small neighborhoods of $\pm 1$. It is worth pointing out that the gain in accuracy and precision from our estimator compared to the DFC estimator is especially promising at the boundary of $x$ in (C3) and (C4) (see panels (c), (f), and (i) in Figures~\ref{Sim4LapLap:box} and \ref{Sim3LapLap:box}). In both cases, data points uniformly distribute over the domain of $m(x)$. Different from (C3) and (C4), in (C1), there are more data points near the boundaries than elsewhere in the domain. Excluding (C2) (since the plotted range of $x$ in Figures~\ref{Sim2GauGau500:box} and \ref{Sim2GauLap500:box} is not the entire observed range), (C1) is the only case among all considered cases here that $\hat m_{\hbox {\tiny DFC}}(x)$ outperforms $\hat m_{\hbox {\tiny HZ}}(x)$ near the boundaries in terms of bias. However, $\hat m_{\hbox {\tiny HZ}}(x)$ is still substantially less variable, and its MSE is lower than that of the competing estimator (see panels (c), (f), and (i) of Figure~\ref{Sim1LapLap500:box}). Finally, contrasting Figure~\ref{Sim2GauGau500:box} and Figure~\ref{Sim2GauLap500:box}, one can see that both estimators are fairly robust to the misspecification of the measurement error distribution. 

When the bandwidth is chosen by the refined CV-SIMEX method, as in Figure~\ref{Sim3LapLapS500:box}, both estimates become more variable, with our estimates better than the DFC estimates over most of the 500 MC replicates. As mentioned earlier, increasing $B$ to a larger value does not substantially change our estimate. More importantly, using a $B$ smaller than ten affects our estimator far less than it affects the DFC estimator.

\subsection{Motorcycle data}
We now apply the two estimation methods to error-contaminated data sets created based on the motorcycle crash data from a simulated motorcycle crash designed to test crash helmets (available under \texttt{R} library \texttt{MASS}). The original data set consists of 133 measurements of head acceleration measured in standard gravity acceleration (gs) at various times in milliseconds after impact. It is of interest to estimate the underlying head acceleration, $Y$, as a function of time after impact, $X$. Having the error-free data in this example allows us to have a reference estimate of the regression function with which the estimates based on error-prone data can be compared. 

Based on the original data, we first obtain the local linear estimate of $m(x)$, denoted by $\hat m(x)$, using the \texttt{R} function \texttt{locpol} in the \texttt{locpol} package, with the bandwidth chosen by cross validation \citep{Wang95} implemented by function \texttt{regCVBwSelC} in the same \texttt{R} package. Compared to the fitted curves using error-prone data, the $\hat m(x)$ can be viewed as the ``ideal" estimate in the sense that one cannot do better than this with error-contaminated data. We use this ideal curve as the reference curve in our follow-up experiments, where we contaminate $X$ with simulated independent Laplace measurement errors to achieve different levels of reliability ratio $\lambda$. At each level of $\lambda$, we use the error-contaminated data to estimate the acceleration curve using the two estimation methods, both assuming Laplace $U$. This experiment of curve estimation is repeated 500 times at each level of $\lambda$. We obtained very similar results when we contaminated $X$ with simulated normal $U$ while assuming Laplace $U$ for estimations.  

\begin{figure}
	\centering
	\setlength{\linewidth}{4.5cm}
	\subfigure[]{ \includegraphics[width=\linewidth]{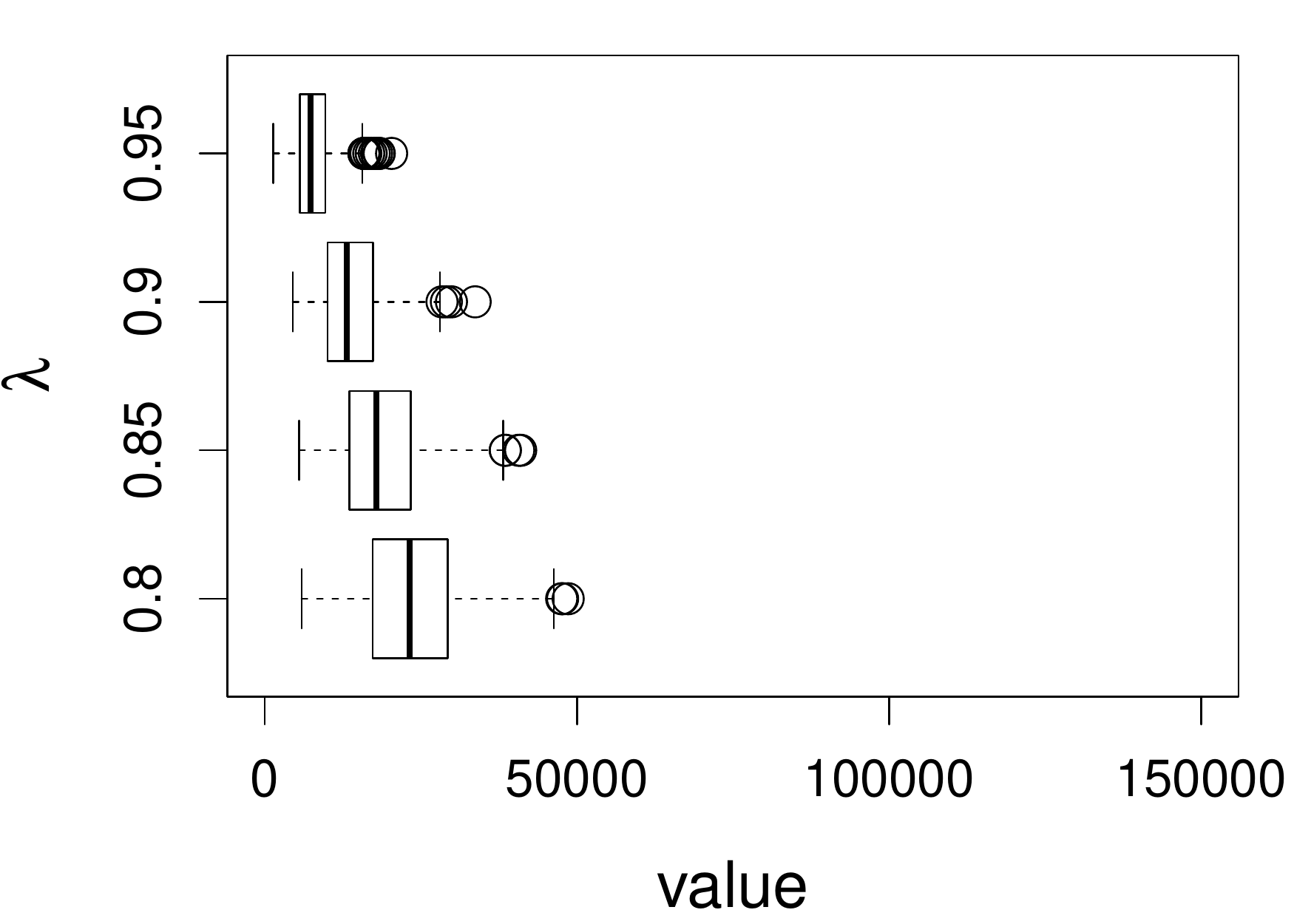} }
	\subfigure[]{ \includegraphics[width=\linewidth]{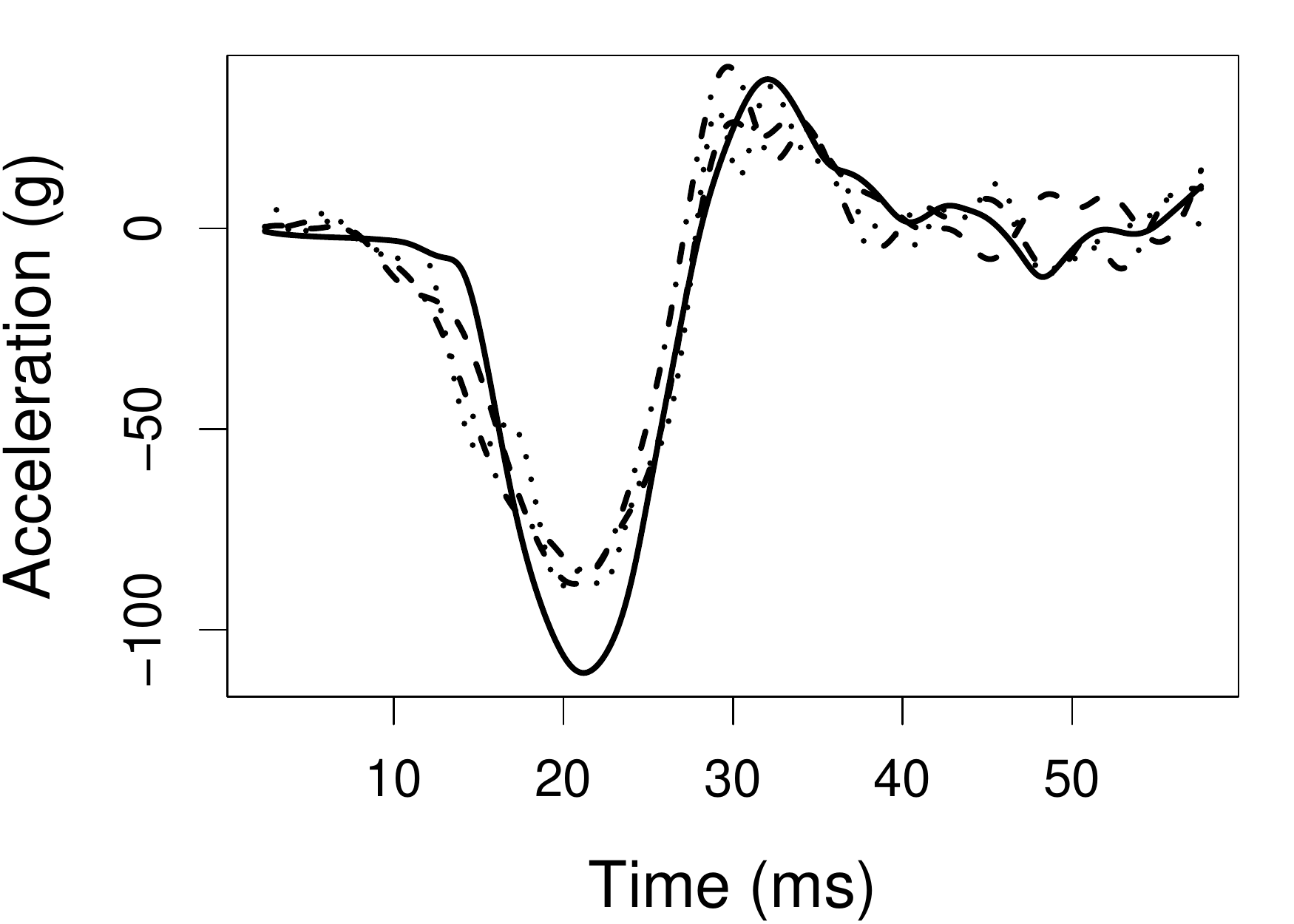} }
	\subfigure[]{ \includegraphics[width=\linewidth]{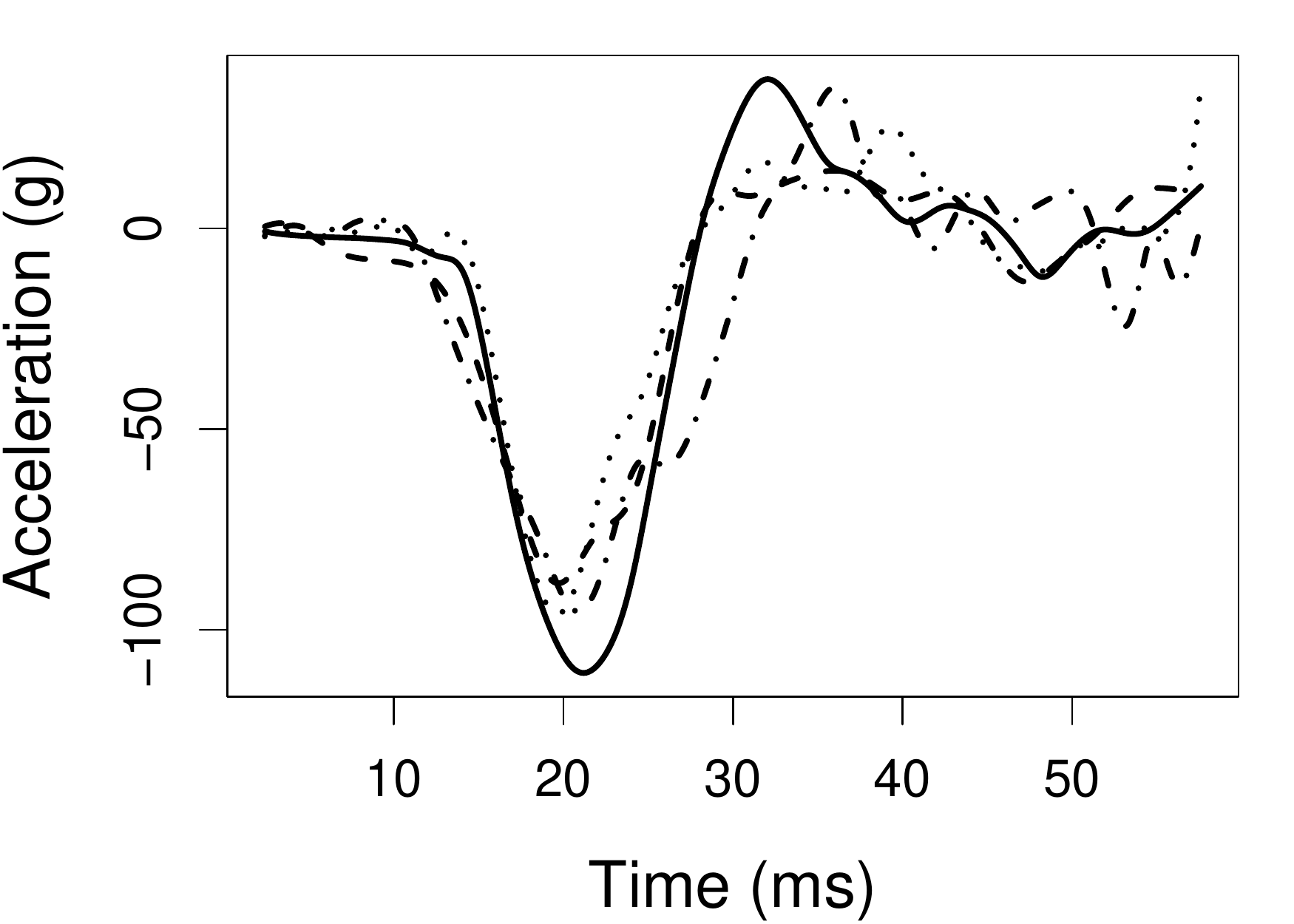}}\\
	\subfigure[]{ \includegraphics[width=\linewidth]{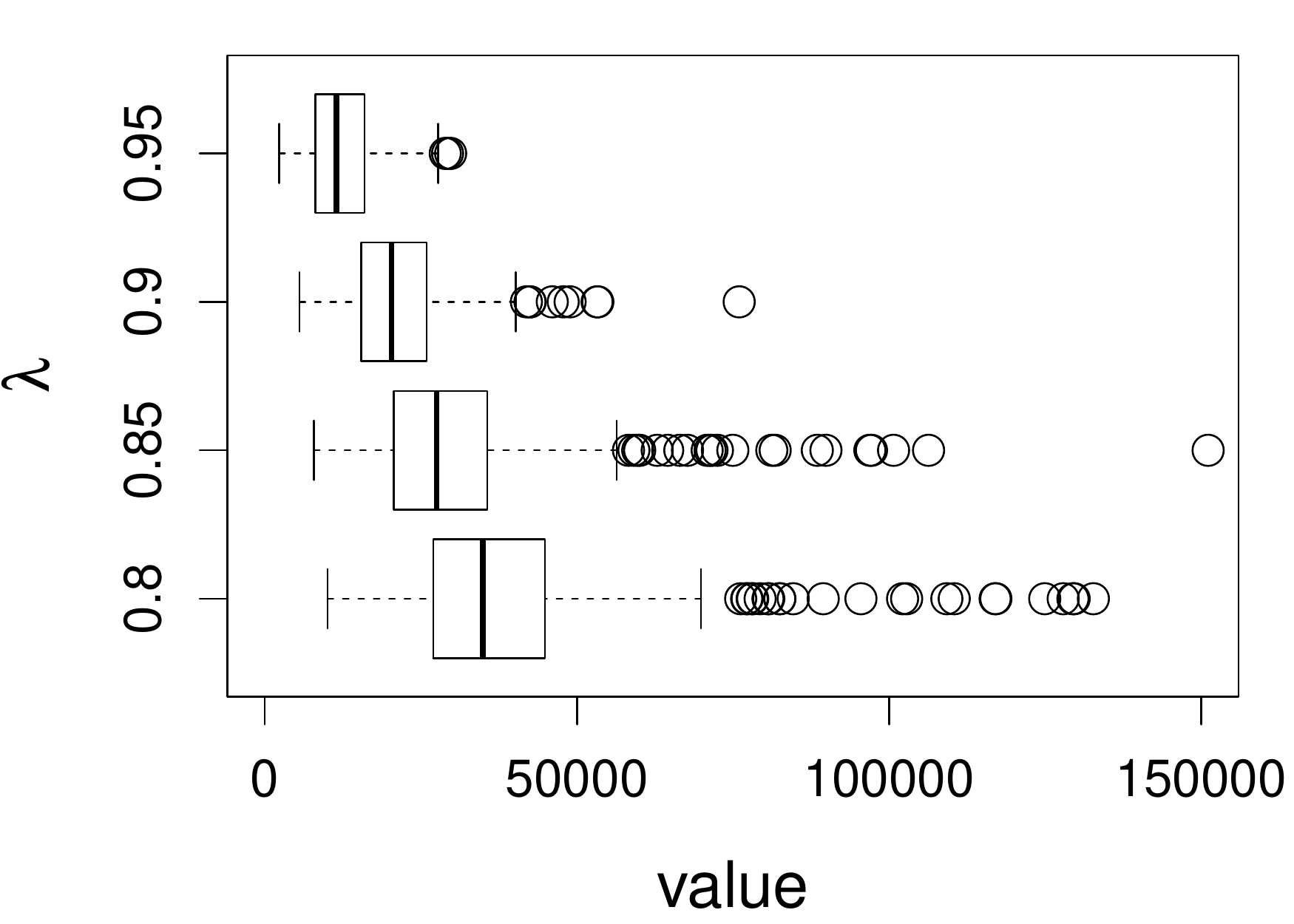} }
	\subfigure[]{ \includegraphics[width=\linewidth]{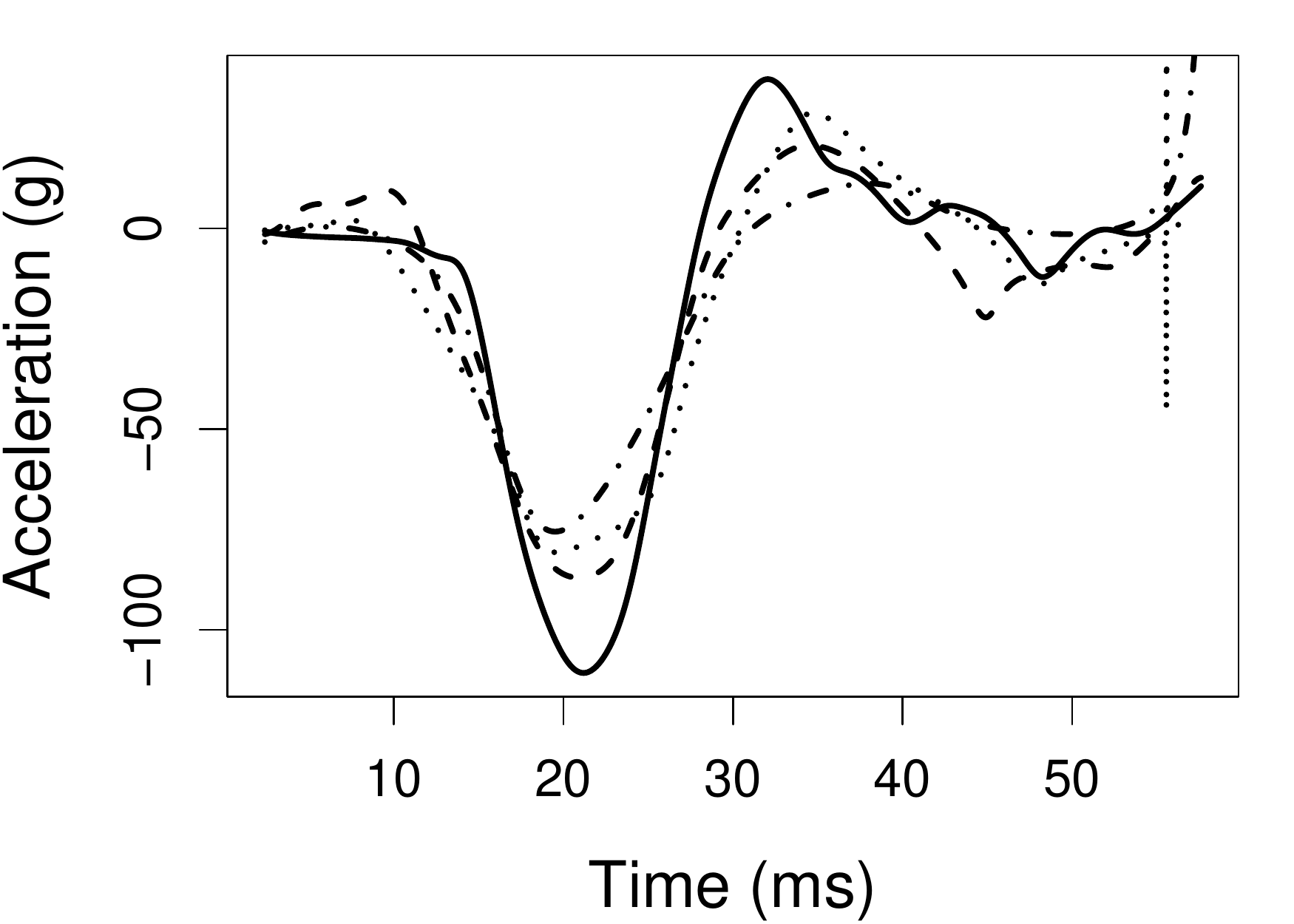} }
	\subfigure[]{ \includegraphics[width=\linewidth]{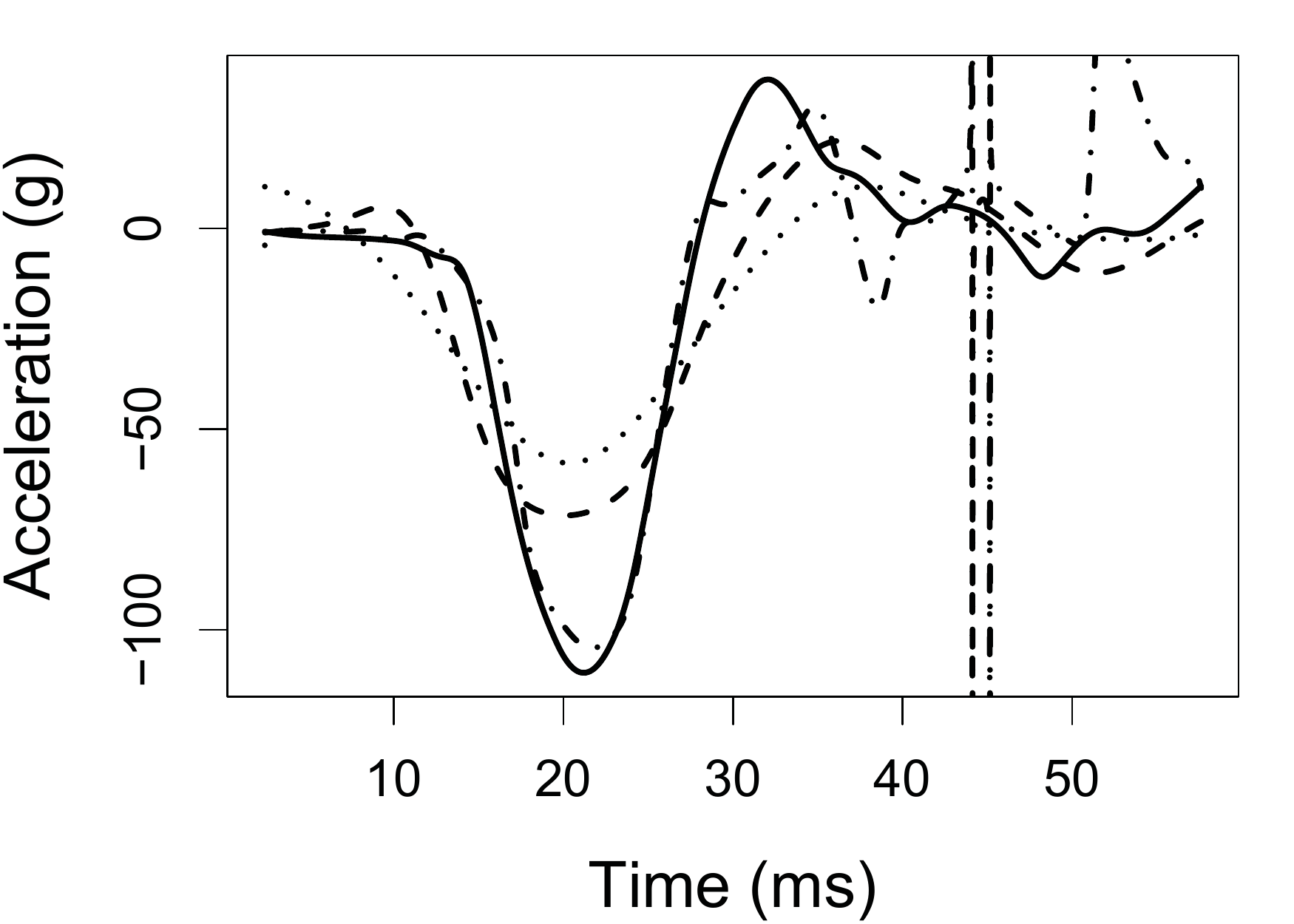}}
\caption{Results for motorcycle data. Panels (a) \& (d): boxplots of ISEs versus $\lambda$ for $\hat m_{\hbox {\tiny HZ}}(x)$ and $\hat m_{\hbox {\tiny DFC}}(x)$, respectively. Panels (b) \& (e):  quantile curves when $\lambda=0.95$ for $\hat m_{\hbox {\tiny HZ}}(x)$ and $\hat m_{\hbox {\tiny DFC}}(x)$, respectively, based on ISEs (dashed lines for the first quartile, dotted lines for the second quartile, and dot-dashed lines for the third quartile, solid lines for the ``truth") when the approximated theoretical optimal $h$ is used. Panels (c) \& (f):  counterpart quantile curves of those in panels (b) \& (e) when $h$ is chosen by the refined CV-SIMEX procedure.} 
	\label{f:motor}
\end{figure}

Figure~\ref{f:motor} depicts the results, including boxplots of ISE at each $\lambda$ level, the fitted curves for $\lambda=0.95$ selected according to quantiles of ISE when the approximated theoretical optimal $h$ is used, and the counterpart fitted curves when the refined CV-SIMEX method is used to select $h$ with $B=10$ and $L=10$. Using the ideal estimate as the ``truth," our estimate appears to be less biased and less variable at all considered levels of error contamination than the DFC estimate. When the refined CV-SIMEX method is used to select $h$, our estimator suffers less numerical instability compared to the competing method. 

\section{Discussion}
\label{s:discussion}
In this study we proposed a local polynomial estimator of the regression function when the covariate is measured with error. The proposed estimator makes direct use of the naive inference, leading to relatively more transparent connections between the properties of the proposed estimator and those of the inference from error-free data. We rigorously derived the asymptotic properties of the proposed estimator in comparison with the estimator proposed by \citet{Delaigle09}. Under very similar regularity conditions, besides the asymptotic normality that both estimators possess, the asymptotic bias and variance of these estimators are carefully compared. Theoretical evidence suggests that the new estimator can be less biased than the competing estimator. Results from extensive simulation study also support this finding.   

To implement the proposed method, we thoughtfully refined the CV-SIMEX bandwidth selection method proposed by \citet{Delaigle08} to narrow the search region of $h$, which in turn allows us to use a much smaller $B$ in the SIMEX implementation without noticeable loss in accuracy. This refinement greatly reduces the computational burden for the otherwise intrinsically cumbersome bandwidth selection procedure. 

In our simulation studies, how the proposed estimator and the DFC estimator compare at the boundary of the support of $X$ depends on the distribution of $X$. Even though the proposed estimator appears to suffer less numerical instability when the refined CV-SIMEX method is used to select $h$, it can still be rather challenging to estimate the curve near the boundary. The properties of our estimator near the boundary deserve further investigation, which may lead to ways to improve its behavior near the boundary. Besides the generalization of the proposed method pointed out earlier in Section~\ref{s:ours} to allow multiple covariates, one can also follow the construction of the proposed estimator in (\ref{eq:mz}) to obtain non-naive estimators of $m(x)$ by starting with a parametric naive estimator $\hat m^*(w)$. For instance, one may naively fit a polynomial regression function to obtain $\hat m^*(w)$, then use it in (\ref{eq:mz}) to achieve a non-naive estimator of $m(x)$ that is not completely nonparametric. However, the obtained estimator of $m(x)$ is usually not of the same functional form as $m^*(w)$. If one wishes to fit a polynomial regression function accounting for measurement error, the method proposed by \citet{Zavala2007} is a more appealing approach than our proposed nonparametric approach. 

\begin{figure}
	\centering
	\setlength{\linewidth}{4.5cm}
	\subfigure[]{ \includegraphics[width=\linewidth]{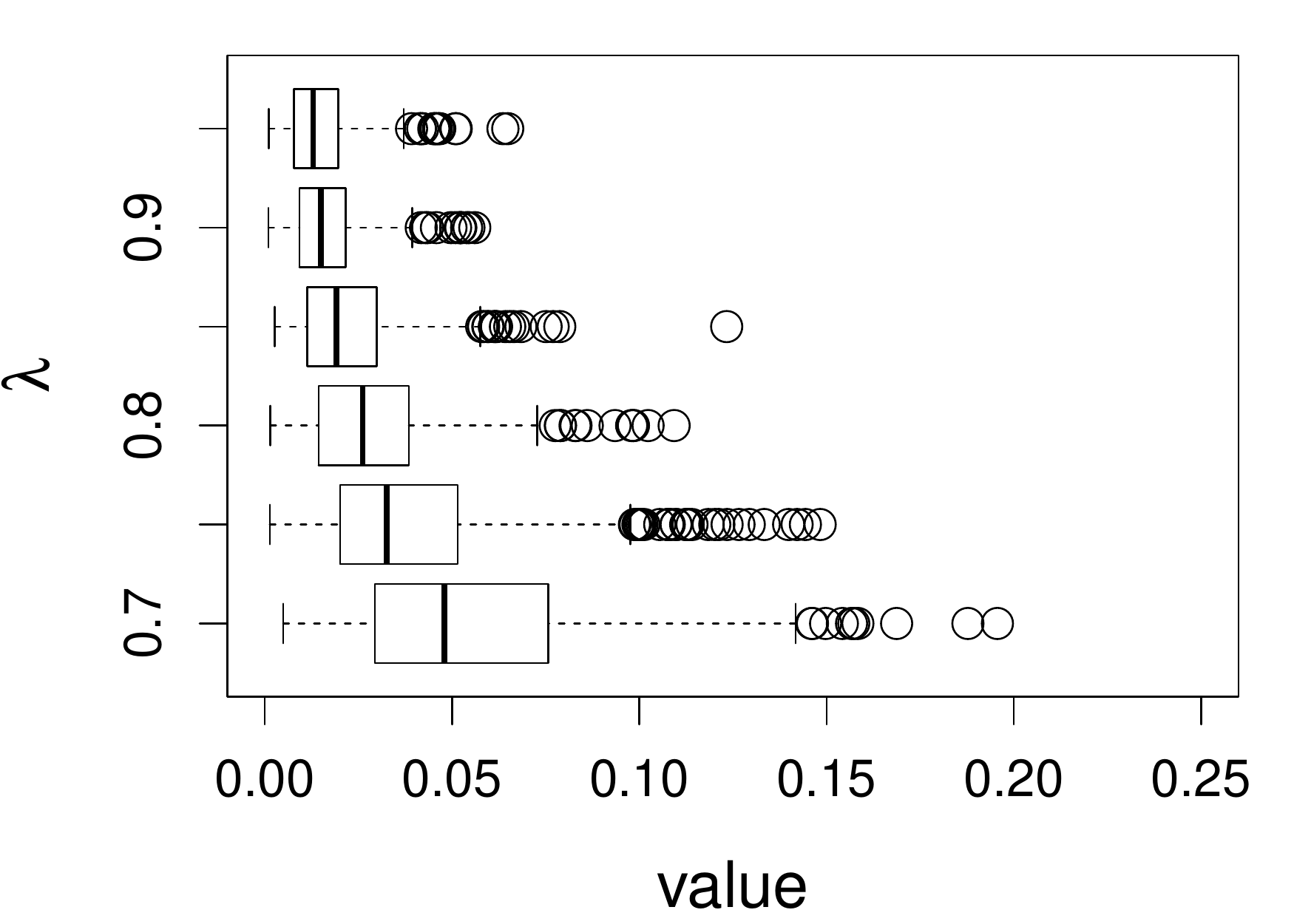} }
	\subfigure[]{ \includegraphics[width=\linewidth]{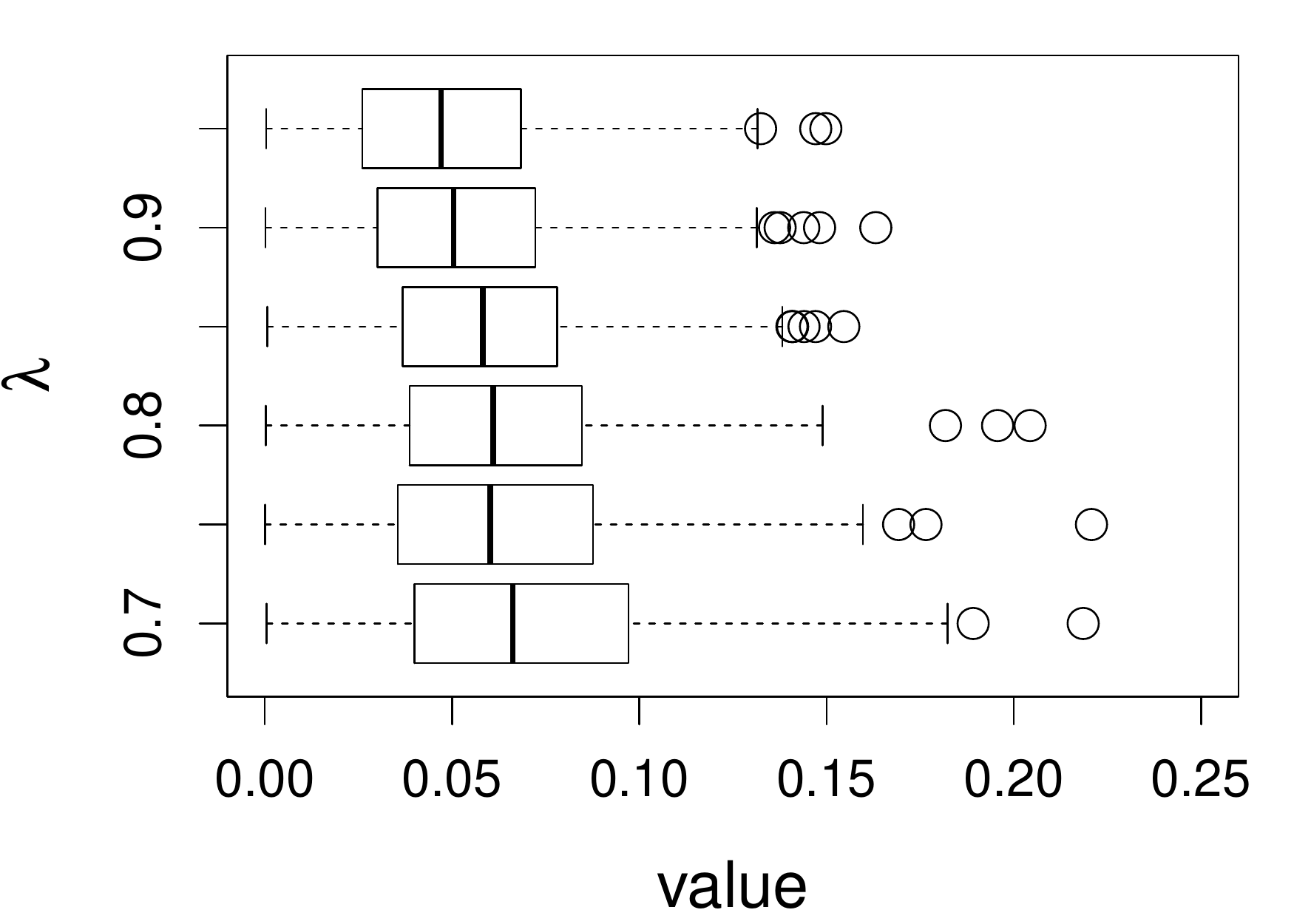} }
	\subfigure[]{ \includegraphics[width=\linewidth]{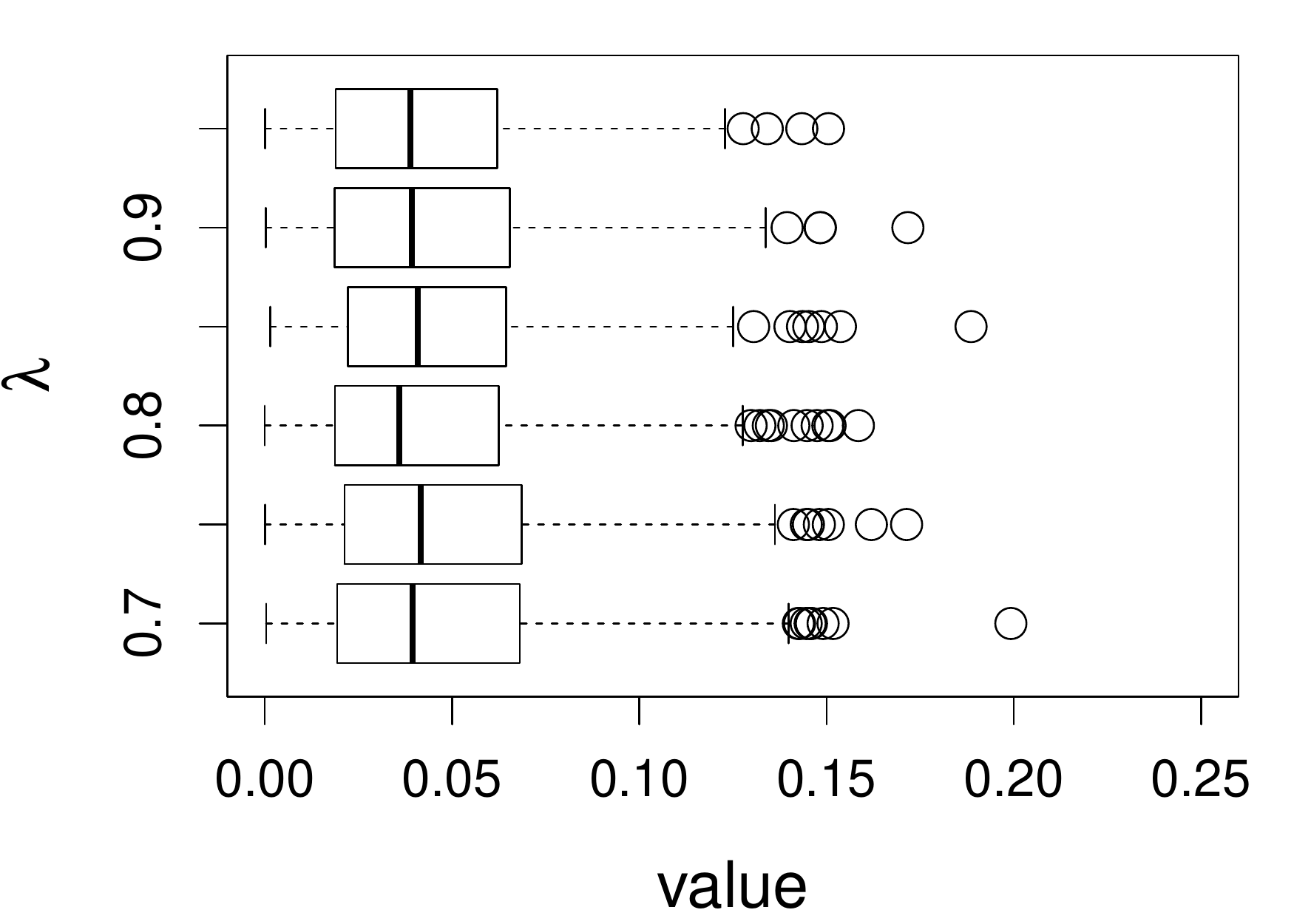} }\\
	\subfigure[]{ \includegraphics[width=\linewidth]{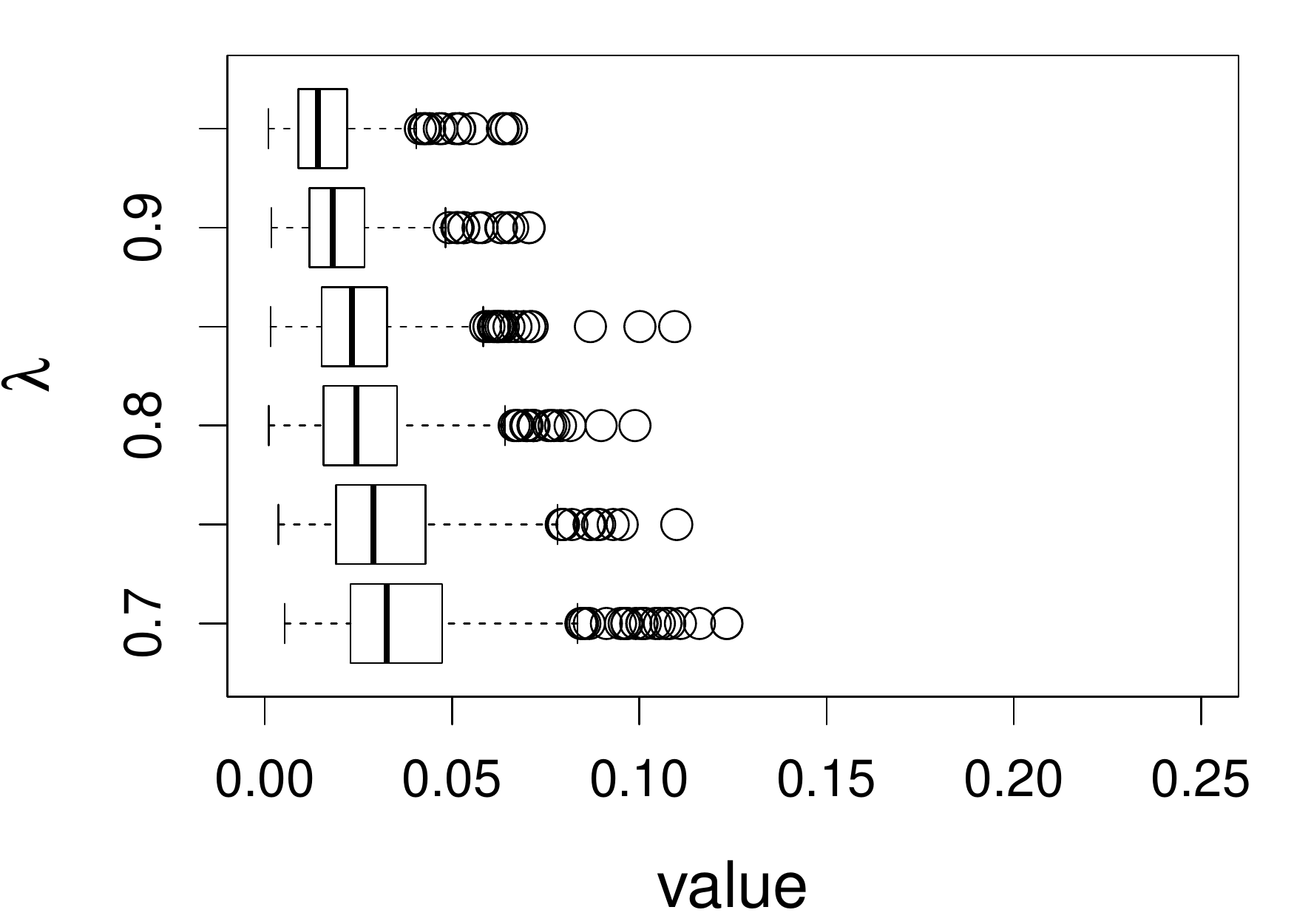} }
	\subfigure[]{ \includegraphics[width=\linewidth]{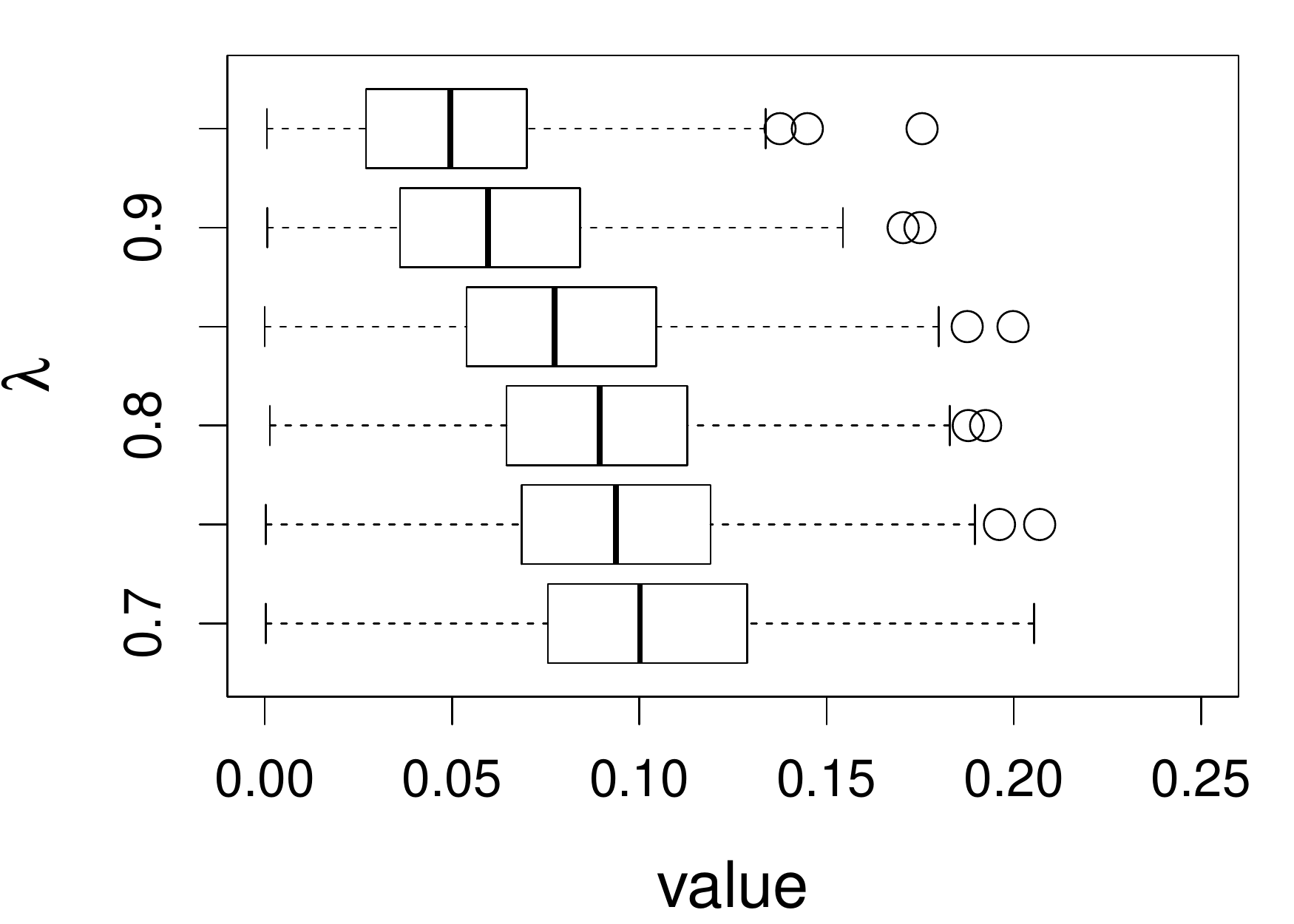} }
	\subfigure[]{ \includegraphics[width=\linewidth]{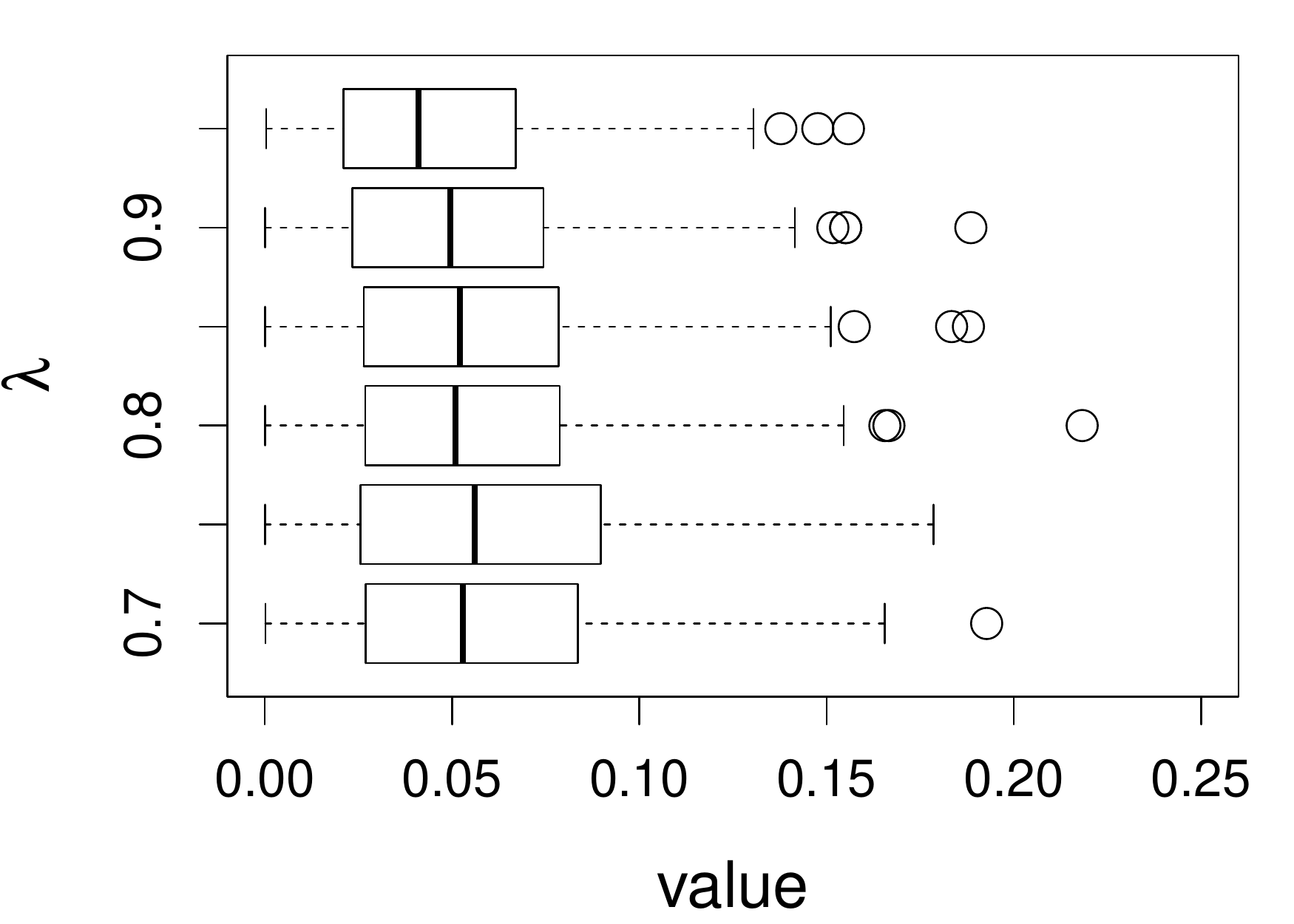} }\\
	\subfigure[]{ \includegraphics[width=\linewidth]{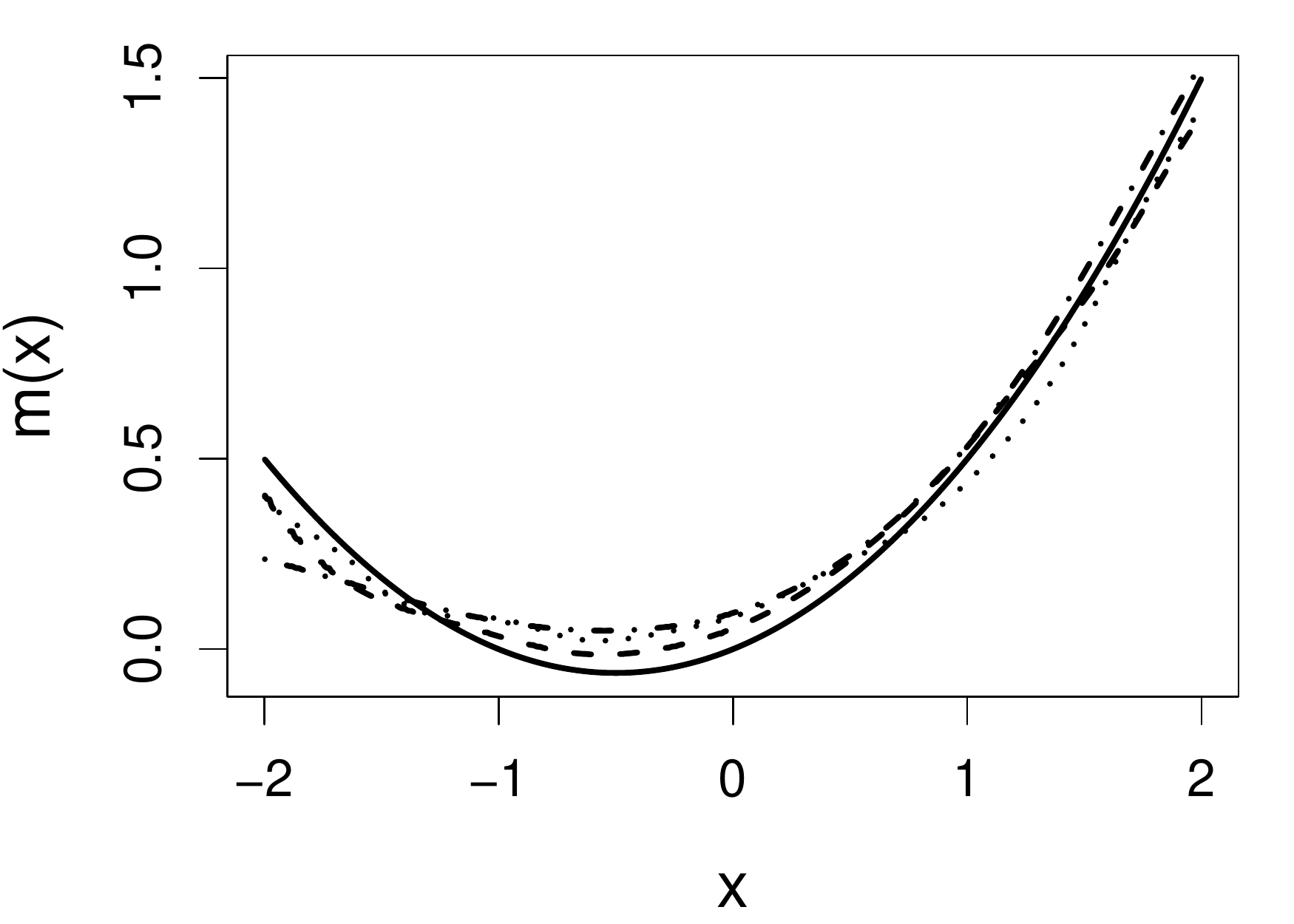} }
	\subfigure[]{ \includegraphics[width=\linewidth]{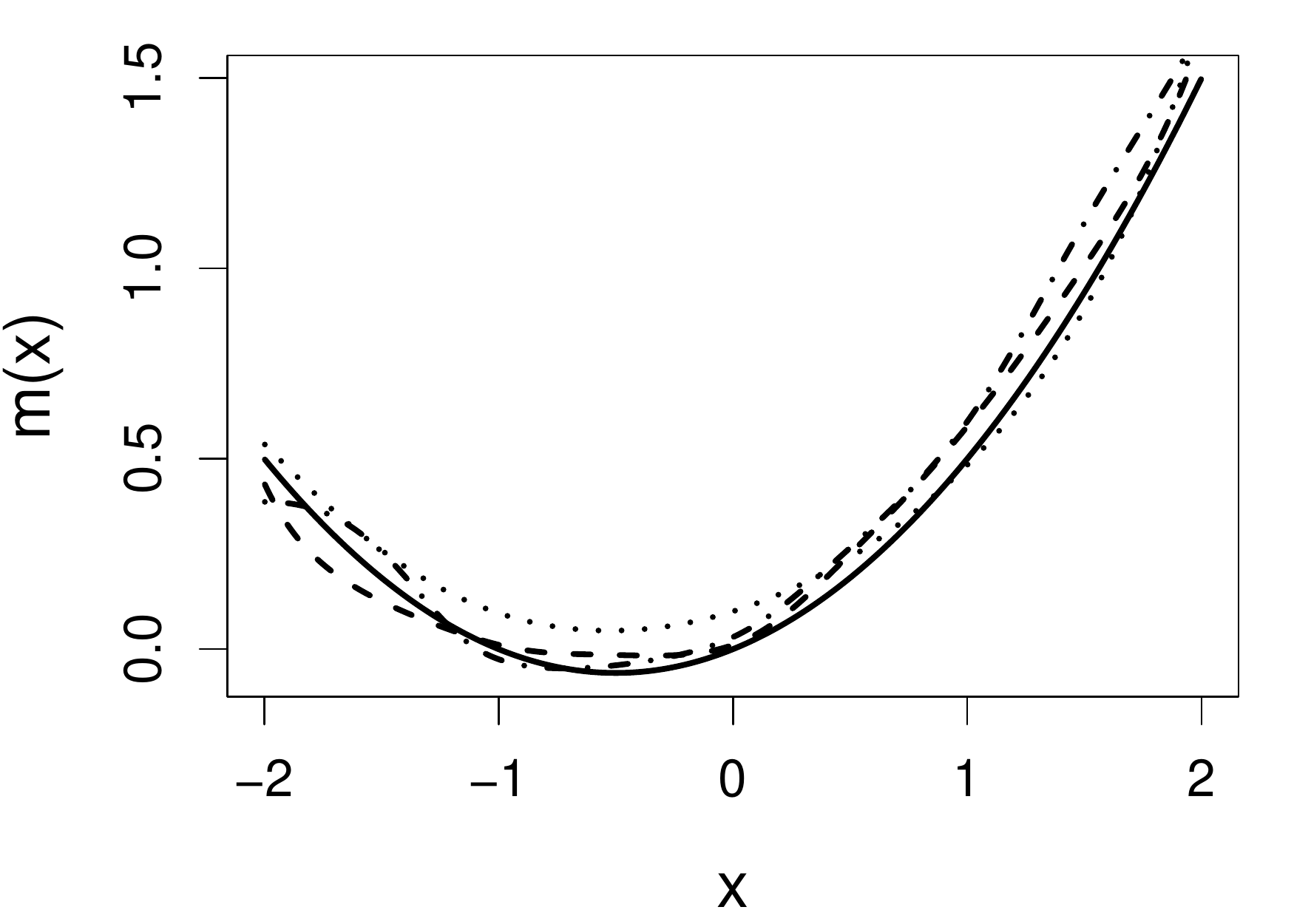} }
	\subfigure[]{ \includegraphics[width=\linewidth]{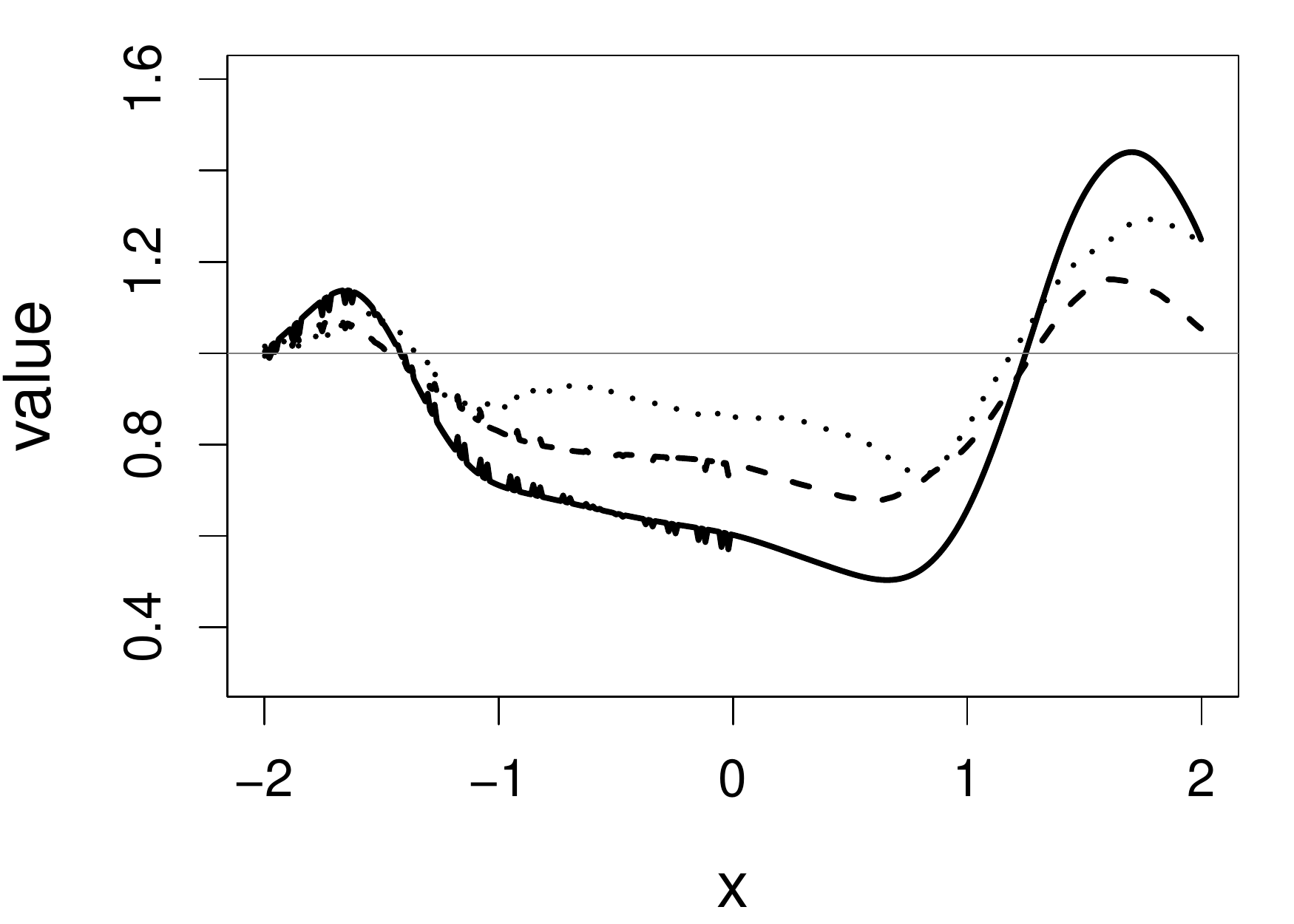} }
	\caption{Simulation results under (C2) using the theoretical optimal $h$, with $U$-distribution misspecified as Laplace and $\sigma^2_u$ estimated using repeated measures. Panels (a) \& (d): boxplots of ISEs versus $\lambda$ for $\hat m_{\hbox {\tiny HZ}}(x)$ and $\hat m_{\hbox {\tiny DFC}}(x)$, respectively. Panels (b) \& (e): boxplots of PAE(0) versus $\lambda$ for $\hat m_{\hbox {\tiny HZ}}(0)$ and $\hat m_{\hbox {\tiny DFC}}(0)$, respectively. Panels (c) \& (f): boxplots of PAE($-1$) versus $\lambda$ for $\hat m_{\hbox {\tiny HZ}}(-1)$ and $\hat m_{\hbox {\tiny DFC}}(-1)$, respectively. Panels (g) \& (h): quantile curves when $\lambda=0.85$ for $\hat m_{\hbox {\tiny HZ}}(x)$ and $\hat m_{\hbox {\tiny DFC}}(x)$, respectively, based on ISEs (dashed lines for the first quartile, dotted lines for the second quartile, and dot-dashed lines for the third quartile, solid lines for the truth). Panel (i): PmAER (dashed line), PsdAER (dotted line), and PMSER (solid line) versus $x$ when $\lambda=0.85$; the horizontal reference line highlights the value 1.}
	\label{Sim2GauLap500:est}
	\end{figure}

The measurement error distribution is assumed be known in the simulation study presented in Section~\ref{s:4cases}, where in one case the distribution is misspecified as a Laplace distribution, and we apprehend little influence of such misspecification on the proposed estimator. This robustness phenomenon is also pointed out in \citet{Delaigle09} for the DFC estimator, and is discussed in \citet{Meister2004} and \citet{Delaigle2008}. Taking advantage of this robustness feature, when the measurement error distribution is unknown, we recommend using the mean-zero Laplace characteristic function, $\phi_{\hbox {\tiny $U$}}(t)=1/\{1+(\sigma^2_u/2) t^2\}$ in the estimator, where $\sigma^2_u$ can be trivially and consistently estimated by equation (4.3) in \citet{LASbook} when repeated measures of each $X_j$ are available. We implement this recommended strategy for the four cases considered in Section~\ref{s:4cases} and observe very similar results as those shown in Figures~\ref{Sim1LapLap500:box}--\ref{Sim2GauLap500:box}. In particular, we generate two replicate measures, $W_{j,k}=X_j+U_{j,k}$, where $U_{j,k}$'s are i.i.d. with variance $2\sigma^2_u$, for $k=1, 2$, $j=1, \ldots, n$. Then we define $W_j=(W_{j,1}+W_{j,2})/2$, for $j=1, \ldots, n$, as the observed covariate values used in $\hat m_{\hbox {\tiny HZ}}(x)$ and $\hat m_{\hbox {\tiny DFC}}(x)$, where the associated measurement error variance is $\sigma_u^2$. Following equation (4.3) in \citet{LASbook}, we estimate $\sigma^2_u$ via $\sum_{j=1}^n\sum_{k=1}^2(W_{j,k}-W_j)^2/(2n)$. Figure~\ref{Sim2GauLap500:est} shows the counterpart results of those shown in Figure~\ref{Sim2GauLap500:box}, from which we can see that using an estimated variance in the misspecified $\phi_{\hbox {\tiny $U$}}(t)$ does not affect the estimates noticeably. Plots parallel to Figures~\ref{Sim1LapLap500:box}, \ref{Sim4LapLap:box}, and \ref{Sim3LapLap:box}, which show estimates obtained using the same strategy under the other three cases, are given in Appendix D.  

\begin{figure}
	\centering
	\setlength{\linewidth}{4.5cm}
	\subfigure[]{ \includegraphics[width=\linewidth]{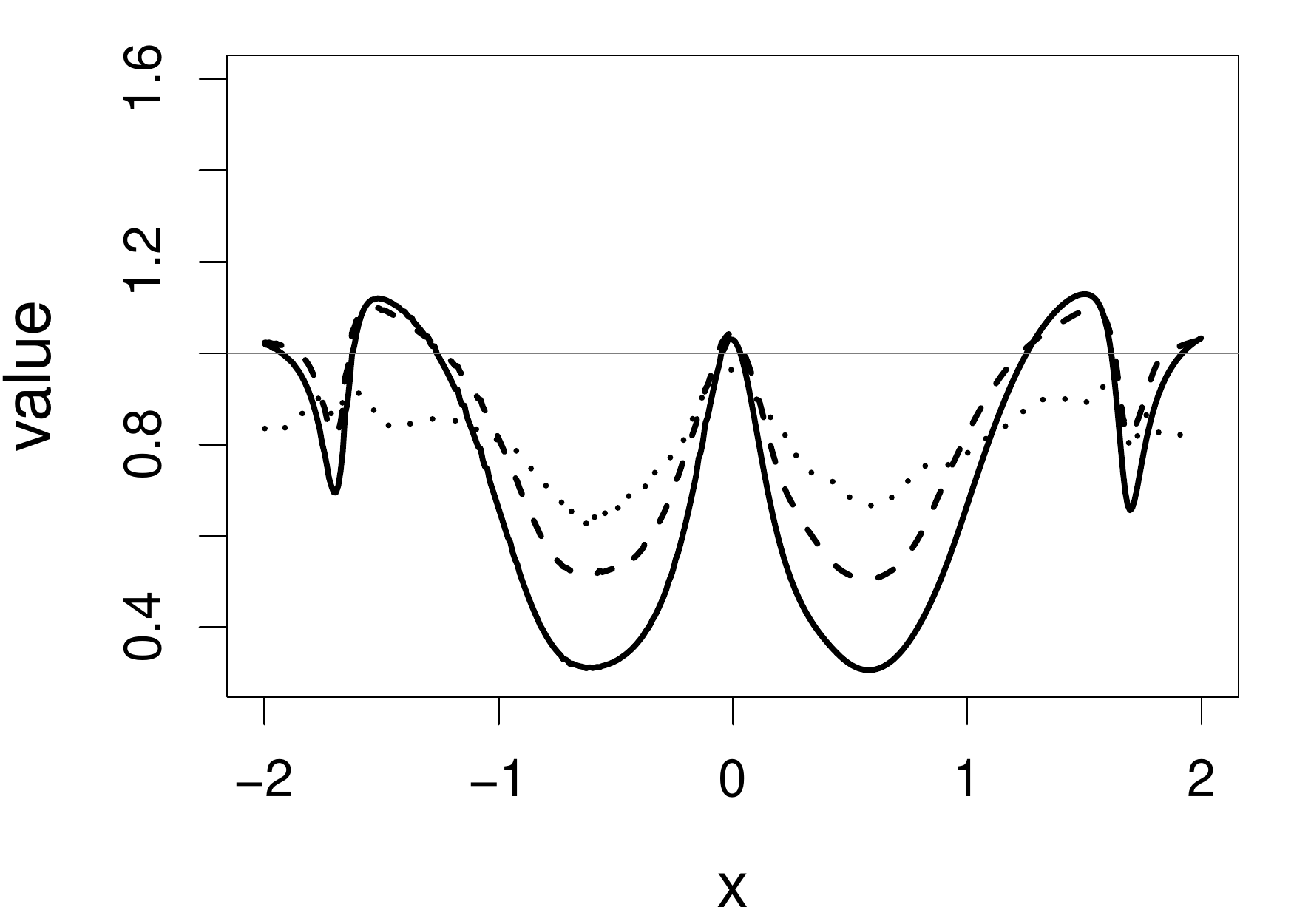} }
	\subfigure[]{ \includegraphics[width=\linewidth]{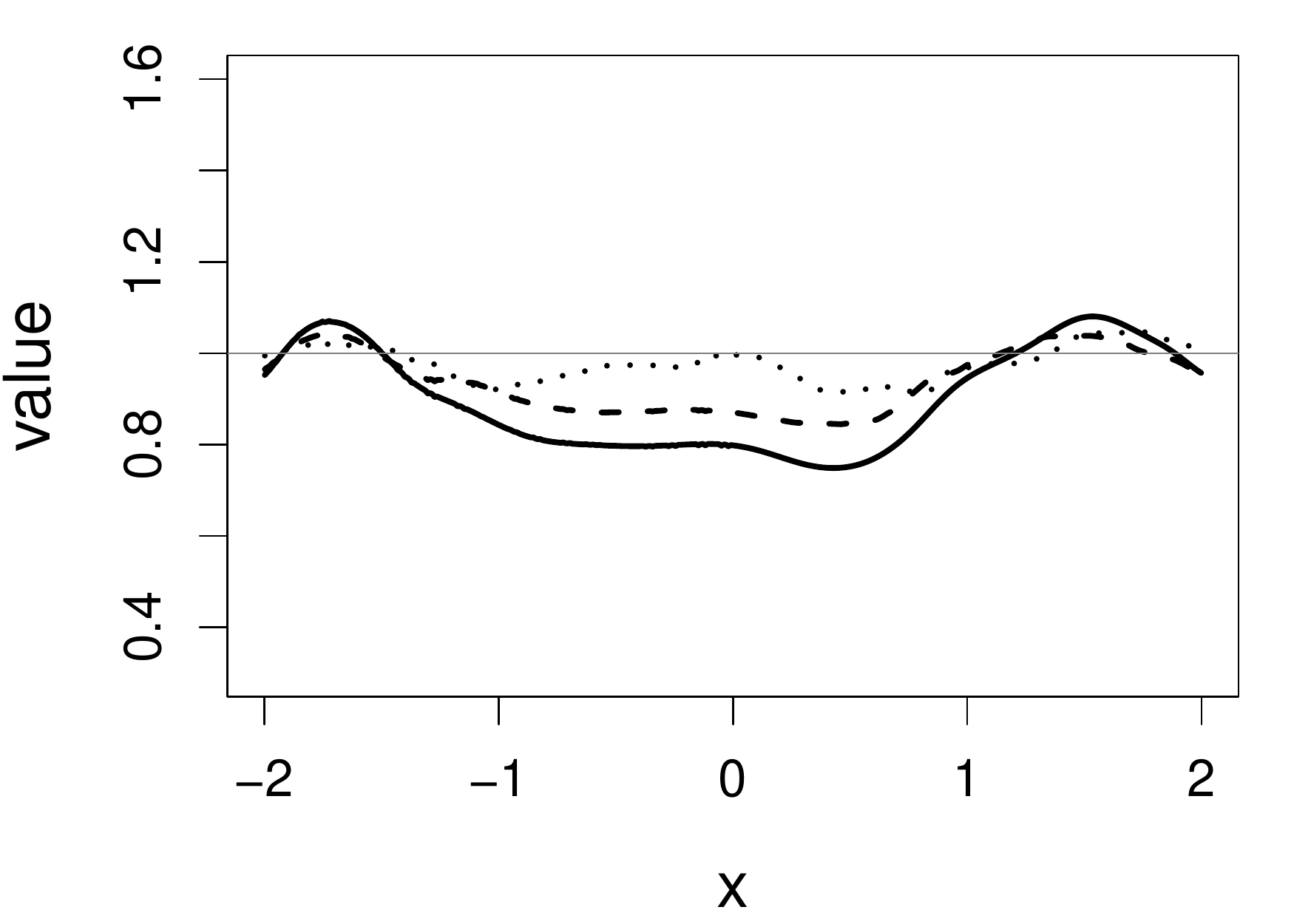} }\\
	\subfigure[]{ \includegraphics[width=\linewidth]{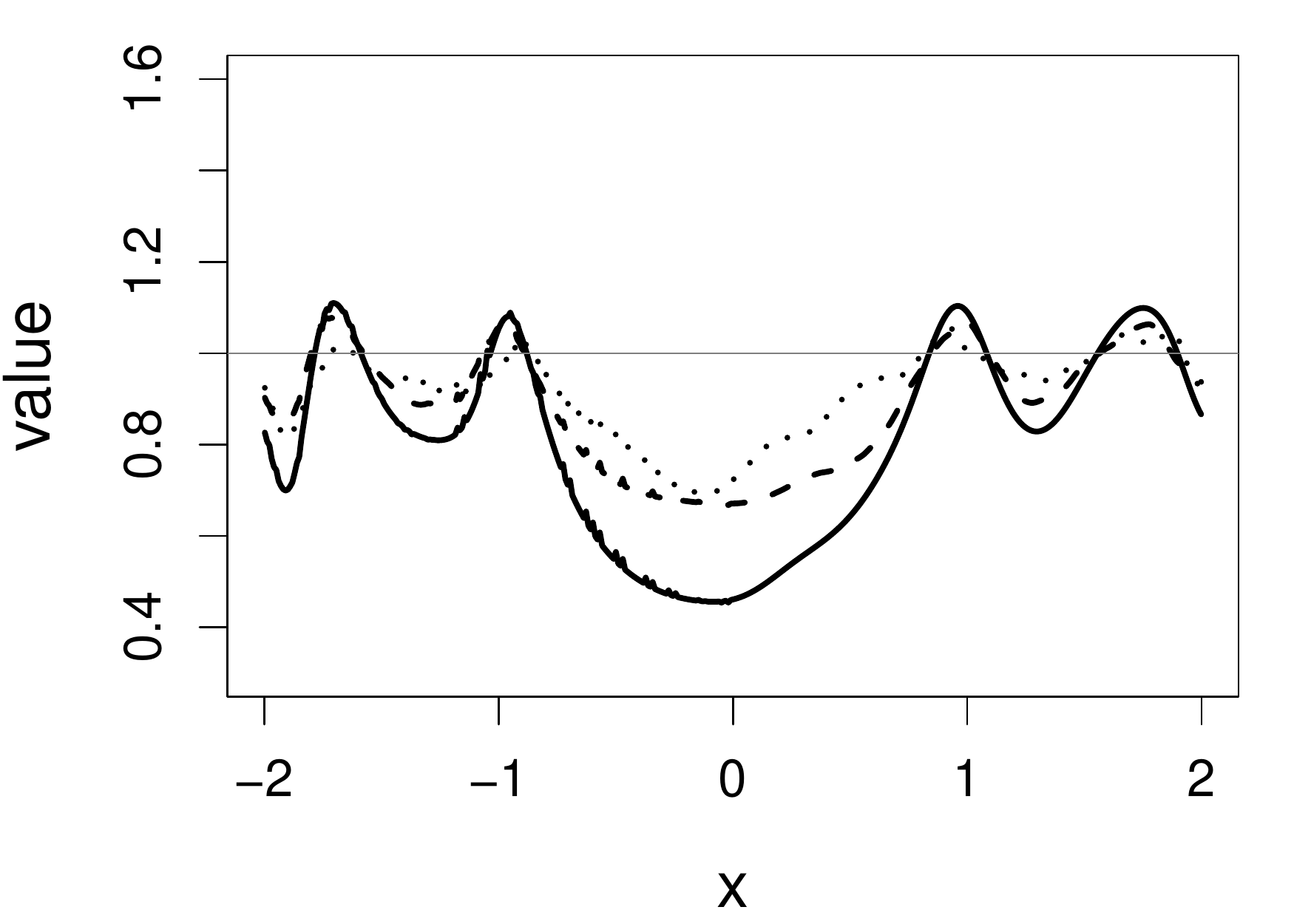} }
	\subfigure[]{ \includegraphics[width=\linewidth]{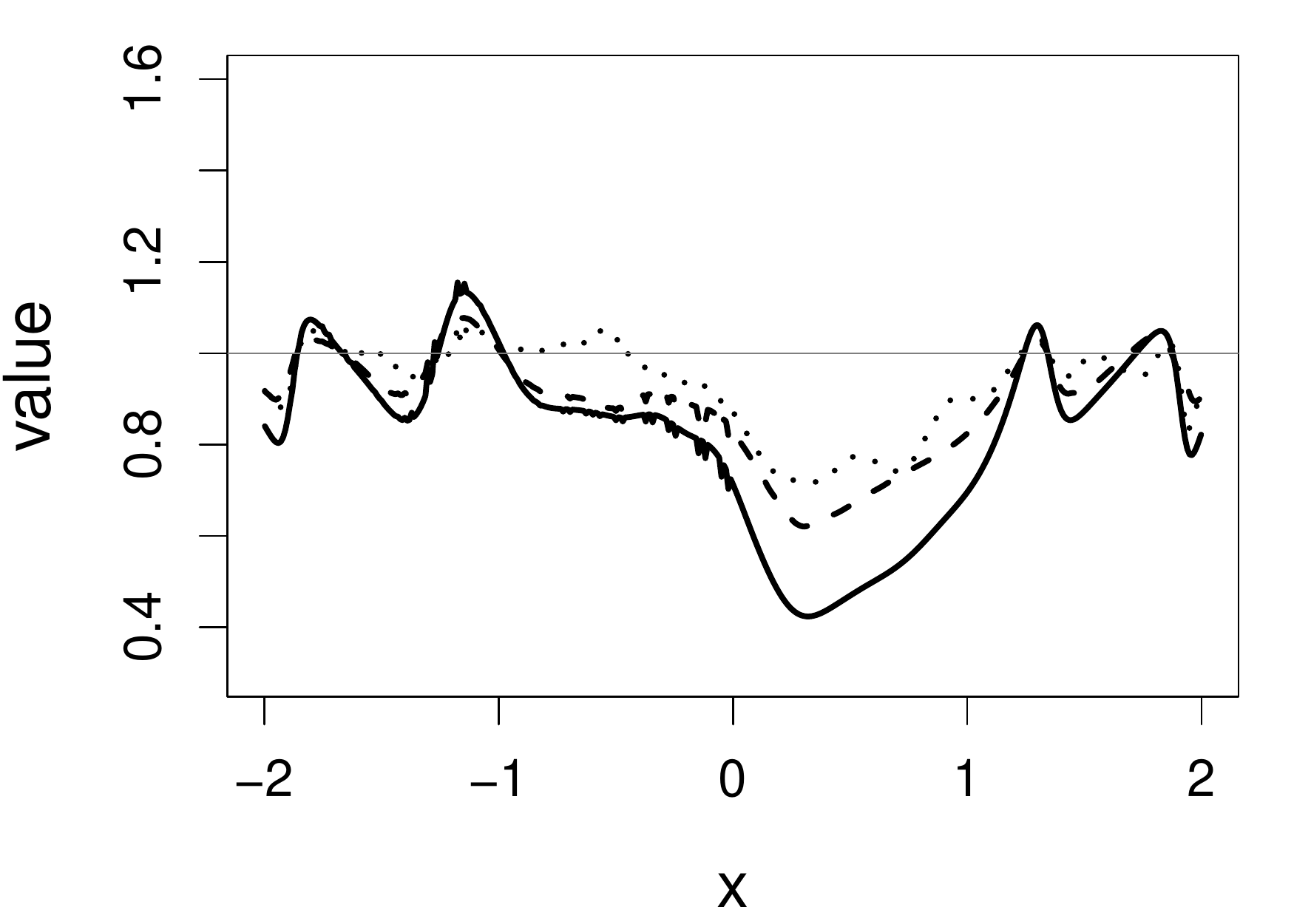} }
\caption{Comparison between $\hat m_{\hbox {\tiny HZ}}(x)$ with $U$-distribution assumed to be Laplace (with $\sigma^2_u$ estimated) and $\hat m_{\hbox {\tiny HZ}}(x)$ with $\phi_{\hbox {\tiny $U$}}(t)$ estimated using repeated measures. Plotted quantities are three ratios, PmAER (dashed line), PsdAER (dotted line), and PMSER (solid line), versus $x$ when $\lambda=0.85$ under (C1) (panel (a)), (C2) (panel (b)), (C3) (panel (c)), and (C4) (panel (d)), respectively. The horizontal reference line highlights the value 1.}
	\label{f:compareours}
	\end{figure}

Alternatively, one may follow the approach proposed by \citet{DHM2008} to estimate $\phi_{\hbox {\tiny $U$}}(t)$ when repeated measures are available, which we also implement in the four cases considered in Section~\ref{s:4cases} using the aforementioned simulated repeated measures. Although this approach frees one from assuming a specific distribution for $U$ and estimating $\sigma^2_u$, the resultant estimates are mostly inferior to the estimates resulting from an assumed Laplace $U$ with $\sigma^2_u$ estimated. Figure~\ref{f:compareours} shows the comparison between these two treatments of $\phi_{\hbox {\tiny $U$}}(t)$ in our proposed estimator in regard to bias, variability, and MSE. In the three ratios, PmAER, PsdAER, and PMSER, depicted in Figure~\ref{f:compareours}, the estimate in the numerators is our estimate assuming Laplace $U$ with an estimated $\sigma_u^2$, and the estimate in the denominator is our estimate with the estimated $\phi_{\hbox {\tiny $U$}}(t)$. The comparison clearly shows that there is no gain from estimating $\phi_{\hbox {\tiny $U$}}(t)$ instead of simply assuming a Laplace $U$ with $\sigma_u^2$ estimated. Obviously, neither $\sigma^2_u$ nor $\phi_{\hbox {\tiny $U$}}(t)$ is identifiable when one does not have repeated measures or other forms of external data that allow one estimate the measurement error distribution. In this case, one can carry out sensitivity analysis with $\sigma^2_u$ varying over a range of practical interest. 

\section*{Supplemental materials}
The supplement to this article contains Appendices A--D referenced in Sections~\ref{s:bias}, \ref{s:variance}, \ref{s:normality}, and \ref{s:discussion}.

\section*{Acknowledgments}
The authors express sincere thanks to the Editor, the Associate Editor and an anonymous referee for their constructive and valuable suggestions on this article, which have led to significant improvements.

\section*{Disclosure statement}
No potential conflict of interest was reported by the authors.

\section*{Funding}
The first author gratefully acknowledges support from the NSF under grant number DMS-1006222.

\end{document}


\title{An Alternative Local Polynomial Estimator for the Error-in-Variables Problem\\Supplementary Materials}
\author{
\name{Xianzheng Huang\textsuperscript{a}$^{\ast}$\thanks{$^\ast$Corresponding author. Email: huang@stat.sc.edu}
and Haiming Zhou\textsuperscript{b}}
\affil{\textsuperscript{a}Department of Statistics, University of South Carolina, Columbia, South Carolina, U.S.A.;
\textsuperscript{b}Division of Statistics, Northern Illinois University, DeKalb, Illinois, U.S.A.}
\received{v4.0 released June 2015}
}

\date{}
\maketitle 

\noindent 
\setcounter{equation}{0}
\setcounter{figure}{0}
\renewcommand{\theequation}{A.\arabic{equation}}
\renewcommand{\thefigure}{A.\arabic{figure}}
\renewcommand{\thesection}{A.\arabic{section}}

\noindent 
\setcounter{equation}{0}
\setcounter{figure}{0}
\renewcommand{\theequation}{A.\arabic{equation}}
\renewcommand{\thefigure}{A.\arabic{figure}}
\renewcommand{\thesection}{A.\arabic{section}}

\section*{Appendix A: Asymptotic bias of $\hat m_{\hbox {\tiny HZ}}(x)$}
\label{s:appA}
\section{Dominating asymptotic bias of $\hat m_{\hbox {\tiny HZ}}(x)$}
Define $\mathbb{W}=(W_1, \ldots, W_n)$. Under the regularity conditions that guarantees the consistency of $\hat f_{\hbox {\tiny $X$}}(x)$, the dominating terms in $E\{\hat m_{\hbox {\tiny HZ}}(x)-m(x)|\mathbb{W}\}$ are the same as those in 
\begin{equation} 
E[\{\hat m_{\hbox {\tiny HZ}}(x)-m(x)\}\hat f_{\hbox {\tiny $X$}}(x)/f_{\hbox {\tiny $X$}}(x)|\mathbb{W}]=[E\{\mathcal{B}(x)|\mathbb{W} \}-m(x)\hat f_{\hbox {\tiny $X$}}(x)]/f_{\hbox {\tiny $X$}}(x).\label{eq:equiv}
\end{equation}
The majority of the following derivation is to elaborate $E\{\mathcal{B}(x)|\mathbb{W}\}$. 

By the relationship between $\mathcal{B}(x)$ and $\mathcal{A}(w)$ indicated by equation (8) in the main article, we immediate have $E\{\mathcal{B}(x)|\mathbb{W}\}=\{E(\mathcal{A}|\mathbb{W})*D\}(x)$. This motivates us to first look into $E\{\mathcal{A}(w)|\mathbb{W}\}$, that is, 
\begin{equation}
  E\left\{\hat m^* (w) \hat f_{\hbox {\tiny $W$}}(w)|\mathbb{W}\right\} 
 =  E\left\{\hat m^* (w)|\mathbb{W}\right\} \hat f_{\hbox {\tiny $W$}}(w) .\nonumber 
\end{equation}
The following two results for kernel-based estimators in the absence of measurement error can be used to derive $E\{\mathcal{A}(w)|\mathbb{W}\}$. The first result is about the local polynomial estimator of the regression function $m^*(w)$. In particular, by Theorem 3.1 in \citet{Fanbook}, if $f'_{\hbox {\tiny $W$}}(\cdot)$ and $m^{*(p+2)}(\cdot)$ are continuous in a neighborhood of $w$ and $nh\to \infty$, then 
\begin{equation}
E\left\{\hat m^*(w) |\mathbb{W}\right\}  =  m^*(w)+\be_1^\T \bS^{-1} \bc_p \frac{1}{(p+1)!} m^{*(p+1)}(w) h^{p+1} +o_{\hbox {\tiny $P$}}(h^{p+1}) \label{eq:Em*podd}
\end{equation} 
when $p$ is odd, and 
\begin{eqnarray}
E\left\{\hat m^*(w) |\mathbb{W}\right\} & = & m^*(w)+ \be_1^\T \bS^{-1} \tilde \bc_p \frac{1}{(p+2)!} \Big\{m^{*(p+2)}(w)\nonumber \\
& & +(p+2)m^{*(p+1)}(w) \frac{f'_{\hbox {\tiny $W$}}(w)}{f_{\hbox {\tiny $W$}}(w)} \Big\}h^{p+2} +o_{\hbox {\tiny $P$}}(h^{p+2}) \label{eq:Em*peven}
\end{eqnarray} 
when $p$ is even, where $\bS=(\mu_{\ell_1+\ell_2})_{0\le \ell_1, \ell_2\le p}$, $\bc_p=(\mu_{p+1}, \ldots, \mu_{2p+1})^\T$, $\tilde\bc_p=(\mu_{p+2}, \ldots, \mu_{2p+2})^\T$, $\mu_\ell=\int u^\ell K(u)du$, and $\be_1$ is the $(p+1)\times 1$ vector with the first entry being 1 and the remaining $p$ entries being 0. The second result is about the kernel-based density estimator, $\hat f_{\hbox {\tiny $W$}}(w)$. By the definition of $\hat f_{\hbox {\tiny $W$}}(w)$ and using Taylor expansion around $h=0$, one has 
\begin{equation*}
E\left\{\hat f_{\hbox {\tiny $W$}}(w)\right\} = f_{\hbox {\tiny $W$}}(w)+ \sum_{\ell=1}^{p+2} f^{(\ell)}_{\hbox {\tiny $W$}}(w) \mu_\ell h^\ell/\ell!+o(h^{p+2}).  
\end{equation*}
Furthermore, $\textrm{Var}\{\hat f_{\hbox {\tiny $W$}}(w)\}=O\{1/(nh)\}$. It follows that 
\begin{eqnarray}
\hat f_{\hbox {\tiny $W$}}(w) & = & E\left\{\hat f_{\hbox {\tiny $W$}}(w)\right\}+O_{\hbox {\tiny $P$}}\left[\sqrt{\textrm{Var}\{\hat f_{\hbox {\tiny $W$}}(w)\}}\right] \nonumber \\
& = & f_{\hbox {\tiny $W$}}(w)+ \sum_{\ell=1}^{p+2} f^{(\ell)}_{\hbox {\tiny $W$}}(w) \mu_\ell h^\ell/\ell!+o_{\hbox {\tiny $P$}}(h^{p+2})+O_{\hbox{\tiny $P$}}(1/\sqrt{nh}).  
\label{eq:Efw}
\end{eqnarray}

Using the two sets of results in (\ref{eq:Em*podd})--(\ref{eq:Efw}), provided that $nh^{2p+3}\to \infty$ when $p$ is odd and $nh^{2p+5}\to \infty$ when $p$ is even, we have
\begin{equation}
E\left\{\mathcal{A}(w)|\mathbb{W}\right\}=
\left\{
\begin{array}{ll}
A(w)+N_p(w)+M_p(w)h^{p+1}/(p+1)!+o_{\hbox {\tiny $P$}}(h^{p+1}), & \textrm{ if $p$ is odd,}\\
A(w)+N_p(w)+M_p(w)h^{p+2}/(p+2)!+o_{\hbox {\tiny $P$}}(h^{p+2}), & \textrm{ if $p$ is even,}
\end{array}
\right.\nonumber 
\end{equation}
where $N_p(w)=m^*(w)\sum_{\ell=1}^p f_{\hbox {\tiny $W$}}^{(\ell)}(w) \mu_{\ell}h^\ell/\ell!$ and 
\begin{equation}
M_p(w)= m^*(w)f_{\hbox {\tiny $W$}}^{(p+1)}(w) \mu_{p+1}
+ m^{*(p+1)}(w)f_{\hbox {\tiny $W$}}(w)\be_1^\T \bS^{-1}\bc_p,\nonumber 
\end{equation}
if $p$ is odd, and 
\begin{equation}
M_p(w)= m^*(w)f_{\hbox {\tiny $W$}}^{(p+2)}(w) \mu_{p+2} + \left\{ m^{*(p+2)}(w)f_{\hbox {\tiny $W$}}(w) 
+(p+2)m^{*(p+1)}(w) f'_{\hbox {\tiny $W$}}(w)\right\}\be_1^\T \bS^{-1}\tilde\bc_p, \nonumber 
\end{equation}
if $p$ is even.

With $E\{\mathcal{A}(w)|\mathbb{W}\}$ derived, we have 
\begin{eqnarray}
& & E\left\{ \mathcal{B}(x) |\mathbb{W}\right\}  = \left\{ E(\mathcal{A}|\mathbb{W})*D \right\}(x) \nonumber \\
& = & 
\left\{
\begin{array}{ll}
B(x)+(N_p*D)(x)+(M_p*D)(x)h^{p+1}/(p+1)!+o_{\hbox {\tiny $P$}}(h^{p+1}), & \textrm{ if $p$ is odd,}\\
B(x)+(N_p*D)(x)+(M_p*D)(x)h^{p+2}/(p+2)!+o_{\hbox {\tiny $P$}}(h^{p+2}), & \textrm{ if $p$ is even.}
\end{array}
\right.\nonumber \\
\label{eq:EBx}
\end{eqnarray}

For the deconvolution kernel density estimator $\hat f_{\hbox {\tiny $X$}}(x)$, by equation (1.9) in \citet{Stefanski90}, one has
$$E\left\{\hat f_{\hbox {\tiny $X$}}(x)\right\} = f_{\hbox {\tiny $X$}}(x)+ \sum_{\ell=1}^{p+2} f^{(\ell)}_{\hbox {\tiny $X$}}(x) \mu_\ell h^\ell/\ell!+o(h^{p+2}).$$
Under the conditions in Theorem 2.1 in \citet{Stefanski90}, $\textrm{Var}\{\hat f_{\hbox {\tiny $X$}}(x)\}$ is bounded from above by $(nh)^{-1}\sup_{x} f_{\hbox {\tiny $X$}}(x) \int K^2(t)dt$. It follows that 
\begin{eqnarray}
\hat f_{\hbox {\tiny $X$}}(x) & = & E\left\{\hat f_{\hbox {\tiny $X$}}(x)\right\} + O_{\hbox {\tiny $P$}}\left[\sqrt{\textrm{Var}\{\hat f_{\hbox {\tiny $X$}}(x)   \}}\right] \nonumber \\
& = & f_{\hbox {\tiny $X$}}(x)+ \sum_{\ell=1}^{p+2} f^{(\ell)}_{\hbox {\tiny $X$}}(x) \mu_\ell h^\ell/\ell!+o_{\hbox {\tiny $P$}}(h^{p+2})+O_{\hbox {\tiny $P$}}(1/\sqrt{nh}). \nonumber
\end{eqnarray}
Therefore, 
\begin{equation}
m(x)\hat f_{\hbox {\tiny $X$}}(x) = B(x)+ m(x)\sum_{\ell=1}^{p+2} f^{(\ell)}_{\hbox {\tiny $X$}}(x) \mu_\ell h^\ell/\ell!+o_{\hbox {\tiny $P$}}(h^{p+2})+O_{\hbox {\tiny $P$}}(1/\sqrt{nh}). \label{eq:mEfx} 
\end{equation}

By (\ref{eq:EBx}) and (\ref{eq:mEfx}), provided that $nh^{2p+5}\to \infty$ when $p$ is even and $nh^{2p+3}\to \infty$ when $p$ is odd, (\ref{eq:equiv}) is equal to $\{f_{\hbox {\tiny $X$}} (x)\}^{-1}$ times 
\begin{eqnarray}
& & (N_p*D)(x)+\left\{(M_p*D)(x)-m(x) f^{(p+1)}_{\hbox {\tiny $X$}} (x) \mu_{p+1}\right\}h^{p+1}/(p+1)!\nonumber \\
& - & m(x)\sum_{\ell=1}^p f^{(\ell)}_{\hbox {\tiny $X$}} (x)\mu_\ell h^\ell/\ell!+o_{\hbox {\tiny $P$}}(h^{p+1}) \nonumber 
\end{eqnarray}
when $p$ is odd, and 
\begin{eqnarray}
& & (N_p*D)(x)+\left\{(M_p*D)(x)-m(x) f^{(p+2)}_{\hbox {\tiny $X$}} (x) \mu_{p+2}\right\}h^{p+2}/(p+2)!\nonumber \\
& - & m(x)\sum_{\ell=1}^p f^{(\ell)}_{\hbox {\tiny $X$}} (x)\mu_\ell h^\ell/\ell!+o_{\hbox {\tiny $P$}}(h^{p+2}) \nonumber 
\end{eqnarray}
when $p$ is even. This gives the dominating bias for $\hat m_{\hbox {\tiny HZ}}(x)$ of order $h^{p+1}$ when $p$ is odd and that of order $h^{p+2}$ when $p$ is even. It is worth noting that, although the derivation of the asymptotic bias of $\hat m_{\hbox {\tiny HZ}}(x)$ is conditional on $\mathbb{W}$, the leading terms of the asymptotic bias do not depend on $\mathbb{W}$, and thus these leading terms can be interpreted as the unconditional dominating asymptotic bias. This is in line with the remarks in \citet[][Remark 1 on page 1351]{Ruppert94} regarding their asymptotic bias and variance of the nonparametric estimator of $m(x)$ with $\mathbb{X}=(X_1, \ldots, X_n)$ observed.  

\section{Dominating asymptotic bias of $\hat m_{\hbox {\tiny HZ}}(x)$ when $m(x)$ is a polynomial}
Under the assumptions that $m(x)=\sum_{k=0}^r \beta_k x^k$, where $r\ge 2$, $X\sim N(0, \, 1)$, $U\sim N(0, \, \sigma_u^2)$, and $X\perp U$, we show in this section that
\begin{equation}
f^{-1}_{\hbox {\tiny $X$}}(x)\left\{(M*D)(x)-m(x)f^{(2)}_{\hbox {\tiny $X$}}(x) \right\}, 
\label{eq:key}
\end{equation}
is a polynomial of order $r$. For notational brevity, $\textrm{Low}_k(t)$ is used in the sequel to stand for a generic polynomial in $t$ of order lower than $k$, for $k > 0$. 

With $f_{\hbox {\tiny $X$}}(x)=\exp(-x^2/2)/\sqrt{2\pi}$, straightforward induction reveals that, 
\begin{equation}
f^{(k)}_{\hbox {\tiny $X$}}(x)=f_{\hbox {\tiny $X$}}(x)\left\{(-1)^k x^k+(-1)^{k-1} {{k}\choose{2}} x^{k-2}+\textrm{Low}_{k-2}(x)\right\}.
\label{eq:fkder}
\end{equation}
It follows that
$$m(x)f^{(2)}_{\hbox {\tiny $X$}}(x)=f_{\hbox {\tiny $X$}}(x)\left\{\beta_r x^{r+2}+\beta_{r-1}x^{r+1}+(\beta_{r-2}-\beta_r)x^r+\textrm{Low}_r(x)\right\}.$$ This solves half of the ``mystery" in (\ref{eq:key}), that is, we have 
\begin{equation}
f^{(-1)}_{\hbox {\tiny $X$}}(x)m(x)f^{(2)}_{\hbox {\tiny $X$}}(x)=\beta_r x^{r+2}+\beta_{r-1}x^{r+1}+(\beta_{r-2}-\beta_r)x^r+\textrm{Low}_r(x).
\label{eq:secondhalf}
\end{equation}
The other half of the mystery is about $f^{-1}_{\hbox {\tiny $X$}}(x)(M*D)(x)$ in (\ref{eq:key}). In what follows, we will show that this half is equal to 
\begin{equation}
\beta_r x^{r+2}+\beta_{r-1}x^{r+1}+\left[\beta_{r-2}+\beta_r \left\{2r(\lambda-1)-1\right\}\right]x^r+\textrm{Low}_r(x), 
\label{eq:firsthalf}
\end{equation}
where $\lambda=1/(1+\sigma^2_u)$. Once this is established, subtracting (\ref{eq:secondhalf}) from (\ref{eq:firsthalf}) reveals that (\ref{eq:key}) is equal to $2r(\lambda-1)\beta_r x^r +\textrm{Low}_r(x)$, i.e., a polynomial of order $r$ as long as $\lambda \ne 1$. 

Recall that $M(w)=m^*(w)f^{(2)}_{\hbox {\tiny $W$}}(w)+m^{*(2)}(w)f_{\hbox {\tiny $W$}}(w)$, which involves $f_{\hbox {\tiny $W$}}(w)$ and  $m^*(w)$. 
Because $X\sim N(0, \, 1)$ is independent of $U\sim N(0, \, \sigma_u^2)$, one has $f_{\hbox {\tiny $W$}}(w)=(1+\sigma_u^2)^{-1/2} \phi(w/\sqrt{1+\sigma_u^2})$, and thus $f^{(2)}_{\hbox {\tiny $W$}}(w)=\lambda f_{\hbox {\tiny $W$}}(w) (\lambda w^2-1)$, where $\phi(\cdot)$ denotes the pdf of the standard normal. Given the current $m(x)$, the naive regression function $m^*(w)$ is equal to 
\begin{eqnarray}
E(Y|W=w) & = & E\{E(Y|X)|W=w\} \nonumber \\
& = & f^{-1}_{\hbox {\tiny $W$}}(w) \int m(x) f_{\hbox {\tiny $X$}}(x) f_{\hbox {\tiny $U$}}(w-x)dx \nonumber\\
& = & f^{-1}_{\hbox {\tiny $W$}}(w) \sum_{k=0}^r \beta_k \int x^k \phi(x) \sigma_u^{-1} \phi\{(w-x)/\sigma_u\}dx \nonumber\\
& = & f^{-1}_{\hbox {\tiny $W$}}(w) f_{\hbox {\tiny $W$}}(w)\sum_{k=0}^r \beta_k \int x^k \frac{\sqrt{1+\sigma_u^2}}{\sigma_u}\phi\left\{\frac{\sqrt{1+\sigma_u^2}}{\sigma_u}\left(x-\frac{w}{1+\sigma^2_u}\right)\right\} dx \nonumber\\
& = & \sum_{k=0}^r \beta_k \times \textrm{the $k$th moment of $N(\lambda w, \, 1-\lambda)$} \nonumber\\
& = & \sum_{k=0}^r \beta_k \lambda^k \sum_{\ell=0}^{\floor{k/2}}{k\choose{2\ell}}(2\ell-1)!!(1-\lambda)^\ell\lambda^{-2\ell}w^{k-2\ell}, 
\label{eq:examplemw}
\end{eqnarray}
where $!!$ is the double factorial symbol, with $(-1)!!$ defined to be 1. It follows that 
$$m^{*(2)}(w) = \sum_{k=0}^r \beta_k \lambda^k \sum_{\ell=0}^{\floor{k/2-1}}{k\choose{2\ell}}(2\ell-1)!!(1-\lambda)^\ell\lambda^{-2\ell}(k-2 \ell)(k-2\ell-1)w^{k-2\ell-2}.$$
Putting $f_{\hbox {\tiny $W$}}(w)$, $f^{(2)}_{\hbox {\tiny $W$}}(w)$, $m^*(w)$, and $m^{*(2)}(w)$ back in $M(w)$, one can see that $M(w)$ is equal to $f_{\hbox {\tiny $W$}}(w)$ times a polynomial in $w$ of order $r+2$. Hence, the key to deriving 
\begin{equation}
(M*D)(x)=\frac{1}{2\pi}\int e^{-itx} \frac{\phi_{M}(t)}{\phi_{\hbox {\tiny $U$}}(t)}dt, 
\label{eq:MDx}
\end{equation}
is to understand, for $k\ge 0$, $(2\pi)^{-1}\int e^{-itx} \phi_{w^k f_{\hbox {\tiny $W$}}(w)}(t)/\phi_{\hbox {\tiny $U$}}(t)dt$.

It is straightforward to show that $\phi_{w^k f_{\hbox {\tiny $W$}}}(t)=i^{-k}\phi^{(k)}_{\hbox {\tiny $W$}}(t)$, where $\phi_{\hbox {\tiny $W$}}(t)$ is the characteristic function of $W$. With a normal $W$, using induction one can show that 
\begin{equation}
\phi^{(k)}_{\hbox {\tiny $W$}}(t)=\{(-1)^k \lambda^{-k}t^k+\textrm{Low}_{k-1}(t)\}\phi_{\hbox {\tiny $W$}}(t).
\label{eq:phiwk}
\end{equation}
Noting that $\phi_{\hbox {\tiny $W$}}(t)=\phi_{\hbox {\tiny $X$}}(t)\phi_{\hbox {\tiny $U$}}(t)$, and using the result that, for $k=0, 1, \ldots$, 
\begin{equation}
(-i)^k (2\pi)^{-1}\int e^{-itx} t^k \phi_{\hbox {\tiny $X$}}(t) dt=f^{(k)}_{\hbox {\tiny $X$}}(x),
\label{eq:tofxkder}
\end{equation}
we now have
\begin{eqnarray}
&  & (2\pi)^{-1}\int e^{-itx} \phi_{w^k f_{\hbox {\tiny $W$}}(w)}(t)/\phi_{\hbox {\tiny $U$}}(t)dt \nonumber \\
& = & i^{-k}(2\pi)^{-1}\int e^{-itx} \{(-1)^k \lambda^{-k}t^k+\textrm{Low}_{k-1}(t)\}\phi_{\hbox {\tiny $W$}}(t)/\phi_{\hbox {\tiny $U$}}(t)dt  \nonumber \\
& = & i^{-k}(-1)^k \lambda^{-k} (-i)^{-k} (-i)^k (2\pi)^{-1}\int e^{-itx} t^k \phi_{\hbox {\tiny $X$}} (t) dt+
\int e^{-itx }\textrm{Low}_{k-1}(t) \phi_{\hbox {\tiny $X$}}(t) dt \nonumber \\
& = & (-1)^k \lambda^{-k} f^{(k)}_{\hbox {\tiny $X$}}(x) + \textrm{ (some coefficient free of $x$)} \times f^{(k-2)}_{\hbox {\tiny $X$}}(x), \textrm{ by (\ref{eq:tofxkder})} \nonumber \\
& = & (-1)^k \lambda^{-k} (-1)^k x^k f_{\hbox {\tiny $X$}}(x) + \textrm{Low}_{k-1}(x)f_{\hbox {\tiny $X$}}(x), \textrm{ by (\ref{eq:fkder})},\nonumber \\ 
& = & \lambda^{-k} x^k f_{\hbox {\tiny $X$}}(x) + \textrm{Low}_{k-1}(x)f_{\hbox {\tiny $X$}}(x). \label{eq:wkfwD}
\end{eqnarray}

We next focus on the terms in $M(w)$ with the two highest powers of $w$. Tracing the coefficients of $w^{r+2}$ and $w^{r+1}$ in $M(w)$ and applying (\ref{eq:wkfwD}) for $k=r+2$ and $r+1$, one can see that (\ref{eq:MDx}) is equal to 
\begin{eqnarray}
& & \lambda^{r+2} \beta_r \lambda^{-(r+2)}x^{r+2}f_{\hbox {\tiny $X$}}(x)+\lambda^{r+1} \beta_{r-1} \lambda^{-(r+1)}x^{r+1}f_{\hbox {\tiny $X$}}(x)+\textrm{Low}_{r+1}(x)f_{\hbox {\tiny $X$}}(x) \nonumber \\
& = & \{\beta_r x^{r+2} + \beta_{r-1} x^{r+1}+\textrm{Low}_{r+1}(x)\}f_{\hbox {\tiny $X$}}(x), \nonumber
\end{eqnarray}
which proves the first two terms in (\ref{eq:firsthalf}). 
 
To prove the third term in (\ref{eq:firsthalf}), we first find the term in $M(w)$ of the third highest order in $w$, i.e., $w^r$, because this term leads to a term with $x^r$ in $(M*D)(x)$ according to (\ref{eq:wkfwD}). One such term shows up in $m^*(w)f^{(2)}_{\hbox {\tiny $W$}}$ is $\lambda^r \{\beta_{r-2}-\beta_r \lambda+\beta_r (1-\lambda) {r \choose 2}\}w^r f_{\hbox {\tiny $W$}}(w)$. Convoluting this term with $D(\cdot)$ yields 
\begin{equation}
\left\{\beta_{r-2}-\beta_r \lambda+\beta_r (1-\lambda) {r \choose 2}\right\}x^r f_{\hbox {\tiny $X$}}(x).
\label{eq:firstxr}
\end{equation}
Secondly, note that $m^*(w)f^{(2)}_{\hbox {\tiny $W$}}$ contains $\lambda^{r+2}\beta_r w^{r+2}f_{\hbox {\tiny $W$}}(w)$, which, after convoluting with $D(\cdot)$ produces $f_{\hbox {\tiny $X$}}^{(r+2)}(x)$, which itself contributes $x^r$ according to (\ref{eq:fkder}). More specifically, this term is 
\begin{equation}
-\beta_r {{r+2} \choose 2} x^r f_{\hbox {\tiny $X$}}(x).
\label{eq:secondxr}
\end{equation}
Thirdly, also due to the involvement of $\lambda^{r+2}\beta_r w^{r+2}f_{\hbox {\tiny $W$}}(w)$ in $M(w)$, which, when convoluting with $D(\cdot)$, yields $\phi_{\hbox {\tiny $W$}}^{(r+2)}(t)$, which contains a terms with $t^r$ according to (\ref{eq:phiwk}). This eventually translates to 
\begin{equation}
\lambda\beta_r x^r {{r+2} \choose 2}f_{\hbox {\tiny $X$}}(x).
\label{eq:thirdxr}
\end{equation}
Summing (\ref{eq:firstxr}), (\ref{eq:secondxr}), and (\ref{eq:thirdxr}) together gives the third term in (\ref{eq:firsthalf}). 

Now (\ref{eq:firsthalf}) is established, and combining it with (\ref{eq:secondhalf}), we show that (\ref{eq:key}) is equal to a polynomial in $x$ of order $r$. 

In the case with $r=2$, we have 
\begin{eqnarray}
\phi_{m^*f^{(2)}_{\hbox {\tiny $W$}}}(t)& = &\{\beta_2 t^4-i \beta_1 t^3-(\beta_0+\beta_2+4\lambda \beta_2)t^2+i 2\lambda \beta_1 + 2 \lambda^2 \beta_2 \}\phi_{\hbox {\tiny $W$}}(t), \nonumber \\
\phi_{m^{*(2)}f_{\hbox {\tiny $W$}}}(t) & = & 2 \lambda^2 \beta_2 \phi_{\hbox {\tiny $W$}}(t). \nonumber 
\end{eqnarray}
It follows that (\ref{eq:MDx}) is equal to the inverse Fourier transform of 
$\{\beta_2 t^4-i \beta_1 t^3-(\beta_0+\beta_2+4\lambda \beta_2)t^2+i 2\lambda \beta_1 + 4 \lambda^2 \beta_2 \}\phi_{\hbox {\tiny $X$}}(t).$
Using (\ref{eq:tofxkder}), one can show that this inverse Fourier transform is equal to 
\begin{eqnarray}
& & \beta_2 f^{(4)}_{\hbox {\tiny $X$}}(x)-\beta_1 f^{(3)}_{\hbox {\tiny $X$}}(x)+(\beta_0+\beta_2+4\lambda \beta_2)f^{(2)}_{\hbox {\tiny $X$}}(x)-2\lambda \beta_1 f'_{\hbox {\tiny $X$}}(x)+4\lambda^2 \beta_2 f_{\hbox {\tiny $X$}}(x) \nonumber \\
& = & \{   \beta_2 x^4 +\beta_1 x^3 +(\beta_0-5\beta_2+4\lambda \beta_2) x^2 +(2\lambda-3)\beta_1 x+ \nonumber \\
& & 4 \lambda^2 \beta_2-4\lambda \beta_2-\beta_0+2\beta_2 \}f_{\hbox {\tiny $X$}}(x). \label{eq:lhs}
\end{eqnarray}
Subtracting $m(x) f^{(2)}_{\hbox {\tiny $X$}}(x)=f_{\hbox {\tiny $X$}}(x)(x^2-1)(\beta_0+\beta_1x+\beta_2x^2)$ from (\ref{eq:lhs}) reveals that (\ref{eq:key}) reduces to $2\{(\lambda-1)(\beta_1+2\beta_2x)x+\beta_2(2\lambda^2-2\lambda+1)\}$.

\noindent 
\setcounter{section}{0}
\setcounter{equation}{0}
\setcounter{figure}{0}
\renewcommand{\theequation}{B.\arabic{equation}}
\renewcommand{\thefigure}{B.\arabic{figure}}
\renewcommand{\thesection}{B.\arabic{section}}

\section*{Appendix B: Detailed derivations for $\textrm{Var}\{\mathcal{B}(x)|\mathbb{W}\}$}
\label{s:appB}
The five steps of the road map outlined in Section 4.1 in the main article are elaborated in this section. 
\section{Step 1: Relating $\textrm{Var}\{\mathcal{B}(x)|\mathbb{W}\}$ to $\textrm{Cov}\{\mathcal{A}(w_1), \, \mathcal{A}(w_1)|\mathbb{W}\}$}
\hspace{1.7pc}Assuming interchangeability of expectation and integration, one has
\begin{equation}
E\{\mathcal{B}^2(x)|\mathbb{W}\}  =  E\left\{\int D(v_1) \mathcal{A}(x-v_1) \mathcal{B}(x) dv_1|\mathbb{W}\right\} 
 =  \int D(v_1) E\left\{\mathcal{A}(x-v_1) \mathcal{B}(x)|\mathbb{W} \right\} dv_1, \nonumber 
\end{equation}
where 
\begin{eqnarray}
E\left\{\mathcal{A}(x-v_1) \mathcal{B}(x)|\mathbb{W}\right\} & = & 
E\left[\mathcal{A}(x-v_1) \int \mathcal{A}(x-v_2)D(v_2) dv_2|\mathbb{W}\right] \nonumber \\
& = & \int D(v_2) E\left\{\mathcal{A}(x-v_1) \mathcal{A}(x-v_2) |\mathbb{W}\right\} dv_2. \nonumber
\end{eqnarray}
Thus
\begin{equation}
E\{\mathcal{B}^2(x)|\mathbb{W}\}= \int D(v_1) \int D(v_2) E\left\{\mathcal{A}(x-v_1) \mathcal{A}(x-v_2)|\mathbb{W} \right\} dv_2 dv_1. \label{eq:first}
\end{equation}
In addition,
\begin{eqnarray}
\left[E\{\mathcal{B}(x)|\mathbb{W}\}\right]^2 
& = & \int D(v_1) E\{\mathcal{A}(x-v_1)|\mathbb{W}\} dv_1 \int D(v_2) E\{\mathcal{A}(x-v_2)|\mathbb{W}\}dv_2 \nonumber \\
& = & \int D(v_1) \int D(v_2) E\{\mathcal{A}(x-v_1)|\mathbb{W}\} E\{\mathcal{A}(x-v_2)|\mathbb{W}\} dv_2 dv_1. \label{eq:second}
\end{eqnarray}

Subtracting (\ref{eq:second}) from (\ref{eq:first}) gives
\begin{eqnarray}
\textrm{Var}\{\mathcal{B}(x)|\mathbb{W}\} 
& = & \int D(v_1) \int D(v_2) \textrm{Cov}\left\{\mathcal{A}(x-v_1), \mathcal{A}(x-v_2) |\mathbb{W}\right\} dv_2 dv_1 \nonumber \\
& = &  \int D(x-w_1) \int D(x-w_2) \textrm{Cov}\left\{\mathcal{A}(w_1), \mathcal{A}(w_2)|\mathbb{W} \right\} dw_2 dw_1, \label{eq:varB}
\end{eqnarray}
where 
\begin{equation}
\textrm{Cov}\left\{\mathcal{A}(w_1), \, \mathcal{A}(w_2) |\mathbb{W}\right\} 
 = \textrm{Cov}\{\hat m^*(w_1), \,  \hat m^*(w_2)|\mathbb{W}\}f_{\hbox {\tiny $W$}}(w_1)f_{\hbox {\tiny $W$}}(w_2)\{1+o_{\hbox {\tiny $P$}}(1)\}. 
\label{eq:covAA}
\end{equation}
The next two steps are devoted to deriving $\textrm{Cov}\{\hat m^*(w_1), \, \hat m^*(w_2)|\mathbb{W} \}$. 

\section{Step2: Approximating $\textrm{Cov}\{\hat m^*(w_1), \, \hat m^*(w_2)|\mathbb{W}\}$}
Naive estimation of $m(x)$ based on error-contaminated data, $\{(Y_j, W_j)\}_{j=1}^n$, entails implementing the weighted least squares estimation in \citet[][Section 3.1]{Fanbook} with $X_j$ and $x$ there replaced by $W_j$ and $w$, respectively, for $j=1, \ldots, n$. In particular, one may consider the naive regression, $Y_j=m^*(W_j)+\nu(W_j)\epsilon^*_j$, where $E(\epsilon^*)=0$, $\textrm{Var}(\epsilon^*)=1$, and $W_j$ and $\epsilon^*_j$ are independent. Then a set of estimators of $m^{*(\ell)}(w_k)$, for $k=1, 2$, and $\ell=0, 1, \ldots, p$, can be obtained by minimizing the following weighted sum of squares, 
\begin{equation}
\sum_{j=1}^n \left\{Y_j-\sum_{\ell=0}^p \beta^*_{\ell k} (W_j-w_k)^\ell \right\}^2 K_h(W_j-w_k),
\label{eq:objfun}
\end{equation}
where $\beta^*_{\ell k}=m^{*(\ell)}(w_k)/\ell !$, for $\ell=0, \ldots, p$, and $K_h(t)=K(t/h)/h$. Denote by $\bbeta^*_k=(\beta^*_{0k}, \beta^*_{1k}, \ldots, \beta^*_{pk})^\T$ and by $\hat\bbeta^*_k$ the minimizer of (\ref{eq:objfun}), for $k=1, 2$. Then $\textrm{Cov}\{\hat m^*(w_1),\, \hat m^*(w_2)|\mathbb{W}\}$ is the $[1, 1]$ element of the $(p+1)\times (p+1)$ variance-covariance matrix $\textrm{Cov}(\hat\bbeta^*_1, \hat\bbeta^*_2|\mathbb{W})$. 

As in equation (3.5) in \citet{Fanbook}, the minimizer of (\ref{eq:objfun}) is, for $k=1, 2$, $\hat \bbeta_k^*=(\bG_k^\T \bW_k \bG_k)^{-1} \bG_k^\T \bW_k \bY$, where $\bW_k = \diag \{ K_h(W_1-w_k), \ldots, K_h(W_n-w_k) \}$ and 
\begin{equation}
\bG_k = 
\begin{bmatrix}
1 & (W_1-w_k) & \ldots & (W_1-w_k)^p \cr 
\vdots & \vdots & \ldots & \vdots \cr
1 & (W_n-w_k) & \ldots & (W_n-w_k)^p
\end{bmatrix}. \nonumber 
\end{equation}
It follows that, since $\textrm{Var}(Y|W=w)=\nu^2(w)$ under the naive regression, 
\begin{equation}
\textrm{Cov}(\hat \bbeta^*_1, \, \hat \bbeta^*_2|\mathbb{W})=\left\{\bS^{(1)}_{n\hbox {\tiny $W$}} \right\}^{-1} \bS^*_{n\hbox {\tiny $W$}}\left\{\bS^{(2)}_{n\hbox {\tiny $W$}} \right\}^{-1}.
\label{eq:ourcov2}
\end{equation}
where $\bS^{(k)}_{n\hbox {\tiny $W$}} = \bG_k^\T \bW_k \bG_k = (S^{(k)}_{n\hbox {\tiny $W$}, \ell_1+\ell_2})_{0\le \ell_1, \ell_2\le p}$,  for $k=1, 2$,
in which, for $\ell=0, 1, \ldots, 2p$,
$S^{(k)}_{n\hbox{\tiny $W$}, \ell} = \sum_{j=1}^n K_h(W_j-w_k)(W_j-w_k)^\ell$; and 
\begin{equation}
\bS^*_{n\hbox {\tiny $W$}} = \bG_1^\T \bSigma_{12} \bG_2 = (S^*_{n\hbox {\tiny $W$}, \ell_1,\ell_2})_{0\le \ell_1, \ell_2\le p}, 
\label{eq:Sn*}
\end{equation}
in which, $\bSigma_{12}=\diag\{K_h(W_1-w_1)K_h(W_1-w_2)\nu^2(W_1), \ldots, K_h(W_n-w_1)K_h(W_n-w_2)\nu^2(W_n) \}$, and, for $\ell_1, \ell_2=0, 1, \ldots, p$, 
\begin{equation}
S^*_{n\hbox {\tiny $W$}, \ell_1,\ell_2} = \sum_{j=1}^n (W_j-w_1)^{\ell_1}(W_j-w_2)^{\ell_2}K_h(W_j-w_1)K_h(W_j-w_2)\nu^2(W_j). 
\label{eq:Snl*} 
\end{equation}
Finally, extracting the $[1, 1]$ element of (\ref{eq:ourcov2}) gives $\textrm{Cov}\{\hat m^*(w_1), \, \hat m^*(w_2)|\mathbb{W}\}$.

Now, to derive a large-sample approximation of $\textrm{Cov}\{\hat m^*(w_1), \, \hat m^*(w_2)|\mathbb{W}\}$, we need to approximate $\bS^{(k)}_{n \hbox {\tiny $W$}}$, for $k=1, 2$, and $\bS^*_{n\hbox {\tiny $W$}}$. Both approximations follow the same spirit as those in \citet[][page 101]{Fanbook} that lead to their (3.54) and (3.55). 

\subsection{Approximate $\bS^{(k)}_{n \hbox {\tiny $W$}}$}
For $k=1, 2$ and $\ell=0, 1, \ldots, 2p$,
\begin{eqnarray}
S^{(k)}_{n\hbox{\tiny $W$}, \ell} & = & E\left\{S^{(k)}_{n\hbox{\tiny $W$}, \ell}\right\}+O_{\hbox {\tiny $P$}} \left[\sqrt{\textrm{Var} \left\{ S^{(k)}_{n\hbox{\tiny $W$}, \ell}  \right\}}\right] \nonumber \\
& = & n\int K_h(w-w_k)(w-w_k)^\ell f_{\hbox {\tiny $W$}}(w)dw + O_{\hbox {\tiny $P$}} \left[\sqrt{n\textrm{Var} \left\{K_h(W_1-w_k)(W_1-w_k)^\ell \right\}}\right] \nonumber \\
& = & n\int K(u) h^\ell u^\ell f_{\hbox {\tiny $W$}}(hu+w_k)du + O_{\hbox {\tiny $P$}} 
\left[
\sqrt{ n E\left\{K^2_h(W_1-w_k)(W_1-w_k)^{2\ell }\right\} } 
\right]\nonumber \\
& = & n h^\ell \left\{f_{\hbox {\tiny $W$}}(w_k)+o_{\hbox {\tiny $P$}}(1)\right\} \int K(u)  u^\ell du + 
nh^\ell O_{\hbox {\tiny $P$}}(1/\sqrt{nh})\nonumber \\
& = & nh^\ell f_{\hbox {\tiny $W$}}(w_k)\mu_\ell \{1+o_{\hbox {\tiny $P$}}(1)\}. \nonumber
\end{eqnarray}
Hence, 
\begin{equation}
\bS^{(k)}_{n\hbox{\tiny $W$}}=nf_{\hbox {\tiny $W$}}(w_k)\bH \bS \bH\{1+o_{\hbox {\tiny $P$}}(1)\},\quad \textrm{for $k=1, 2$},
\label{eq:Skv2}
\end{equation}
where $\bH=\diag(1, h, \ldots, h^p)$ and $\bS=(\mu_{\ell_1+\ell_2})_{0\le \ell_1, \ell_2 \le p}$.

\subsection{Approximate $\bS^*_{n\hbox {\tiny $W$}}$}
For $\ell_1, \ell_2=0, 1, \ldots, p$, 
$S^*_{n\hbox {\tiny $W$},\ell_1,\ell_2}  =  E\left(S^*_{n\hbox {\tiny $W$},\ell_1,\ell_2}\right)+O_{\hbox {\tiny $P$}}\left\{\sqrt{\textrm{Var}\left(S^*_{n\hbox {\tiny $W$},\ell_1,\ell_2}\right)}\right\}$,
where 
\begin{eqnarray}
& & E\left(S^*_{n\hbox {\tiny $W$},\ell_1,\ell_2}\right) \nonumber \\
& = & n \int (w-w_1)^{\ell_1}(w-w_2)^{\ell_2}K_h(w-w_1)K_h(w-w_2)\nu^2(w) f_{\hbox {\tiny $W$}}(w) dw \nonumber \\
& = &  nh^{\ell_1+\ell_2-1} \int \left(u-\frac{w_1-w_2}{2h}\right)^{\ell_1} \left(u+\frac{w_1-w_2}{2h}\right)^{\ell_2} K\left(u-\frac{w_1-w_2}{2h}\right) K\left(u+\frac{w_1-w_2}{2h}\right) \times \nonumber \\
& & \nu^2\left(hu+\frac{w_1+w_2}{2}\right) f_{\hbox {\tiny $W$}}\left(hu+\frac{w_1+w_2}{2}\right)du \nonumber \\
& = &  nh^{\ell_1+\ell_2-1} \{\nu^2_{\hbox {\tiny $W$}}\left(\frac{w_1+w_2}{2}\right) f_{\hbox {\tiny $W$}}\left(\frac{w_1+w_2}{2}\right)\times \nonumber \\
& & \int \left(u-\frac{w_1-w_2}{2h}\right)^{\ell_1} \left(u+\frac{w_1-w_2}{2h}\right)^{\ell_2} K\left(u-\frac{w_1-w_2}{2h}\right) K\left(u+\frac{w_1-w_2}{2h}\right)du + o(1) \} \nonumber \\
& = & nh^{\ell_1+\ell_2-1} \left\{\nu^2\left(\frac{w_1+w_2}{2}\right) f_{\hbox {\tiny $W$}}\left(\frac{w_1+w_2}{2}\right)
\xi_{\ell_1, \ell_2}\left(\frac{w_1-w_2}{2}, h \right) + o(1) \right\}, \label{eq:es*long}
\end{eqnarray}
in which 
\begin{equation}
\xi_{\ell_1, \ell_2}(w, h)=\int (u-w/h)^{\ell_1} (u+w/h)^{\ell_2} K(u-w/h) K(u+w/h)du;
\label{eq:tauh}
\end{equation}
and
\begin{eqnarray}
& & \textrm{Var}\left( S^*_{n {\hbox {\tiny $W$}},\ell_1,\ell_2} \right) \nonumber \\
& = & n E\left\{(W_1-w_1)^{2\ell_1} (W_1-w_2)^{2\ell_2} K^2_h(W_1-w_1) K^2_h(W_1-w_2) \nu^4(W_1) \right\}-\nonumber \\
& &  n \left[ E\left\{(W_1-w_1)^{\ell_1} (W_1-w_2)^{\ell_2} K_h(W_1-w_1) K_h(W_1-w_2) \nu^2(W_1) \right\}\right]^2   \nonumber \\
& = & n \int (w-w_1)^{2\ell_1} (w-w_2)^{2\ell_2} K^2_h(w-w_1) K^2_h(w-w_2) \nu^4(w) f_{\hbox {\tiny $W$}}(w) dw -  \nonumber \\
& & n\left[h^{\ell_1+\ell_2-1} \left\{\nu^2\left(\frac{w_1+w_2}{2}\right) f_{\hbox {\tiny $W$}}\left(\frac{w_1+w_2}{2}\right)
\xi^{(h)}_{\ell_1, \ell_2}\left(\frac{w_1-w_2}{2}\right) + o(1) \right\} \right]^2 \nonumber \\
& = & n h^{2(\ell_1+\ell_2-1)-1} \int \left(u-\frac{w_1-w_2}{2h}\right)^{2\ell_1} \left(u+\frac{w_1-w_2}{2h}\right)^{2\ell_2} K^2\left(u-\frac{w_1-w_2}{2h}\right) \times \nonumber \\
& & K^2\left(u+\frac{w_1-w_2}{2h}\right) \nu^4\left(hu+\frac{w_1+w_2}{2}\right) f_{\hbox {\tiny $W$}}\left(hu+\frac{w_1+w_2}{2}\right) du \nonumber \\
& &  +o\left\{nh^{2(\ell_1+\ell_2-1)-1}  \right\}\label{eq:intermvar} \\
& = & n h^{2(\ell_1+\ell_2-1)-1} \zeta_{\ell_1, \ell_2}\left(\frac{w_1-w_2}{2}\right) \Big\{ \nu^4\left(\frac{w_1+w_2}{2}\right) f_{\hbox {\tiny $W$}}\left(\frac{w_1+w_2}{2}\right) \nonumber \\
& & +o(1)\Big\} +  o\left\{n h^{2(\ell_1+\ell_2-1)-1}  \right\}, \nonumber 
\end{eqnarray}
in which  
\begin{equation}
\zeta_{\ell_1, \ell_2}(w, h) = \int (u-w/h)^{2\ell_1} (u+w/h)^{2\ell_2} K^2(u-w/h) K^2(u+w/h) dw. 
\label{eq:xih}
\end{equation}
Note that (\ref{eq:intermvar}) is reached under the assumptions that $\nu^2(w)$ is bounded, and $\xi_{\ell_1, \ell_2}(w, h)$ is bounded for all $w$, $h>0$, and $0\le \ell_1, \ell_2 \le p$. 

Now we see that, if $\zeta_{\ell_1, \ell_2}(w, h)$ is bounded for all $w$, $h>0$, and $0\le \ell_1, \ell_2 \le p$,  
\begin{equation}
O_{\hbox {\tiny $P$}}\left\{\sqrt{\textrm{Var}\left( S^*_{n {\hbox {\tiny $W$}},\ell_1,\ell_2} \right)} \right\} 
= nh^{\ell_1+\ell_2-1} O_{\hbox {\tiny $P$}}(1/\sqrt{nh}). \label{eq:Opvar}
\end{equation}
Combining (\ref{eq:es*long}) and (\ref{eq:Opvar}), we have 
\begin{equation}
S^*_{n\hbox {\tiny $W$},\ell_1,\ell_2}=nh^{\ell_1+\ell_2-1} f_{\hbox {\tiny $W$}}\left(\frac{w_1+w_2}{2}\right) \nu^{2}\left(\frac{w_1+w_2}{2}\right)\xi_{\ell_1, \ell_2}\left(\frac{w_1-w_2}{2}, h\right)\{1+o_{\hbox {\tiny $P$}}(1) \}, \label{eq:ourS*}
\end{equation}
which is similar to (3.56) in \citet{Fanbook} although they have $\nu_{\ell_1+\ell_2}$ in the place of $\xi_{\ell_1, \ell_2}\{(w_1-w_2)/2, h\}$ above, where $\nu_\ell=\int u^\ell K^2(u)du$. We shall point out that their $\nu_\ell$ is free of $h$, whereas our $\xi_{\ell_1, \ell_2}\{(w_1-w_2)/2, h\}$ depends on $h$. In fact, the dependence of $\xi_{\ell_1, \ell_2}\{(w_1-w_2)/2, h\}$ on $h$ is crucial in the follow-up derivations. 

Putting (\ref{eq:ourS*}) inside the matrix in (\ref{eq:Sn*}), we have 
\begin{equation}
\bS^*_{n\hbox {\tiny $W$}} =  nh^{-1} f_{\hbox {\tiny $W$}}\left(\frac{w_1+w_2}{2}\right) \nu^{2}\left(\frac{w_1+w_2}{2}\right)\bH \bS^*_{\hbox {\tiny $W$}, h} \bH \{1+o_{\hbox {\tiny $P$}}(1) \} \label{eq:Sn*v2},
\end{equation}
where 
\begin{equation}
\bS^*_{\hbox {\tiny $W$},h}=\left(\xi_{\ell_1, \ell_2}\left(\frac{w_1-w_2}{2}, h\right)\right)_{0\le \ell_1, \ell_2 \le p}. 
\label{eq:mySw*}
\end{equation}
The result in (\ref{eq:Sn*v2}) is the counterpart of (3.57) in \citet{Fanbook}

\section{Step 3: Go from $\textrm{Cov}\left\{\mathcal{A}(w_1), \, \mathcal{A}(w_2) |\mathbb{W}\right\}$ to $\textrm{Var}\left\{\mathcal{B}(x)|\mathbb{W} \right\}$}
Substituting (\ref{eq:Skv2}) and (\ref{eq:Sn*v2}) in (\ref{eq:ourcov2}) yields 
\begin{equation}
\text{Cov}(\hat\bbeta_1^*, \, \hat\bbeta_2^*|\mathbb{W})=\frac{\nu^2\{(w_1+w_2)/2\}}{nh}\frac{f_{\hbox {\tiny $W$}}\{(w_1+w_2)/2\}}{f_{\hbox {\tiny $W$}}(w_1)f_{\hbox {\tiny $W$}}(w_2)} \bH^{-1} \bS^{-1}\bS^*_{\hbox {\tiny $W$},h}\bS^{-1}\bH^{-1}\{1+o_{\hbox {\tiny $P$}}(1)\}, 
\label{eq:ourcov3}
\end{equation}
which is the counterpart of (3.58) in \citet{Fanbook}. Hence
\begin{equation}
\textrm{Cov}\{\hat m^*(w_1), \, \hat m^*(w_2) |\mathbb{W}\}=\frac{\nu^2\{(w_1+w_2)/2\}}{nh}\frac{f_{\hbox {\tiny $W$}}\{(w_1+w_2)/2\}}{f_{\hbox {\tiny $W$}}(w_1)f_{\hbox {\tiny $W$}}(w_2)}\be_1^\T \bS^{-1}\bS^*_{\hbox {\tiny $W$},h}\bS^{-1}\be_1+o_{\hbox {\tiny $P$}}\left(\frac{1}{nh}\right), 
\label{eq:covmm2}
\end{equation}
as a counterpart of (3.7) in \citet{Fanbook}. Finally, by (\ref{eq:covAA}), we have 
\begin{equation}
\textrm{Cov}\left\{\mathcal{A}(w_1), \, \mathcal{A}(w_2) |\mathbb{W}\right\} =  \frac{\gamma\{(w_1+w_2)/2\}}{nh}\be_1^\T \bS^{-1}\bS^*_{\hbox {\tiny $W$},h}\bS^{-1}\be_1+o_{\hbox {\tiny $P$}}\left(\frac{1}{nh}\right), 
\label{eq:covAA2}
\end{equation}
where $\gamma(w)=\nu^2(w)f_{\hbox {\tiny $W$}}(w)$. 

Plugging (\ref{eq:covAA2}) in (\ref{eq:varB}) gives
\begin{eqnarray}
\textrm{Var}\left\{\mathcal{B}(x) |\mathbb{W}\right\} & = &\int D(x-w_1) \int D(x-w_2)
 \Big[\frac{\gamma\{(w_1+w_2)/2\}}{nh}\be_1^\T \bS^{-1}\bS^*_{\hbox {\tiny $W$},h}\bS^{-1}\be_1 \nonumber \\
 & & +o_{\hbox {\tiny $P$}}\left(\frac{1}{nh}\right)\Big] dw_2 dw_1. 
\label{eq:varB2}
\end{eqnarray}
Given the definition of $\bS^*_{\hbox {\tiny $W$},h}$ in (\ref{eq:mySw*}) and the definition of its entries in (\ref{eq:tauh}), we shall elaborate the following integral, 
\begin{equation}
\int D(x-w_1) \int D(x-w_2) \xi_{\ell_1, \ell_2}\left(\frac{w_1-w_2}{2}, h\right) \gamma\left(\frac{w_1+w_2}{2}\right) dw_2 dw_1. 
\label{eq:varB2en}
\end{equation}
The next step tackles this integral in detail. 

\section{Step 4: Elaborate (\ref{eq:varB2en})}
First, substituting $\xi_{\ell_1, \ell_2}(\cdot, \cdot)$ (\ref{eq:varB2en}) with its definition in (\ref{eq:tauh}) yields
\begin{eqnarray}
&  & \int D(x-w_1) \int D(x-w_2)  \gamma\left(\frac{w_1+w_2}{2}\right)  \int \left(u-\frac{w_1-w_2}{2h}\right)^{\ell_1} \times \nonumber \\
& & \left(u+\frac{w_1-w_2}{2h}\right)^{\ell_2} K\left(u-\frac{w_1-w_2}{2h}\right) K\left(u+\frac{w_1-w_2}{2h}\right)du dw_2 dw_1. \nonumber \\
\label{eq:3layers}
\end{eqnarray}
Using multivariate change-of-variable and letting $s_1=u-(w_1-w_2)/(2h)$ and $s_2=u+(w_1-w_2)/(2h)$, (\ref{eq:3layers}) becomes 
\begin{eqnarray}
& & h\int D(x-w_2) \int s_1^{\ell_1} K(s_1) \int D\{x-w_2-h(s_2-s_1)\}\gamma\{w_2+h(s_2-s_1)/2\}\times \nonumber \\
&  & s_2^{\ell_2} K(s_2)ds_2ds_1dw_2 \nonumber \\
& = & h\int D(x-w_2) \{\gamma(w_2)+O(h)\} \int s_1^{\ell_1} K(s_1) \int D\{x-w_2-h(s_2-s_1)\}\times \nonumber \\
& & s_2^{\ell_2} K(s_2)ds_2ds_1dw_2. \label{eq:ds2ds1dw2}
\end{eqnarray}

Second, zooming on the inner integral (with respect to $s_2$) in (\ref{eq:ds2ds1dw2}), we have 
\begin{eqnarray}
& & \int D\{x-w_2-h(s_2-s_1)\}s_2^{\ell_2} K(s_2)ds_2 \nonumber \\
& = & (2\pi)^{-1} \int  e^{-it\{x-w_2-h(s_2-s_1)\}}\{\phi_{\hbox {\tiny $U$}}(t)\}^{-1} \int e^{iths_2} s_2^{\ell_2}  K(s_2) ds_2 dt. \nonumber 
\nonumber 
\end{eqnarray}
Using the fact that $\phi_{\hbox {\tiny $K$}}^{(\ell)}(t)=i^\ell \int e^{itv} v^\ell K(v)dv$, the preceding expression is equal to
\begin{eqnarray}
&  &   i^{-\ell_2} (2\pi)^{-1} \int e^{-it(x-w_2+hs_1)} \frac{\phi^{(\ell_2)}_{\hbox {\tiny $K$}}(th)}{\phi_{\hbox {\tiny $U$}}(t)} dt  \nonumber \\
& = & h^{-1} i^{-\ell_2}(2\pi)^{-1}\int e^{-is(x-w_2+hs_1)/h} \frac{\phi^{(\ell_2)}_{\hbox {\tiny $K$}}(s)}{\phi_{\hbox {\tiny $U$}}(s/h)} ds \nonumber \\
& = & h^{-1} K_{\hbox {\tiny $U$}, \ell_2}\{(x-w_2)/h+s_1\}, \label{eq:KUl1}
\end{eqnarray}
where we use equation (5) in the main article to introduce the ``transformed kernel" in \citet{Delaigle09}, $K_{\hbox {\tiny $U$}, \ell}(x)$. 

Third, putting (\ref{eq:KUl1}) back in (\ref{eq:ds2ds1dw2}) to deal with the remaining two-dimensional integral (with respect to $s_1$ and $w_2$), we have 
\begin{eqnarray*}
& & \int D(x-w_2) \{\gamma(w_2)+O(h)\} \int s_1^{\ell_1}K(s_1) K_{\hbox {\tiny $U$}, \ell_2}\left(\frac{x-w_2}{h}+s_1\right) ds_1 dw_2. 
\end{eqnarray*}
Letting $v=(s-w_2)/h+s_1$, the above integral is equal to 
\begin{eqnarray}
& & h \int K_{\hbox {\tiny $$U}, \ell_2} (v) \int D\{(v-s_1)h\}[\gamma\{x-(v-s_1)h\}+O(h)] s_1^{\ell_1}K(s_1) ds_1dv \nonumber \\
& = & h \{\gamma(x)+O(h)\} \int K_{\hbox {\tiny $$U}, \ell_2} (v) \int (2\pi)^{-1} e^{-ithv} \frac{1}{\phi_{\hbox {\tiny $U$}}(t)} \int e^{iths_1}s_1^{\ell_1}K(s_1) ds_1 dt dv \nonumber \\
& = & \{\gamma(x)+O(h)\} \int K_{\hbox {\tiny $$U}, \ell_2} (v)  i^{-\ell_1}(2\pi)^{-1}\int e^{-isv} \frac{\phi_{\hbox {\tiny $K$}}^{(\ell_1)}(s)}{\phi_{\hbox {\tiny $U$}}(s/h)}ds dv \nonumber\\
& =  & \{\gamma(x)+O(h)\} \int K_{\hbox {\tiny $$U}, \ell_1} (v) K_{\hbox {\tiny $$U}, \ell_2} (v) dv. \label{eq:KK}
\end{eqnarray}

\section{Step 5: Lemmas needed for elaborating (\ref{eq:KK})}
To elaborate (\ref{eq:KK}), as related in Section 4.1 in the main article, we use directly Lemma B.4, Lemma B.6 (for ordinary smooth $U$) and Lemma B.9 (for super smooth $U$) in \citet{Delaigle09}. For completeness, these lemmas are restated next. 

\begin{description}
\item[Lemma B.4:] Assume that, for $\ell=\ell_1, \ell_2$, $\|\phi_{\hbox {\tiny $K$}}^{(\ell)}\|_{\infty}<\infty$, $\|\phi_{\hbox {\tiny $K$}}^{(\ell+1)}\|_{\infty}<\infty$, $\|\phi'_{\hbox {\tiny $U$}}\|_\infty <\infty$, $\int (|t|^b+|t|^{b-1})\{|\phi_{\hbox {\tiny $K$}}^{(\ell)}|+|\phi_{\hbox {\tiny $K$}}^{(\ell+1)}|\}dt<\infty$, and $\int |t|^b |\phi_{\hbox {\tiny $K$}}^{(\ell)}|dt<\infty$, then, for a bounded function $g$,  
\begin{eqnarray}
& & \lim_{n\to \infty} h^{2b} \int K_{\hbox {\tiny $U$},\ell_1}(v)K_{\hbox {\tiny $U$},\ell_2}(v)g(x-hv)dv \nonumber \\
& = & i^{-\ell_1-\ell_2}(-1)^{-\ell_2}\frac{g(x)}{c^2}\frac{1}{2\pi}\int |t|^{2b}\phi_{\hbox {\tiny $K$}}^{(\ell_1)}(t)
\phi_{\hbox {\tiny $K$}}^{(\ell_2)}(t)dt.\nonumber 
\end{eqnarray}
\item[Lemma B.6:] Suppose, for $\ell=\ell_1, \ell_2$, $\|\phi^{(\ell)}_{\hbox {\tiny $K$}}(t) \|_{\infty}<\infty$ and $\int |t|^{2b}|\phi^{(\ell)}_{\hbox {\tiny $K$}}(t) |^2dt<\infty $. Then $|\int_{-\infty}^\infty K_{\hbox {\tiny $U$}, \ell_1}(v) K_{\hbox {\tiny $U$}, \ell_2}(v) dv| \le Ch^{-2b}$ for some finite positive constant $C$.
\item[Lemma B.9:] Suppose that $\phi_{\hbox {\tiny $K$}}(t)$ is supported on $[-1, 1]$, and, for $\ell=\ell_1$ and $\ell_2$, $\|\phi^{(\ell)}_{\hbox {\tiny $K$}}(t) \|_{\infty}<\infty$. Then $|\int_{-\infty}^\infty K_{\hbox {\tiny $U$}, \ell_1}(v) K_{\hbox {\tiny $U$}, \ell_2}(v) dv| \le Ch^{2b_2}\exp(2h^{-b}/d_2)$, where $b_2=b_0I(b_0<1/2)$.
\end{description}
The conditions required in Lemma B.6 and Lemma B.9 are included in or implied by {\bf Condition O} (for ordinary smooth $U$) and {\bf Condition S} (for super smooth $U$), respectively.

\section{Elaboration of $\gamma(\cdot)$}
Define $\tau^2(x)=\textrm{Var}(Y|X=x)$, then 
\begin{eqnarray}
\nu^2(w) &=&  \textrm{Var}(Y|W=w)\nonumber \\
& = & E\left\{\textrm{Var}(Y|X)|W=w\right\}+\textrm{Var}\left\{E(Y|X)|W=w\right\} \nonumber \\
& = & E\left\{\tau^2(X)|W=w\right\}+\textrm{Var}\left\{m(X)|W=w\right\} \nonumber \\
& = & \left\{f_{\hbox {\tiny $W$}}(w) \right\}^{-1} \int \tau^2(x)f_{\hbox {\tiny $X$}}(x) f_{\hbox {\tiny $U$}}(w-x)dx + \nonumber \\
& &	E\left\{m^2(X)|W=w \right\}-\left[E\left\{m(X)|W=w \right\} \right]^2 \nonumber \\
& =  & \frac{\int \left\{ \tau^2(x)+m^2(x) \right\} f_{\hbox {\tiny $X$}}(x) f_{\hbox {\tiny $U$}}(w-x)dx}{f_{\hbox {\tiny $W$}}(w)}
-\frac{\left\{\int m(x)f_{\hbox {\tiny $X$}}(x) f_{\hbox {\tiny $U$}}(w-x)dx  \right\}^2}{\left\{f_{\hbox {\tiny $W$}}(w) \right\}^2} \nonumber \\
& = & \frac{\left\{\left(\tau^2+m^2 \right)f_{\hbox {\tiny $X$}}\right\}*f_{\hbox {\tiny $U$}}(w)}{f_{\hbox {\tiny $W$}}(w)}
-\frac{\left\{(m f_{\hbox {\tiny $X$}})*f_{\hbox {\tiny $U$}}(w)\right\}^2}{\left\{f_{\hbox {\tiny $W$}}(w) \right\}^2}. \nonumber 
\end{eqnarray}
In the above elaboration, we use the following identity according to \citet[][Theorem 34.4]{Billingsley79}, 
$$ E\{g(Y)|W\} = E\left[E\{g(Y)|X,W\} |W\right],$$
where $g(\cdot)$ a generic function such that the relevant expectations exist. Under the assumption of nondifferential measurement error, the right-hand side of this identity is equal to $E[E\{g(Y)|X\} |W]$.

It follows that 
\begin{eqnarray}
\gamma(w) & = & \left\{\left(\tau^2+m^2 \right)f_{\hbox {\tiny $X$}}\right\}*f_{\hbox {\tiny $U$}}(w)
-\left\{f_{\hbox {\tiny $W$}}(w) \right\}^{-1}\left\{(m f_{\hbox {\tiny $X$}})*f_{\hbox {\tiny $U$}}(w)\right\}^2 \nonumber \\
& = & \left\{\left(\tau^2+m^2 \right)f_{\hbox {\tiny $X$}}\right\}*f_{\hbox {\tiny $U$}}(w)
-\frac{\left\{(m f_{\hbox {\tiny $X$}})*f_{\hbox {\tiny $U$}}(w)\right\}^2}{(f_{\hbox {\tiny $X$}}*f_{\hbox {\tiny $U$}})(w)}. \nonumber
\end{eqnarray}

\noindent 
\setcounter{section}{0}
\setcounter{equation}{0}
\setcounter{figure}{0}
\renewcommand{\theequation}{C.\arabic{equation}}
\renewcommand{\thefigure}{C.\arabic{figure}}
\renewcommand{\thesection}{C.\arabic{section}}

\section*{Appendix C: Asymptotic normality of $\hat m_{\hbox {\tiny HZ}}(x)$}
\section{A sufficient condition for asymptotic normality}
To show $\hat m_{\hbox {\tiny HZ}}(x)$ converges in distribution to a normal distribution as $n\to \infty$, by Slutsky's Theorem and the fact that $\hat f_{\hbox {\tiny $X$}}(x)\stackrel{p}{\to} f(x)$ \citep[][Theorem 2.1]{Stefanski90}, it suffices to show the asymptotic normality for the difference $\mathcal{B}(x)-B(x)$, that is,  
\begin{equation}
\hat m_{\hbox {\tiny HZ}}(x)\hat f_{\hbox {\tiny $X$}}(x)-m(x)f_{\hbox {\tiny $X$}}(x) 
= \frac{1}{2\pi}\int e^{-itx}\frac{\phi_{\hat m^* \hat f_{\hbox {\tiny $W$}}- m^* f_{\hbox {\tiny $W$}}}(t)}{\phi_{\hbox {\tiny $U$}}(t)}dt=\{(\mathcal{A}-A)*D\}(x).
\label{eq:mfdiff}
\end{equation}

To show the asymptotic normality of (\ref{eq:mfdiff}), we first show that (\ref{eq:mfdiff}) can be approximated by an average, 
$n^{-1}\sum_{j=1}^n \tilde U_{n, j}(x)$, for some independent and identically distributed (i.i.d) random variables (at each fixed $x$) $\{\tilde U_{n, j}(x)\}_{j=1}^n$, each of which depends on $n$ via its dependence on $h$. Then we show that, for some positive constant $\eta$, 
\begin{equation}
\lim_{n\to \infty} \frac{E|\tilde U_{n,1}|^{2+\eta}}{n^{\eta/2}\{E(\tilde U_{n,1}^2)\}^{(2+\eta)/2}}=0, 
\label{eq:suffcond}
\end{equation}
which is a sufficient condition for 
\begin{equation}
\frac{\sum_{j=1}^n \tilde U_{n,j} -n E(\tilde U_{n,j})}{\sqrt{n\textrm{Var}(\tilde U_{n,j})}}\stackrel{\mathcal{L}}{\to} N(0, 1). \nonumber 
\end{equation}

Because convolution is a linear operator, approximating (\ref{eq:mfdiff}) via an average of i.i.d random variables can be realized by approximating $\mathcal{A}(w)-A(w)$ via an average of another set of $n$ i.i.d random variables at a fixed $w$ in the support of $W$. We achieve this goal following four steps described next. 

\section{Step 1: Re-express $\mathcal{A}(w)-A(w)$ as a summation}
\label{s:A-A}
Assuming $\hat m^*(w)$ bounded, we have 
\begin{eqnarray}
\mathcal{A}(w)-A(w) & = & \hat m^*(w) \hat f_{\hbox {\tiny $W$}}(w)-m^*(w)f_{\hbox {\tiny $W$}}(w) \nonumber \\
& = & \left\{\sum_{\ell=0}^p S^{0,\ell}_{n\hbox {\tiny $W$}} (w) T_{n\hbox {\tiny $W$},\ell} (w)-m^*(w)\right\}f_{\hbox {\tiny $W$}}(w)+o_{\hbox {\tiny $P$}}(1), \label{eq:stmdiff}
\end{eqnarray}
where 
$T_{n\hbox {\tiny $W$}, \ell}=\sum_{j=1}^n Y_j (W_j-w)^\ell K_h(W_j-w), \textrm{ for $\ell=0, 1, \ldots, p$,}$
and $S^{0,\ell}_{n\hbox {\tiny $W$}} (w)$ is the $[1, \ell+1]$ element of $\bS^{-1}_{n\hbox {\tiny $W$}}(w)$, with $\bS_{n\hbox {\tiny $W$}}(w)=(S_{n\hbox {\tiny $W$}, \ell_1+\ell_2}(w))_{0\le \ell_1, \ell_2\le p}$ and
\begin{equation}
S_{n\hbox {\tiny $W$}, \ell}=\sum_{j=1}^n (W_j-w)^\ell K_h(W_j-w), \textrm{ for $\ell=0, 1, \ldots, 2p$.} \nonumber
\end{equation}

Because
$\sum_{\ell=0}^p S_{n\hbox {\tiny $W$}}^{0, \ell}(w) S_{n\hbox {\tiny $W$},\ell+ \ell'}(w)=I(\ell'=0),$ where $I(\cdot)$ is the indicator function, 
inside the curly brackets in (\ref{eq:stmdiff}) we have
\begin{eqnarray}
&  & \sum_{\ell=0}^p S^{0,\ell}_{n\hbox {\tiny $W$}} (w) T_{n\hbox {\tiny $W$},\ell} (w)-m^*(w)\nonumber \\
& =&  \sum_{\ell=0}^p S^{0,\ell}_{n\hbox {\tiny $W$}} (w) T_{n\hbox {\tiny $W$},\ell} (w)- \sum_{\ell'=0}^p h^{\ell'} \frac{m^{*(\ell')}(w)}{\ell'!}\sum_{\ell=0}^p S_{n\hbox {\tiny $W$}}^{0, \ell}(w) S_{n\hbox {\tiny $W$},\ell+ \ell'}(w) \nonumber \\
& = & \sum_{\ell=0}^p S^{0,\ell}_{n\hbox {\tiny $W$}} (w) \left\{ T_{n\hbox {\tiny $W$},\ell} (w)- \sum_{\ell'=0}^p h^{\ell'} \frac{m^{*(\ell')}(w)}{\ell'!}S_{n\hbox {\tiny $W$},\ell+ \ell'}(w) \right\} \nonumber \\
& = & \sum_{\ell=0}^p S^{0,\ell}_{n\hbox {\tiny $W$}} (w)T^*_{n\hbox {\tiny $W$},\ell} (w), \label{eq:sumpst}
\end{eqnarray}
where $T^*_{n\hbox {\tiny $W$},\ell} (w)=T_{n\hbox {\tiny $W$},\ell} (w)- \sum_{\ell'=0}^p h^{\ell'} \{m^{*(\ell')}(w)/\ell'!\}S_{n\hbox {\tiny $W$},\ell+ \ell'}(w)$, for $\ell=0, 1, \ldots, p$.

In what follows, we show that (\ref{eq:sumpst}) is equivalent to 
\begin{eqnarray}
& & \sum_{\ell=0}^p [\textrm{a nonrandom function of $w$ as an approximation of $nS^{0,\ell}_{n\hbox {\tiny $W$}} (w)$}]\nonumber \\
& \times & \left\{n^{-1}T^*_{n\hbox {\tiny $W$},\ell} (w)\right\}+O_{\hbox {\tiny $P$}}(h^{\textrm{some positive power}}). \nonumber 
\end{eqnarray}
This is accomplished in two steps. First, studying $n^{-1}T^*_{n\hbox {\tiny $W$},\ell} (w)$ to understand its order in $h$. Second, approximating $nS^{0,\ell}_{n\hbox {\tiny $W$}} (w)$. 

\section{Step 2: The order of $n^{-1}T^*_{n\hbox {\tiny $W$},\ell} (w)$}
\label{s:ninvT}
Because
$n^{-1}T^*_{n\hbox {\tiny $W$},\ell} (w)=E\{n^{-1}T^*_{n\hbox {\tiny $W$},\ell} (w)\}+O_{\hbox {\tiny $P$}}[\sqrt{\textrm{Var}\{n^{-1}T^*_{n\hbox {\tiny $W$},\ell} (w)\}}]$,
we study the order of the expectation and variance separately in this section. 

For the expectation, we have $E\{n^{-1}T^*_{n\hbox {\tiny $W$},\ell} (w)\}$ equal to 
\begin{eqnarray}
& & E\left\{n^{-1}T_{n\hbox {\tiny $W$},\ell} (w)\right\}- \sum_{\ell'=0}^p h^{\ell'} 
\frac{m^{*(\ell')}(w)}{\ell'!}E\left\{n^{-1}S_{n\hbox {\tiny $W$},\ell+\ell'} (w)\right\} \nonumber \\
& = & E\left\{Y(W-w)^\ell K_h(W-w)\right\}-\sum_{\ell'=0}^p h^{\ell'} 
\frac{m^{*(\ell')}(w)}{\ell'!} E\left\{(W-w)^{\ell+\ell'} K_h(W-w)\right\} \nonumber \\
& = & \int m(v_2) \int h^\ell v_1^\ell K(v_1)f_{\hbox {\tiny $U$}}(hv_1-v_2+w)dv_1 f_{\hbox {\tiny $X$}}(v_2)dv_2-\nonumber \\
& & \sum_{\ell'=0}^p h^{\ell'} \frac{m^{*(\ell')}(w)}{\ell'!} \int \int h^{\ell+\ell'} v_1^{\ell+\ell'} K(v_1)f_{\hbox {\tiny $U$}}(hv_1-v_2+w)dv_1 f_{\hbox {\tiny $X$}}(v_2)dv_2.\nonumber
\end{eqnarray}
Using the first-order Taylor expansion of $f_{\hbox {\tiny $U$}}(hv_1-v_2+w)$ around $h=0$, i.e., 
\begin{equation}
f_{\hbox {\tiny $U$}}(hv_1-v_2+w)=f_{\hbox {\tiny $U$}}(w-v_2)+hf'_{\hbox {\tiny $U$}}(w-v_2)v_1+O(h^2),
\label{eq:taylorfu}
\end{equation}
in the above integral, we have $E\{n^{-1}T^*_{n\hbox {\tiny $W$},\ell} (w)\}$ equal to 
\begin{eqnarray}
& & h^\ell \Big\{\mu_\ell \int m(v_2) f_{\hbox {\tiny $U$}}(w-v_2)f_{\hbox {\tiny $X$}}(v_2)dv_2+h\mu_{\ell+1}\int m(v_2) f'_{\hbox {\tiny $U$}}(w-v_2)f_{\hbox {\tiny $X$}}(v_2)dv_2 \nonumber \\
& & +O(h^2) \Big\} - \sum_{\ell'=0}^p h^{\ell+2\ell'} \frac{m^{*(\ell')}(w)}{\ell'!} \Big\{ \mu_{\ell+\ell'} f_{\hbox {\tiny $W$}}(w) + \nonumber \\ 
& & h\mu_{\ell+\ell'+1}\int  f'_{\hbox {\tiny $U$}}(w-v_2)f_{\hbox {\tiny $X$}}(v_2)dv_2 +O(h^2) \Big\} \nonumber \\
& = & h^\ell \mu_\ell f_{\hbox {\tiny $W$}}(w) m^*(w) +h^{\ell+1}\mu_{\ell+1}\int m(v_2) f'_{\hbox {\tiny $U$}}(w-v_2)f_{\hbox {\tiny $X$}}(v_2)dv_2
 - h^\ell \mu_\ell f_{\hbox {\tiny $W$}}(w)  m^*(w) \nonumber \\
& & - h^{\ell+2}m^{*'}(w) \mu_{\ell+1}f_{\hbox {\tiny $W$}}(w)-h^{\ell+1} m^*(w) \mu_{\ell+1}\int f'_{\hbox {\tiny $U$}}(w-v_2)f_{\hbox {\tiny $X$}}(v_2)dv_2  +O(h^{\ell+2})\nonumber \\
& = & h^{\ell+1} \mu_{\ell+1} \int \left\{m(v_2)-m^*(w)\right\} f'_{\hbox {\tiny $U$}}(w-v_2)f_{\hbox {\tiny $X$}}(v_2)dv_2 - h^{\ell+2}m^{*'}(w) \mu_{\ell+1}f_{\hbox {\tiny $W$}}(w)\nonumber \\
& & +O(h^{\ell+2}) \nonumber \\
& = &  \left\{
\begin{array}{ll}
O(h^{\ell+1}) & \textrm{if $\ell$ is odd} \\
O(h^{\ell+2}) & \textrm{if $\ell$ is even}
\end{array}
\right. .\label{eq:et*}
\end{eqnarray}

For the variance, we have $\textrm{Var}\{n^{-1}T^*_{n\hbox {\tiny $W$},\ell} (w)\}$ equal to 
\begin{eqnarray}
& & \textrm{Var}\left\{n^{-1}T_{n\hbox {\tiny $W$},\ell} (w)-n^{-1}\sum_{\ell'=0}^p h^{\ell'}\frac{m^{*(\ell')}(w)}{\ell'!} S_{n\hbox {\tiny $W$}, \ell+\ell'}(w) \right\}\nonumber \\
& = & \textrm{Var}\left\{n^{-1}T_{n\hbox {\tiny $W$},\ell} (w)-n^{-1}m^*(w)S_{n\hbox {\tiny $W$},\ell} (w)-n^{-1}\sum_{\ell'=1}^p h^{\ell'}\frac{m^{*(\ell')}(w)}{\ell'!} S_{n\hbox {\tiny $W$}, \ell+\ell'}(w) \right\}\nonumber \\
& = & O\left[\textrm{Var}\left\{n^{-1}T_{n\hbox {\tiny $W$},\ell} (w)-n^{-1}m^*(w)S_{n\hbox {\tiny $W$},\ell} (w)\right\}\right]+O\left[\sum_{\ell'=1}^p \textrm{Var}\left\{n^{-1} S_{n\hbox {\tiny $W$}, \ell+\ell'}(w) \right\}\right]. \nonumber \\
\label{eq:twovar}
\end{eqnarray}

Looking into the first variance in (\ref{eq:twovar}), we have 
\begin{eqnarray}
& & \textrm{Var}\left\{n^{-1}T_{n\hbox {\tiny $W$},\ell} (w)-n^{-1}m^*(w)S_{n\hbox {\tiny $W$},\ell} (w)\right\} \nonumber \\
& = & \textrm{Var}\left[n^{-1}\sum_{j=1}^n\left\{Y_j-m^*(w) \right\}(W_j-w)^\ell K_h(W_j-w) \right] \nonumber \\
& \le & \frac{1}{nh^2} E\left[ \left\{Y-m^*(w)\right\}^2 (W-w)^{2\ell} K^2\left(\frac{W-w}{h}\right) \right] \nonumber \\
& = &  \frac{1}{nh^2} \int E\left[\left\{Y-m^*(w)\right\}^2|X=v_2\right]\int h^{2\ell+1}v_1^{2\ell} K^2(v_1)f_{\hbox {\tiny $U$}}(hv_1-v_2+w)dv_1f_{\hbox {\tiny $X$}}(v_2) dv_2. \nonumber
\end{eqnarray}
Define $\tilde \kappa (w, X)=E\left[\left\{Y-m^*(w)\right\}^2|X\right]$ and $\delta_\ell=\int v^\ell K^2(v)dv$ for $\ell=0, 1, \ldots, 2p$, using (\ref{eq:taylorfu}), the preceding expression becomes 
\begin{eqnarray}
& & \frac{h^{2\ell}}{nh} \int \tilde \kappa (w, v_2) \left\{\delta_{2\ell} f_{\hbox {\tiny $U$}}(w-v_2)+h\delta_{2\ell+1} f'_{\hbox {\tiny $U$}}(w-v_2)+O(h^2) \right\} f_{\hbox {\tiny $X$}}(v_2) dv_2 \nonumber \\
& = & \frac{h^{2\ell}}{nh} \delta_{2\ell} E\left\{\tilde \kappa (w, X)f_{\hbox {\tiny $U$}}(w-X)\right\} +  \frac{h^{2\ell+1}}{nh} \delta_{2\ell+1}E\left\{\tilde \kappa (w, X)f'_{\hbox {\tiny $U$}}(w-X)\right\}+ \frac{1}{nh}O(h^{2\ell+2}) \nonumber \\
& = & \frac{h^{2\ell}}{nh} \delta_{2\ell} E\left\{\tilde \kappa (w, X)f_{\hbox {\tiny $U$}}(w-X)\right\} +\frac{1}{nh}O(h^{2\ell+2}), \textrm{ as $\delta_k=0$ when $k$ is odd.} \nonumber \\
& = & O\left(\frac{h^{2\ell}}{nh}\right), \textrm{ assuming $\delta_{2\ell}E\left\{\tilde \kappa (w, X)f_{\hbox {\tiny $U$}}(w-X)\right\}$ bounded and nonzero.} \nonumber 
\end{eqnarray}
Hence, $\textrm{Var}\left\{n^{-1}T_{n\hbox {\tiny $W$},\ell} (w)-n^{-1}m^*(w)S_{n\hbox {\tiny $W$},\ell} (w)\right\}$ is bounded from above by some nonrandom quantity of order  $(nh)^{-1}O\left(h^{2\ell}\right)$. 

Similarly, for the second variance in (\ref{eq:twovar}), we show that 
$$\textrm{Var}\{n^{-1} S_{n\hbox {\tiny $W$}, \ell+\ell'}(w) \} \le h^{2(\ell+\ell')}/(nh) \delta_{2k}f_{\hbox {\tiny $W$}}(w) +(nh)^{-1}O(h^{2k+2}).$$ 
Because $\ell+\ell'\ge \ell+1$ in (\ref{eq:twovar}), assuming $\delta_{2k}f_{\hbox {\tiny $W$}}(w)$ bounded and nonzero, we see that $\sum_{\ell'=1}^p \textrm{Var}\left\{n^{-1} S_{n\hbox {\tiny $W$}, \ell+\ell'}(w) \right\}$ is bounded from above by some nonrandom quantity of order $(nh)^{-1}O(h^{2\ell+2})$, which converges to 0 faster than the first variance in (\ref{eq:twovar}). Therefore,  $\textrm{Var}\{n^{-1}T^*_{n\hbox {\tiny $W$},\ell} (w)\}\le C h^{2\ell}/(nh)$, for some positive constant $C$. If $1/\sqrt{nh}=O(h^2)$, i.e., $h=O(n^{-1/5})$, then $\sqrt{\textrm{Var}\{n^{-1}T^*_{n\hbox {\tiny $W$},\ell} (w)\}}\le \sqrt{C} h^{\ell+2}$, which tends to 0 at least as fast as $E\{n^{-1}T^*_{n\hbox {\tiny $W$},\ell} (w)\}$ according to (\ref{eq:et*}). 

In conclusion, we establish that 
\begin{equation}
n^{-1}T^*_{n\hbox {\tiny $W$},\ell} (w)=\left\{
\begin{array}{ll}
h^\ell\left\{ C h+ O_{\hbox {\tiny $P$}}(h^2)\right\} & \textrm{if $\ell$ is odd} \\
h^\ell\left\{ C' h^2+ O_{\hbox {\tiny $P$}}(h^3)\right\}& \textrm{if $\ell$ is even}
\end{array}
\right., \label{eq:t*order}
\end{equation}
for some finite nonzero nonrandom quantities $C$ and $C'$ that depend on $w$ (but not on $n$). This completes the first task stated in Section~\ref{s:A-A}. 

\section{Step 3: Approximate $nS^{0,\ell}_{n\hbox {\tiny $W$}} (w)$}
\label{s:nSWw}
Since $S^{0,\ell}_{n\hbox {\tiny $W$}}(w)$ is an element of $\bS^{-1}_{n\hbox {\tiny $W$}}(w)$, we may first study the elements in $\bS_{n\hbox {\tiny $W$}}(w)$, namely $S_{n\hbox {\tiny $W$},\ell}(w)$. Following similar strategies used in Section~\ref{s:ninvT}, we begin with 
$n^{-1}S_{n\hbox {\tiny $W$},\ell}(w)=E\{n^{-1}S_{n\hbox {\tiny $W$},\ell}(w)\}+O_{\hbox {\tiny $P$}}[\sqrt{\textrm{Var}\{n^{-1}S_{n\hbox {\tiny $W$},\ell}(w) \}}]$

For the expectation above, we have $E\{n^{-1}S_{n\hbox {\tiny $W$},\ell}(w)\}$ equal to 
\begin{eqnarray}
&  & E\left\{(W-w)^\ell K_h(W-w)\right\} \nonumber \\
& = & \int \int (u+v_2-w)^\ell h^{-1}K\left(\frac{u+v_2-w}{h}\right)f_{\hbox {\tiny $U$}}(u)du f_{\hbox {\tiny $X$}}(v_2)dv_2 \nonumber \\
& = & h^\ell \int \int v_1^\ell K(v_1)f_{\hbox {\tiny $U$}}(hv_1-v_2+w)dv_1 f_{\hbox {\tiny $X$}}(v_2)dv_2 \nonumber \\ 
& = & h^\ell \int \int v_1^\ell K(v_1)\left\{f_{\hbox {\tiny $U$}}(w-v_2)+hf'_{\hbox {\tiny $U$}}(w-v_2)v_1+O(h^2)\right\}dv_1 f_{\hbox {\tiny $X$}}(v_2)dv_2 \nonumber \\ 
& = & h^\ell \left[\mu_\ell f_{\hbox {\tiny $W$}}(w) + h \mu_{\ell+1}E\left\{f'_{\hbox {\tiny $U$}}(w-X) \right\}+O(h^2)\right]. \label{eq:es*}
\end{eqnarray}

For the variance, we have $\textrm{Var}\{n^{-1}S_{n\hbox {\tiny $W$},\ell}(w)\}$ equal to 
\begin{eqnarray}
&  & n^{-1}\textrm{Var}\left\{(W-w)^\ell K_h(W-w)\right\} \nonumber \\
& \le & \frac{1}{nh^2}E\left\{(W-w)^{2\ell}K^2\left(\frac{W-w}{h}\right)\right\} \nonumber \\
& = & \frac{1}{nh^2} \int \int h^{2\ell+1} v_1^{2\ell} K^2(v_1)f_{\hbox {\tiny $U$}}(hv_1-v_2+w) dv_1 f_{\hbox{\tiny $X$}}(v_2)dv_2 \nonumber \\
& = & \frac{1}{nh^2} \int \int h^{2\ell+1} v_1^{2\ell} K^2(v_1)\left\{f_{\hbox {\tiny $U$}}(w-v_2)+hf'_{\hbox {\tiny $U$}}(w-v_2)v_1+O(h^2)\right\} dv_1 f_{\hbox{\tiny $X$}}(v_2)dv_2 \nonumber \\
& = & \frac{1}{nh^2} h^{2\ell+1}\left[\delta_{2\ell} f_{\hbox {\tiny $W$}}(w)+h \delta_{2\ell+1}E\left\{f'_{\hbox {\tiny $U$}}(w-X)\right\}+O(h^2) \right] \nonumber \\
& = & \frac{h^{2\ell}}{nh} \left\{\delta_{2\ell} f_{\hbox {\tiny $W$}}(w)+O(h^2) \right\}. \nonumber 
\end{eqnarray}
Therefore, $\sqrt{\textrm{Var}\{n^{-1}S_{n\hbox {\tiny $W$},\ell}(w)\}}\le C h^\ell/\sqrt{nh}$,
for some positive constant $C$ that depends on $w$ but not on $n$. And if $h=O(n^{-1/5})$, 
$\sqrt{\textrm{Var}\left\{n^{-1}S_{n\hbox {\tiny $W$},\ell}(w) \right\}}\le C' h^{\ell+2}$,
for some positive constant $C'$ that depends on $w$ but not on $n$. Hence, the dominating terms in $n^{-1}S_{n\hbox {\tiny $W$},\ell}(w)$ are in the  expectation elaborated in (\ref{eq:es*}). 

Define $\widetilde\bS=(\mu_{\ell_1+\ell_2+1})_{0\le \ell_1, \ell_2 \le p}$. By (\ref{eq:es*}), we now have 
\begin{equation*}
n^{-1}\bS_{n\hbox {\tiny $W$}}(w)=\bH\left[f_{\hbox {\tiny $W$}}(w)\bS+ h E\left\{f'_{\hbox {\tiny $U$}}(w-X)\right\}\widetilde\bS +O_{\hbox {\tiny $P$}}(h^2)\right]\bH. 
\end{equation*}
It follows that 
$$\bH \left\{n\bS^{-1}_{n\hbox {\tiny $W$}}(w)\right\} \bH 
 =  \left[\bI_{p+1}+hE\left\{f'_{\hbox {\tiny $U$}}(w-X)\right\} f^{-1}_{\hbox {\tiny $W$}}(w)\bS^{-1} \widetilde\bS \right]^{-1}\bS^{-1} f^{-1}_{\hbox {\tiny $W$}}(w)+ O_{\hbox {\tiny $P$}}(h^2). $$
Using the first order Taylor expansion of $\left[\bI_{p+1}+hE\left\{f'_{\hbox {\tiny $U$}}(w-X)\right\} f^{-1}_{\hbox {\tiny $W$}}(w)\bS^{-1}\widetilde\bS\right]^{-1}$ around $h=0$, the above expression is equal to 
\begin{eqnarray}
&  &  \left[\bI_{p+1}-h E\left\{f'_{\hbox {\tiny $U$}}(w-X)\right\} f^{-1}_{\hbox {\tiny $W$}}(w)\bS^{-1}\widetilde\bS+O(h^2) \right] \bS^{-1} f^{-1}_{\hbox {\tiny $W$}}(w)+ O_{\hbox {\tiny $P$}}(h^2) \nonumber \\
& = & \bS^{-1} f^{-1}_{\hbox {\tiny $W$}}(w)-h E\left\{f'_{\hbox {\tiny $U$}}(w-X)\right\} f^{-2}_{\hbox {\tiny $W$}}(w)\bS^{-1}\widetilde\bS \bS^{-1}+ O_{\hbox {\tiny $P$}}(h^2). \nonumber 
\end{eqnarray}
Hence, 
\begin{equation}
n\bS^{-1}_{n\hbox {\tiny $W$}}(w)=\bH^{-1}\left[\bS^{-1} f^{-1}_{\hbox {\tiny $W$}}(w)-h E\left\{f'_{\hbox {\tiny $U$}}(w-X)\right\} f^{-2}_{\hbox {\tiny $W$}}(w)\bS^{-1}\widetilde\bS \bS^{-1}+ O_{\hbox {\tiny $P$}}(h^2)\right]\bH^{-1}.
\label{eq:Snwinv}
\end{equation}

Denote by $S_{\hbox {\tiny $Z$}}^{0, \ell}(w)$ the $[1, \ell+1]$ element of the matrix $\bS^{-1} f^{-1}_{\hbox {\tiny $W$}}(w)$, and by $\breve S_{\hbox {\tiny $W$}}^{0, \ell}(w)$ the $[1, \ell+1]$ element of the matrix $E\{f'_{\hbox {\tiny $U$}}(w-X)\} f^{-2}_{\hbox {\tiny $W$}}(w)\bS^{-1}\widetilde\bS \bS^{-1}$, for $\ell=1, \ldots, p$. Then (\ref{eq:Snwinv}) indicates that 
\begin{equation}
nS_{n\hbox {\tiny $W$}}^{0,\ell}(w) =  h^{-\ell}\left\{S_{\hbox {\tiny $Z$}}^{0, \ell}(w)-h\breve S_{\hbox {\tiny $W$}}^{0, \ell}(w) + O_{\hbox {\tiny $P$}}(h^2) \right\} 
 =  h^{-\ell}\left\{R^{0,\ell}_{\hbox {\tiny $W$}}(w) + O_{\hbox {\tiny $P$}}(h^2)\right\},  \label{eq:Snwinv1}
\end{equation}
where $R^{0,\ell}_{\hbox {\tiny $W$}}(w) = -h\breve S_{\hbox {\tiny $W$}}^{0, \ell}(w)$ if $\ell$ is odd, and 
$R^{0,\ell}_{\hbox {\tiny $W$}}(w)=S_{\hbox {\tiny $Z$}}^{0, \ell}(w)$ if $\ell$ is even. The definition of $R^{0,\ell}_{\hbox {\tiny $W$}}(w)$ comes from the observation that the locations in $\bS$ where the elements are 0 remain to be 0 in the same locations in $\bS^{-1}$, and the locations in $\widetilde \bS$ where the elements are 0 remain to be 0 in the same locations in $\bS^{-1}\widetilde \bS \bS^{-1}$. More specifically, with an even kernel $K(v)$, for $\ell=0, 1, \ldots, p$, the $[1, \ell+1]$ element of $\bS$, $\mu_{\ell}$, is equal to 0 if $\ell$ is odd, and thus the $[1, \ell+1]$ element of $\bS^{-1}f^{-1}_{\hbox {\tiny $W$}}(w)$, $S_{\hbox {\tiny $Z$}}^{0,\ell}(w)$, is also 0 if $\ell$ is odd. Similarly, the $[1, \ell+1]$ element of $\widetilde\bS$, $\mu_{\ell+1}$, is equal to 0 if $\ell$ is even, and thus the $[1, \ell+1]$ element of $E\left\{f'_{\hbox {\tiny $U$}}(w-X)\right\} f^{-2}_{\hbox {\tiny $W$}}(w)\bS^{-1}\widetilde\bS \bS^{-1}$, $\breve S_{\hbox {\tiny $W$}}^{0,\ell}(w)$, is also 0 if $\ell$ is even. This completes the second task stated in Section~\ref{s:A-A}.

\section{Step 4: Reexpress (\ref{eq:sumpst})}
\label{s:rexpD5}
By (\ref{eq:Snwinv1}), the summand in (\ref{eq:sumpst}) is equal to 
\begin{eqnarray}
& & \left\{nS_{n\hbox {\tiny $W$}}^{0,\ell}(w)\right\}\left\{n^{-1}T^*_{n\hbox {\tiny $W$},\ell}(w)\right\} \nonumber \\
& = & h^{-\ell}\left\{R^{0,\ell}_{\hbox {\tiny $W$}}(w) + O_{\hbox {\tiny $P$}}(h^2)\right\}\left\{n^{-1}T^*_{n\hbox {\tiny $W$},\ell}(w)\right\} \nonumber \\
& = & h^{-\ell}R^{0,\ell}_{\hbox {\tiny $W$}}(w)\left\{n^{-1}T^*_{n\hbox {\tiny $W$},\ell}(w)\right\} +h^{-\ell}O_{\hbox {\tiny $P$}}(h^2)\left\{n^{-1}T^*_{n\hbox {\tiny $W$},\ell}(w)\right\}, \nonumber
\end{eqnarray}
where the second term above is $O_{\hbox {\tiny $P$}}(h^3)$ if $\ell$ is odd and $O_{\hbox {\tiny $P$}}(h^4)$ if $\ell$ is even according to (\ref{eq:t*order}). Assuming $p\ge 1$ (so that $\ell$ is odd at least once in (\ref{eq:sumpst})), we deduce that (\ref{eq:sumpst}) is equal to 
\begin{eqnarray}
& & \sum_{\ell=0}^ph^{-\ell}R^{0,\ell}_{\hbox {\tiny $W$}}(w)\left\{n^{-1}T^*_{n\hbox {\tiny $W$},\ell}(w)\right\} + O_{\hbox {\tiny $P$}}(h^3) \nonumber\\
& = & \sum_{\ell=0}^ph^{-\ell}R^{0,\ell}_{\hbox {\tiny $W$}}(w) n^{-1}\nonumber  \Bigg\{\sum_{j=1}^n Y_j(W_j-w)^\ell K_h(W_j-w) -\nonumber \\
& & \sum_{\ell'=0}^p h^{\ell'}\frac{m^{*(\ell')}(w)}{\ell'!}\sum_{j=1}^n (W_j-w)^{\ell+\ell'} K_h(W_j-w) \Bigg\} + O_{\hbox {\tiny $P$}}(h^3) \nonumber \\
& = & n^{-1} \sum_{j=1}^n \Bigg[\sum_{\ell=0}^p h^{-\ell}R^{0,\ell}_{\hbox {\tiny $W$}}(w)Y_j(W_j-w)^\ell K_h(W_j-w)-   \nonumber \\
& & \sum_{\ell=0}^p\sum_{\ell'=0}^p h^{\ell'-\ell}R^{0,\ell}_{\hbox {\tiny $W$}}(w) \frac{m^{*(\ell')}(w)}{\ell'!} (W_j-w)^{\ell+\ell'} K_h(W_j-w)\Bigg] + O_{\hbox {\tiny $P$}}(h^3)  \nonumber \\
& = & n^{-1} \sum_{j=1}^n \Bigg[\sum_{\ell=0}^p h^{-\ell}R^{0,\ell}_{\hbox {\tiny $W$}}(w)\{Y_j-m^*(w)\}(W_j-w)^\ell K_h(W_j-w) \nonumber \\
& & -\sum_{\ell=0}^p\sum_{\ell'=1}^p h^{\ell'-\ell}R^{0,\ell}_{\hbox {\tiny $W$}}(w) \frac{m^{*(\ell')}(w)}{\ell'!} (W_j-w)^{\ell+\ell'} K_h(W_j-w)\Bigg] + O_{\hbox {\tiny $P$}}(h^3),  \nonumber \\
\label{eq:UPQ}
\end{eqnarray}
which is finally in the form of an average of $n$ i.i.d. random variables for a fixed $w$ plus $O_{\hbox {\tiny $P$}}(h^3)$. Denote by $U_{\hbox {\tiny $W$},j}(w)$ the summand inside the square brackets in (\ref{eq:UPQ}) and decompose it as $U_{\hbox {\tiny $W$},j}(w)=P_{\hbox {\tiny $W$},j}(w)+Q_{\hbox {\tiny $W$},j}(w)$, where 
\begin{eqnarray}
P_{\hbox {\tiny $W$},j}(w) & = & \sum_{\ell=0}^p h^{-\ell}R^{0,\ell}_{\hbox {\tiny $W$}}(w)\{Y_j-m^*(w)\}(W_j-w)^\ell K_h(W_j-w), \nonumber \\
Q_{\hbox {\tiny $W$},j}(w) & = & -\sum_{\ell=0}^p\sum_{\ell'=1}^p h^{\ell'-\ell}R^{0,\ell}_{\hbox {\tiny $W$}}(w) \frac{m^{*(\ell')}(w)}{\ell'!} (W_j-w)^{\ell+\ell'} K_h(W_j-w). \nonumber 
\end{eqnarray}

Now (\ref{eq:stmdiff}) reduces to 
\begin{eqnarray}
\hat m^*(w) \hat f_{\hbox {\tiny $W$}}(w)-m^*(w)f_{\hbox {\tiny $W$}}(w)=n^{-1}\sum_{j=1}^n U_{\hbox {\tiny $W$},j}(w)f_{\hbox {\tiny $W$}}(w)\left\{1+O_{\hbox {\tiny $P$}}(1)\right\}. \label{eq:mfwdiff}
\end{eqnarray}
After repeating the exercise already seen in Sections~\ref{s:ninvT} and \ref{s:nSWw}, by looking into $\hat f_{\hbox {\tiny $W$}}(w)=E\{\hat f_{\hbox {\tiny $W$}}(w)\}+O_{\hbox {\tiny $P$}}[\sqrt{\textrm{Var}\{\hat f_{\hbox {\tiny $W$}}(w)\}}]$, we show that $\hat f_{\hbox {\tiny $W$}}(w)=f_{\hbox {\tiny $W$}}(w)+O_{\hbox {\tiny $P$}}(h^2)$. So the $O_{\hbox {\tiny $P$}}(h^3)$ in (\ref{eq:UPQ}) is dominated by (or absorbed in) this $O_{\hbox {\tiny $P$}}(h^2)$. Hence, we actually have 
$$\hat m^*(w) \hat f_{\hbox {\tiny $W$}}(w)-m^*(w)f_{\hbox {\tiny $W$}}(w)=n^{-1}\sum_{j=1}^n U_{\hbox {\tiny $W$},j}(w)f_{\hbox {\tiny $W$}}(w)+O_{\hbox {\tiny $P$}}(h^2)$$ before concluding (\ref{eq:mfwdiff}). We reach (\ref{eq:mfwdiff}) by showing that $n^{-1}\sum_{j=1}^n U_{\hbox {\tiny $W$},j}(w)f_{\hbox {\tiny $W$}}(w)$ is of order $O_{\hbox {\tiny $P$}}(h^2)$ or tends to zero at a slower rate than $h^2$.

Finally, plugging (\ref{eq:mfwdiff}) in (\ref{eq:mfdiff}), we obtain the following desired form, 
\begin{eqnarray}
\hat m_{\hbox {\tiny HZ}}(x)\hat f_{\hbox {\tiny $X$}}(x)-m(x) f_{\hbox {\tiny $X$}}(x) & = & \left\{n^{-1}\sum_{j=1}^n \frac{1}{2\pi}\int e^{-itx}
\frac{\phi_{U_{\hbox {\tiny $W$},j}f_{\hbox {\tiny $W$}}}(t)}{\phi_{\hbox {\tiny $U$}}(t)} dt \right\} \left\{1+O_{\hbox {\tiny $P$}}(1)\right\} \nonumber \\
& = & \left\{n^{-1}\sum_{j=1}^n \tilde U_{n,j} (x)\right\}\left\{1+O_{\hbox {\tiny $P$}}(1)\right\}, \label{eq:mfdiff2}
\end{eqnarray}
where, for $j=1, \ldots, n$, $\tilde U_{n,j} (x) = \tilde P_{n,j} (x)+\tilde Q_{n,j} (x)$, 
with 
$$
\tilde P_{n,j} (x) =  \frac{1}{2\pi}\int e^{-itx}
\frac{\phi_{P_{\hbox {\tiny $W$},j}f_{\hbox {\tiny $W$}}}(t)}{\phi_{\hbox {\tiny $U$}}(t)} dt, \quad
\tilde Q_{n,j} (x) = \frac{1}{2\pi}\int e^{-itx}
\frac{\phi_{Q_{\hbox {\tiny $W$},j}f_{\hbox {\tiny $W$}}}(t)}{\phi_{\hbox {\tiny $U$}}(t)} dt.$$ 

\section{The order (in $h$) of $E|\tilde P_{n,j} (x)|^{2+\eta}$, $E|\tilde Q_{n,j} (x)|^{2+\eta}$, and $E|\tilde U_{n,j} (x)|^2$}
In order to show (\ref{eq:suffcond}), we need to study the orders (in $h$) of $E|\tilde P_{n,j} (x)|^{2+\eta}$, $E|\tilde Q_{n,j} (x)|^{2+\eta}$, and $E|\tilde U_{n,j} (x)|^2$. The orders of these quantities mainly depend on two facts. First, the orders of $E|\phi_{P_{\hbox {\tiny $W$},j}f_{\hbox {\tiny $W$}}}(t)|^{2+\eta}$ and $E|\phi_{Q_{\hbox {\tiny $W$},j}f_{\hbox {\tiny $W$}}}(t)|^{2+\eta}$; second, the smoothness of $U$. We first look into the first factor in the upcoming subsection, which leads to the intermediate results needed for showing asymptotic normality. In this section, we use $s\asymp t$ to indicate that $s$ and $t$ are of the same order in $h$ as $n\to \infty$. 

\subsection{Intermediate results}
\label{s:interres}
First, by the definition of $P_{\hbox {\tiny $W$},j}(w)$,
\begin{eqnarray}
P_{\hbox {\tiny $W$},j}(w)f_{\hbox {\tiny $W$}}(w) 
& = & f_{\hbox {\tiny $W$}}(w)\sum_{\ell=0}^p h^{-\ell}R^{0,\ell}_{\hbox {\tiny $W$}}(w)\{Y_j-m^*(w)\}(W_j-w)^\ell K_h(W_j-w) \nonumber \\
& = & \sum_{\ell=0}^p h^{-\ell}\tilde R^{0,\ell}_{\hbox {\tiny $W$}}(w)\{Y_j-m^*(w)\}(W_j-w)^\ell K_h(W_j-w) \nonumber, 
\end{eqnarray}
where $\tilde R^{0,\ell}_{\hbox {\tiny $W$}}(w)=f_{\hbox {\tiny $W$}}(w)R^{0,\ell}_{\hbox {\tiny $W$}}(w)$. By the definition of $R^{0,\ell}_{\hbox {\tiny $W$}}(w)$ given after (\ref{eq:Snwinv1}), when $\ell$ is odd, $\tilde R^{0,\ell}_{\hbox {\tiny $W$}}(w)$ is equal to $-hE\{f_{\hbox {\tiny $U$}}(w-X)\}f_{\hbox {\tiny $W$}}^{-1}(w)$ times the $[1, \ell+1]$ entry of $\bS^{-1}\widetilde{S}\bS^{-1}$, and, when $\ell$ is even, $\tilde R^{0,\ell}_{\hbox {\tiny $W$}}(w)$ is equal to the $[1, \ell+1]$ entry of $\bS^{-1}$. 

Define $\kappa_{\hbox {\tiny $W$}}(w, W)=E\{|Y-m^*(w)|^{2+\eta}|W\}$. Assuming $\tilde R^{0,\ell}_{\hbox {\tiny $W$}}(w)$ and $\|\kappa_{\hbox {\tiny $W$}}(w, W)\|_\infty$ bounded, we have, for some positive finite constants $C$ and $C'$, 
\begin{eqnarray}
&   & E|\phi_{P_{\hbox {\tiny $W$},j}f_{\hbox {\tiny $W$}}}(t)|^{2+\eta} \nonumber \\
& = & h^{-2-\eta}E\left| \int e^{-itw} \{Y_j-m^*(w)\} \sum_{\ell=0}^p \tilde R^{0,\ell}_{\hbox {\tiny $W$}}(w)\left(\frac{W_j-w}{h}\right)^\ell K\left(\frac{W_j-w}{h}\right) dw\right|^{2+\eta} \nonumber \\
& \le & C h^{-2-\eta} E\left|\sum_{\ell=0}^p \int e^{-itw} \{Y-m^*(w)\} \left(\frac{W_j-w}{h}\right)^\ell K\left(\frac{W_j-w}{h}\right) dw \right|^{2+\eta} \nonumber \\
& \asymp & C h^{-2-\eta} \sum_{\ell=0}^p E\left| \int e^{-itw} \{Y-m^*(w)\} \left(\frac{W_j-w}{h}\right)^\ell K\left(\frac{W_j-w}{h}\right) dw \right|^{2+\eta} \nonumber \\
& \le & C' h^{-2-\eta} \sum_{\ell=0}^p E\left| \int e^{-itw} \left(\frac{W_j-w}{h}\right)^\ell K\left(\frac{W_j-w}{h}\right) dw \right|^{2+\eta} \nonumber \\
& = & C' h^{-2-\eta} \sum_{\ell=0}^p \int \int \left | e^{-itw} \left(\frac{v-w}{h}\right)^\ell K\left(\frac{v-w}{h}\right) dw \right |^{2+\eta} f_{\hbox {\tiny $U$}}(v-v_2) dv f_{\hbox {\tiny $X$}}(v_2) dv_2 \nonumber \\
& = & C' h^{-2-\eta} \sum_{\ell=0}^p \int \left |-h e^{-itv} \int e^{ithv_1} v_1^\ell K(v_1)dv_1 \right |^{2+\eta} f_{\hbox {\tiny $U$}}(v-v_2) dv f_{\hbox {\tiny $X$}}(v_2) dv_2 \nonumber \\
& = & C' h^{-2-\eta} \sum_{\ell=0}^p \int \left |-h e^{-itv} i^{-\ell} \phi_{\hbox {\tiny $K$}}^{(\ell)}(th) \right |^{2+\eta} f_{\hbox {\tiny $U$}}(v-v_2) dv f_{\hbox {\tiny $X$}}(v_2) dv_2 \nonumber \\
& = & C' \sum_{\ell=0}^p \left |\phi_{\hbox {\tiny $K$}}^{(\ell)}(th) \right |^{2+\eta} .\nonumber
\end{eqnarray}
Therefore, $E|\phi_{P_{\hbox {\tiny $W$},j}f_{\hbox {\tiny $W$}}}(t)|^{2+\eta}$ is bounded by a non-random quantity of the same order in $h$ as $\sum_{\ell=0}^p \left |\phi_{\hbox {\tiny $K$}}^{(\ell)}(th) \right |^{2+\eta}.$

Second, by the definition of $Q_{\hbox {\tiny $W$},j}(w)$, we have 
$$Q_{\hbox {\tiny $W$},j}(w)f_{\hbox {\tiny $W$}}(w)= -\sum_{\ell=0}^p\sum_{\ell'=1}^p h^{\ell'-\ell}\tilde R^{0,\ell}_{\hbox {\tiny $W$}}(w) \frac{m^{*(\ell')}(w)}{\ell'!} (W_j-w)^{\ell+\ell'} K_h(W_j-w).$$ 
It follows that
\begin{eqnarray}
& & E|\phi_{Q_{\hbox {\tiny $W,j$}}f_{\hbox {\tiny $W$}}}(t)|^{2+\eta} \nonumber \\
& \asymp & \sum_{\ell=0}^p \sum_{\ell'=1}^p E\left|\int e^{itw}h^{\ell'-\ell}\tilde R^{0,\ell}_{\hbox {\tiny $W$}}(w) \frac{m^{*(\ell')}(w)}{\ell'!} (W-w)^{\ell+\ell'} K_h(W-w)   \right|^{2+\eta} \nonumber \\
& \le & C \sum_{\ell=0}^p \sum_{\ell'=1}^p h^{(\ell'-\ell)(2+\eta)}E\left|\int e^{itw} (W-w)^{\ell+\ell'}K_h(W-w)dw \right|^{2+\eta} \nonumber \\
& = & C \sum_{\ell=0}^p \sum_{\ell'=1}^p h^{2\ell'(2+\eta)}\left|\int e^{ithv} (-1)^{1+\ell+\ell'}v^{\ell+\ell'} K(v) dv \right|^{2+\eta} \nonumber \\
& = & C \sum_{\ell=0}^p \sum_{\ell'=1}^p h^{2\ell'(2+\eta)}\left| i^{-(\ell+\ell')}\phi_{\hbox {\tiny $K$}}^{(\ell+\ell')}(th) \right|^{2+\eta} \nonumber \\
& = & C \sum_{\ell=0}^p \sum_{\ell'=1}^p h^{2\ell'(2+\eta)}\left|\phi_{\hbox {\tiny $K$}}^{(\ell+\ell')}(th) \right|^{2+\eta}.
\end{eqnarray}
Therefore, $E|\phi_{Q_{\hbox {\tiny $W$},j}f_{\hbox {\tiny $W$}}}(t)|^{2+\eta}$ is bounded by a non-random quantity of the same order in $h$ as $\sum_{\ell=0}^p \sum_{\ell'=1}^p h^{2\ell'(2+\eta)}\left|\phi_{\hbox {\tiny $K$}}^{(\ell+\ell')}(th) \right|^{2+\eta}$.

\subsection{Normality with ordinary smooth $U$}
\label{s:normord}
Now we are ready to tackle the orders of $E|\tilde P_{n,j} (x)|^{2+\eta}$ and $E|\tilde Q_{n,j} (x)|^{2+\eta}$. For ordinary smooth measurement error, 
\begin{eqnarray}
& & E|\tilde P_{n,j} (x)|^{2+\eta} \nonumber \\
& = & E\left| \frac{1}{2\pi}\int e^{-itx}
\frac{\phi_{P_{\hbox {\tiny $W$},j}f_{\hbox {\tiny $W$}}}(t)}{\phi_{\hbox {\tiny $U$}}(t)} dt  \right|^{2+\eta} \nonumber \\
& \le & E\left\{ \left(\frac{1}{2\pi}\right)^{2+\eta}\int 
\frac{|\phi_{P_{\hbox {\tiny $W$},j}f_{\hbox {\tiny $W$}}}(t)|^{2+\eta}}{|\phi_{\hbox {\tiny $U$}}(t)|^{2+\eta}} dt  \right\} \nonumber \\
& = & \left(\frac{1}{2\pi}\right)^{2+\eta}\int 
\frac{E|\phi_{P_{\hbox {\tiny $W$},j}f_{\hbox {\tiny $W$}}}(t)|^{2+\eta}}{|\phi_{\hbox {\tiny $U$}}(t)|^{2+\eta}} dt  \nonumber \\
& = & \left(\frac{1}{2\pi}\right)^{2+\eta}\left\{\int_{|t|\le M} 
\frac{E|\phi_{P_{\hbox {\tiny $W$},j}f_{\hbox {\tiny $W$}}}(t)|^{2+\eta}}{|\phi_{\hbox {\tiny $U$}}(t)|^{2+\eta}} dt
+\int_{|t|> M} 
\frac{E|\phi_{P_{\hbox {\tiny $W$},j}f_{\hbox {\tiny $W$}}}(t)|^{2+\eta}}{|\phi_{\hbox {\tiny $U$}}(t)|^{2+\eta}} dt\right\}  \nonumber \\
& \le & \left(\frac{1}{2\pi}\right)^{2+\eta}\left[\left\{\inf_{|t|\le M}|\phi_{\hbox {\tiny $U$}}(t)|^{2+\eta}\right\}^{-1}  \int_{|t|\le M} 
E|\phi_{P_{\hbox {\tiny $W$},j}f_{\hbox {\tiny $W$}}}(t)|^{2+\eta} dt
+\int_{|t|> M} 
\frac{E|\phi_{P_{\hbox {\tiny $W$},j}f_{\hbox {\tiny $W$}}}(t)|^{2+\eta}}{\left(\frac{c}{2}|t|^{-b}\right)^{2+\eta}} dt\right]. \nonumber \\
\label{eq:breakEP}
\end{eqnarray}
Using the result in Section~\ref{s:interres}, we have that (\ref{eq:breakEP}) is bounded from above by 
\begin{equation}
C \int_{|t|\le M} \sum_{\ell=0}^p \left |\phi_{\hbox {\tiny $K$}}^{(\ell)}(th) \right |^{2+\eta} dt+  C' \int_{|t|>M} |t|^{b(2+\eta)}\sum_{\ell=0}^p \left |\phi_{\hbox {\tiny $K$}}^{(\ell)}(th) \right |^{2+\eta} dt, \nonumber 
\end{equation}
where the first term in the sum is 
\begin{equation}
C h^{-1}\sum_{\ell=0}^p \int_{|s|\le Mh} \left | \int e^{isu} u^{\ell} K(u) du \right | ds 
 \le  C h^{-1}\sum_{\ell=0}^p \int_{|s|\le Mh} \int \left |u^{\ell} K(u) \right | du dt, \nonumber
\end{equation}
where the integral is of order $O(h)$, and thus the first term is bounded by a finite constant. And the second term is equal to 
$$ C' h^{-b(2+\eta)-1} \int_{|s|>Mh} |s|^{b(2+\eta)}\sum_{\ell=0}^p \left |\phi_{\hbox {\tiny $K$}}^{(\ell)}(s) \right |^{2+\eta} ds,$$
which is of order $h^{-b(2+\eta)-1}$ under the assumption that $\int \{|s|^b |\phi_{\hbox {\tiny $K$}}^{(\ell)}(s) |\}^{2+\eta} ds< \infty$, for $\ell=0, 1, \ldots, p$. Hence, $E|\tilde P_{n,j} (x)|^{2+\eta}$ is bounded by a quantity of order $h^{-b(2+\eta)-1}$. 

As for $E|\tilde Q_{n,j}(x)|^{2+\eta}$, we have 
\begin{eqnarray}
& & E|\tilde Q_{n,j}(x)|^{2+\eta} \nonumber \\
& \le & C \left\{\int_{|t|\le M} \frac{E|\phi_{Q_{\hbox {\tiny $W,j$}}f_{\hbox {\tiny $W$}}}(t)|^{2+\eta}}{|\phi_{\hbox {\tiny $U$}}(t)|^{2+\eta}}dt+
\int_{|t|> M} \frac{E|\phi_{Q_{\hbox {\tiny $W,j$}}f_{\hbox {\tiny $W$}}}(t)|^{2+\eta}}{|\phi_{\hbox {\tiny $U$}}(t)|^{2+\eta}}dt  \right\} \nonumber \\
& \le & C_1 \int_{|t|\le M} E|\phi_{Q_{\hbox {\tiny $W,j$}}f_{\hbox {\tiny $W$}}}(t)|^{2+\eta}dt+C_2 \int_{|t|> M} |t|^{b(2+\eta)}E|\phi_{Q_{\hbox {\tiny $W$}}f_{\hbox {\tiny $W$}}}(t)|^{2+\eta}dt \nonumber \\
& \asymp & h^{2(2+\eta)+1}+C_3\sum_{\ell=0}^p\sum_{\ell'=1}^p h^{2\ell'(2+\eta)} \int_{|s|\le Mh} h^{-b(2+\eta)-1} |s|^{b(2+\eta)} |\phi_{\hbox {\tiny $K$}}(s)^{(\ell+\ell')}|^{2+\eta}dw, \nonumber 
\end{eqnarray}
which is of the order $h^{(2-b)(2+\eta)-1}$ assuming that $\int |t|^{b(2+\eta)}|\phi_{\hbox {\tiny $K$}}^{(k)}(t)|^{2+\eta}dt<\infty$ for $k=1,\ldots, 2p$.  

Finally, the order of $E(\tilde U^2_{n,j})$ is the same as that of the variance of $\hat m(x)_{\hbox {\tiny HZ}}$, which is $h^{-2b-1}$. Combing the above three parts of the derivations, we conclude that $E|\tilde U_{n,j}|^{2+\eta}=O(h^{-b(2+\eta)-1})$ and $E(\tilde U^2_{n,j})=Ch^{-2b-1}\{1+o(1)\}$. Therefore, if $\eta \ge 2$ and $(nh)^{-\eta/2}\to 0$ as $n\to \infty$, (\ref{eq:suffcond}) holds. This completes the proof of the asymptotic normality of $\hat m(x)_{\hbox {\tiny HZ}}$ when the density of $U$ is ordinary smooth. 

\subsection{Normality with super smooth $U$}
When $U$ is super smooth, we assume $\phi_{\hbox {\tiny $K$}}(t)$ supported on $[-1, 1]$. The main change from the derivations in Section~\ref{s:normord} is how to partition the range of integrations. 

For the order of $E|\tilde P_{n,j}(x)|^{2+\eta}$, we have 
\begin{eqnarray}
&& E|\tilde P_{n,j}(x)|^{2+\eta} \nonumber \\
& \le & \left(\frac{1}{2\pi}\right)^{2+\eta} \left\{\int_{|t|\le M} \frac{E|\phi_{P_{\hbox {\tiny $W,j$}} f_{\hbox {\tiny $W$}}}(t)|^{2+\eta}}{|\phi_{\hbox {\tiny $U$}}(t)|^{2+\eta}} dt+\int_{M<|t|\le 1/h} \frac{E|\phi_{P_{\hbox {\tiny $W,j$}} f_{\hbox {\tiny $W$}}}(t)|^{2+\eta}}{|\phi_{\hbox {\tiny $U$}}(t)|^{2+\eta}} dt   \right\} \nonumber \\
& \le & \left(\frac{1}{2\pi}\right)^{2+\eta}  \Big[\left\{\inf_{|t|\le M} |\phi_{\hbox {\tiny $U$}}(t)|^{2+\eta} \right\}^{-1} \int_{|t|\le M} E|\phi_{P_{\hbox {\tiny $W,j$}} f_{\hbox {\tiny $W$}}}(t)|^{2+\eta}dt\nonumber \\
& & +  \int_{M<|t|\le 1/h} \frac{E|\phi_{P_{\hbox {\tiny $W,j$}} f_{\hbox {\tiny $W$}}}(t)|^{2+\eta}}{|d_0|t|^{b_0}\exp(-|t|^b/d_2)/2|^{2+\eta}} dt\Big], \nonumber 
\end{eqnarray}
which is bounded from above by, using the result in Section~\ref{s:interres}, 
\begin{eqnarray}
& & C\int_{|t|\le M}\sum_{\ell=0}^p |\phi^{(\ell)}_{\hbox {\tiny $K$}}(th)|^{2+\eta}dt \nonumber \\
& & + C\int_{M<|t|\le 1/h}  |t|^{-b_0(2+\eta)}\exp\{(2+\eta)|t|^b/d_2 \}\sum_{\ell=0}^p |\phi^{(\ell)}_{\hbox {\tiny $K$}}(th)|^{2+\eta}dt \nonumber \\
& = & C h^{-1} \int_{|s|\le Mh}\sum_{\ell=0}^p |\phi^{(\ell)}_{\hbox {\tiny $K$}}(s)|^{2+\eta}dt \nonumber \\
& & + C h^{b_0(2+\eta)-1} \int_{Mh<|s|\le 1}  |s|^{-b_0(2+\eta)}\exp\{(2+\eta)|s|^b/(h^b d_2)\}\sum_{\ell=0}^p |\phi^{(\ell)}_{\hbox {\tiny $K$}}(s)|^{2+\eta}dt, \nonumber
\end{eqnarray}
of which the first term is $O(1)$ under the assumption that $\|\phi^{\ell}_{\hbox {\tiny $K$}}(t)\|_{\infty}<\infty$, for $\ell=0, 1, \ldots, p$, and the second term is bounded from above by, under the same assumption,  
\begin{eqnarray}
& & Ch^{b_0(2+\eta)-1}\exp\{(2+\eta)h^{-b}/d_2\}\int_{Mh < |s| \le 1} |s|^{-b_0(2+\eta)}ds. \nonumber \\
& = & \left\{
\begin{array}{ll}
Ch^{b_0(2+\eta)-1} \exp\{(2+\eta)h^{-b}/d_2\} & \textrm{ if $b_0<1/(2+\eta)$,} \\
Ch^{-1} \exp\{(2+\eta)h^{-b}/d_2\} & \textrm{ if $b_0=1/(2+\eta)$,} \\
C \exp\{(2+\eta)h^{-b}/d_2\} & \textrm{ if $b_0>1/(2+\eta)$.} 
\end{array} 
\right.\nonumber 
\end{eqnarray} 

Similarly, one can show that 
\begin{equation}
E|\tilde Q_{n,j}(x)|^{2+\eta} \le 
\left\{
\begin{array}{ll}
Ch^{(2+b_0)(2+\eta)-1} \exp\{(2+\eta)h^{-b}/d_2\} & \textrm{ if $b_0<1/(2+\eta)$,} \\
Ch^{2(2+\eta)-1} \exp\{(2+\eta)h^{-b}/d_2\} & \textrm{ if $b_0=1/(2+\eta)$,} \\
C h^{2(2+\eta)}\exp\{(2+\eta)h^{-b}/d_2\} & \textrm{ if $b_0>1/(2+\eta)$.} 
\end{array} 
\right.
\nonumber 
\end{equation} 
Hence, 
\begin{eqnarray}
E|\tilde U_{n,j}(x)|^{2+\eta} & \le &
\left\{
\begin{array}{ll}
Ch^{b_0(2+\eta)-1} \exp\{(2+\eta)h^{-b}/d_2\} & \textrm{ if $b_0<1/(2+\eta)$,} \\
Ch^{-1} \exp\{(2+\eta)h^{-b}/d_2\} & \textrm{ if $b_0=1/(2+\eta)$,} \\
C \exp\{(2+\eta)h^{-b}/d_2\} & \textrm{ if $b_0>1/(2+\eta)$,} 
\end{array} 
\right.\nonumber \\
& \le & C h^{(2+\eta)b_3-1}\exp\{(2+\eta)h^{-b}/d_2\}, \nonumber 
\end{eqnarray} 
where $b_3=b_0I(b_0<0.5)$. 

Using the variance result for $\hat m_{\hbox {\tiny HZ}}(x)$ for the super smooth $U$, we have $E|\tilde U^2_{n,j}(x)|\le Ch^{2b_3-2}\exp(2h^{-b}/d_2)$. Putting these together, we have 
\begin{equation}
\frac{E|\tilde U_{n,j}(x)|^{2+\eta}}{n^{\eta/2}E|\tilde U^2_{n,j}(x)|^{(2+\eta)/2}}\asymp h^{1+\eta/2}/n^{\eta/2}\to 0,\nonumber
\end{equation}
as $n \to \infty$, for any $\eta>0$. Hence, the normality of $\hat m_{\hbox {\tiny HZ}}(x)$ for the case with super smooth $U$ is proved. 

\noindent 
\setcounter{section}{0}
\setcounter{equation}{0}
\setcounter{figure}{0}
\renewcommand{\theequation}{D.\arabic{equation}}
\renewcommand{\thefigure}{D.\arabic{figure}}
\renewcommand{\thesection}{D.\arabic{section}}

\section*{Appendix D: Additional simulation studies without assuming measurement error distribution known}
\label{s:appD}

To address the practical scenario where the measurement error distribution is unknown, we consider two strategies in Section 7 in the main article. The first strategy, which we recommend, is to assume Laplace measurement error with characteristic function given by $\phi_{\hbox {\tiny $U$}}(t)=1/\{1+(\sigma^2_u/2) t^2\}$, in which $\sigma^2_u$ is estimated using repeated measures following equation (4.3) in \citet{Carroll06}. The second strategy, which is inferior to the first strategy according to Figure 9 in the main article, is to estimate $\phi_{\hbox {\tiny $U$}}(t)$ following the approach in \citet{DHM2008}. Suppose there are two repeated measures, $W_{j,1}$ and $W_{j,2}$, for each true covariate value $X_j$, for $j=1, \ldots, n$, then, assuming a symmetric measurement error distribution, this approach yields an estimated characteristic function of the measurement error associated with $W_{j,k}$ ($k=1,2$) given by $\hat \phi_{\hbox {\tiny $U_1$}}(t)=\sqrt{\sum_{j=1}^n \cos \{it (W_{j,1}-W_{j,2})\}/n}$. Then we define $W_j=(W_{j,1}+W_{j,2})/2$ as the error-contaminated surrogate of $X_j$, for $j=1, \ldots, n$, and the estimated characteristic function associated with the measurement error in $W_j=X_j+U_j$ is given by $\hat \phi_{\hbox {\tiny $U$}}(t)=\{\hat \phi_{\hbox {\tiny $U_1$}}(t/2)\}^2$.

\begin{figure}
	\centering
	\setlength{\linewidth}{4.5cm}
	\subfigure[]{ \includegraphics[width=\linewidth]{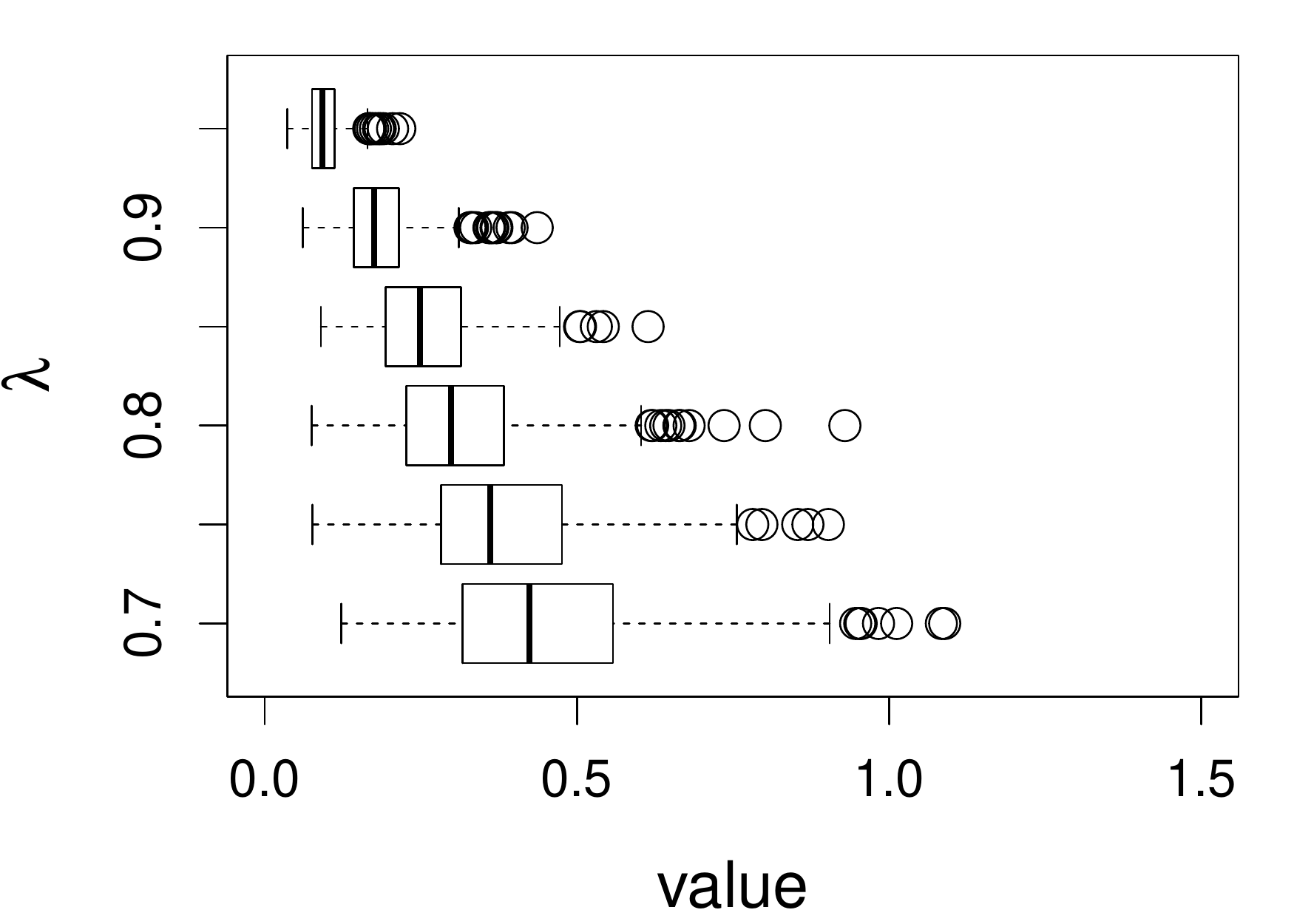} }
	\subfigure[]{ \includegraphics[width=\linewidth]{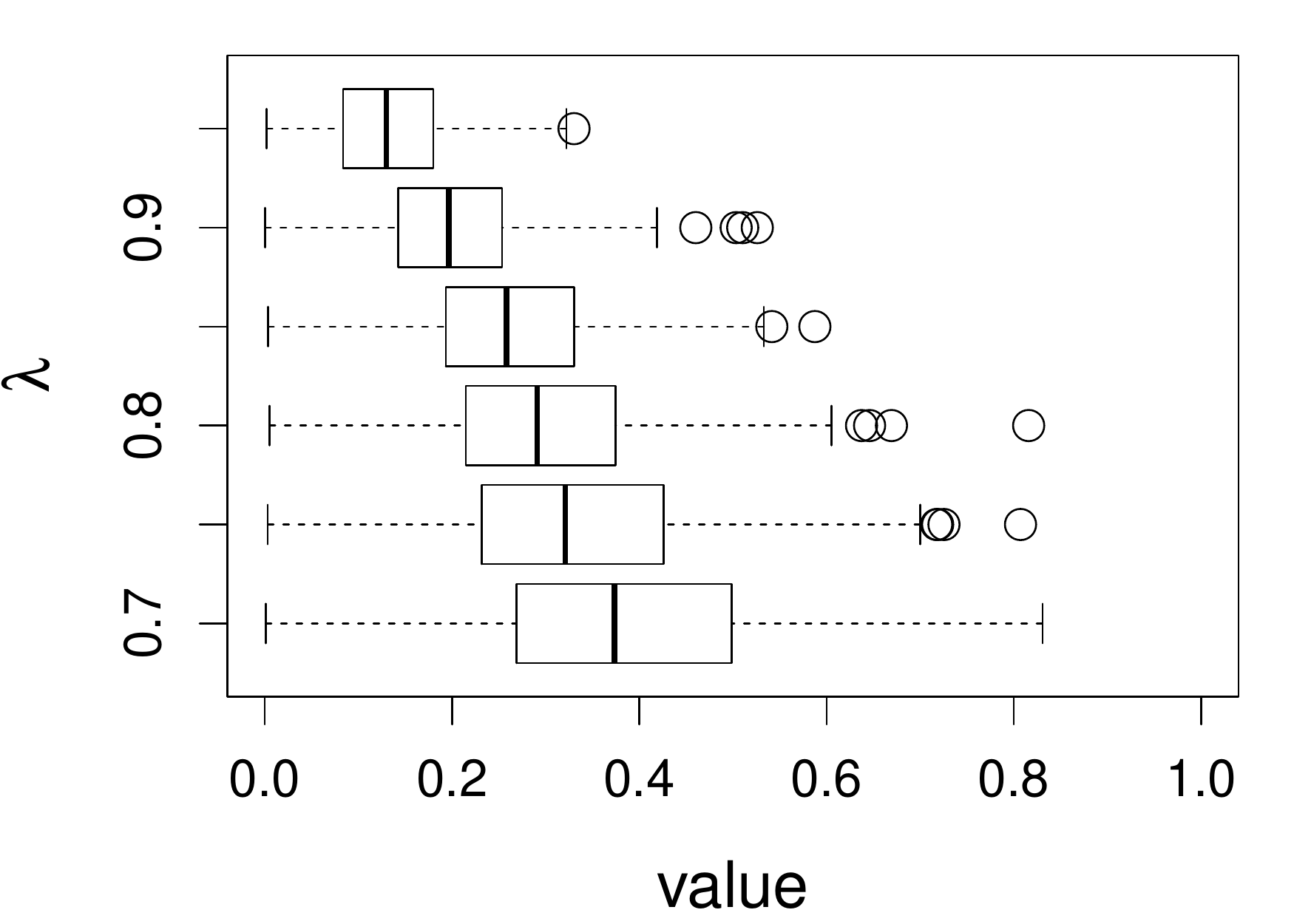} }
	\subfigure[]{ \includegraphics[width=\linewidth]{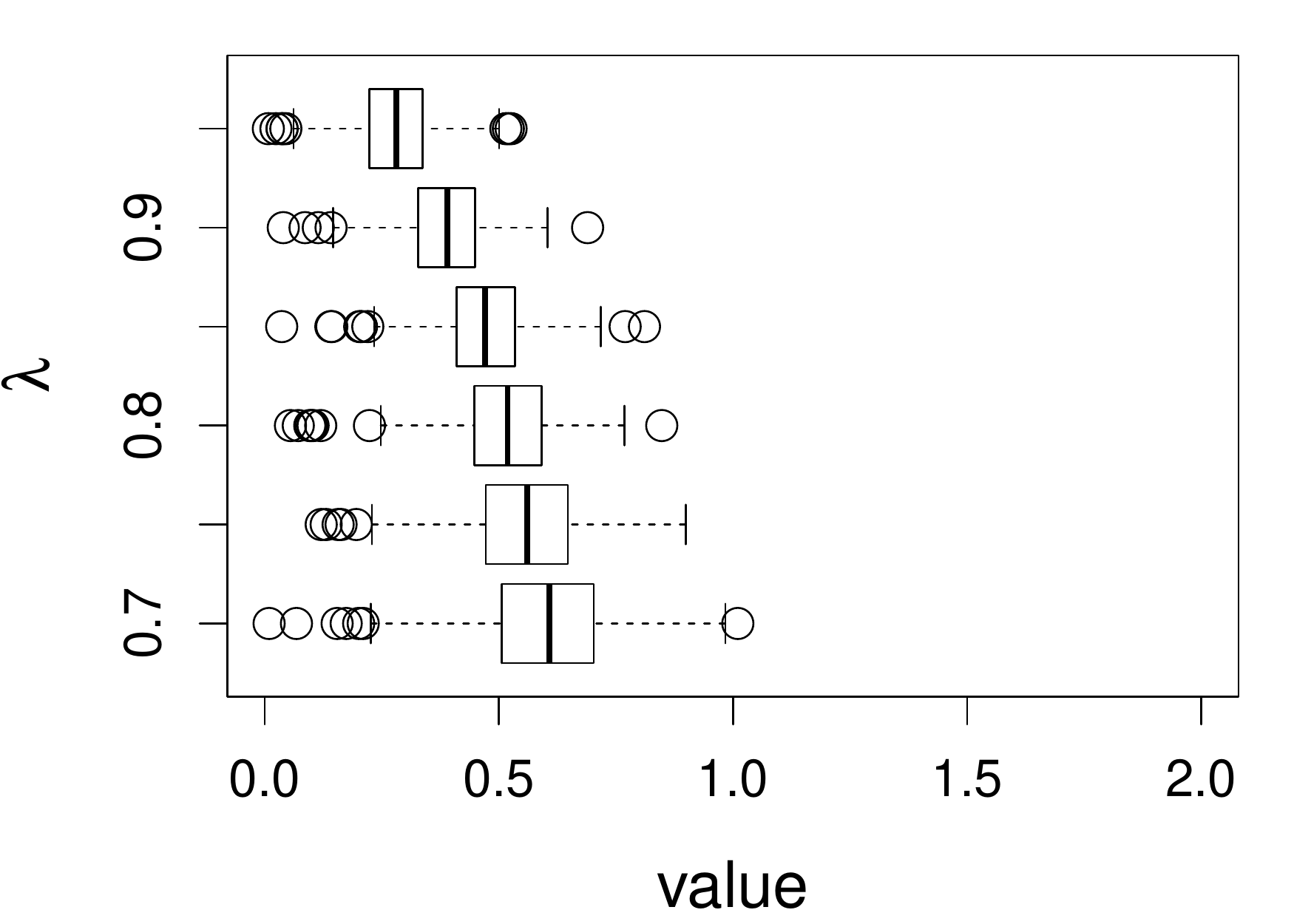} }\\
	\subfigure[]{ \includegraphics[width=\linewidth]{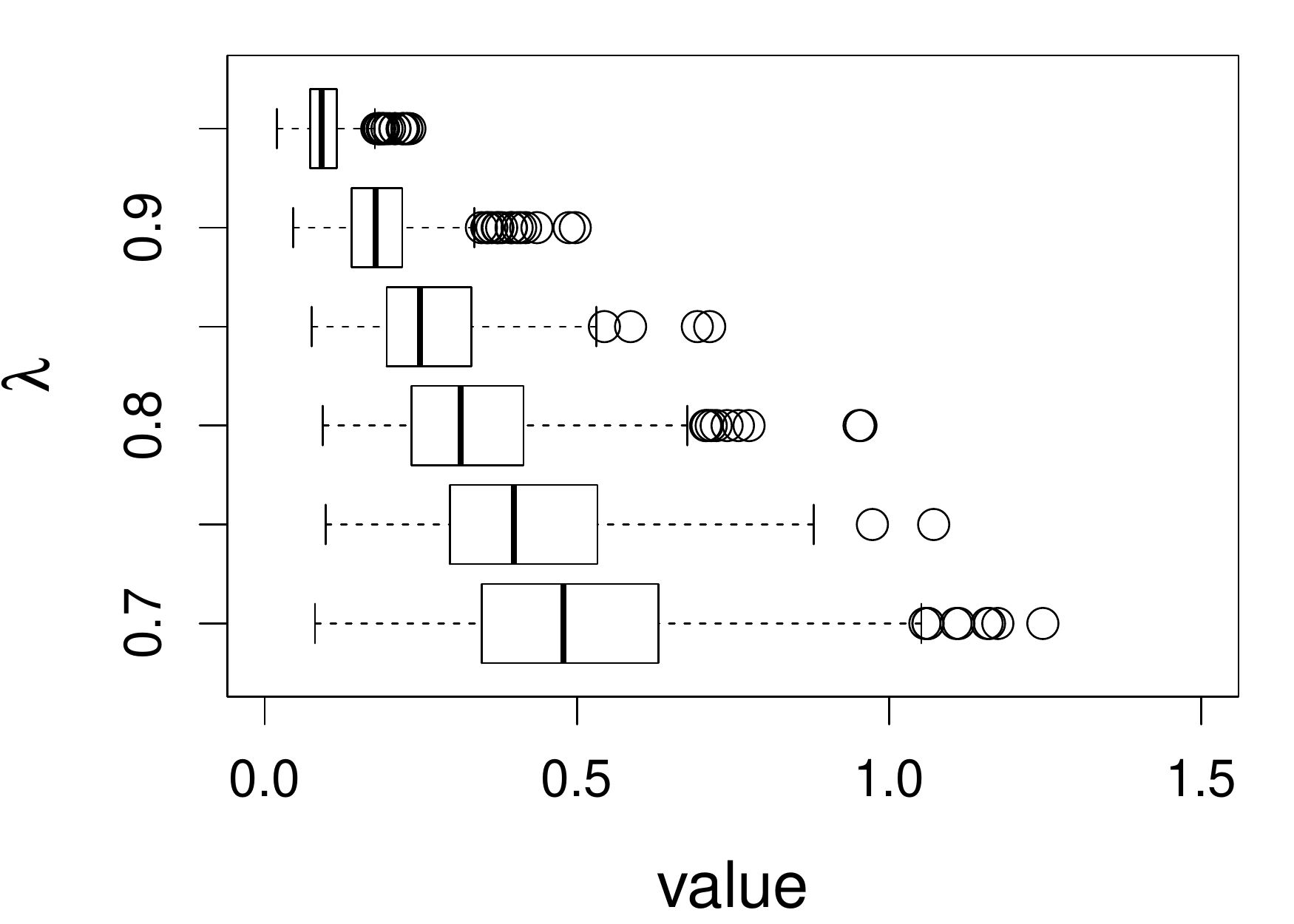} }
	\subfigure[]{ \includegraphics[width=\linewidth]{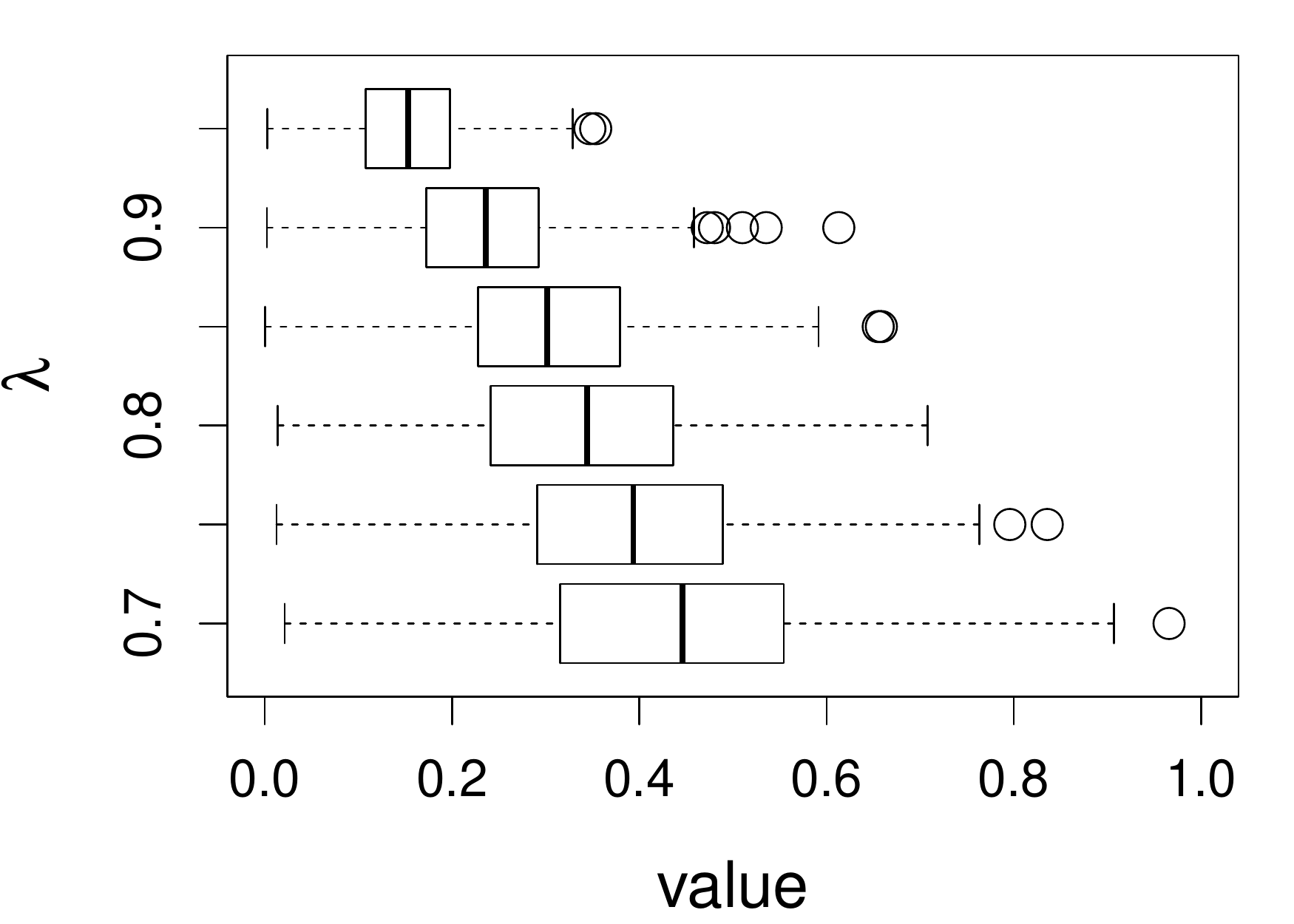} }
	\subfigure[]{ \includegraphics[width=\linewidth]{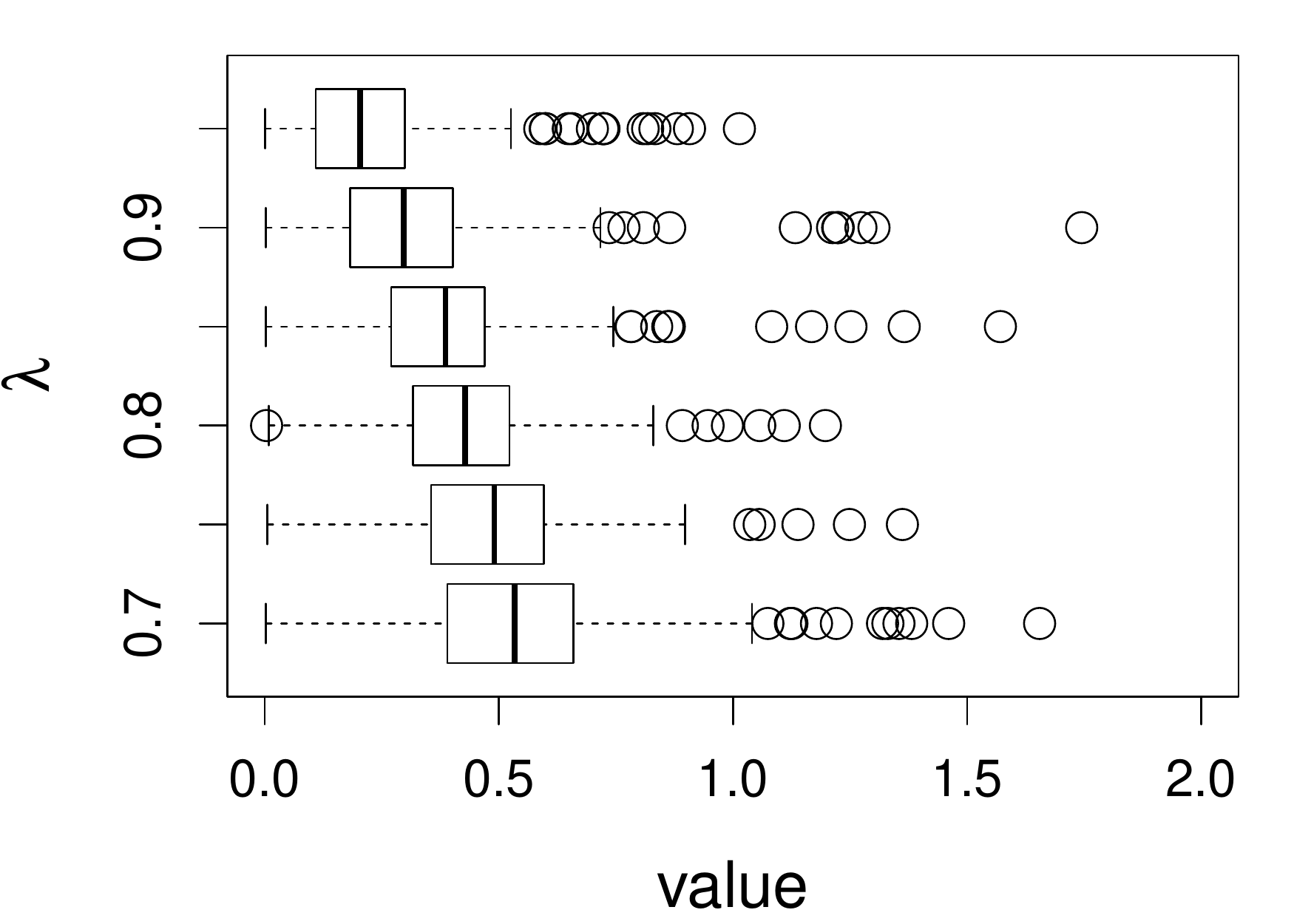} }\\
	\subfigure[]{ \includegraphics[width=\linewidth]{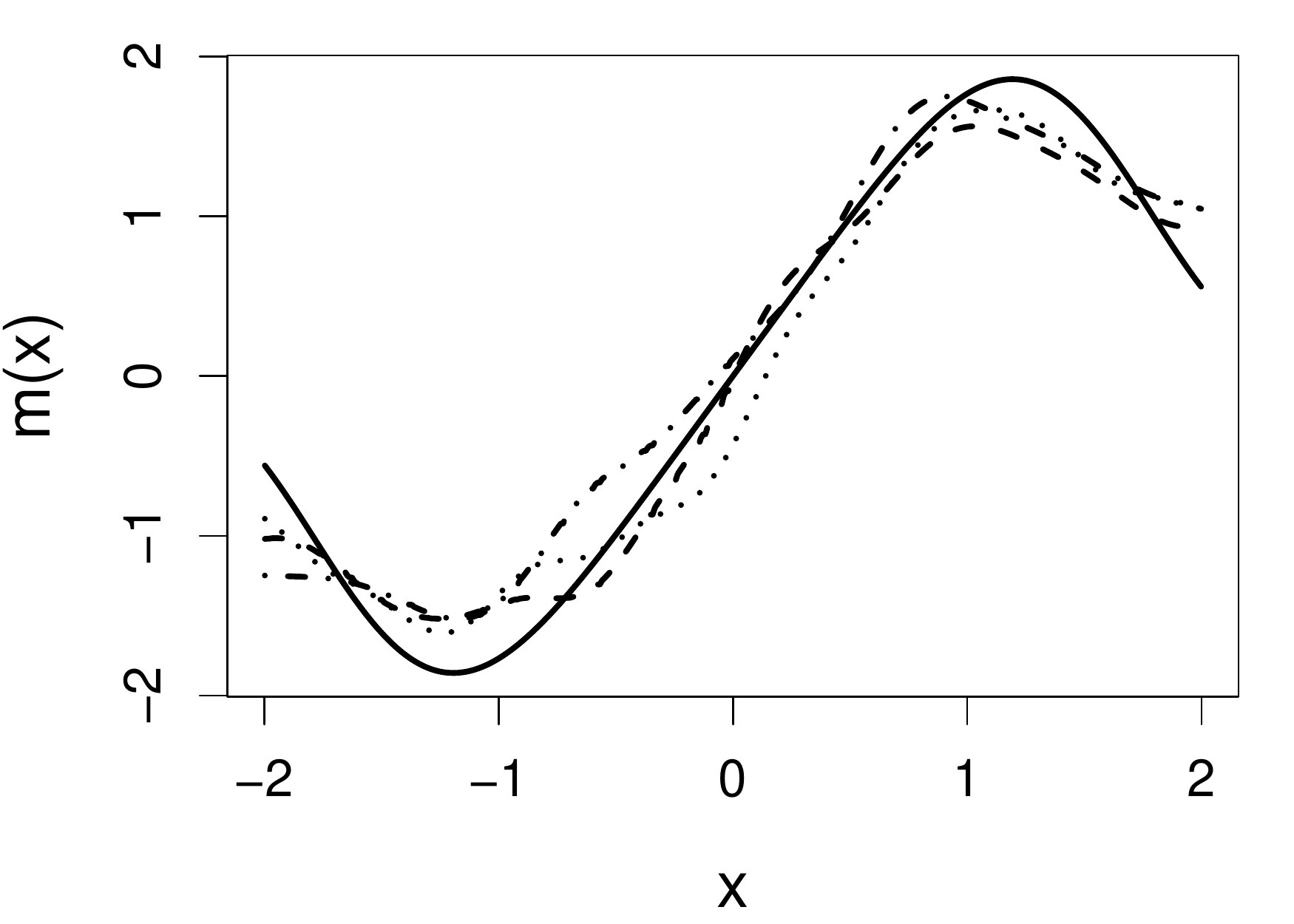} }
	\subfigure[]{ \includegraphics[width=\linewidth]{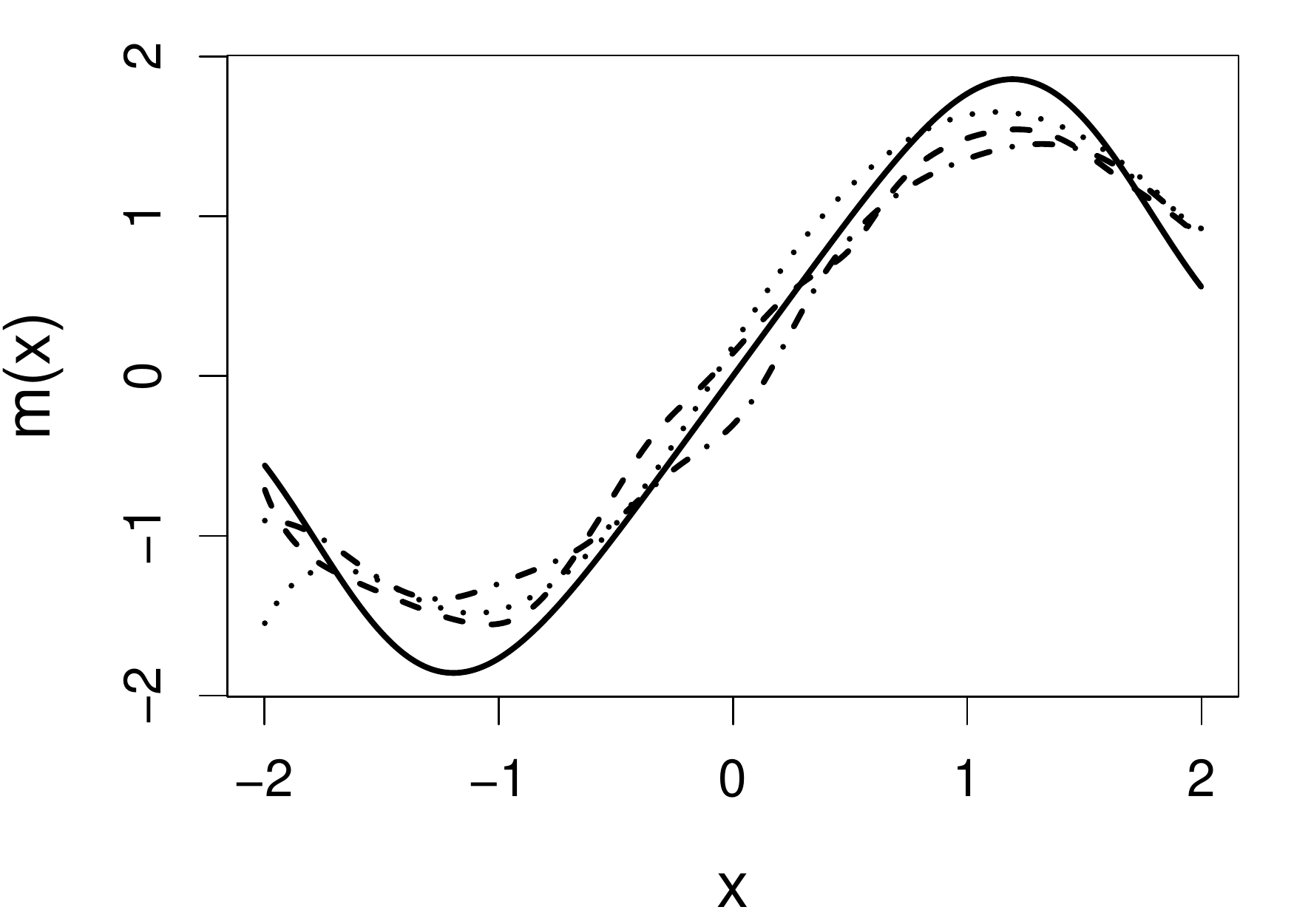} }
	\subfigure[]{ \includegraphics[width=\linewidth]{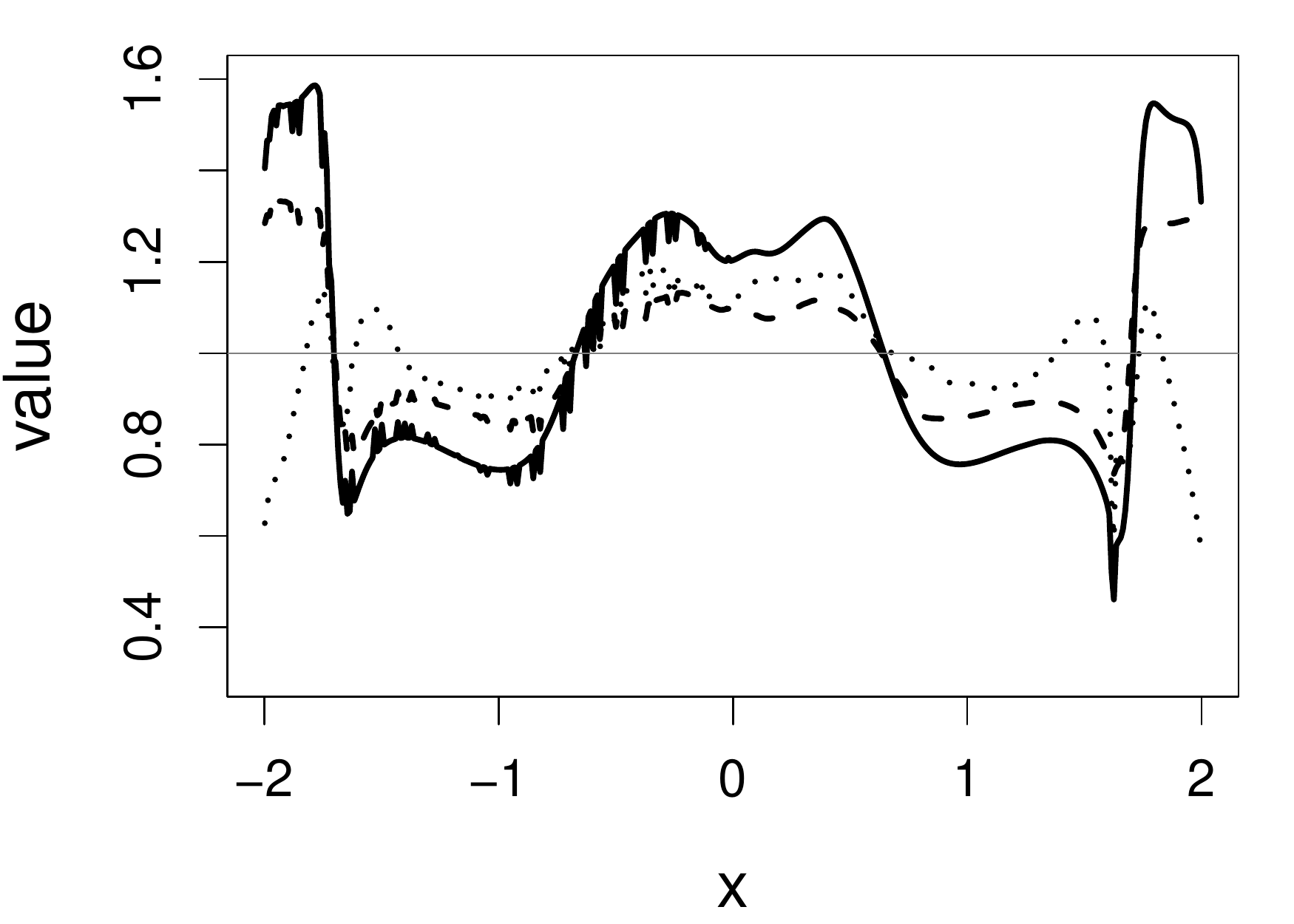} }
\caption{Simulation results under (C1) using the theoretical optimal $h$, assuming Laplace $U$ with $\sigma^2_u$ estimated using repeated measures. Panels (a) \& (d): boxplots of ISEs versus $\lambda$ for $\hat m_{\hbox {\tiny HZ}}(x)$ and $\hat m_{\hbox {\tiny DFC}}(x)$, respectively. Panels (b) \& (e): boxplots of PAE(1) versus $\lambda$ for $\hat m_{\hbox {\tiny HZ}}(1)$ and $\hat m_{\hbox {\tiny DFC}}(1)$, respectively. Panels (c) \& (f): boxplots of PAE(2) versus $\lambda$ for $\hat m_{\hbox {\tiny HZ}}(2)$ and $\hat m_{\hbox {\tiny DFC}}(2)$, respectively. Panels (g) \& (h): quantile curves when $\lambda=0.85$ for $\hat m_{\hbox {\tiny HZ}}(x)$ and $\hat m_{\hbox {\tiny DFC}}(x)$, respectively, based on ISEs (dashed lines for the first quartile, dotted lines for the second quartile, and dot-dashed lines for the third quartile, solid lines for the truth). Panel (i): PmAER (dashed line), PsdAER (dotted line), and PMSER (solid line) versus $x$ when $\lambda=0.85$; the horizontal reference line highlights the value 1.} 
	\label{Sim1LapLap500:box}
\end{figure}

\begin{figure}
	\centering
	\setlength{\linewidth}{4.5cm}
	\subfigure[]{ \includegraphics[width=\linewidth]{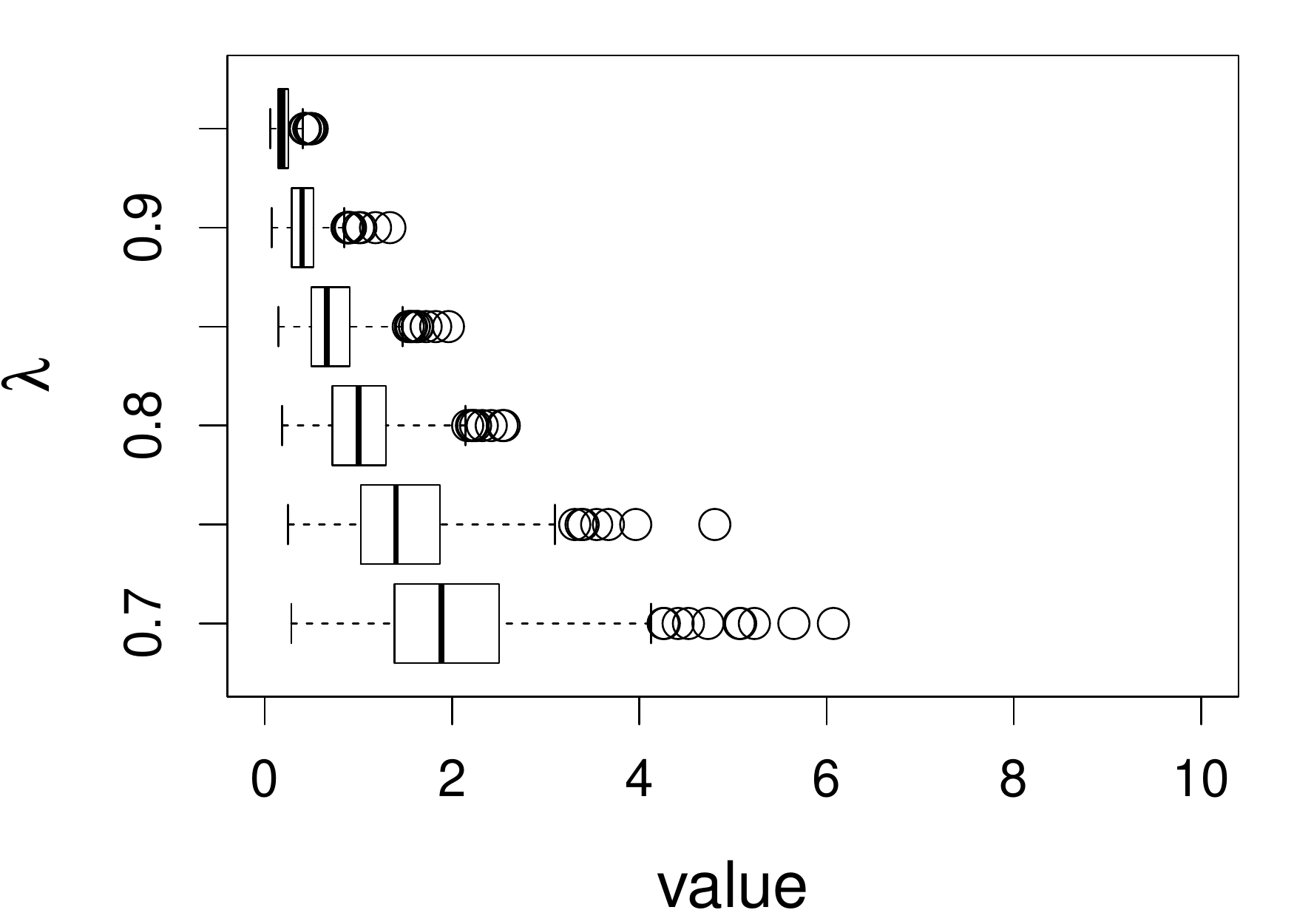} }
	\subfigure[]{ \includegraphics[width=\linewidth]{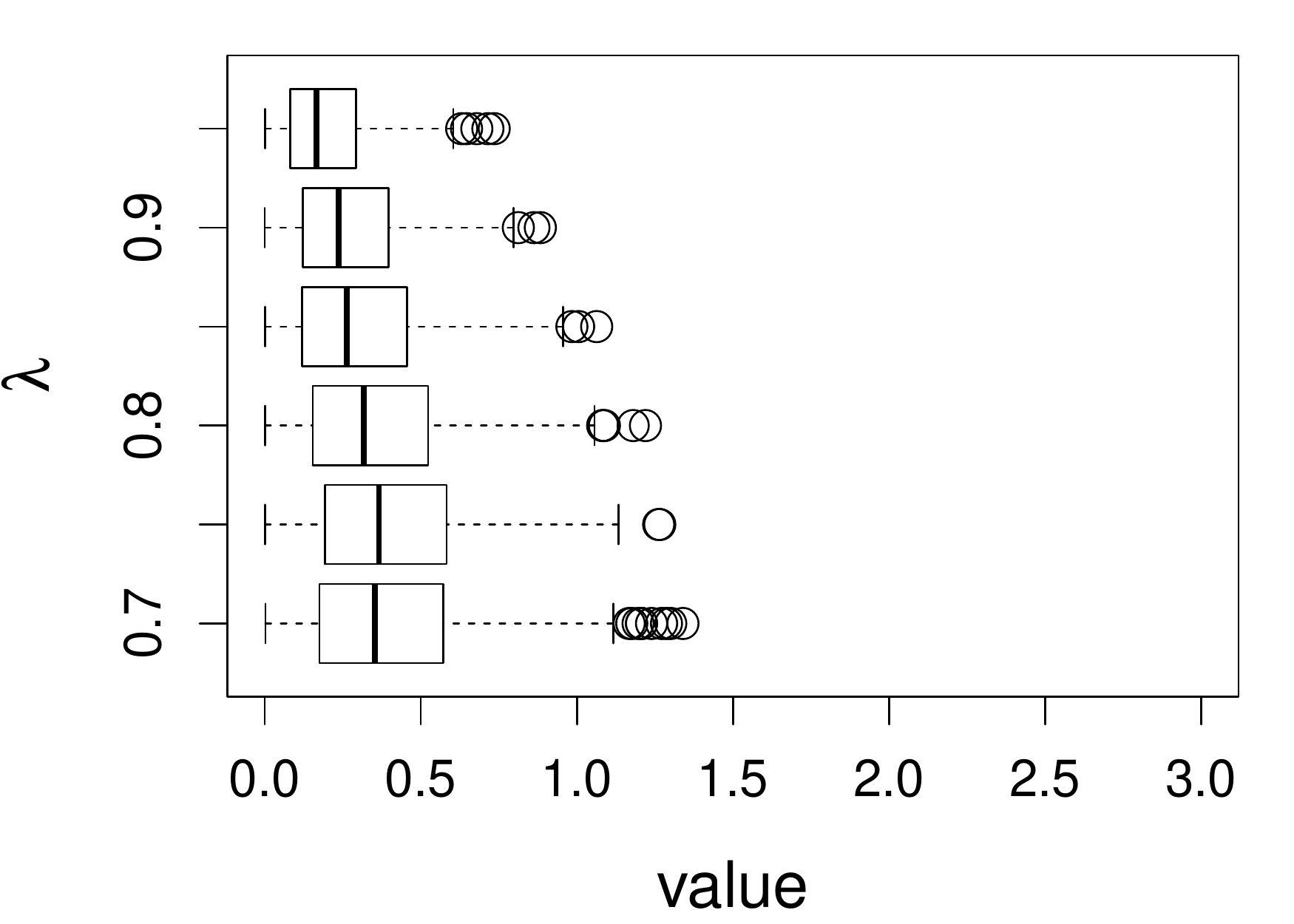} }
	\subfigure[]{ \includegraphics[width=\linewidth]{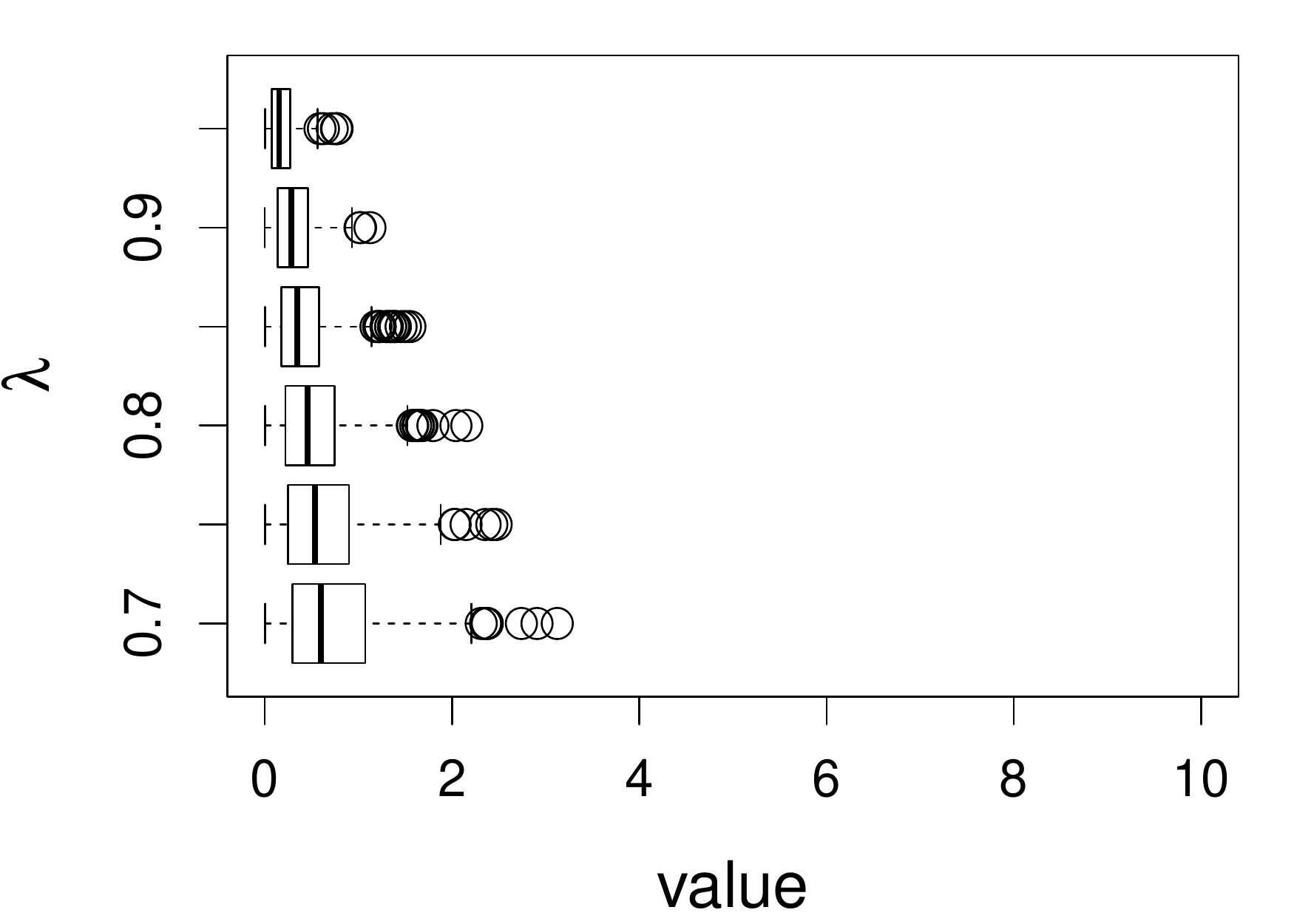} }\\
	\subfigure[]{ \includegraphics[width=\linewidth]{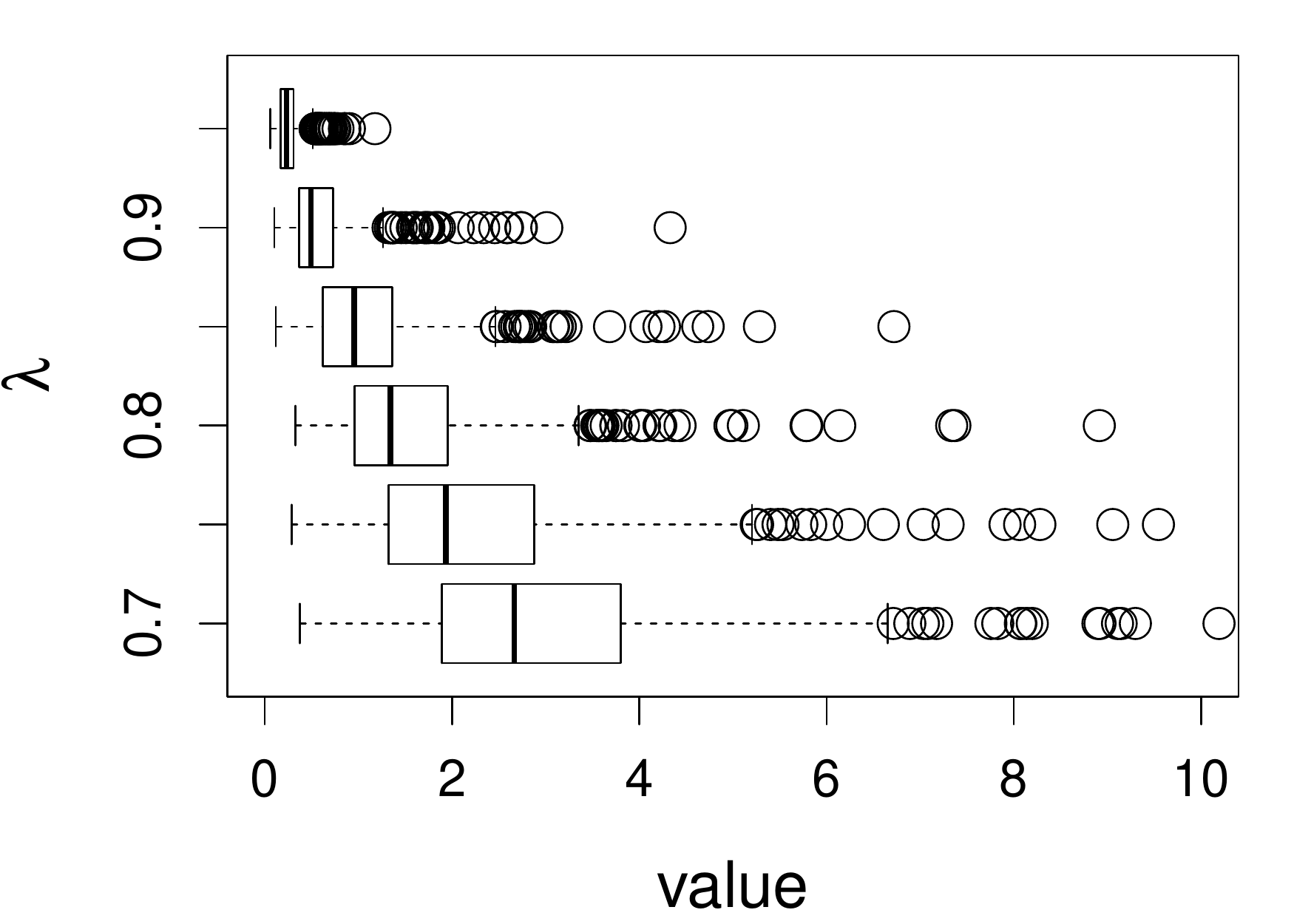} }
	\subfigure[]{ \includegraphics[width=\linewidth]{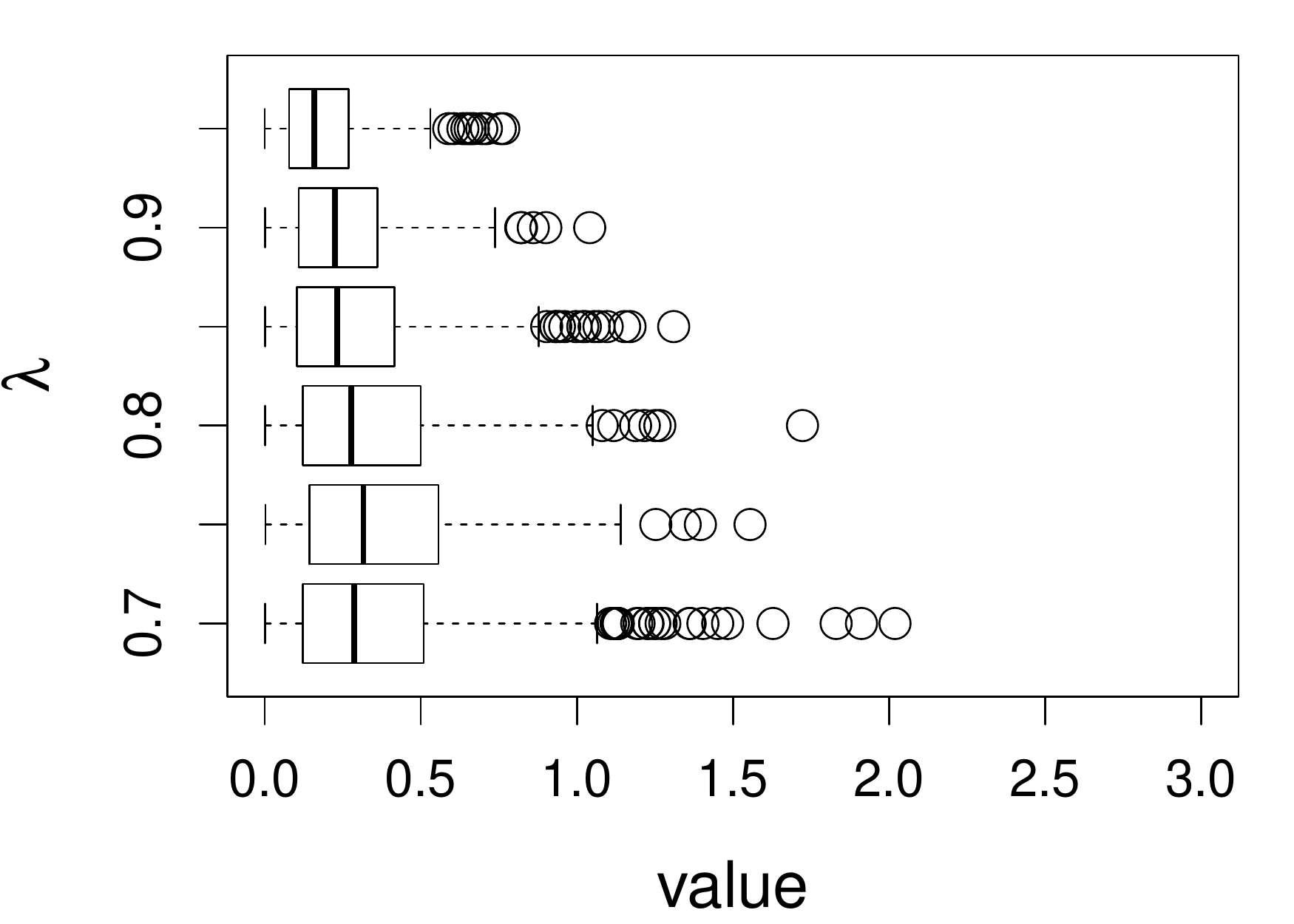} }
	\subfigure[]{ \includegraphics[width=\linewidth]{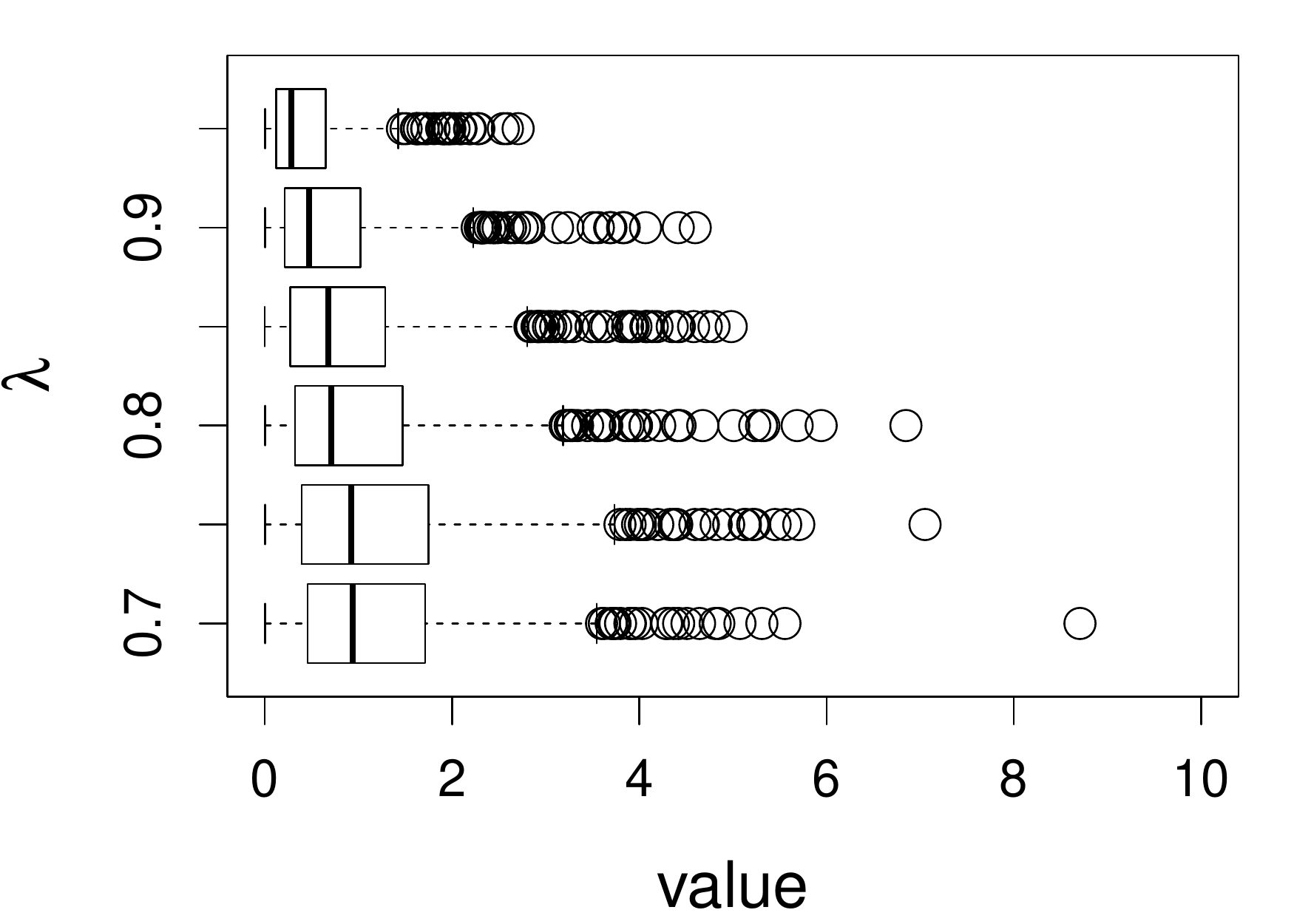} }\\
	\subfigure[]{ \includegraphics[width=\linewidth]{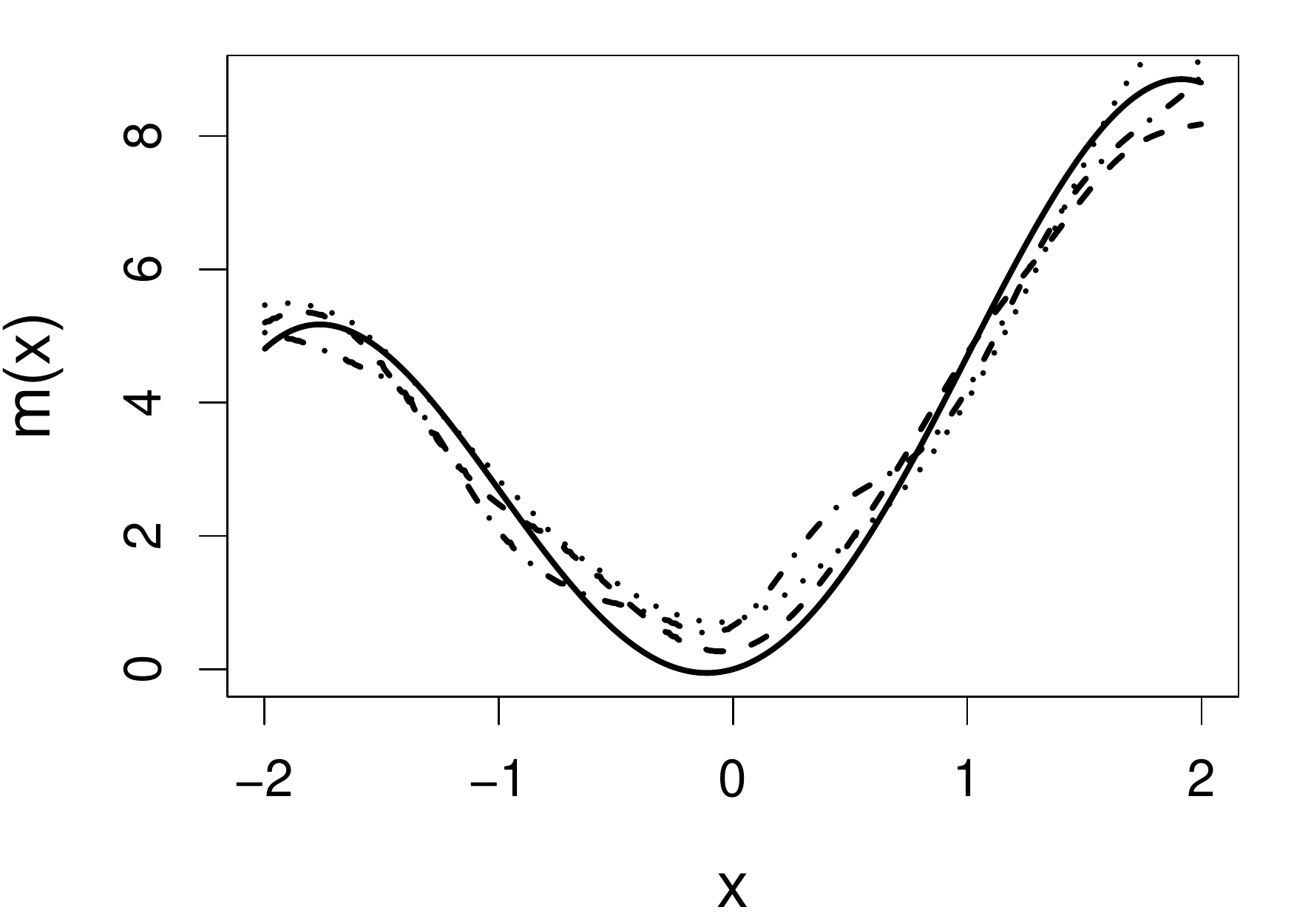} }
	\subfigure[]{ \includegraphics[width=\linewidth]{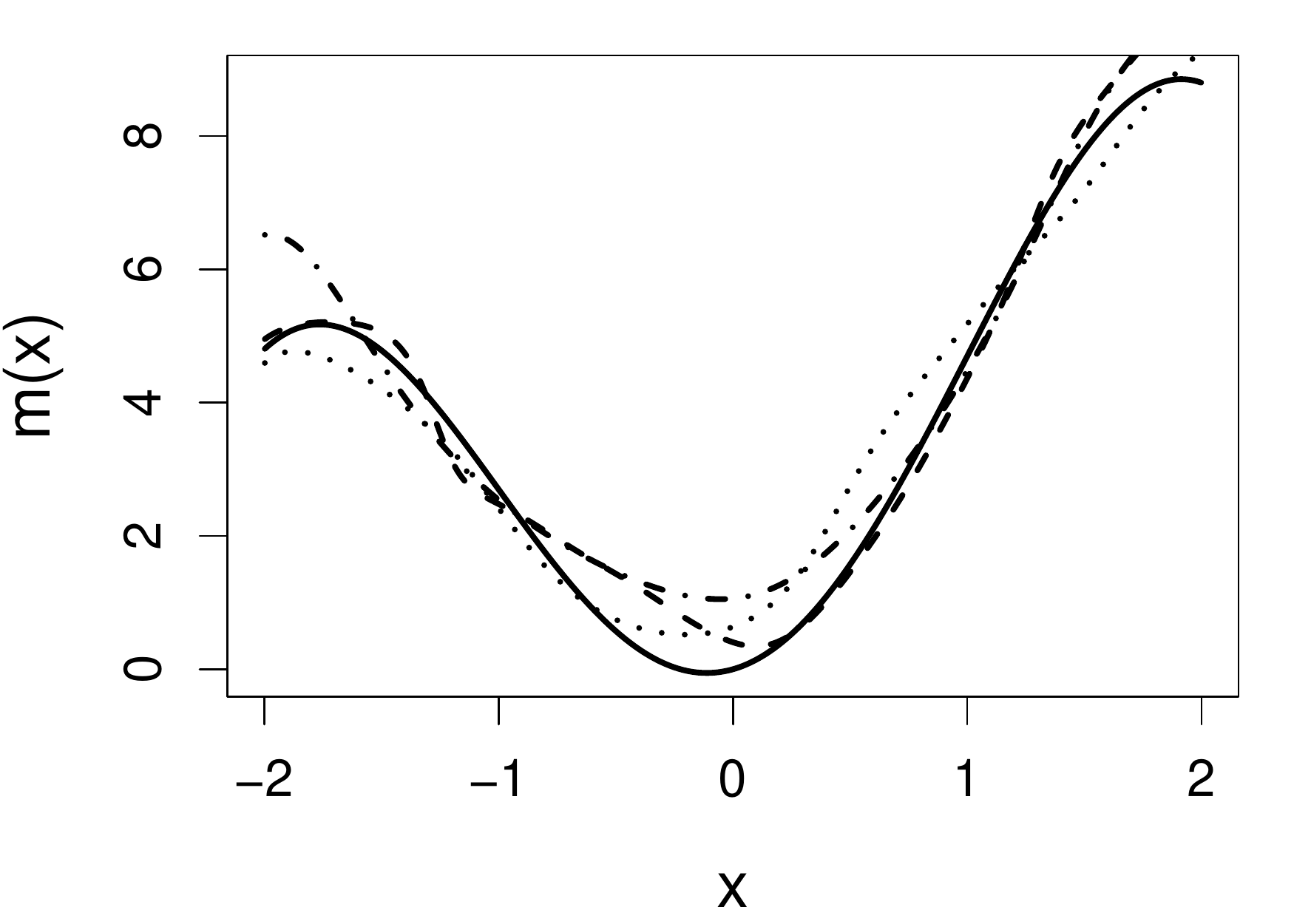} }
	\subfigure[]{ \includegraphics[width=\linewidth]{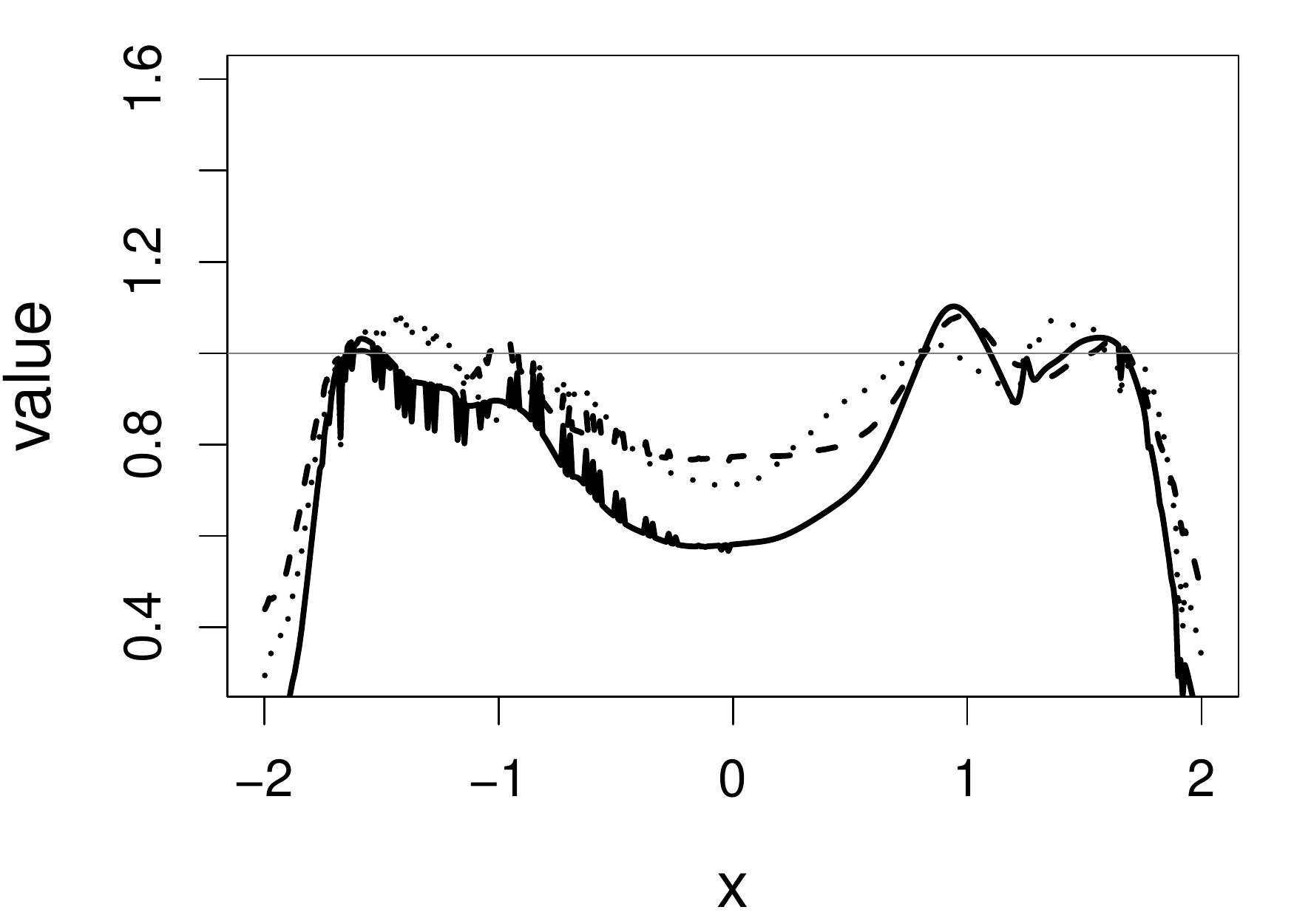} }
\caption{Simulation results under (C3) using the theoretical optimal $h$, assuming Laplace $U$ with $\sigma^2_u$ estimated using repeated measures. Panels (a) \& (d): boxplots of ISEs versus $\lambda$ for $\hat{m}_{\hbox {\tiny HZ}}(x)$ and $\hat{m}_{\hbox {\tiny DFC}}(x)$, respectively. Panels (b) \& (e): boxplots of PAE(1) versus $\lambda$ for  $\hat{m}_{\hbox {\tiny HZ}}(1)$ and $\hat{m}_{\hbox {\tiny DFC}}(1)$, respectively. Panels (c) \& (f): boxplots of PAE(2) versus $\lambda$ for  $\hat{m}_{\hbox {\tiny HZ}}(2)$ and $\hat{m}_{\hbox {\tiny DFC}}(2)$, respectively. Panels (g) \& (h): quantile curves when $\lambda=0.8$ for $\hat{m}_{\hbox {\tiny HZ}}(x)$ and $\hat{m}_{\hbox {\tiny DFC}}(x)$, respectively, based on ISEs (dashed lines for the first quartile, dotted lines for the second quartile, dot-dashed lines for the third quartile, and solid lines for the truth). Panel (i): PmAER (dashed line), PsdAER (dotted line), and PMSER (solid line) versus $x$ when $\lambda=0.8$; the horizontal reference line highlights the value 1.} 
	\label{Sim3LapLap500:box}
\end{figure}

\begin{figure}
	\centering
	\setlength{\linewidth}{4.5cm}
	\subfigure[]{ \includegraphics[width=\linewidth]{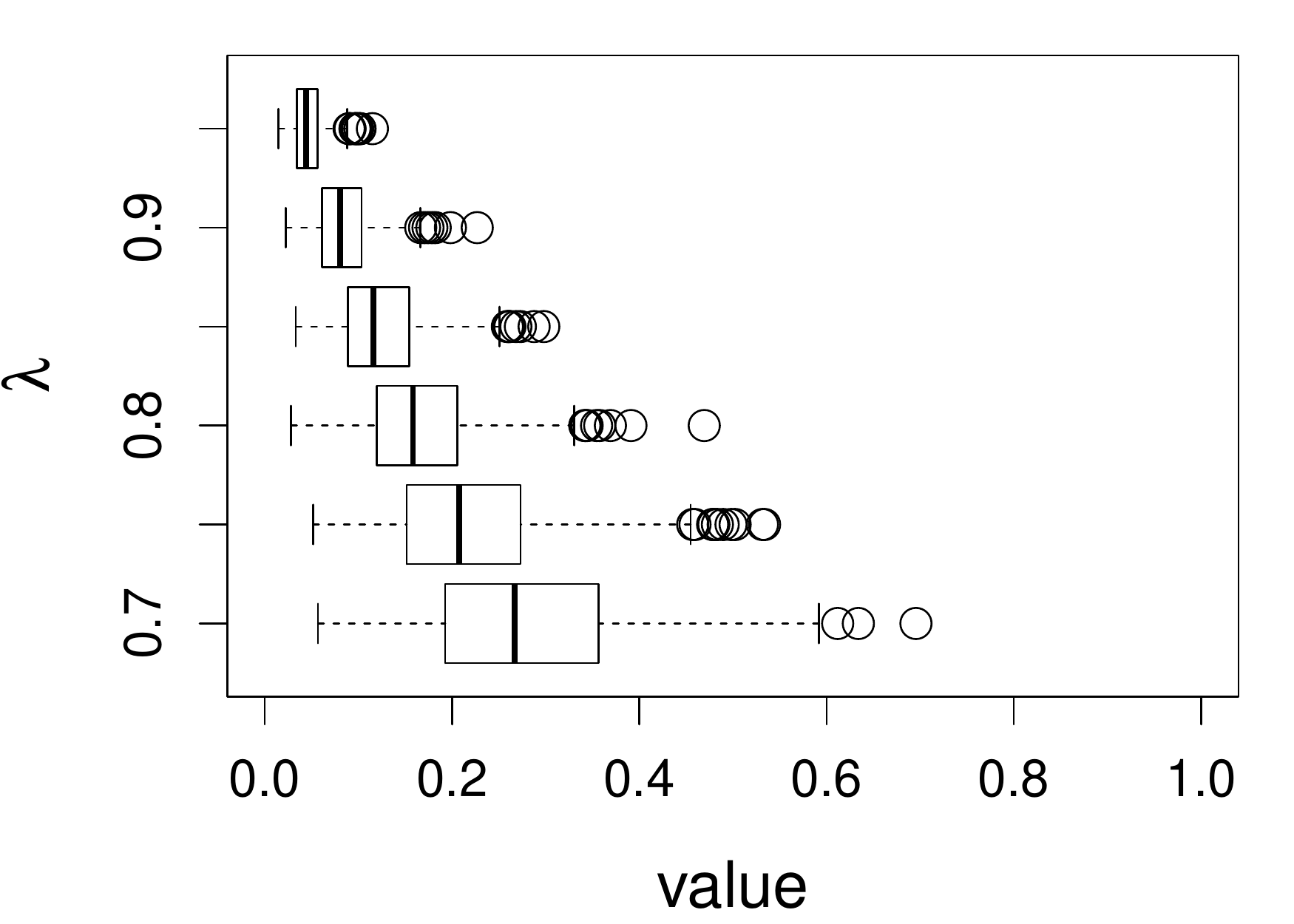} }
	\subfigure[]{ \includegraphics[width=\linewidth]{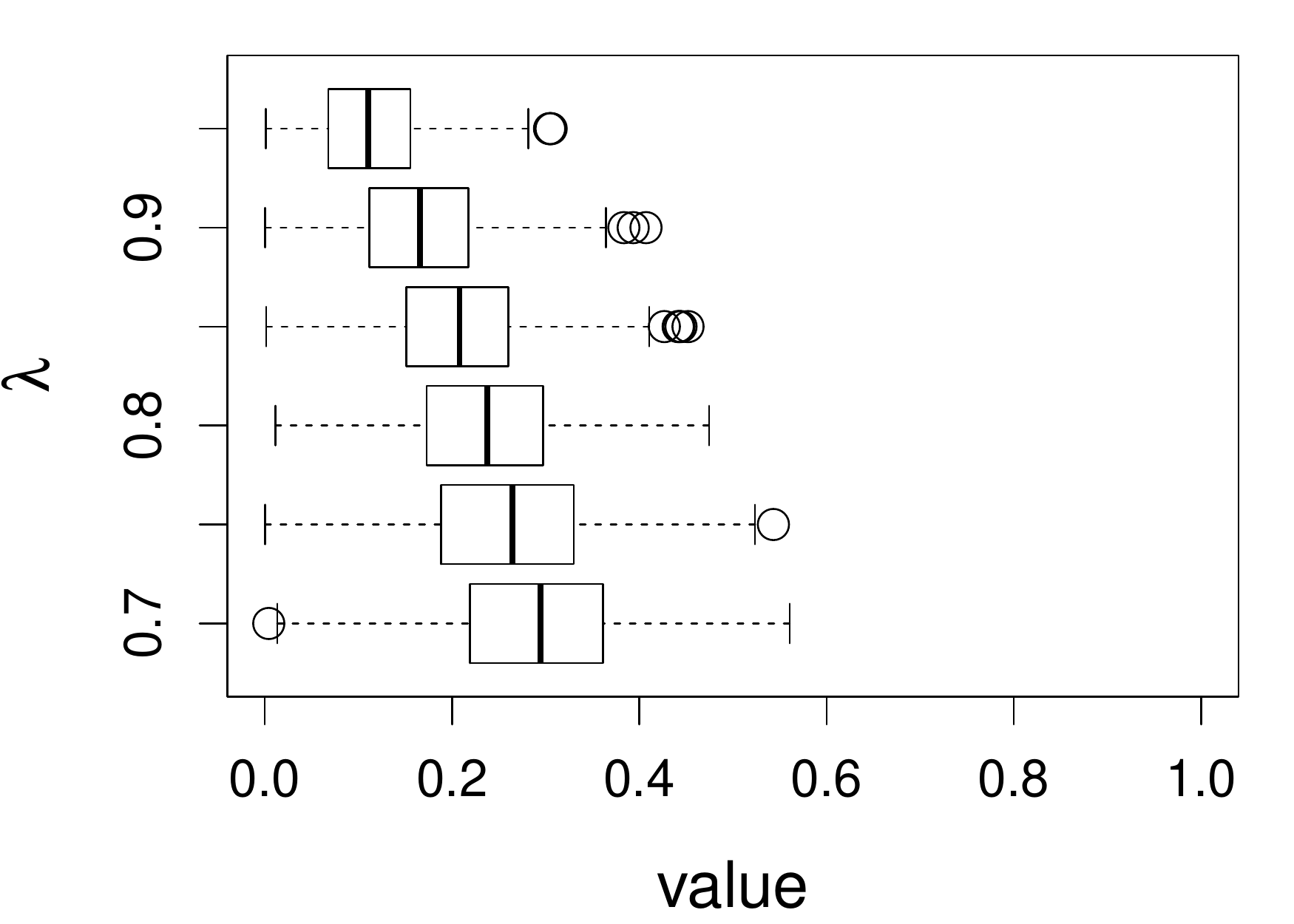} }
	\subfigure[]{ \includegraphics[width=\linewidth]{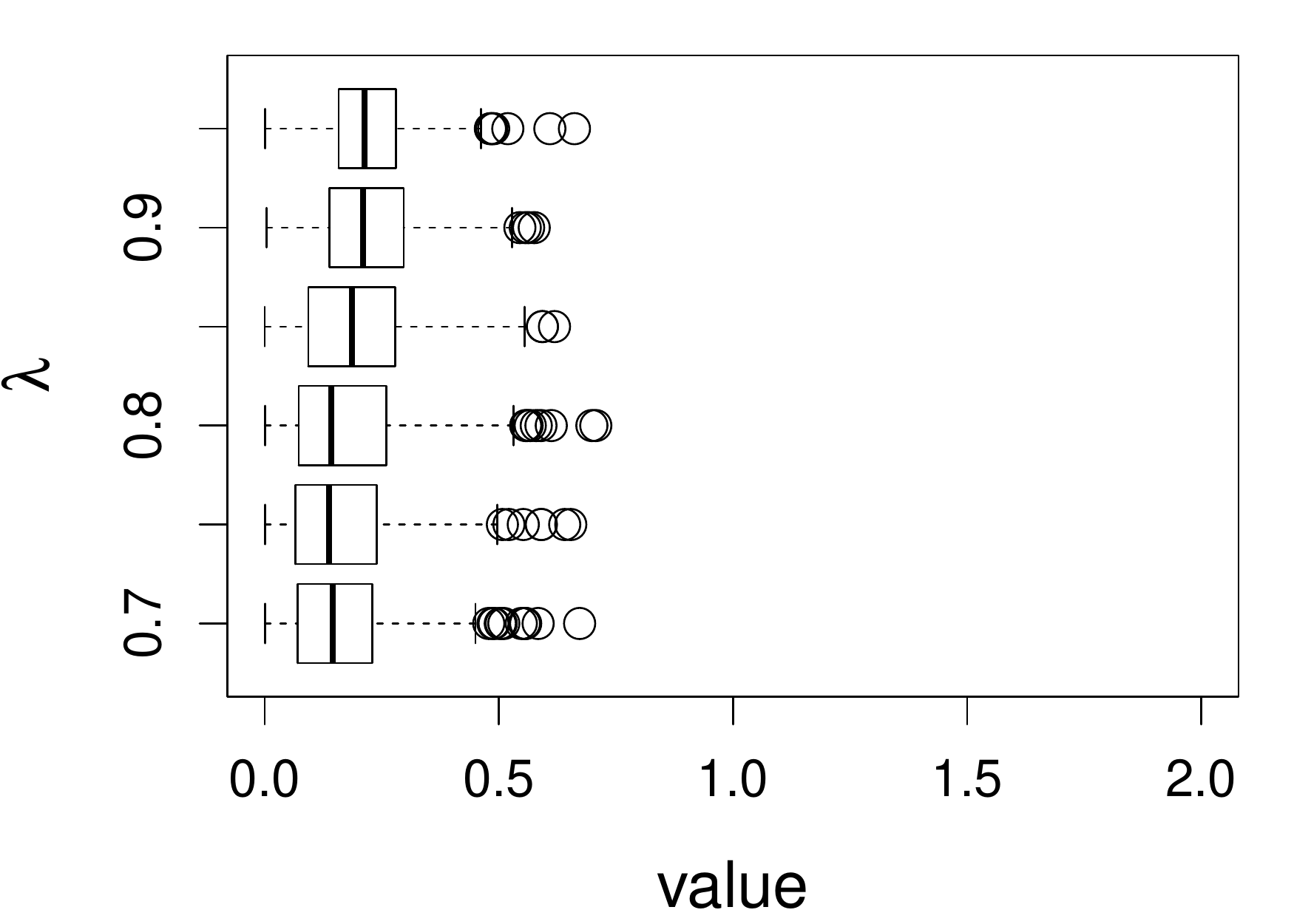} }\\
	\subfigure[]{ \includegraphics[width=\linewidth]{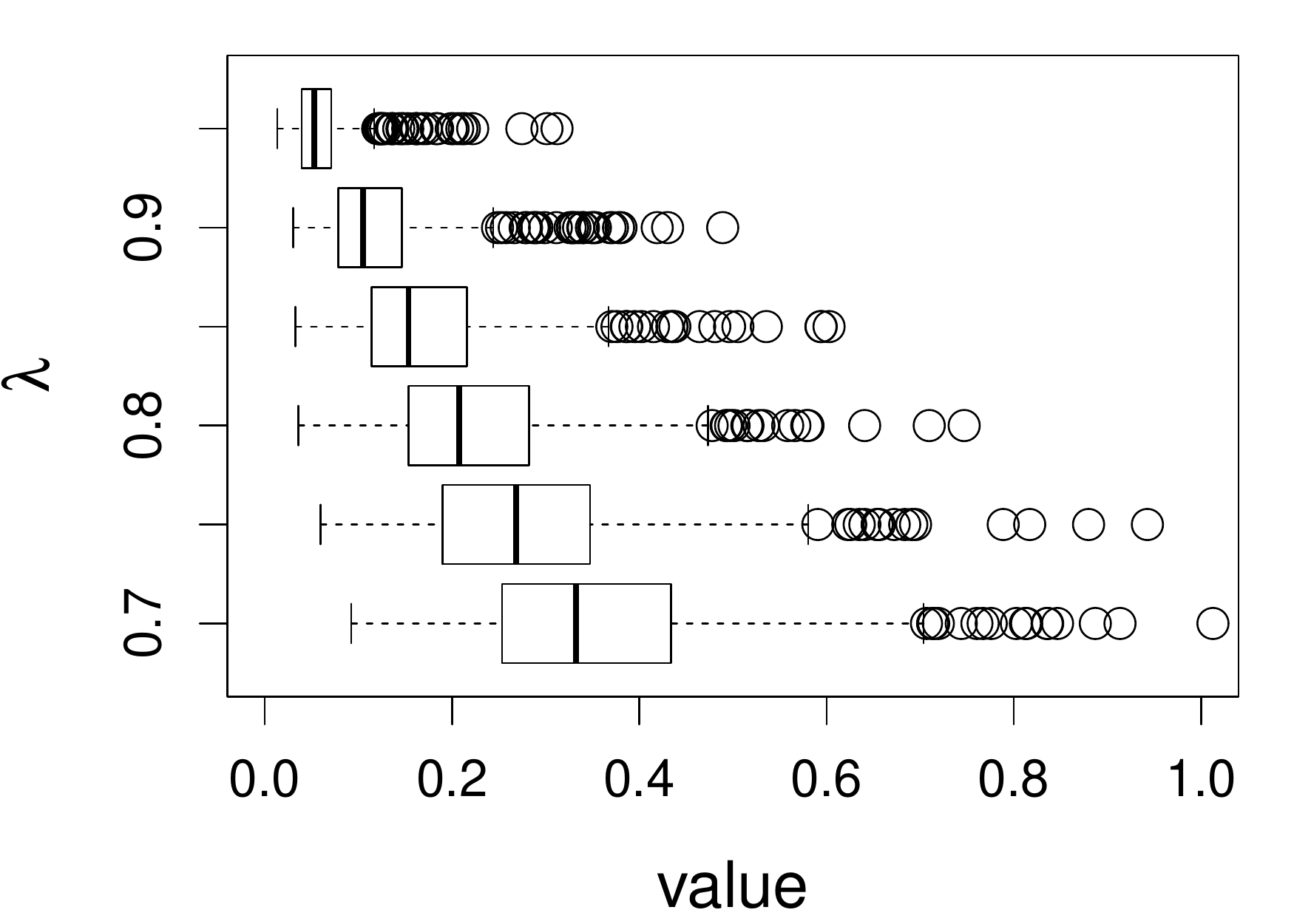} }
	\subfigure[]{ \includegraphics[width=\linewidth]{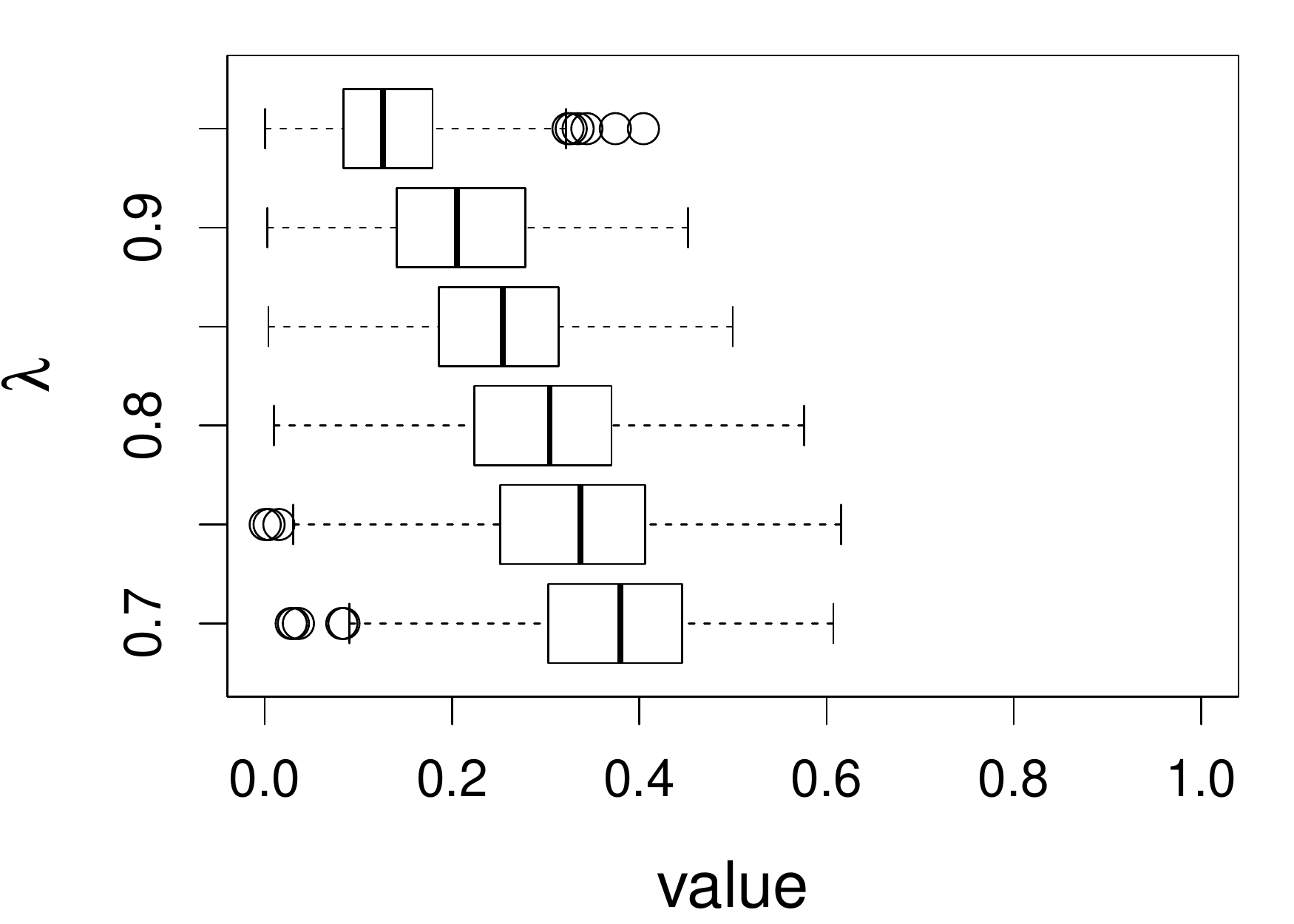} }
	\subfigure[]{ \includegraphics[width=\linewidth]{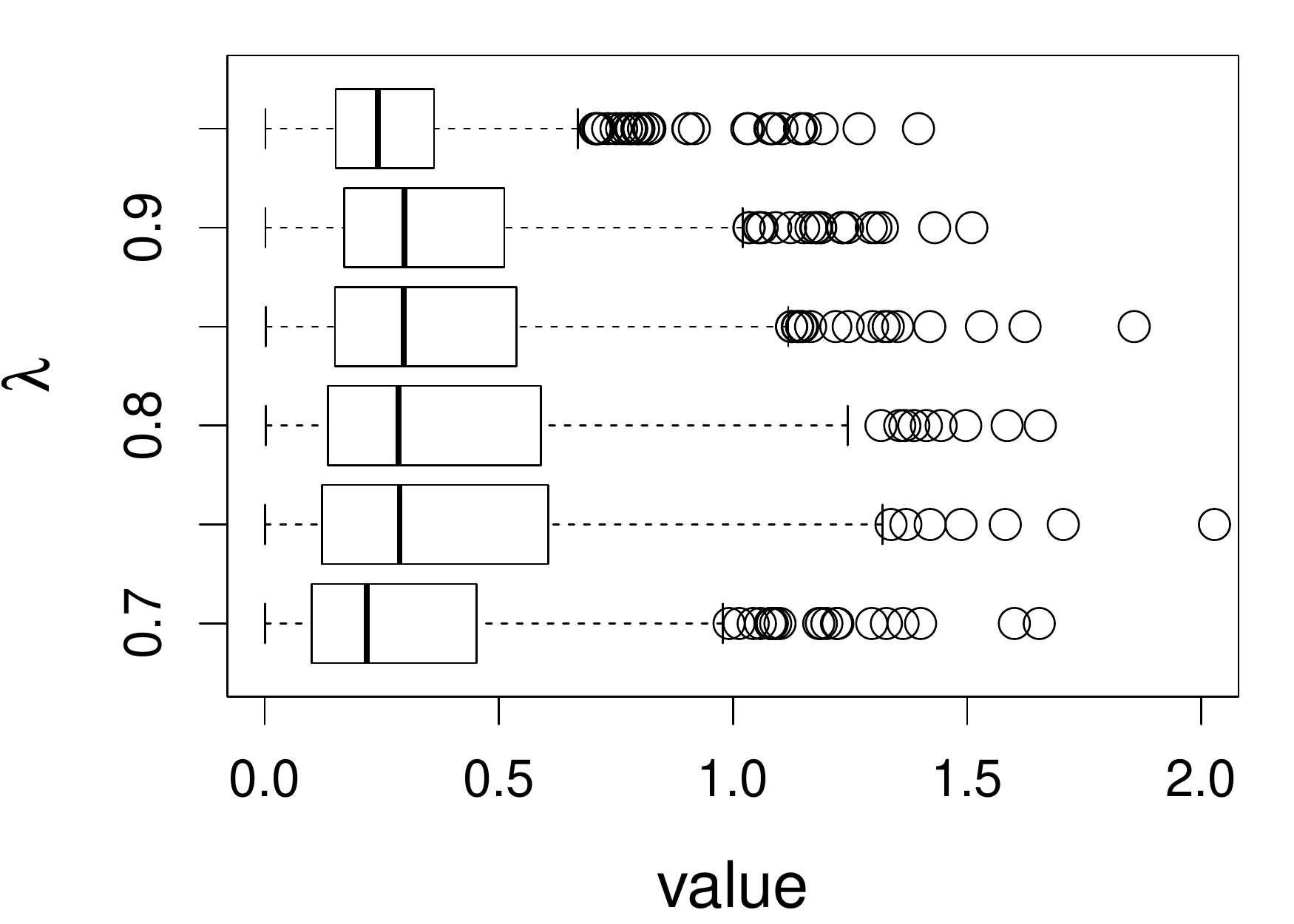} }\\
	\subfigure[]{ \includegraphics[width=\linewidth]{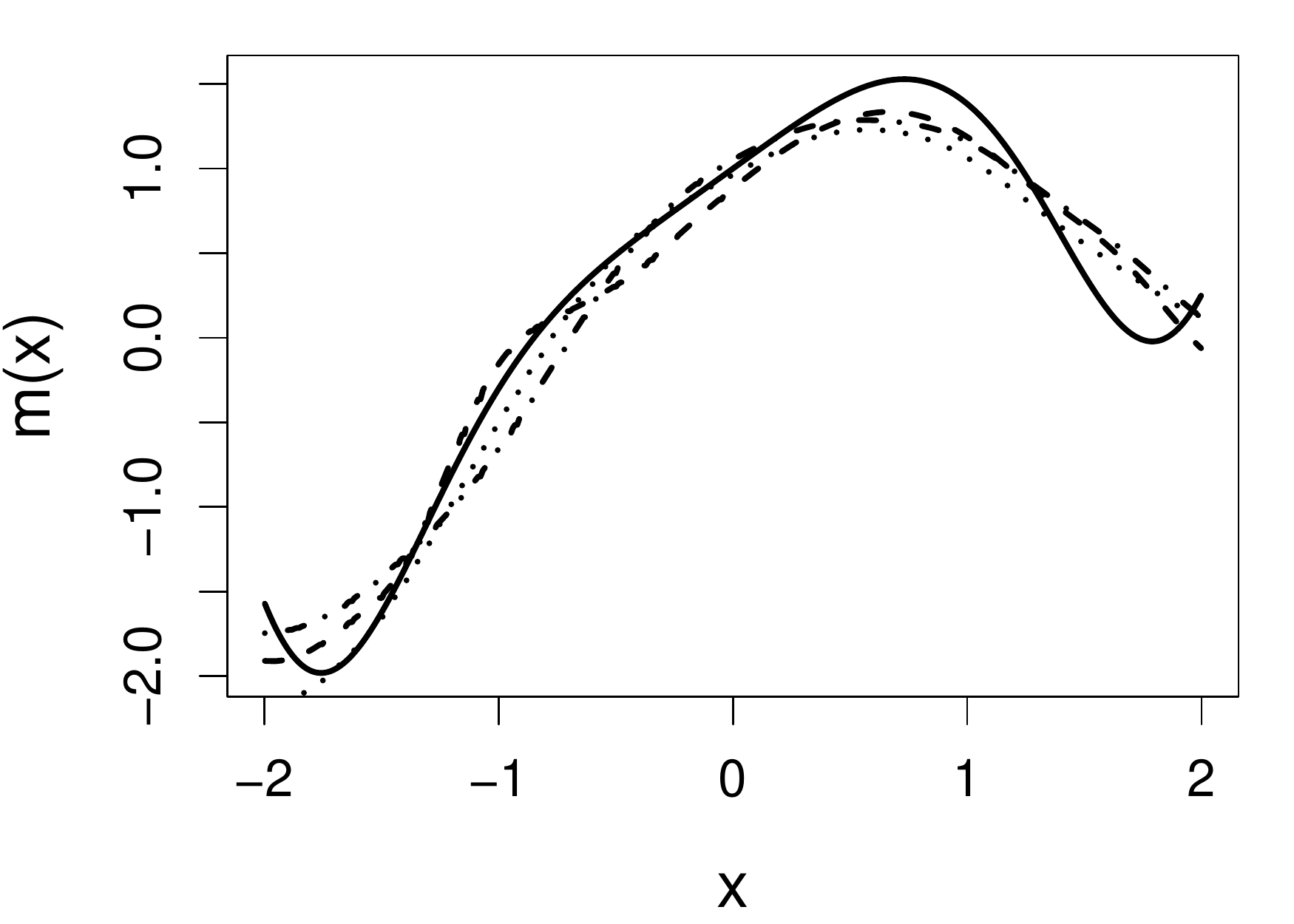} }
	\subfigure[]{ \includegraphics[width=\linewidth]{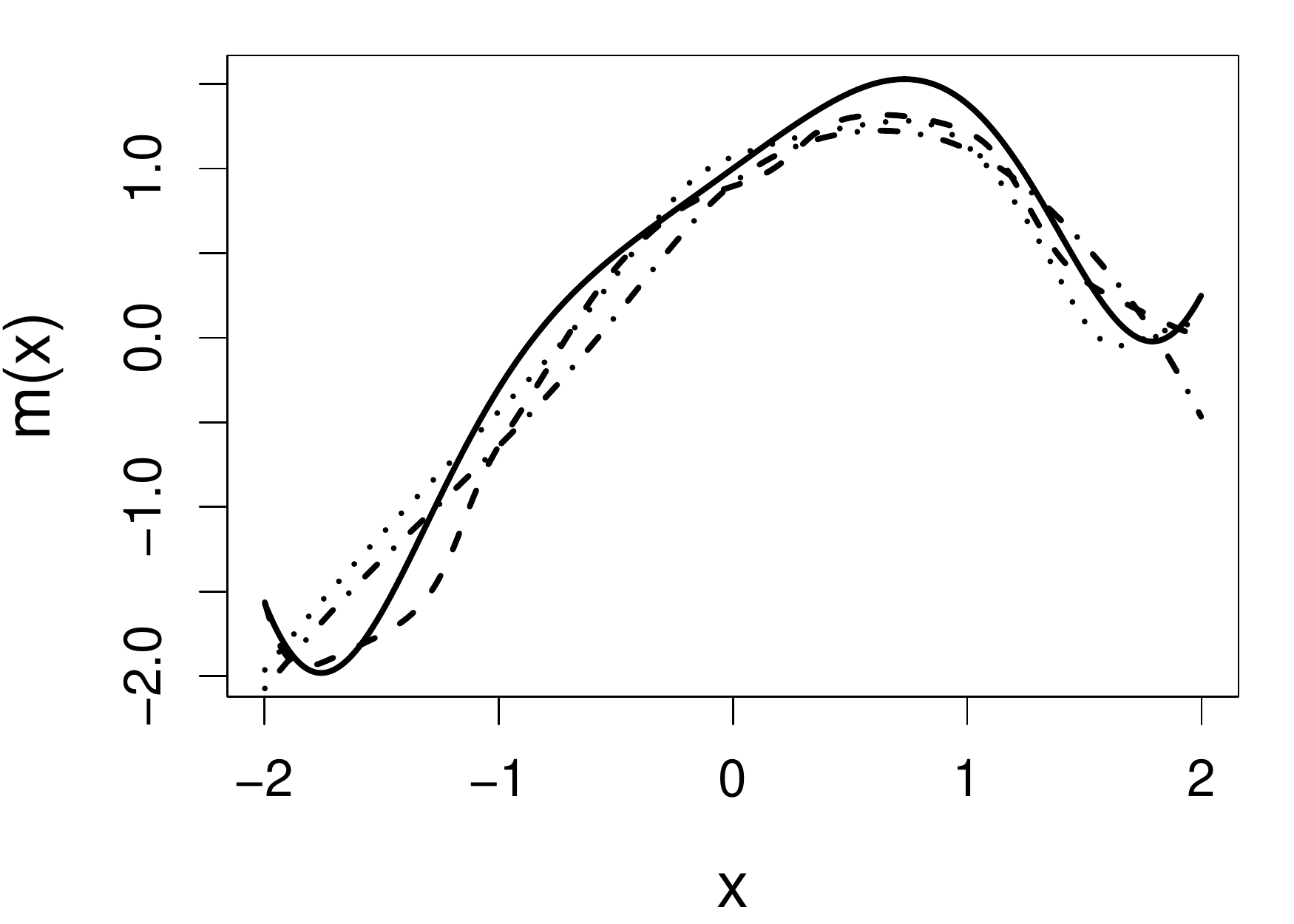} }
	\subfigure[]{ \includegraphics[width=\linewidth]{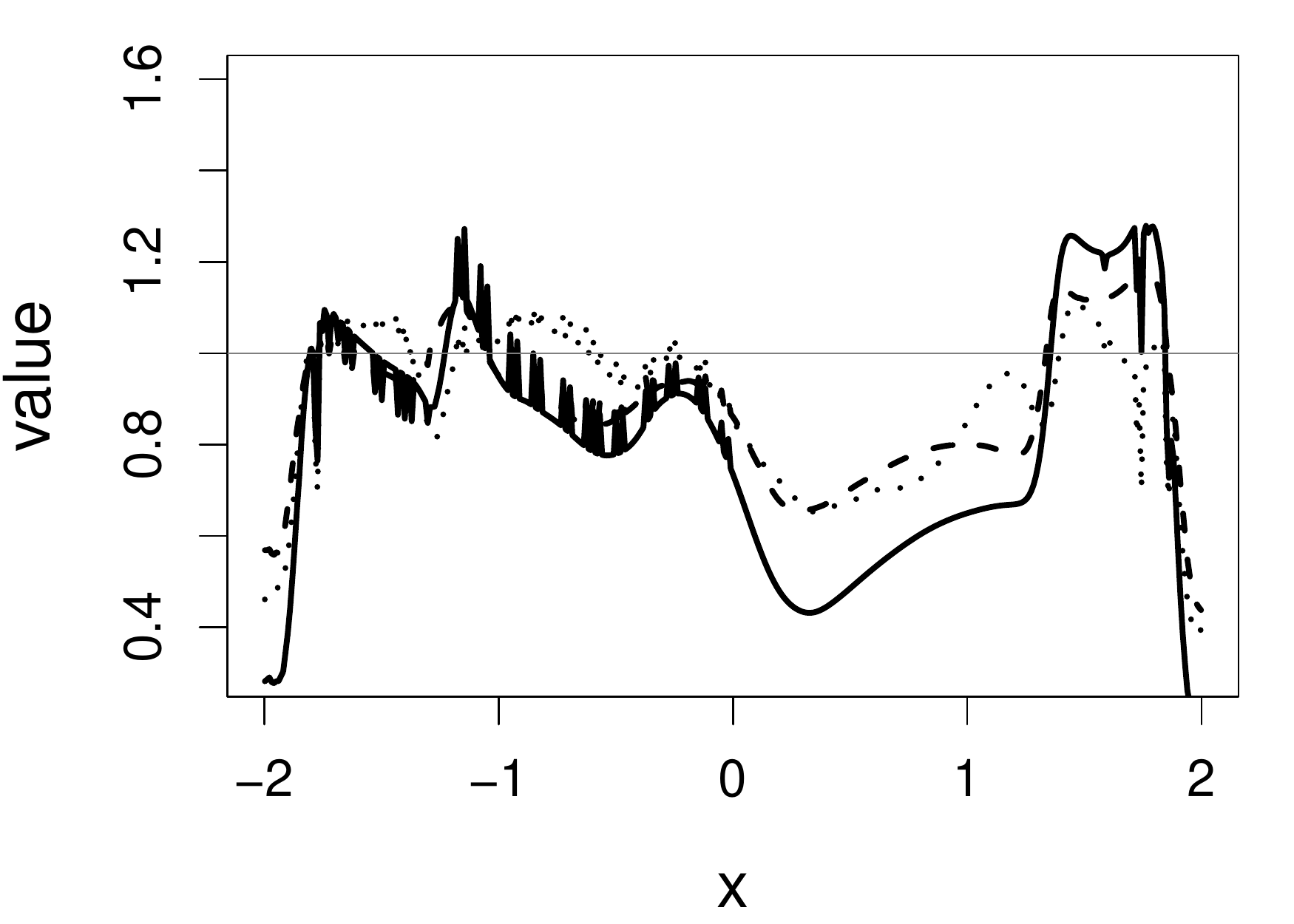} }

	\caption{Simulation results under (C4) using the theoretical optimal $h$, assuming Laplace $U$ with $\sigma^2_u$ estimated using repeated measures. Panels (a) \& (d): boxplots of ISEs versus $\lambda$ for $\hat{m}_{\hbox {\tiny HZ}}(x)$ and $\hat{m}_{\hbox {\tiny DFC}}(x)$, respectively. Panels (b) \& (e): boxplots of PAE(1) versus $\lambda$ for  $\hat{m}_{\hbox {\tiny HZ}}(1)$ and $\hat{m}_{\hbox  {\tiny DFC}}(1)$, respectively. Panels (c) \& (f): boxplots of PAE(2) versus $\lambda$ for  $\hat{m}_{\hbox {\tiny HZ}}(2)$ and $\hat{m}_{\hbox {\tiny DFC}}(2)$, respectively. Panels (g) \& (h): quantile curves when $\lambda=0.8$ for $\hat{m}_{\hbox {\tiny HZ}}(x)$ and $\hat{m}_{\hbox {\tiny DFC}}(x)$, respectively, based on ISEs (dashed lines for the first quartile, dotted lines for the second quartile, dot-dashed lines for the third quartile, and solid lines for the truth). Panel (i): PmAER(dashed line), PsdAER (dotted line), and PMSER (solid line) versus $x$ when $\lambda=0.8$; the horizontal reference line highlights the value 1.} 
	\label{Sim4LapLap500:box}
\end{figure}

Adopting the first strategy, Figures~\ref{Sim1LapLap500:box}--\ref{Sim4LapLap500:box} provide the results for our estimator and the DFC estimator under cases (C1), (C3), and (C4) considered in Section 6.3 in the main article, respectively. These are parallel to Figures 1, 3, and 4 in the main article, where one assumes a known measurement error distribution. Contrasting these two sets of figures, one can see that estimating $\sigma^2_u$ has very little impact on the estimates. 

\begin{figure}
\setlength{\linewidth}{4.5cm}
$\begin{array}{ccc}
	\includegraphics[width=\linewidth]{./Sim1LapLap-500-box-NEW-est.pdf}   & 
	\includegraphics[width=\linewidth]{./Sim1LapLap-500-box1-NEW-est.pdf}  & 
	\includegraphics[width=\linewidth]{./Sim1LapLap-500-box2-NEW-est.pdf} \\
	\includegraphics[width=\linewidth]{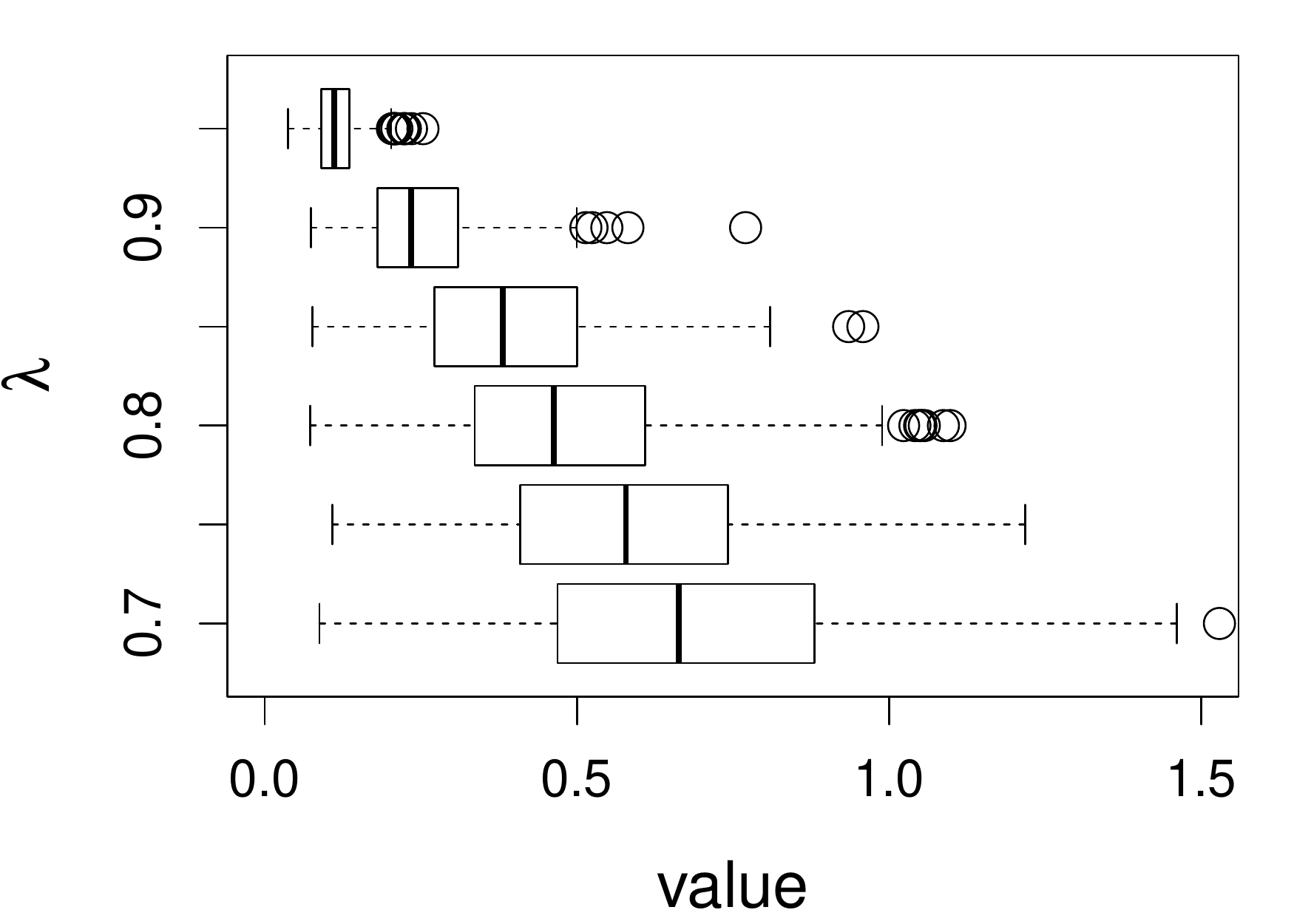}  &
	\includegraphics[width=\linewidth]{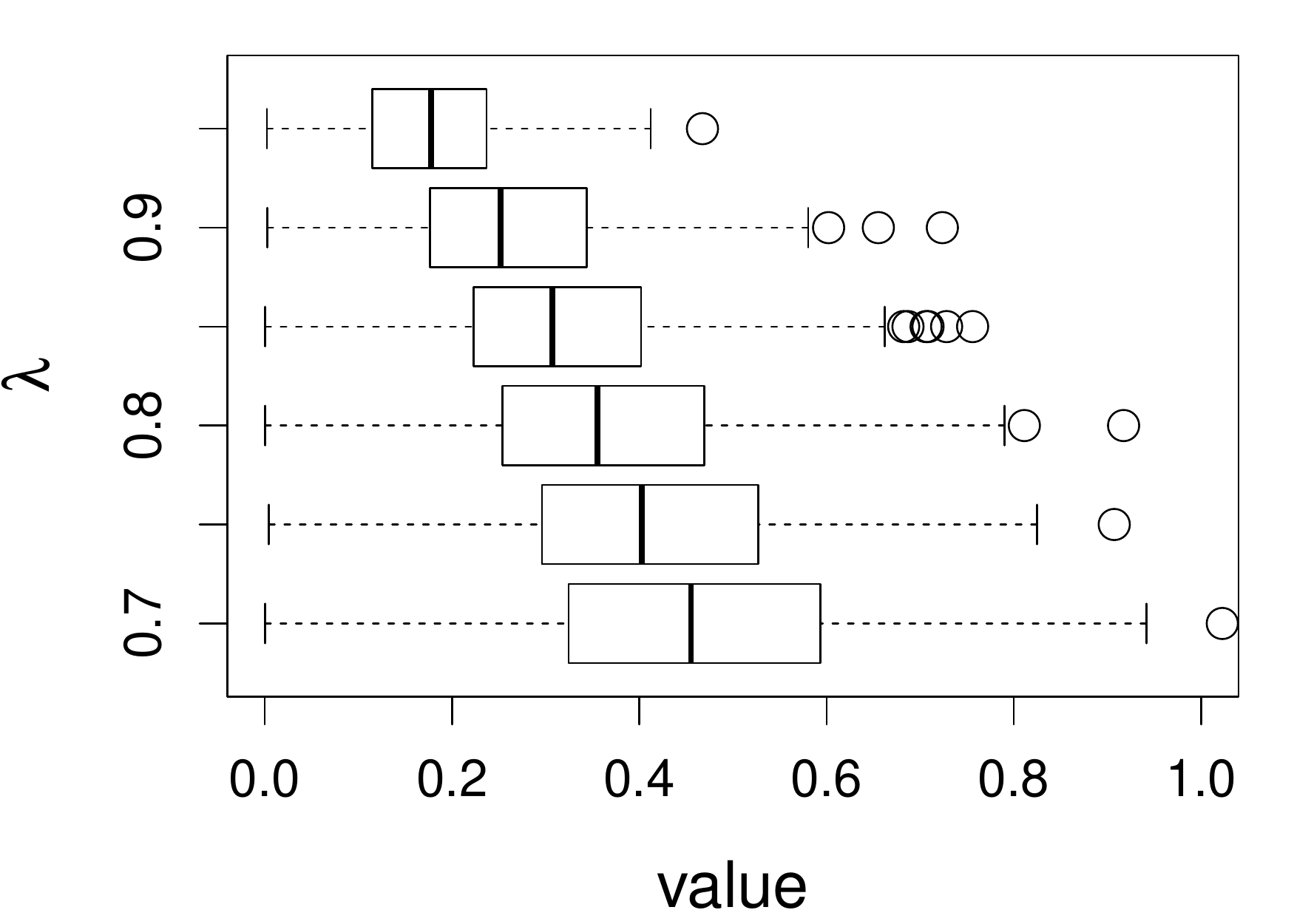} & 
	\includegraphics[width=\linewidth]{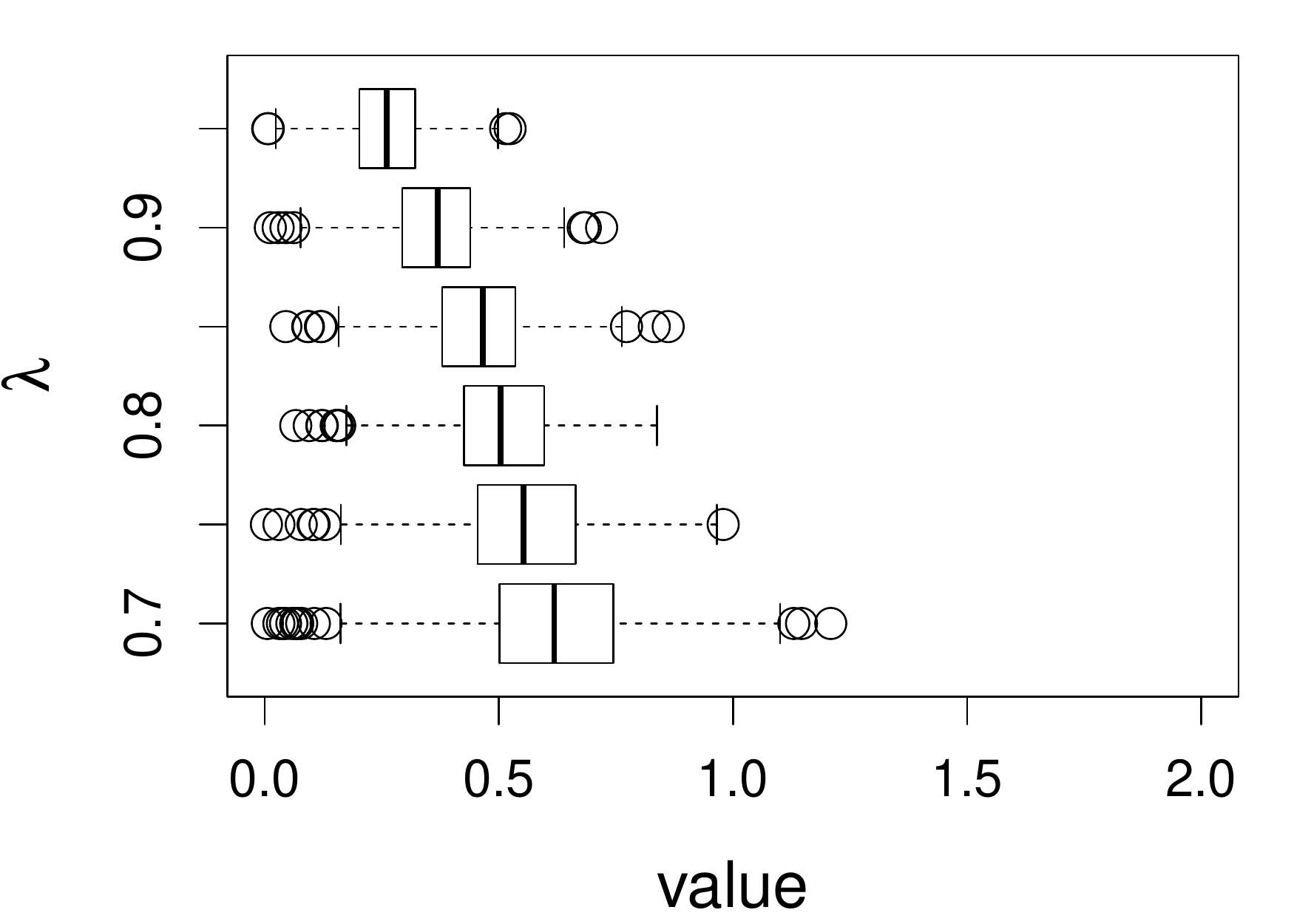} \\
	\includegraphics[width=\linewidth]{./Sim1LapLap-500-curve-NEW-85-est.pdf} &
	\includegraphics[width=\linewidth]{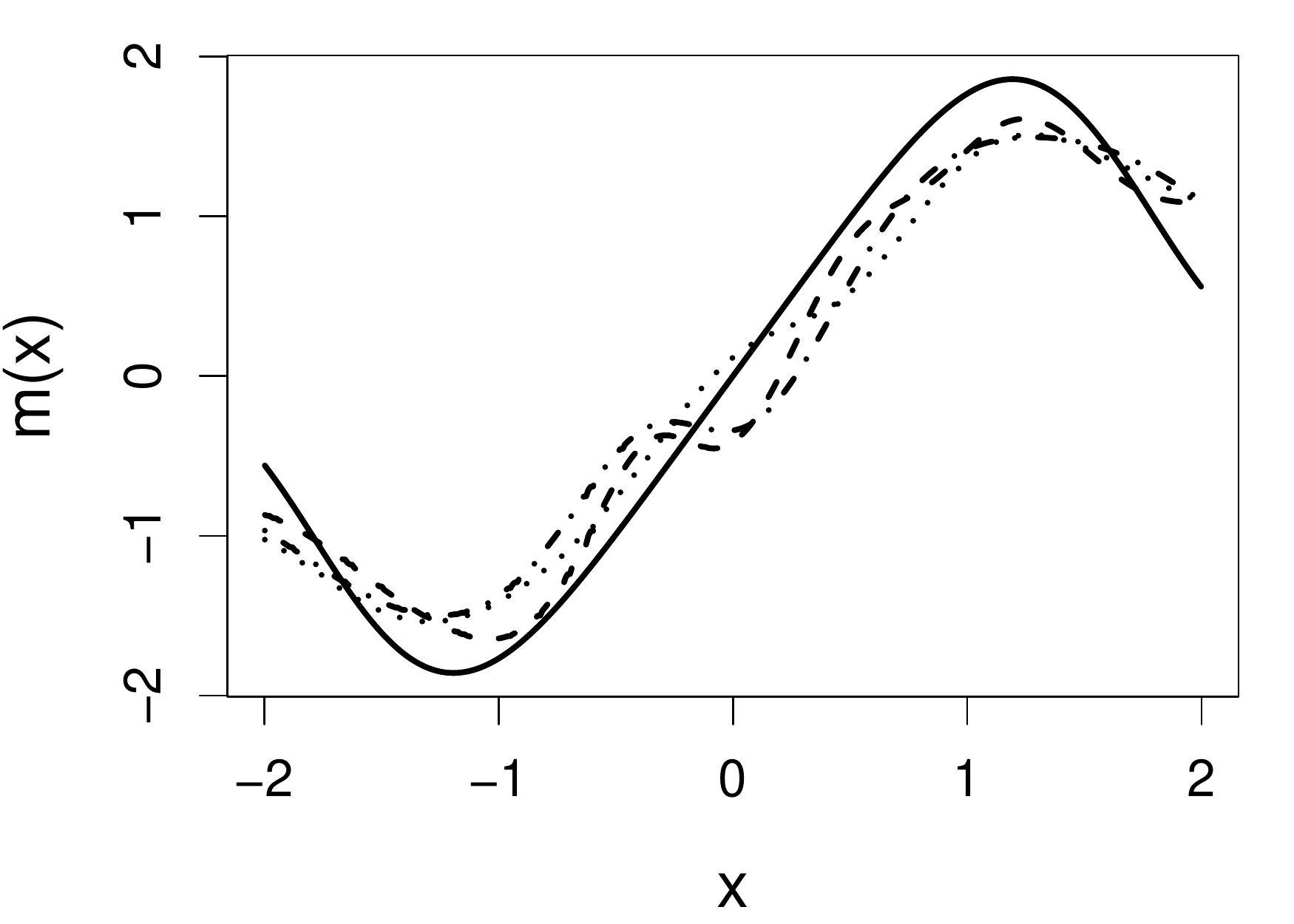} &
\end{array}$	
\caption{Simulation results under (C1) using the theoretical optimal $h$ without assuming measurement error distribution known. The first row is identical to the first row in Figure~\ref{Sim1LapLap500:box}, presenting (from left to right) the boxplots of ISE,  PAE(1), and PAE(2), respectively, associated with $\hat m_{\hbox {\tiny HZ}}(x)$ when one assumes Laplace $U$ with $\sigma^2_u$ estimated. The second row presents the counterpart boxplots when one estimates $\phi_{\hbox {\tiny $U$}}(t)$ as described in the second strategy. The third row contains the quantile curves when $\lambda=0.85$ for $\hat m_{\hbox {\tiny HZ}}(x)$ resulting from the first strategy (same as panel (g) in Figure~\ref{Sim1LapLap500:box}) on the left, and the counterpart quantile curves resulting from the second strategy of accounting for unknown $\phi_{\hbox {\tiny $U$}}(t)$ on the right.}
	\label{f:twostrategies}

\end{figure}

For illustration purpose, we demonstrate in Figure~\ref{f:twostrategies} our estimate resulting from the first strategy and our estimate employing the second strategy to account for an unknown measurement error distribution under (C1). This comparison shows that using $\hat\phi_{\hbox {\tiny $U$}}(t)$ in the estimate usually leads to more biased estimates with higher variability than using an assumed Laplace characteristic function with $\sigma^2_u$ estimated. 

